# New vacua for Yang-Mills theory on a 3–torus

# New vacua for Yang-Mills theory on a 3–torus

PROEFSCHRIFT

DOOR

## Johannes Arnoldus Hermanus Keurentjes

GEBOREN TE BLADEL EN NETERSEL IN 1969

*'I know what you're thinking about'*, said Tweedledum,*'but it isn't so, nohow.'*
*'Contrariwise'*, continued Tweedledee, *'if it was so, it might be;*
*and if it were so, it would be; but as it isn't, it ain't. That's logic.'*
Alice's adventures in Wonderland - Lewis Carroll

# Contents

# 1 Introduction

## 1.1 Gauge theory in a finite volume

All the known forces in nature, electromagnetism, the weak and strong nuclear forces, and gravity appear to be well described by gauge theories. In these theories there are roughly two kinds of particles: The particles that are subject to the forces, commonly denoted as "matter", and particles that exert the forces, the gauge-particles. In a gauge theory there are, apart from the variables describing the properties of particles, like momentum and energy, extra variables parametrising a certain internal space. These variables are acted upon by a symmetry group. The gauge principle states that the physics should be invariant under local transformations by elements of this symmetry group. This means that measurable quantities are not affected by transformations by elements of this symmetry group, even if the transformation is not the same at each point in space-time. This does not mean however that the gauge symmetry has no consequences. Interactions between different particles are constrained to be invariant under the gauge symmetry group. As another consequence, it is not possible to give mass to the gauge particles in most cases, as it is not possible to write down a mass term for the theory that respects the gauge symmetry.

Probably the simplest example of a gauge theory is the theory of electromagnetism. In this case the internal space is one dimensional, and it appears to be a circle. That it is a circle and not an infinite line is already a profound idea, leading for example to the quantisation of electrical charge. The symmetry group is the group of translations along this circle, which mathematicians denote as $U(1)$. The internal space describes the phases of the particles, which cannot be measured. Different particles may transform in different ways under symmetry transformations. In a sense one can think of this as the internal circle attached to each particle being of a different size, the (inverse) size representing the electrical charge of the particle. Mathematically, the charge labels a representation of the group. The gauge particle that exerts the electromagnetic force, the photon, has no electrical charge, and therefore does not (directly) "feel" the force. In the mathematical description of the theory this can be traced back to the fact that the group $U(1)$ is Abelian, which means that the order in which group transformations are done is not relevant.

The two nuclear forces, weak and strong, are described by non-Abelian gauge groups, non-Abelian meaning that the order of the symmetry transformations is relevant. One of the consequences of this is that the gauge particles are not invariant under the symmetry, which again means that they must interact with themselves. This self-interaction leads to non-linearities in the theory, which make that even a theory of gauge particles alone (called a Yang-Mills theory, so named the discoverers) can exhibit complicated physics. In contrast, a theory of electromagnetism without any charged particles is a trivial theory of free particles.

The weak nuclear force is responsible for the radioactive decay of nuclear particles. The full description of the weak nuclear force includes electromagnetism, and this theory is said to describe the "electroweak" force. In this description, the $U(1)$ of electromagnetism is part of a bigger group called $SU(2) \times U(1)$. This bigger group is not the symmetry that is measured at low energies, as it is "spontaneously broken". This means that, although the theory is invariant under the symmetry, it is not manifest in the low energy theory, because there is no starting point for perturbation theory that respects the symmetry. In the spontaneously broken theory the gauge particles associated to the weak interaction are no longer massless and only at energies much higher than the scale of symmetry breaking is the symmetry recovered. Long distance effects are under control because the mass of the gauge bosons gives the weak force a finite range. The remaining massless gauge boson in the electroweak theory is identified with the photon, as its effects fall off with the distance as is familiar from electromagnetism.

The strong nuclear force is responsible for the binding of atomic nuclei, and in these, the bindings of quarks into protons and neutrons. In this case the gauge group is called $SU(3)$ and unlike the case for the weak force, it is not spontaneously broken, and the gauge particles, called gluons, are massless. Therefore the strong force has long range effects, but unlike in electromagnetism, the strong force does *not* fall off powerlike with the distance between the particles. This property is a consequence of the self-interaction of the gluons. In the short distance limit, or at high energies, one has "asymptotic freedom" and the particles that feel the strong force, called quarks, behave approximately as free particles. At low energies, or long distances, the quarks form bound states, called hadrons. The hadrons are subdivided into baryons, like the proton and neutron, which are thought to be bound states of three quarks, and mesons, like the pions, which are bound states of a quark and anti-quark. In experiments, quarks always occur in these bound states and never as single particles. It is believed that the quarks are permanently "locked up" in bound states, and this property is called confinement. One of the great challenges for the theory of the strong force, called Quantumchromodynamics (QCD) is to explain confinement from the microscopic theory, and this has not been achieved yet in all rigour.

Confinement is a long distance effect, and a particular way to study such effects is to formulate the theory in a finite volume. The finiteness of the volume provides a maximum distance, and one may attempt to study effects in the theory as a function of this distance. This is one of the motivations for studying gauge theories in a finite volume. Another important motivation is given by the attempt to study gauge theories on a computer. As memory size is necessarily finite, one needs to approximate the theory, that has infinitely may degrees of freedom by a theory that has finitely many degrees of freedom. This is achieved by approximating the space-time as a finite set of points, lying on a lattice. The finite size of the lattice does not only imply a minimum length (being the smallest lattice distance), but also a maximum length, and hence a finite volume.

A third motivation can be found from string theory. String theory is a candidate "Theory of everything", a theory that may be able to unify all forces, and in particular it provides a description of quantum gravity. Although at all accessible scales gravity is well described by the theory of General Relativity (which is a classical field theory), there is still a need for a microscopic description of gravity, to cope with the paradoxes and questions that arise when one tries to study the theory of gravity at small distances. At small distances we expect the classical field theory to loose its validity, and quantummechanical effects to enter. The recipe that has been successful in formulating the quantum field theories of electromagnetism and the nuclear forces however fails on the theory of General Relativity, due to the fact that the resulting quantum theory is not renormalisable, which means that at short distances it is ill-behaved. This suggests that we need a modification to the theory of gravity at short distances. String theory proposes that this modification is due to the fact that at truly microscopic scales, particles are no longer pointlike, but extended one-dimensional objects called strings. Apart from the variables that describe the motion of the string in space (which are also encountered in the description of point particles), there are also internal degrees of freedom: a string has several vibrational modes that can be excited. The different vibrational modes are interpreted as different particles in a macroscopic description.

String theory can be successfully quantised, but only in a highly restricted set-up. Naive quantisation of strings leads to anomalies and loss of Lorentz invariance, and/or to instabilities (tachyons) of the theory. The only known way to avoid these problems is to turn to strings that incorporate supersymmetry, the symmetry that relates bosons to fermions, and to assume that space-time is really 10 dimensional (9 space dimensions and 1 time dimension). Another feature is the possibility to include non-Abelian gauge symmetries in string theory. The resulting low energy theory has then a sector that describes a non-Abelian gauge theory. That this is a 10-dimensional gauge theory appears to frustrate the hope that it may be related to a description of our 4-dimensional universe. There is however no contradiction if we assume that 6 out of the 10 dimensions of string theory are very small, and hence a subspace of the 10 dimensional space-time actually has a finite (hyper-)volume. This again motivates the study of gauge theories in a finite volume.

The simplest way to get to a finite volume theory is to take $n$ of the dimensions of finite length, and impose periodic boundary conditions. The result is a product of $n$ circles, or an $n$–torus (in this thesis $n$ will most of the time have the value 3). The $n$-torus is flat, meaning that for sufficiently large tori the physics is the same as in a flat but non-compact space-time. The main difference with flat non-compact space-time is the presence of non-contractible loops. For the case of supersymmetric field or string theory, another relevant fact about tori is that they preserve all supersymmetries.

Quantum theory tells us that the fields fluctuate around the values allowed by the classical field equations. If the fluctuations are small, one may use a semiclassical approach to study the fluctuations on top of the classical background. If the fluctuations are large then this approach looses its validity. In this thesis we will study the possible backgrounds for gauge theories on a 3–torus, and hardly deal with the fluctuations. As a further restriction we study only backgrounds that carry no energy. When a gauge theory is formulated on a torus, there is the possibility of setting the background gauge fields to values that cannot be gauged away. A way of parametrising these background fields is to study the non-Abelian phases that particles pick up when transported around a closed loop. These are called holonomies and are elements of the gauge group. The requirement of zero energy leads to the requirement of zero field strength, which implies that the holonomy around a contractible loop equals the identity. Holonomies around non-contractible loops are far less restricted. The zero field strength condition implies that holonomies around different non-contractible loops commute, which only depends on the homotopy class, but they are otherwise unrestricted. The classification of possible zero field strength backgrounds on an $n$-torus then amounts to classifying all sets of $n$ commuting elements of the group. The main part of this thesis is devoted to studying this problem for $n = 3$.

A motivation for studying gauge theories on the 3–torus came from a problem with counting the number of vacua for the supersymmetric extension of four-dimensional gauge theories. In the next section we will give some background on supersymmetry, and supersymmetric gauge theories in particular. One of the insights that led to a solution of the problem came from a discussion on gauge symmetries inspired by string theory. The full solution to the problem lead to new results for gauge theories with exceptional and orthogonal groups compactified on tori. The fact that some of these groups appear in string theories suggests to apply the new results in string theory. We will therefore also devote a section to give some background on gauge symmetries in string theory. This chapter concludes with an outline of the topics discussed in the remainder of this thesis.

## 1.2   Supersymmetric gauge theories

The original motivation for this work came from an argument in supersymmetric gauge theories, but we stress that the results are applicable to gauge theories in general. Supersymmetry plays a role in various topics we will discuss. It is also a crucial ingredient in string theories, although it will not often enter our analysis.

Supersymmetry is a continuous symmetry, whose generators satisfy anti-commutation instead of commutation relations. One may nevertheless obtain the symmetry transformation by exponentiating the generators, provided one uses a Grassmannian parameter, instead of an ordinary parameter. An infinitesimal supersymmetry transformation multiplies a bosonic commuting field with a Grassmanian parameter, resulting in an anti-commuting fermionic field. Similarly, fermionic fields turn into bosonic fields under infinitesimal supersymmetry transformations.

Supersymmetry is a remarkable idea, with several attractive features. Probably the most remarkable aspect of realisation of supersymmetry on physical models is that the Hamiltonian can be expressed in the supersymmetry charges. This leads to strong constraints on the dynamics. The non-trivial relation between the symmetry generators and dynamical quantities does not only forbid certain interactions, it also protects certain quantities and relations between them against quantum corrections. Examples are the relation between mass and charges of special representations of supersymmetry (BPS-states), and non-renormalisation theorems for interaction parameters. Therefore many properties of supersymmetric physical systems may actually be deduced from symmetry considerations alone, making complicated calculations easier or even superfluous.

Because of its special properties, supersymmetry might be helpful (or even crucial) in solving certain riddles in high energy physics, such as the hierarchy problem and the cosmological constant problem. It also is a crucial ingredient in some approaches to unify gravity with the other known elementary forces, like supergravity and superstring theories.

Supersymmetry is not observed in nature, so if it is realised in some way it should be broken. The most attractive option is spontaneous breaking, since this conserves many of the attractive features of supersymmetry. A criterion for the possibility of spontaneous breaking of supersymmetry was developed by Witten [52]. He introduced a quantity that became known as the "Witten index", and argued that spontaneous breaking of supersymmetry is only possible if this index is zero. He also argued that, under suitable conditions, the value of the index should not depend on the details of the computation to establish its value. The index may be computed from counting the ground states of the theory.

In this thesis we study non-Abelian gauge theories. For non-Abelian gauge theories supersymmetric extensions exist. The simplest of these theories are the supersymmetric Yang-Mills theories. This name is motivated by the fact that these are extensions of Yang-Mills theories. They are however not Yang-Mills theories in the usual sense, as supersymmetry requires the introduction of an extra field which is not a gauge field. This spinor field is called the gluino, a name motivated by the fact that it is the superpartner of the gluon.

The Lagrangian for supersymmetric Yang-Mills is

$$\mathcal{L} = -\frac{1}{4g^2} F^a_{\mu\nu} F^{\mu\nu a} + i\bar{\psi}^a \slashed{D}_{ab} \psi^b \qquad (1.2.1)$$

The indices $a$, $b$ run over the number of group generators, and $\slashed{D}_{ab}$ is the contraction of the covariant derivative with the Dirac $\gamma$-matrices, $\slashed{D}_{ab} = \gamma_\mu(\partial^\mu \delta_{ab} - iA^{\mu c} f_{abc})$, with $f_{abc}$ the structure constants. It is very similar to the Lagrangian for QCD with one massless quark. Note however that supersymmetry requires the gluino's $\psi^a$ to transform in the same representation as the gauge field, being the adjoint. The Lagrangian is invariant under the supersymmetry transformations

$$\delta_\xi F^a_{\mu\nu} = 0 \qquad (1.2.2)$$
$$\delta_\xi \psi^a = \sigma^{\mu\nu} \xi F^a_{\mu\nu}, \qquad (1.2.3)$$

where $\xi$ is an Grassmannian parameter, and $\sigma^{\mu\nu} = \frac{i}{2}[\gamma^\mu, \gamma^\nu]$. The resemblance with QCD suggests that many interesting non-perturbative effects, such as confinement and chiral symmetry breaking that occur in QCD may also occur in supersymmetric Yang-Mills theory. One question one may ask is whether such non-perturbative effects may spontaneously break supersymmetry. To get an idea one might calculate Witten's index for supersymmetric Yang-Mills theory. This was done by Witten [52] who computed the index to be a non-zero number depending on the rank of the gauge group. The calculation requires a count of the number of ground states, which is hard to do because in perturbation theory the theory has a continuous spectrum with massless particles. A way to achieve a well defined calculation is to formulate the theory in a finite volume, and because the compactification should not break supersymmetry, compactification on a spatial 3–torus is the logical thing to do.

The fact that the index is non-zero indicates that supersymmetry cannot be broken spontaneously. It is nevertheless an interesting result as the invariance of the index suggests that the finite number of vacuum states may also be identified in other limits, such as the large volume limit, when the theory becomes strongly coupled. A simple argument based on the index theorem for adjoint fermion zero-modes in an instanton background, and the assumption that the gluino's will condense, suggests that indeed it is possible to obtain a finite number in the infinite volume limit. Comparison of the numbers obtained from the two different calculations now however leads to a surprise. For supersymmetric Yang-Mills theories with unitary or symplectic groups the answers agree. For theories with orthogonal and exceptional groups the numbers disagree however. Apparently one of the calculations is incorrect, or the Witten index is not invariant. There seems to be however no easy argument explaining the discrepancy between the results for orthogonal and exceptional groups, that does not affect the computations for unitary and symplectic groups.

Subsequent work lent further support to the assumptions made for the infinite volume computation [1][35][45]. In these works the powerful constraints imposed by supersymmetry were used to facilitate a computation for the value of the gluino condensate. According to these computations, the gluino's indeed condense. At the same time there were no new insights in the calculation for the finite volume, sharpening the paradox.

It took more than 15 years before it was found out that, in the computation for the index for supersymmetric Yang-Mills theories on a 3–torus with orthogonal groups something was overlooked [55]. It turns out that there exist ground states for gauge theories with orthogonal groups on a 3–torus, that were not included in the previous computation of the index. It is then not hard to show which assumption in the old computation is incorrect. The fatal assumption is however true for unitary and symplectic groups, explaining why for these cases there was never a discrepancy.

The discovery of the new vacua was achieved using new techniques coming from developments in string theory, in particular D-branes and orientifolds. These techniques however are inapplicable to gauge theories with exceptional groups. To understand these, one had to understand the existence of the new vacua from a group theoretic point of view, and extend the analysis to exceptional groups. This is the main topic of this thesis. It turns out that also in the case of exceptional groups there exist new ground states, and that for exceptional groups the structure is even much richer than for orthogonal groups. Including the new ground states for the theory on the torus, all index computations agree. Summarising, the argument that Witten's index is an invariant for supersymmetric Yang-Mills theories appears to be correct. Supersymmetry is not spontaneously broken in supersymmetric Yang-Mills theory, but the development that lead to this insight gave new results for gauge theories on 3–tori. The analysis can be extended to gauge theories on higher dimensional tori, and also there new ground states not previously considered are found.

## 1.3   Gauge symmetry in string theory

We mentioned that the new vacua for gauge theories were found from an analysis motivated by string theory. The extension of the analysis to all groups also indicates the occurrence of new vacua in the exceptional groups, and in particular in the exceptional group $E_8$. $E_8$ is a group that plays an important role in string theory, suggesting that the newly found ground states for this group may also play a role in string theory. This is indeed true. In this section we give some background on non-Abelian gauge symmetries in string theory [20] [44].

Consider a theory of open strings living in $d+1$ dimensions. An open string may vibrate in spatial directions transverse to its extension, which give it $d-1$ transverse degrees of freedom. This may remind the reader of the fact that a gauge field in $d+1$ dimensions also fluctuates in $d-1$ transverse directions. Indeed if one checks the spectrum of excitations of the open string, it turns out that there is a state that is massless, and caries a vector index. This suggest to identify this state as a $U(1)$ gauge boson. The gauge boson is described by a $d+1$-component vector field. Not all these components are independent, as gauge symmetry eliminates one degree of freedom, and Gauss' law a second one. A similar scenario is true for the string, which vibrates in $d+1$ dimensions, but one degree of freedom is eliminated by the possibility of equivalent embedding in space time (which is a gauge symmetry), and a second one by a constraint that follows from the equations of motion.

In an open string theory one may add degrees of freedom to the endpoints of the string. Consider the possibility that each endpoint of the string may be in $n$ states. To describe these, one needs $n$ labels for each endpoint of the open string, leading to a total of $n^2$ different species of open strings, labelled by two indices. All the states occurring by exciting such strings, are labelled with these two indices. Choosing a basis for the $n \times n$ matrices, one may expand each state as

$$|\phi_{ij}>= \lambda^a_{ij}|\phi^a> \qquad (1.3.1)$$

with the $\lambda^a$ hermitian matrices, with $a$ taking $n^2$, and $i$ and $j$ each $n$ different values. This is the Chan-Paton construction. The $\lambda^a$ may be normalised according to $\text{Tr}(\lambda^a \lambda^b) = \delta^{ab}$. Each state now comes in $n^2$ different species, which we may write as an $n \times n$-matrix. The gauge boson we described previously is promoted by the above construction to a set of $n^2$ gauge bosons, $A_\mu = A^a_\mu \lambda^a$. As strings interact, the states they carry interact and in particular the gauge bosons form an interacting theory. Then the $n^2$ massless gauge bosons cannot transform in an Abelian group, but should form a fully interacting non-Abelian gauge theory. The quantum numbers from the Chan-Paton construction are appropriate for a set of states transforming under $U(n)$, and closer examination of the interactions indeed reveals a $U(n)$ gauge theory.

In the above we implicitly assumed oriented strings, meaning that there is a notion of a left and a right endpoint. One may also consider unoriented open strings, that is strings that are invariant under reversal of the orientation of the worldsheet, commonly denoted as $\Omega$. To make the projection on the states, one has to decide how $\Omega$ acts on these. In particular it is relevant how $\Omega$ acts on the gauge bosons. The mode of the string that represents the gauge bosons turns out to be odd under $\Omega$. One however also need to specify the action of $\Omega$ on the Chan-Paton basis. $\Omega$ exchanges the endpoints of the string and therefore the indices $i$ and $j$. One may also combine the action on the Chan-Paton basis with a unitary transformation. By an appropriate change of Chan-Paton basis, one may reduce the possibilities to two cases, resulting in a gauge theory with $O(n)$ gauge group, or a theory with $Sp(n/2)$ gauge group (the latter is only possible if $n$ is even).

A theory of open strings automatically incorporates closed strings, as the two endpoints of the string can interact and may therefore join. When the space-time in which a closed string moves contains a closed circle, two quantum numbers are relevant. First, there is the momentum in the direction of the circle, which as quantum mechanics teaches us is proportional to $n/R$, with $n$ an integer and $R$ the radius of the circle. A second quantum number is given by the possibility of winding strings around the circle. This is characterised by another integer, the winding number $w$. The length of a winding string is $wR$ and to find the energy associated to the winding one has to multiply this number with the string tension. In the mass formula of the string states there appear several contributions. Roughly, the mass squared is the sum of the momentum squared (giving $(n/R)^2$), a contribution of the energy squared of the winding states $((wR)^2)$, and a contribution of the internal excited modes. For closed strings there is a symmetry of the spectrum, called T-duality, that exchanges $n$ and $w$, and at the same time $R$ and $1/R$. Apparently it is not possible to distinguish a string wrapped around a circle of radius $R$, with momentum $n$ and winding number $w$, from a string wound on a circle of radius $1/R$ with momentum $w$ and winding number $n$.

The closed strings occurring in open string theories also have this symmetry, and it can be extended to the open string sector. Open strings moving freely in space cannot wind, and therefore only carry a momentum proportional to $n$. T-duality suggest that this should be the same as a string with no momentum at all (in the compact direction), and only winding. A string without momentum suggest that its endpoints are stuck somewhere, explaining at the same time why the string can wind. The "stuff" that the string end-points are attached to is called a D-brane, where D comes from "Dirichlet", for the appropriate boundary conditions at the fixed endpoints of the strings. The endpoints of strings attached to D-branes may move along the branes, but cannot come off. This opens up the possibility of interpreting all open string theories as theories of D-branes, if we imagine the existence of D-branes that fill all of space.

For strings with Chan-Paton factors and therefore gauge symmetry, the momentum receives contributions from the background gauge field, in a similar way as for a charged particle moving in an electromagnetic field. The momentum may therefore no longer be integer. Applying the previously described T-duality, one should then also allow for fractional windings. This is achieved by putting not one, but several D-branes on a circle. A string beginning on one D-brane and ending on another traverses a fraction of the circle. In this way the background gauge field in one theory may be made visible in the relative positions of D-branes in the dual theory. This important idea appears at several places in this thesis. Strings have a non-zero tension, and one can imagine that strings stretching from one D-brane to another "pull" on both branes. The D-branes are thus dynamical, interacting objects.

Also the orientation reversal that led to unoriented strings has a place in this picture. In the original theory we had open strings that were free to move. If these strings are unoriented, we have $O(n)$ or $Sp(n/2)$ as its gauge group. In these groups the eigenvalues of the matrix representations are not unrelated, as in $U(n)$, but occur in pairs: if $\phi \in U(1)$ is an eigenvalue, then so is $\phi^*$. In the T-dual description, the eigenvalues for the background gauge field are represented in the position of D-branes. The pairing of eigenvalues for $O(n)$ and $Sp(n/2)$ should then be reflected in a geometrical symmetry of the D-brane configuration. Indeed one may show that in this theory, particular hyperplanes act as "mirrors", reflecting the directions transverse to the plane. These "mirror" planes are called orientifold planes. On top of the reflection of the transverse spatial directions, there is also a projection that reverses the orientation of the string. In the hypervolume of the orientifold fixed plane, a string is mapped to a string on the same position, but with its orientation reflected. This projects onto unoriented strings. Strings away from the orientifold planes are reflected into a string at the mirrored position, but with their orientation reflected. In analogy with the possibility of D-branes filling all of space, one may also imagine an orientifold plane filling all of space. This leads to the unoriented string theories described before. A last point on orientifold planes is that there are no strings ending on them, and that they are therefore not dynamic.

In the above we have not yet mentioned any of the string consistency conditions. A survey of the dynamics of these theories learns that almost all suffer from inconsistencies. The only open string theory that seems free of problems is a supersymmetric theory of unoriented strings, with $SO(32)$ as its gauge group. This theory is named type I string theory.

There is another way to incorporate non-Abelian gauge symmetries in string theory. Consider closed instead of open strings. A closed string does not have endpoints but one may incorporate extra degrees of freedom by adding fields living on the world sheet of the string. Such fields then form internal degrees of freedom. A particular construction involves 16 internal bosonic degrees of freedom. These 16 bosons can be thought of as living on a compact manifold, being a torus $\mathbb{R}^{16}/\Lambda$, where $\Lambda$ is a 16 dimensional lattice. Consistency can be achieved if $\Lambda$ is even and self-dual. There are precisely two such lattices in 16 dimensions. One consists of two copies of the $E_8$-root lattice, the other is the root lattice of the Lie-algebra of $SO(32)$, with one of the spin-weight lattices added. These are the heterotic string theories. The theories with the two different lattices result in a 10-dimensional theory with $E_8 \times E_8$, resp. $SO(32)$ non-Abelian gauge symmetry.

It is also possible to compactify this theory on a $d$-dimensional spatial torus. For sufficiently small size of the $d$-dimensional torus, some degrees of freedom living on the $d$-dimensional spatial torus cannot be distinguished from the bosons living on the internal 16-dimensional torus. The theory then obtains symmetries that mix directions in the internal torus with directions on the spatial torus. This is responsible for interesting phenomena, such as the possibility to enhance the $E_8 \times E_8$ or $SO(32)$ groups to larger groups, by combining them with Kaluza-Klein bosons in a non-trivial way. Another interesting consequence is that the compactified heterotic theories are actually one theory, as it turns out that both the $E_8 \times E_8$-theory and the $SO(32)$-theory can be reached as different decompactification limits of the same compactified theory.

## 1.4 Outline

This thesis treats the construction of gauge field configurations with zero field strength on 3 dimensional (and at some places higher dimensional) tori. We will mainly follow the historic lines, as described in the previous sections.

In chapter 2 we shortly review Witten's index, and the calculation of the index for supersymmetric gauge theories on a 3–torus with periodic boundary conditions. We also describe Witten's analysis for the $SU(n)$ gauge theories with twisted boundary conditions. Then the analysis for the infinite volume case is described and the original paradox outlined. Then we move on to Witten's original construction of the extra vacua for orthogonal groups, and explain the results both in terms of D-branes as in terms of explicit holonomies in various representations. We also indicate how the analysis is almost trivially extended to the exceptional group $G_2$.

In chapter 3 we describe a pragmatic approach to constructing extra vacua with periodic boundary conditions in various groups. This explicit construction yields explicit expressions for the holonomies. We demonstrate that it reproduces the results of chapter 2 and that for the remaining exceptional groups sufficient vacua exist to solve the Witten index problem.

In chapter 4 we briefly review the approach of two other groups [6] [26] on the same subject. This approach is more formal and somewhat more systematic. We will not cover these results in detail, but describe some used techniques and point out the parallels with our work as described in chapter 3. The chapter ends with a description of the calculation of the Chern-Simons invariant, which turns out to take fractional values for the new vacua.

In chapter 5 we study vacuum solutions for gauge theories with classical groups (unitary, orthogonal, symplectic) on the 2–torus and the 3–torus. These results were already studied in [47] and [6], but we reproduce them by an approach that uses D-branes on orientifolds, in a similar setting as the original result from [55]. These results also show how to interpret various orientifold configurations as arising from configurations for classical gauge fields for the 2– or 3–torus.

The results of chapter 5 were obtained completely ignoring the string consistency conditions. In chapter 6 we turn to consistent string theories, starting with the $E_8 \times E_8$ heterotic string theory, compactified on a 3–torus, resulting in a 7–dimensional theory. We describe how only relatively few of the many configurations possible in $E_8 \times E_8$-gauge theory lead to consistent string-theories. One of the configurations turns out to be considered before: It is a version of the CHL-string in disguise. The CHL-string [7] [8] is an $E_8 \times E_8$-string theory compactified on a circle, with a holonomy that interchanges the two $E_8$-factors. It may be compactified on additional tori. The version of the CHL-string we will describe in chapter 6 is formulated in a different way, but is related to the original formulation by string duality. In 8 dimensions there exists a dual formulation in terms of a $Spin(32)/\mathbb{Z}_2$ string with twisted boundary conditions, which together with the standard compactification are the only two possibilities for the $Spin(32)/\mathbb{Z}_2$-heterotic string. String dualities relate the heterotic string to the type I string, and the type I string is related by T-dualities to various D-brane and orientifold configurations. Of these, a few were encountered in chapter 5. The remaining configurations from chapter 5 are inconsistent, for various reasons. Lastly, we briefly address dualities to M-theory and F-theory realisations of the relevant gauge symmetries, which turn out to live on special compactification manifolds that involve the 4-dimensional Calabi-Yau manifold $K3$.

In two appendices, we outline our conventions for Lie-algebra's. A third appendix is devoted to a small derivation of heterotic-heterotic duality, in a formulation that emphasises concepts relevant to this thesis.

# 2 Witten's index and vacua for Yang-Mills theories

*'... They maintain that the operation of counting modifies quantities and converts them from indefinite into definite sums. The fact that several individuals who count the same quantity should obtain the same result is, for the psychologists, an example of association of ideas or a good exercise of memory....'*

"Tlön, Uqbar, Orbis Tertius" - Jorge Luis Borges
(English translation by J. E. Irby)

In this chapter we will review Witten's index [52], and its computation for non-Abelian gauge theories. Previously, for some groups computations of Witten's index in finite and infinite volumes gave different answers. The resolution of this puzzle, for the case of supersymmetric Yang-Mills theory with an orthogonal gauge group, was found by Witten [55]. We will review this solution, and extend the analysis to the exceptional group $G_2$. The full resolution of the puzzle together with a more detailed understanding will be presented in the next two chapters.

## 2.1 Witten's index

Although applicable in a more general situations, we will only need Witten's index for four dimensional relativistic field theories. The relevant symmetry is then four dimensional Poincaré supersymmetry, which is the extension of Poincaré symmetry with fermionic generators. It is possible to introduce more than one supersymmetry charge (up to four for renormalisable theories, up to eight if one drops renormalisability as a criterion), but we will not do so.

The fermionic supersymmetry generators $Q, \overline{Q}$ obey the following algebra (in the conventions of [4]):

$$\{Q_\alpha, \overline{Q}_{\dot\beta}\} = 2\sigma^\mu_{\alpha\dot\beta} p_\mu \qquad (2.1.1)$$
$$[Q_\alpha, p_\mu] = 0 \qquad (2.1.2)$$

where the four momentum $p^\mu$ is $(p_o = E, \vec{p})$ and the four vector $\sigma^\mu$ consists of the 2x2-identity matrix and the 3 Pauli matrices $\sigma^\mu = (\mathbb{1}, \vec{\tau})$. The indices $\alpha$ and $\dot\beta$ are chiral spinor indices, taking the values 1,2. Dotted indices transform oppositely under chirality as undotted ones ($Q$ has also a non-trivial commutator with the generators of the Lorentz group, expressing that it is a spinor). It is easily verified that

$$\{Q_1, \overline{Q}_1\} + \{Q_2, \overline{Q}_2\} = 4E \qquad (2.1.3)$$

so indeed the energy (Hamiltonian) can be expressed in the supersymmetry charges.

From the algebra it is quickly deduced that:

$$E = <\Psi|H|\Psi> = \frac{1}{4}\sum_i ||Q_i|\Psi>||^2 + ||\overline{Q}_i|\Psi>||^2 \geq 0 \qquad (2.1.4)$$

The energy in a supersymmetric theory is bounded from below, and a state with lowest possible energy has $E = 0$, which implies that it is invariant under supersymmetry ($Q|\Psi> = \overline{Q}|\Psi> = 0$).

In the following we wish to count states. This is only well-defined in the case that the theory has a discrete spectrum, which can be achieved by putting it in a finite volume. We assume that appropriate, supersymmetry preserving boundary conditions have been chosen.

We now define an operator, which gives $+1$ on bosonic states and $-1$ on fermionic states; such an operator can be symbolically denoted as $(-)^F$ (where $F$ stands for fermion number) (For some subtleties see [52]). By definition $(-)^F$ anticommutes with the supersymmetry generators, and hence (the trace is over the Hilbert space)

$$\text{Tr}((-)^F 2\sigma^\mu p_\mu) = \text{Tr}((-)^F \{Q,\overline{Q}\}) = 0 \qquad (2.1.5)$$

Since the four-momentum $p$ commutes with the supersymmetry generators, the calculation can be performed at fixed non-zero $p$ giving $\text{Tr}_p(-)^F = 0$. This states that at non-zero $p$ the number of bosonic and fermionic states is equal, which is a consequence of the often-heard statement that supersymmetry maps bosons into fermions, and vice versa. This last statement is however slightly inaccurate, since there can be states (bosonic or fermionic) that are invariant under supersymmetry. From the previous it follows that such states necessarily have momentum equal to zero, and therefore energy equal to zero, and hence

$$\text{Tr}(-)^F = n_B^{E=0} - n_F^{E=0}, \qquad (2.1.6)$$

where $n_B^{E=0}$ ($n_F^{E=0}$) is the number of bosonic (fermionic) states at zero energy. Actually (2.1.6) is not accurate since the left-hand side is ill-defined (it is not absolutely convergent). It is clear from the above that there is an implicit counting prescription, that can be made explicit by replacing $\text{Tr}(-)^F$ by the regulated version $\text{Tr}((-)^F \exp(-\beta E))$. This is convergent, and gives the same result as (2.1.6). This regulated object is what has become known as Witten's index. We will keep on using the inaccurate notation $\text{Tr}(-)^F$ for shortness.

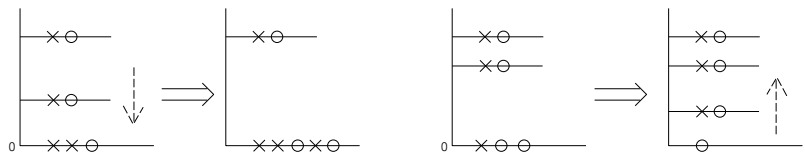

Figure 2-1. Invariance of Witten's index under perturbations, crosses are bosons, circles are fermions

$\text{Tr}(-)^F$ is argued to be invariant under perturbations (provided these respect supersymmetry). Witten's index only depends on the states that have zero energy. There might be a change in the number of zero energy states, for example if some perturbation would cause states to come down from non-zero to zero energy. But by the above states at non-zero energy have to come in bose-fermi pairs, and thus will not contribute to Witten's index when they arrive at zero energy (see fig. 2.1). Similarly, if some states move away from zero energy they can only do so in bose-fermi pairs, again conserving the index.

There is an implicit assumption in the above which is probably best illustrated with the example of the Wess-Zumino model, which is in a sense the supersymmetrisation of $\phi^4$-theory. The bosonic field in this theory is a complex scalar field $\phi$, with a potential of the form $|m\phi|^2|1-g\phi/m|^2$. This potential has two minima, and hence gives rise to two bosonic vacua, one at $\phi = 0$ and one at $\phi = m/g$. The fermionic superpartner of the scalar field is massive and does not contribute to the index, giving $\text{Tr}(-)^F = 2$. This theory may be deformed by varying the parameter $g$, and indeed any small variation of $g$ will not alter the index. If we however take $g \to 0$, then the potential takes the form $|m\phi|^2$ which seems to lead to a unique minimum. The limit $g \to 0$ is however a singular one, as the expectation value of the vacuum at $\phi = m/g$ diverges. Also the height of the barrier between the two vacua diverges, such that the two decouple. It is important to avoid such pathologies when calculating the index. For gauge theories in a finite volume, in which we will be interested, the configuration space is compact and there is no possibility for fields to develop infinite expectation values. In this situation the index is expected to be a constant, even when taking the infinite volume limit.

The importance of Witten's index lies in its relation to spontaneous breaking of supersymmetry. By definition, spontaneous breaking of a symmetry means that the ground state $|0>$ of the system is not invariant under the symmetry. For supersymmetry this implies that $Q|0> \neq 0$, and necessarily that its energy $E_0 > 0$. But if this is the state with lowest energy then this must imply that $\text{Tr}(-)^F = 0$. Hence spontaneous breaking of supersymmetry is impossible in theories with a non-zero index.

## 2.2  Gauge theory on a torus

We wish to calculate Witten's index for supersymmetric Yang-Mills theory. In the previous section it was argued that a sensible calculation can be performed by formulating the theory in a finite volume with supersymmetry preserving boundary conditions. In the following we will discuss gauge theory on a 3–torus, and the possible boundary conditions.

In the Lagrangian formulation the gauge theory is defined on $\mathbb{R} \times T^n$, with $\mathbb{R}$ representing the non-compact time dimension. Then $A_0$ can be gauged away and we will set $A_0 = 0$. An n–torus $T^n$ can be viewed as the quotient of $\mathbb{R}^n$ by a lattice $\Lambda$. For simplicity we will set $\Lambda = L\mathbb{Z}^n$ here, with the length $L$ the same for all directions. We consider a gauge theory on this torus with compact gauge group $G$. The covariant derivative is $D_\mu = \partial_\mu - iA_\mu$. Under gauge transformations, the covariant derivative transforms in the adjoint:

$$D_\mu \to \text{Ad}(g(x))D_\mu = g(x)D_\mu g^{-1}(x) \qquad (2.2.1)$$

A zero field strength solution for the gauge fields implies

$$F_{\mu\nu} = i[D_\mu, D_\nu] = \partial_\mu A_\nu - \partial_\nu A_\mu - i[A_\mu, A_\nu] = 0 \qquad (2.2.2)$$

The (local) solution to this equation is given by

$$A_\mu(x) = iU(x)\partial_\mu(U^{-1}(x)) \tag{2.2.3}$$

where $U$ takes values in $G$, and hence $A_\mu$ takes values in the Lie-algebra of $G$. This is easily rewritten to $\partial_\mu U = iA_\mu U$ which can be integrated around a cycle of the torus to give

$$U(x+L_i) = P\exp\left\{i\int_x^{x+L_i} A_\mu(y)dy^\mu\right\} U(x) \equiv \Omega_i(x)U(x), \tag{2.2.4}$$

where $P$ is for path ordering. By $L_i$ we denote the unit vector $e_i$ in the $i$-the direction, multiplied by the length $L$. It is implicitly assumed that there exists a path form $x$ to $x+L_i$ such that $U$ can be integrated along it. $\Omega_i(x)$ does not depend on the path chosen, as $F_{\mu\nu} = 0$.

We demand periodicity of the gauge field, such that $A_\mu(x+L_i) = A_\mu(x)$. This does not necessarily lead to periodicity of $U$. Instead one can verify that periodicity of $A_\mu$ implies that

$$U^{-1}(x)U(x+L_i) = \omega_i \tag{2.2.5}$$

is independent of $x$ [3]. This implies that $\omega_i\omega_j = \omega_j\omega_i$, and that

$$\Omega_i(x) = U(x)\omega_i U(x)^{-1} \tag{2.2.6}$$

$U(x)$ is only defined up to multiplication by a constant element on the right, and we may use this ambiguity to set $U$ to the identity for a reference point, say $U(0) = \mathbb{1}$, fixing $\Omega_i(0) = \omega_i$.

If there are additional matter fields $\phi(x)$ present, transforming in a representation $R$ of the gauge group, these will pick up a phase when translated around a closed cycle, as

$$\phi(x+L_i) = R(\Omega_i(x))\phi(x), \tag{2.2.7}$$

with $R(g)$ denoting the group element $g$ in the representation $R$. One needs a constraint to avoid ambiguities at $x+L_i+L_j$ which is

$$R(\Omega_i(x+L_j))R(\Omega_j(x)) = R(\Omega_j(x+L_i))R(\Omega_i(x)) \tag{2.2.8}$$

This can be rewritten to $R(\omega_i)R(\omega_j) = R(\omega_j)R(\omega_i)$.

This condition is in general more restrictive than the condition on $A_\mu$. For a theory with gauge fields only, we may take $U$, and therefore $\omega_i$ in the adjoint representation of the group. If the condition on the matter fields is satisfied, then the $\omega_i$ certainly commute in the adjoint representation.

We have not specified the topology of the gauge group $G$, but every compact $G$ has a simply connected cover $\tilde{G}$. Every representation $R$ of $G$ is also a representation of $\tilde{G}$. Defining $\tilde{U}$, $\tilde{\Omega}_i$, and $\tilde{\omega}_i$ to be liftings of $U$, $\Omega_i$, $\omega_i \in G$ to $\tilde{G}$, the above consistency conditions may be translated to

$$\tilde{\omega}_i\tilde{\omega}_j = z_{ij}\tilde{\omega}_j\tilde{\omega}_i \tag{2.2.9}$$

with $z_{ij}$ any element of the centre of $\tilde{G}$ that is projected to the identity in $G$, $R(z_{ij}) = \mathbb{1} \in G$.

Note that $z_{ij} = \mathbb{1} \in \tilde{G}$ is always a solution and this is called periodic boundary conditions. We have seen that a flat connection with periodic boundary conditions gives commuting holonomies. To go in the other direction we need a restriction, namely that the holonomies commute in a simply connected representation. For example, for the 3–torus we have that:

**Theorem** *For any set $\{\Omega_1, \Omega_2, \Omega_3\}$, $\Omega_i \in G$ where $G$ is a simple, connected, and simply connected group, and the $\Omega_i$ mutually commute, a periodic flat connection exists such that $\Omega_i$ are the holonomies.*

This can be proven as follows:

The flat connection is written as $A_i = iU\partial_i U^{-1}$ with $U(x,y,z) \in G$. We demand that $U(x,y,z)$ satisfies the following boundary conditions

$$
\begin{aligned}
U(x+L,y,z) &= U(x,y,z)\Omega_1 \\
U(x,y+L,z) &= U(x,y,z)\Omega_2 \\
U(x,y,z+L) &= U(x,y,z)\Omega_3
\end{aligned}
\tag{2.2.10}
$$

with constant commuting $\Omega_i$ (commutativity of $\Omega_i$ ensures the uniqueness of $U$). Then $A_i(x,y,z)$ is periodic. Choosing $U(0,0,0) = \mathbb{1}$, the matrices $\Omega_i$ are the holonomies. The matrix $U(x,y,z)$ can be constructed in several steps.

At the first step, we define

$$U(x,0,0) = \exp\left\{i\pi T_1 \frac{x}{L}\right\}, \qquad U(0,y,0) = \exp\left\{i\pi T_2 \frac{y}{L}\right\}, \qquad U(0,0,z) = \exp\left\{i\pi T_3 \frac{z}{L}\right\}.$$

where $\Omega_i = \exp\{i\pi T_i\}$ (The choice of $T_i$, once $\Omega_i$ are given is not unique, but it is irrelevant. Take *some* set of the logarithms of holonomies $\Omega_i$). Having done this, $U$ can be extended over all other edges of the 3-cube so that the boundary conditions (2.2.10) are fulfilled. For example, we define

$$U(L,y,0) = \exp\left\{i\pi T_2 \frac{y}{L}\right\}\Omega_1, \qquad U(x,L,0) = \exp\left\{i\pi T_1 \frac{x}{L}\right\}\Omega_2, \qquad \text{etc.}$$

With $U(x,y,z)$ defined on the edges of the cube, $U$ can also be continued to the *faces* of the cube due to the fact that, by assumption, $\pi_1(G) = 0$ i.e. any loop in the group is contractible. First do this for the 3 faces adjacent to the vertex (0,0,0). On the other 3 faces of the cube $U(x,y,z)$ can now be found by imposing the required boundary conditions:

$$U(x,y,L) = U(x,y,0)\Omega_3, \qquad U(x,L,z) = U(x,0,z)\Omega_2, \qquad U(L,y,z) = U(0,y,z)\Omega_1$$

With $U(x,y,z)$ defined on the surface of the cube, it can be continued to the interior using the fact that $\pi_2(G) = 0$ for all simple Lie groups. By construction, $U(x,y,z)$ satisfies the boundary conditions (2.2.10) and hence $A_i(x,y,z)$ is periodic.

This skeleton construction is common in homotopy theory and can be found also in the physics literature (see e.g. [2]). Simply connectedness of the group is an essential ingredient in the construction.

If $G$ itself is not simply connected, then there are additional solutions to the consistency requirements [23] [24]. The $z_{ij}$ parametrise the different boundary conditions, and different $z_{ij}$ can clearly not be deformed into each other. If any of the $z_{ij} \neq \mathbb{1}$, and thus $G$ not simply connected, the above construction fails, as it is impossible to continue to the face with coordinates $x_i$ and $x_j$. Traversing the loop around this face one picks up a factor $z_{ij} \in \tilde{G}$, which projects to $\mathbb{1} \in G$. This signals that the loop is closed in $G$ but not contractible. A flat connection may nevertheless be constructed by covering the face with patches, as in [2]. For the example where one of the $z_{ij} \neq \mathbb{1}$, we cover the relevant face with two patches. We cut out a disk from the face, and set $U = \mathbb{1}$ there. On the complement of the disk, we construct the $U$ with the appropriate holonomies. On the boundary of the disk, the two different definitions of $U$ should be glued together using a transition function. This transition function maps the boundary of the disk, which is a circle, to a closed loop in the group, and therefore carries the homotopy type.

## 2.3  The index for supersymmetric Yang-Mills theories

In this section the original calculation of Witten's index for supersymmetric Yang-Mills theories will be presented. For a well-defined counting procedure, the spatial dimensions will be compactified on a 3–torus. We also have to specify boundary conditions. It turns out that also in the large volume a sensible calculation is possible.

### 2.3.1  Periodic boundary conditions

For periodic boundary conditions (all $z_{ij} = 1$), we have that $A_\mu(x+L_i) = A_\mu$ (with $L_i = Le_i$ with $e_i$ the unit vector in the $i$-th direction as before). Supersymmetry requires us to impose the same boundary conditions on the gluinos, and hence $\psi(x+L_i) = \psi(x)$. A periodic solution for the gauge fields can be obtained by setting (remember $A_0 = 0$)

$$A_i = c_i/L, \tag{2.3.1}$$

where each of the $c_i = 2\pi(h_\alpha)_i$ is an element of the Cartan subalgebra (CSA) (the Cartan subalgebra is parameterised by vectors $\alpha$ taking values in $\mathbb{R}^r$ with $r$ the rank of the Lie group, see appendix A). This fixes the $\Omega_i$ to $\exp\{ic_i\}$. Making this choice does not completely eliminate the gauge freedom, there are two types of gauge transformation that are still allowed.

It is possible to shift an $\alpha$ by a coroot, say $\beta$, by using a local gauge transformation of the form $g = \exp\{2\pi i x^i h_\beta/L\}$. Since $\exp\{2\pi i h_\beta\} = 1$ this is periodic, and because we have to identify under these transformations, this makes the moduli space compact. The second kind of gauge transformations are global discrete gauge transformations that map the Cartan subalgebra to itself. These discrete transformations make up the Weyl group $W$. The moduli space parametrising the vacua for the gauge theory on $T^3$ will then be of the form $S(r)^3/W$, where $S(r)$ is the quotient of $\mathbb{R}^r$ by the coroot lattice of the Lie algebra.

With such a gauge field in the background, the Dirac equation has zero modes of the form

$$\psi = \epsilon \tag{2.3.2}$$

with $\epsilon$ a constant spinor taking a value in the Cartan subalgebra. This is because of the commutator in the Dirac equation, which requires the $\epsilon$ to commute with $c_i$. Another derivation of the same result may be obtained by noting that a zero mode of $\psi$ in the background of the gauge field has to obey $\psi(x+L_i) = \text{Ad}(\Omega_i)\psi(x) = \Omega_i\psi(x)\Omega_i^{-1}$, but on the other hand our boundary conditions require $\psi(x+L_i) = \psi(x)$. The two conditions imply that $\psi$ should commute with $\Omega_i$, and as the $\Omega_i$ are on a maximal torus, $\psi$ can be taken to be an element of the CSA.

The strategy of [52] consists of only quantising the modes (2.3.1) and (2.3.2). The rationale for this is that these are zero energy modes, and that other modes will typically carry an energy of order $2\pi/L$ with $L$ the size of the torus. Of course these modes interact with the zero modes, but the expectation is that for sufficiently small torus size the energy separation is large and the interaction small. Moreover, as motivated in the general discussion, the index is thought to be independent of such calculational details.

The Lagrangian for the slow modes $c_i$ and $\epsilon$ becomes

$$L_{eff} = \int d^3x \mathcal{L}_{eff} = \frac{V}{2h(G)}\text{tr}\left(\frac{1}{g^2}(\frac{\partial c_i}{\partial t})^2 + i\bar{\epsilon}\gamma_0\frac{\partial \epsilon}{\partial t}\right). \quad (2.3.3)$$

Here $h(G)$ is the dual Coxeter number of the group. One now has to expand in the group generators $c_i = c_i^a T^a$, and $\epsilon = \epsilon^a T^a$. The $T^a$ are generators for the adjoint representation, and are normalised according to $\text{tr}(T^a T^b) = h(G)\delta^{ab}$. Quantisation leads to the Hamiltonian

$$H = \frac{g^2}{2V}\left(\frac{\partial}{\partial c_i^a}\right)^2. \quad (2.3.4)$$

The fermions have dropped out because their action is only linear in $t$. The $c_i$ are periodic variables, and therefore have a discrete spectrum. It can therefore be safely asserted that this Hamiltonian has a unique ground state, other states having a finite energy.

We still have to include the fermions. When adding these the residual gauge freedom has to be taken into account. The residual gauge freedom from periodic gauge transformations acts trivial on the fermions, the Weyl group however does not, and only combinations invariant under the Weyl group are allowed. To find these we note that the Weyl group is a discrete subgroup of $O(r)$. This group has an invariant tensor $\delta_{ab}$ leading to the invariant combination $U = \delta_{ab}\varepsilon^{\alpha\beta}\epsilon_\alpha^a\epsilon_\beta^b$ (where one has to antisymmetrise in the spinor indices to make the combination non-zero). Also the products $U^2, \ldots U^r$ are invariant. The operators $U^k$ with $k > r$ are zero because the antisymmetry requirements cannot be fulfilled anymore.

If $|0>$ is the ground state of the bosonic Hamiltonian, then $U|0>, U^2|0>, \ldots$ are also valid ground states. In each of these, an even number of fermionic zero modes is occupied, so these are all bosonic. The Witten index count gives then $\text{Tr}(-)^F = r+1$.

One may argue that the adiabatic approach used to derive this result is questionable. It does not deal with the fact that the moduli space for periodic flat connections has singular points, such as $c_i^a = 0$. At such singular points the group commuting with the $A_i$ is non-Abelian, and the zero-modes are not limited to the CSA. It seems then that the above derivation is too naïve. We will not resolve these matters here, nor is it resolved by the results to be presented later. We stress however that the fact that new vacuum components are found is independent of the question whether the adiabatic approach is applicable.

One may attempt to avoid the problems associated to the singular points in the moduli space by choosing boundary conditions that eliminate the vacuum degeneracy. In some cases this is indeed possible, as will be described in the next section.

### 2.3.2  Twisted boundary conditions

Periodic boundary conditions are not the only supersymmetry preserving boundary conditions for supersymmetric Yang-Mills theories on the 3–torus. In these theories only the adjoint representation is present, and 't Hoofts twisted boundary conditions [23] [24] can be imposed. These lead to a completely different but still supersymmetric theory. With such boundary conditions part of the vacuum degeneracy is removed. Here we will study the case in which it is completely removed, which is only possible for the group $SU(N)$. For the other groups the treatment with twisted boundary conditions is in a sense intermediate to this extreme case and the treatment with periodic boundary conditions. Results for these theories can be found in [6]. In chapter 5 we will study twisted boundary conditions for theories with classical groups from a different viewpoint, and reproducing the analysis of [6] for these groups.

Consider supersymmetric Yang-Mills theory with $SU(N)$ gauge group. This theory is actually a $SU(N)/\mathbb{Z}_N$ gauge theory as all fields transform in the adjoint. We will work in the simply connected group $SU(N)$. Compactify the theory on a spatial 3–torus with non-compact time direction, and gauge away $A_0 = 0$. The centre of $SU(N)$ is formed by elements of the form $\exp(2\pi in/N)\mathbb{1}$. Of these we pick an element $z$ that generates the full centre and set

$$z_{12} = z \quad z_{13} = \mathbb{1} \quad z_{32} = \mathbb{1}. \tag{2.3.5}$$

We calculate the set of $\tilde{\omega}_i$ $i = 1,2,3$ that satisfy the commutation relation (2.2.9). One can choose $\tilde{\omega}_1$ to be diagonal, and it follows from the commutation relations that its diagonal entries are $c\exp(2\pi in/N)$, with $c$ an overall constant to make $\det\tilde{\omega}_1 = 1$. It is defined up to a factor $\exp(2\pi in/N)$ but all choices are equivalent, since they can be obtained by conjugating with $\tilde{\omega}_2$. Choosing a specific diagonal form for $\tilde{\omega}_1$ fixes, by the commutation relations, also the form of $\tilde{\omega}_2$ completely, again up to an overall factor $\exp(2\pi in/N)$. Also here all choices are equivalent, as they can be obtained by conjugation with $\tilde{\omega}_1$. Finally $\tilde{\omega}_3$ should be a constant matrix commuting with $\tilde{\omega}_1$ and $\tilde{\omega}_2$. An element of $SU(n)$ commuting with $\tilde{\omega}_1$ should be diagonal, a diagonal element commuting with $\tilde{\omega}_2$ should have all its diagonal entries equal, and hence the only possibilities for $\tilde{\omega}_3$ are the $N$ elements of the centre of $SU(N)$. We have therefore in total $N$ distinct possibilities for the $\tilde{\omega}_i$, and this leads to $N$ distinct possibilities for the gauge field $A_\mu$.

To investigate the zero-modes for the gluino's we may use again that a zero mode in the $A_\mu$ background has $\psi(x+L_i) = \mathrm{Ad}(\Omega_i(x))\psi(x) = \mathrm{Ad}(U(x)\omega_i U^{-1}(x))\psi(x)$, but on the other hand $\psi(x+L_i) = \psi(x)$ is required by the boundary conditions. This implies that if $\psi$ is a zero-mode, then $\mathrm{Ad}(U^{-1}(x))\psi$ commutes with all the $\omega_i$. One may solve this in the $SU(N)$ cover of $SU(N)/\mathbb{Z}_N$. $\mathrm{Ad}(U^{-1}(x))\psi$ is a certain element of the Lie-algebra of $SU(N)$ which consists of traceless hermitian matrices. We have again that a matrix commuting with both $\tilde{\omega}_1$ and $\tilde{\omega}_2$ has to be diagonal with all diagonal entries equal; together with the condition of tracelessness this implies that $\text{Ad}(U^{-1}(x))\psi$ vanishes, and hence there is no fermionic zero-mode in this background.

The only possible solutions are those specified by the $N$ boundary conditions on the $U$'s. This gives (up to gauge transformations) $N$ purely bosonic vacua, and performing the index count thus gives $\text{Tr}(-)^F = N$. As the rank of $SU(N)$ is $N-1$, this is identical to the result found with periodic boundary conditions.

### 2.3.3 Infinite volume

Both finite volume calculations indicate a vacuum degeneracy for supersymmetric Yang-Mills theory. In view of the discussion on the Witten index, it is natural to ask how this degeneracy can be interpreted in the large volume limit.

To answer this question, observe that in the infinite volume limit the classical theory has a chiral symmetry:

$$\lambda \to e^{i\theta\gamma_5}\lambda \tag{2.3.6}$$

This symmetry does not survive quantisation because of the chiral anomaly. The current associated to the chiral symmetry is not conserved, since the current conservation law is changed into:

$$\partial_\mu J_5^\mu = \frac{h(G)}{16\pi^2} {}^*F^{\mu\nu a} F^a_{\mu\nu} \tag{2.3.7}$$

with $h(G)$ the dual Coxeter number of the gauge group. This relation states that in a background of non-zero ${}^*F^{\mu\nu} F_{\mu\nu}$, hence in the presence of instantons, the $U(1)$ of chiral symmetry is broken to a discrete group, being $\mathbb{Z}_{2h(G)}$.

It is expected that dynamical effects will generate a non-vanishing gluino condensate $<\lambda\lambda> \neq 0$. This expectation is confirmed by various calculations [1] [11] [35] [45] (Note however that these groups do not agree on the value of the gluino condensate, but all find a non-zero result.) The condensate is not invariant under $\mathbb{Z}_{2h(G)}$ but breaks it further to $\mathbb{Z}_2$, giving $h(G)$ degenerate vacua.

The reasoning behind the Witten index suggests that these $h(G)$ vacua could be the $r+1$ vacua of the finite volume theory. To support this $h(G)$ and $r+1$ should be equal. A glance at the following table shows however that this is in general not the case.

|       | $SU(N)$ | $Sp(N)$ | $SO(2N)$ | $SO(2N+1)$ | $G_2$ | $F_4$ | $E_6$ | $E_7$ | $E_8$ |
|-------|---------|---------|----------|------------|-------|-------|-------|-------|-------|
| $r+1$ | $N$     | $N+1$   | $N+1$    | $N+1$      | 3     | 5     | 7     | 8     | 9     |
| $h$   | $N$     | $N+1$   | $2N-2$   | $2N-1$     | 4     | 9     | 12    | 18    | 30    |

Table 2-1. rank $r+1$, and dual Coxeter number $h$ for all simple compact Lie groups

Remarkable is however that for the infinite number of $SU(N)$-theories all calculations do agree, as well as for $Sp(N)$. The number of groups where the results disagree is however also infinite. Note that in case of disagreement $h$ is always larger than $r+1$.

## 2.4 Extra vacua from orientifolds

The puzzle on the disagreement of the Witten index computation in various limits lasted for over 15 years. It was again Witten who pointed out what the flaw in the original argument is [55]. The rather innocent looking but wrong assumption is, that on the 3–torus all classical solutions to the Yang-Mills vacuum equations can be put in the form (2.3.1), that is globally constant gauge fields taking values in the Cartan subalgebra. This assumption can be found on various places in the literature, sometimes accompanied by "proofs". Unfortunately all these proofs contain flaws, as we will see a counterexample shortly. But let us first discuss to what extent the assumption is really wrong.

It is not hard to show that if the Yang-Mills gauge group is $SU(N)$ or $Sp(N)$, then every solution to the vacuum equations can be transformed to a globally constant one taking values in the CSA. But these are precisely the groups for which all Witten index computations agreed. For orthogonal and exceptional gauge groups, the assumption can not be proven easily, and as we will see, it is actually incorrect. One can also show that if one considers a gauge theory with simply connected and simple gauge group on a circle or a 2–torus then (2.3.1) is always correct (for *any* gauge group). It is only for 3– and higher dimensional tori, and theories with orthogonal or exceptional gauge groups that there are extra solutions.

In [55] Witten focussed on the orthogonal groups (the exceptional groups cannot be treated within the formalism presented there). The orthogonal group can be realised as a gauge group in string theory by means of the Chan-Paton construction [20] [44]. One attaches charges to the endpoints of open strings, which have to be unoriented to give orthogonal symmetry. The strings are allowed to move freely through space. Now this theory is compactified on a 3–torus. By using a T-duality transformation in all 3 directions of the torus, this string theory is transformed to another theory with the same physical content. This dual theory has open strings whose endpoint are confined to hypersurfaces called D-branes [43]. These D-branes are transverse to the dual 3–torus, intersecting it only in a point. The positions of D-branes on this dual torus parametrise the holonomies around the three cycles of the original torus. The dual torus is not exactly a torus however. The original theory was unoriented, which means that only states invariant under reflection of the string worldsheet are kept. In the dual theory this translates into an orientifold $T^3/\mathbb{Z}_2$, where the $\mathbb{Z}_2$ acts as a reflection on all three coordinates of the torus simultaneously, and reflects the string world sheet. The D-brane configuration is required to be invariant under the orientifold projection. This requires all D-branes to occur in pairs on the double cover of $T^3/\mathbb{Z}_2$, unless they are located at a fixed point of the orientifold projection. An $O(n)$ theory on a 3–torus is thus translated in a collection of $n$ D-branes, living on an orientifold.

In this way one can parametrise a compactification of an $O(n)$-gauge theory on a 3–torus. We require however a more restrictive set-up. First of all, we demand that the gauge theory can be restricted to the connected component of $O(n)$, being $SO(n)$. Second, we demand periodic boundary conditions, which translates to the requirement that the $SO(n)$-configuration lifts to a configuration in the simply connected double cover $Spin(n)$. There exists mathematical machinery to tackle this problem in the form of characteristic classes. The special case of investigating the liftings from $O(n) \to SO(n) \to Spin(n)$ occurs often in physics and in mathematics and the characteristic classes associated with these are known as Stiefel-Whitney classes.

The first Stiefel-Whitney class $w_1$ is a $\mathbb{Z}_2$-valued one-form. It measures the obstruction to lifting an $O(n)$ configuration to an $SO(n)$ configuration: If the class is non-trivial then lifting is impossible. Similarly, the second Stiefel-Whitney class $w_2$ is a $\mathbb{Z}_2$ valued two-form measuring the obstruction of lifting an $SO(n)$ configuration to $Spin(n)$.

The task is then to compute the Stiefel-Whitney classes for a given orientifold-D-brane-configuration. First of all, note that a smooth deformation of the flat connection corresponds to smoothly changing the position of some D-branes. The isolated D-branes at the orientifold fixed points cannot be moved, as this would be inconsistent with the orientifold projection. Pairs of D-branes can be moved on the orientifold, in particular to the origin of the orientifold (corresponding to the eigenvalue 1 in the holonomy). The pairs can therefore never be the source of an obstruction to the liftings we consider, and we may as well take the number of D-branes at each point modulo 2. This is of course in keeping with the $\mathbb{Z}_2$ nature of the Stiefel-Whitney classes. The remaining branes are the isolated ones which are located at the fixed points on the double cover of the orientifold. Except for the origin, all fixed points correspond to one or more eigenvalues $-1$ in the holonomies. Therefore, one can associate to each of the branes a formal class $1 + a_x \delta x + a_y \delta y + a_z \delta z$. Here the $a_i$ are either 0 or 1 (we are using an additive representation of $\mathbb{Z}_2$). We set $a_i$ to 1 if the corresponding brane contributes an eigenvalue $-1$ to the holonomy $\Omega_i$. The total Stiefel-Whitney class $w = 1 + w_1 + w_2 + \ldots$ can be obtained by multiplying the contributions of all D-branes (note that, since $-1 = 1 \bmod 2$ there is no ordering ambiguity).

A way to proceed now is to calculate the Stiefel-Whitney class of a general configuration of D-branes, and demand that the first and second Stiefel-Whitney class vanish. One can also make an ansatz based on the knowledge of the moduli space for $Spin(n \leq 6)$, since these all correspond to unitary and symplectic groups which are known to admit only one class of solutions. From this one concludes that one needs at least 7 D-branes for a non-trivial solution, and the only non-trivial configuration of 7 D-branes that does not correspond to anything one should encounter in lower dimensional $Spin$-groups, is to distribute these 7 D-branes over all orientifold fixed points apart from the origin. Computation of the Stiefel-Whitney class of this configuration shows that $w$ is trivial, and hence this can be lifted to $SO(7)$ and $Spin(7)$. Moreover, by the theorem below formula (2.2.9), it corresponds to a solution of the vacuum equations.

The new vacuum corresponds to the $SO(7)$-holonomies

$$\begin{aligned} \Omega_1 &= \mathrm{diag}(\ 1,-1,-1,-1,\ 1,\ 1,-1) \\ \Omega_2 &= \mathrm{diag}(\ 1,-1,\ 1,\ 1,-1,-1,-1) \\ \Omega_3 &= \mathrm{diag}(-1,\ 1,\ 1,-1,\ 1,-1,-1) \end{aligned} \qquad (2.4.1)$$

Other choices for the set $\{\Omega_i\}$ differing from (2.4.1) by permutations of the columns are possible, since the orientifold construction does not specify the ordering of these. They all can be obtained from each other by global $SO(7)$ rotations.

That the holonomies can be lifted to $Spin(7)$ can also be shown explicitly. The spinor representation of $Spin(7)$ is 8 dimensional, and can actually be chosen to be real. The easiest way to construct $Spin(7)$ is to find a representation of the 7-dimensional Euclidean $\Gamma$–matrices, satisfying the Clifford algebra $\{\Gamma_i, \Gamma_j\} = -2\delta_{ij}$. A particular choice for these $\Gamma$–matrices is

$$\begin{aligned}&\Gamma_1 = i\sigma^2 \otimes \sigma^2 \otimes \sigma^2; \quad \Gamma_2 = -i\mathbb{1} \otimes \sigma^1 \otimes \sigma^2; \quad \Gamma_3 = -i\mathbb{1} \otimes \sigma^3 \otimes \sigma^2; \quad \Gamma_4 = i\sigma^1 \otimes \sigma^2 \otimes \mathbb{1}; \\ &\Gamma_5 = -i\sigma^3 \otimes \sigma^2 \otimes \mathbb{1}; \quad \Gamma_6 = -i\sigma^2 \otimes \mathbb{1} \otimes \sigma^1; \quad \Gamma_7 = -i\sigma^2 \otimes \mathbb{1} \otimes \sigma^3,\end{aligned} \quad (2.4.2)$$

with $\sigma_i$ the $2 \times 2$ Pauli-matrices, and $\mathbb{1}$ the $2 \times 2$ identity matrix. Notice that these $\Gamma$-matrices are all real.

The generators of Spin(7) $T_{ij}^S$ can be simply constructed as

$$T_{ij}^S = \frac{1}{4}[\Gamma_i, \Gamma_j] \quad (2.4.3)$$

These obey the same commutation relations as the 7 dimensional matrices

$$\left(T_{ij}^V\right)_{kl} = \delta_{il}\delta_{jk} - \delta_{ik}\delta_{jl} \quad (2.4.4)$$

which are the generators of the vector representation.

To construct the liftings of the holonomies (2.4.1), one writes these holonomies as exponentials of generators $T_{ij}^V$, and replaces these by the generators of the spinor representation $T_{ij}^S$. There are multiple ways to obtain the $\Omega_i$ from exponentiating generators $T_{ij}^V$, but this does not matter, as any particular choice only affects an overall sign (this reflects the twofold ambiguity one has in lifting from $SO(7)$ to $Spin(7)$). The set of $Spin(7)$ holonomies corresponding to the set (2.4.1) of $SO(7)$ holonomies is

$$\Omega_1^{spin} = \pm\sigma^3 \otimes \mathbb{1} \otimes \sigma^3; \quad \Omega_2^{spin} = \pm\sigma^3 \otimes \sigma^3 \otimes \mathbb{1}; \quad \Omega_3^{spin} = \pm\sigma^3 \otimes \mathbb{1} \otimes \mathbb{1} \quad (2.4.5)$$

It is easy to see that $[\Omega_i^{spin}, \Omega_j^{spin}] = 0$. As $\pi_1[Spin(7)] = 0$, non–trivial periodic flat connections with the holonomies (2.4.5)(2.4.1) exist. Important is also that the 8 different liftings of the holonomies are all equivalent. To see this, notice that $\exp\{\pi T_{ij}^S\} = 2T_{ij}^S$, so $2T_{ij}^S$ is an element of $Spin(7)$. It is easily checked that for example $2T_{25}^S$ changes the sign of $\Omega_1^{spin}$ leaving the other two invariant, and $2T_{16}^S$ and $2T_{56}^S$ do the same for $\Omega_2^{spin}$ and $\Omega_3^{spin}$. Hence all signs are related by gauge transformations.

The new vacuum does not correspond to a situation where the gauge potentials $A_\mu$ are constant commuting matrices. This can be shown explicitly. Suppose one could choose the $A_\mu$ to be constants, then we could write $\Omega_i = \exp\{i\pi S_i\}$. The condition $[S_i, S_j] = 0 \;\forall i, j$ implies that $[S_i, \Omega_j] = 0 \;\forall i, j$. But, as one can easily check, a matrix that commutes with all three $\Omega_j$ has to be diagonal. The generators of $SO(7)$ however are antisymmetric, so no generator of $SO(7)$ commutes with all $\Omega_i$. This proves that the assumption $[S_i, S_j] = 0$ must be wrong.

This new vacuum is disconnected from standard vacua for $Spin(7)$-theory, which correspond to one D-brane at the origin and 3 pairs of D-branes placed at arbitrary points, representing a configuration where the gauge potentials $A_\mu$ can be chosen to be constant commuting matrices. This can be easily seen since there is no continuous way to deform from one configuration to the other; the orientifold condition prevents either of the 7 D-branes to come of their fixed point for one vacuum, or to split any pair of D-branes for the other vacua. It can also be deduced from group theory. Try to perturb the $\Omega_i$

$$\Omega'_i = \Omega_i(1 + \alpha_i^a T^a) \qquad (2.4.6)$$

and require that the $\Omega'_i$ still commute. This implies the conditions

$$\alpha_i^a \Omega_i [T^a, \Omega_j] = \alpha_j^a \Omega_j [T^a, \Omega_i] \qquad (2.4.7)$$

To solve these, we note that:

- With the given $\Omega^i$ and the standard basis $T^a$ of the $so(7)$ Lie algebra (denoted by $T_{ij}^V$ in the above), either $[\Omega^i, T^a] = 0$ or $\{\Omega^i, T^a\} = 0$
- $\alpha_i^a = 0$ if $[T^a, \Omega_i] = 0$ (since $[T^a, \Omega_j] \neq 0$ for some $i \neq j$)
- $\alpha_i^a = \alpha_j^a$ if both $[T^a, \Omega_i] \neq 0$ and $[T^a, \Omega_j] \neq 0$

From these observations it follows that we can rewrite (2.4.6) as

$$\Omega'_i = \Omega_i + \beta^a [\Omega_i, T^a] \qquad (2.4.8)$$

with $\beta^a$ independent of $i$. This is a global group rotation, and not a nontrivial deformation. The new vacuum does not admit deformations, and hence is isolated.

Being a new vacuum component, its contribution to the Witten index of $SO(7)$ supersymmetric Yang-Mills theory was not included in previous calculations. The new vacuum breaks all continuous gauge symmetries. Quantisation shows again a unique bosonic vacuum, but now, in the absence of continuous gauge symmetries, there are no fermion zero-modes. So its contribution to the Witten index is 1. This should be added to the Witten index from the previous calculation, which is $r_{SO(7)} + 1 = 4$, giving $\text{Tr}(-)^F = 4 + 1 = 5 = h(SO(7))$.

For any $SO(2N+7)$ theory on a 3–torus, extra vacua can be obtained by adding $N$ pairs of branes in the orientifold representation. A non-trivial $SO(8)$ vacuum can be obtained by adding to the non-trivial $SO(7)$-vacuum one single brane, by placing it at the orientifold fixed plane at the origin. Vacua for $SO(2N+8)$-theories can be obtained by adding $N$ pairs of branes to orientifolds representing $SO(8)$-vacua.

For $SO(7)$, the new component in the moduli space of classical vacua presents just a single point. The same is true for $SO(8)$: up to a global gauge transformation, any set of commuting $SO(8)$ matrices whose logarithms do not commute can be presented in the form $\Omega_i^{SO(8)} = \text{diag}(\Omega_i^{SO(7)}, 1)$ with $\Omega_i^{SO(7)}$ given by Eq.(2.4.1). Consider still higher orthogonal groups. For $SO(N > 8)$, an additional freedom appears associated with Cartan rotations in extra dimensions; any set $\Omega_i^{SO(N)} = \text{diag}(\Omega_i^{SO(7)}, \omega_i^{SO(N-7)})$ with logarithms of $\omega_i^{SO(N-7)}$ belonging to the Cartan subalgebra of $SO(N-7)$ gives rise to a nontrivial $SO(N)$ connection. The extra component of the moduli space is not an isolated point anymore, but represents a manifold. Its dimension is $3r_{SO(N-7)}$. There are $r_{SO(N-7)} + 1$ eigenstates of the corresponding Born–Oppenheimer Hamiltonian. All together we have $(r_{SO(N)} + 1) + (r_{SO(N-7)} + 1) = N - 2$ vacuum states [55] which is also the value of the dual Coxeter number for the group $SO(N)$.

Two theories have a minor subtlety. In the $SO(9)$-case, the non-trivial vacuum has the continuous unbroken symmetry group is $SO(2)$, which is Abelian. The index for $SO(2)$-theory is $\text{Tr}(-1)^F = 0$ [52], which does not seem to solve the problem for $SO(9)$-theory. This is however resolved as follows. Apart from the continuous $SO(2)$, there are also some discrete symmetries unbroken. An example of such a discrete symmetry is represented by the matrix $\text{diag}(1,1,1,1,1,1,-1,1,-1)$. This matrix commutes with the holonomies $\text{diag}(\Omega_i^{SO(7)},1,1)$, and acts as $\text{diag}(1,-1)$ in the unbroken $SO(2)$-subgroup. It is a gauge symmetry, so we have to demand invariance under this symmetry. In this way, the unbroken $SO(2)$ is enhanced to $O(2)$, and we need $\text{Tr}(-1)^F$ for $O(2)$-theory, not $SO(2)$. To calculate the index one can simply repeat the analysis for $SO(2)$ from [52], with the requirement of invariance under the extra symmetry $\text{diag}(1,-1)$. One finds that, of the four states mentioned in [52], the two states with one fermion are not invariant under the extra symmetry, while the two bosonic states (two fermions or none) are invariant. In this way we find $\text{Tr}(-1)^F = 2$ for $O(2)$-theory, in contrast to the zero result of $SO(2)$-theory. Hence for $SO(9)$ this results in $\text{Tr}(-)^F = (r_{SO(9)}+1)+2 = 7$, the correct number.

Another seemingly problematic case is resolved in a similar way (this case is not mentioned in [55] [27]). In $SO(11)$, the connected part of the unbroken gauge group for the non-trivial vacuum is $SO(4)$. Its double cover is $Spin(4) = SU(2) \times SU(2)$, which is not simple. We have not treated the calculation of Witten's index for non-simple groups, but the extension of the analysis is straightforward. $SU(2) \times SU(2)$ theory has a unique bosonic vacuum, and both $SU(2)$-factors allow a bilinear combination of fermion zero-modes. The index for $Spin(4)$ theory is then 4, coming from one empty bosonic vacuum, two vacua in which one of the respective bilinear combinations of zero-modes for the $SU(2)$ factors is occupied, and one in which all fermion zero-modes are occupied. Again this naively leads to the wrong value for the Witten index. The resolution lies again in an extra parity symmetry, as the unbroken subgroup is $O(4)$ instead of $SO(4)$. Parity in $SO(4)$ acts by exchanging the two spin representations. Imposing this as a gauge symmetry, we see that of the four vacua for $SO(4)$-theory, two are actually equivalent (related by a gauge transformation) in $O(4)$. The index for $O(4)$-theory is therefore 3, and not 4, leading to the right answer for the Witten index.

The extra parity symmetry is actually present for all higher orthogonal groups, the unbroken symmetry group being $O(N-7)$ (for the $Spin$ groups this is $Pin(N-7)$, the double cover of $O(N-7)$). However, the extra symmetry does not affect the analysis for the other cases, and the previous results remain valid.

In the orientifold representation, it is clear that there are no more than two vacua; due to the fact that there are only 8 non-trivial fixed points, one can have at most 8 isolated branes; for these, only two configurations lead to a trivial Stiefel-Whitney-class. Of course the same result follows from group theory. For $N \geq 14$, one might want to try to write down holonomies of the form $\text{diag}(\Omega_i^{SO(7)},\Omega_i^{SO(7)})$. These are however contained in the trivial component of the moduli space. One can write explicitly

$$\Omega_1^{SO(14)} = \text{diag}(1,-1,-1,-1,1,1,-1;1,-1,-1,-1,1,1,-1) =$$
$$\exp\{\pi[T_{2,9} + T_{3,10} + T_{4,11} + T_{7,14}]\} \qquad (2.4.9)$$

and similarly for $\Omega_2, \Omega_3$. The $\log \Omega_i^{SO(14)}$ defined according to the prescription (2.4.9) com-mute and can be put into the Cartan subalgebra of $SO(14)$. So nothing new is obtained. Also for still higher groups nothing new happens. Hence for $SO(N)$, there are never more than two components in the moduli space, no matter how large $N$ gets.

## 2.5 Extension to $G_2$

As was observed in [27] the above results are trivial to extend to the smallest of the exceptional groups $G_2$, which is a subgroup of $SO(7)$ and can be defined in various ways.

A particular definition defines it as the subgroup of $SO(7)$ leaving invariant the combination $f_{ijk}Q^i P^j R^k$ where $Q^i, P^j, R^k$ are 3 arbitrary 7–vectors and $f_{ijk}$ is a certain antisymmetric tensor. One particular convention for $f_{ijk}$ is

$$f_{165} = f_{341} = f_{523} = f_{271} = f_{673} = f_{475} = f_{246} = 1 \tag{2.5.1}$$

and all other non-zero components are recovered using antisymmetry. It is easy to see now that the matrices (2.4.1) do belong to the $G_2$ subgroup of $SO(7)$.

Another way to see the same is to define $G_2$ as a subgroup of Spin(7) leaving a particular spinor invariant. The matrices (2.4.5) leave invariant the spinor

$$\eta = \begin{pmatrix} 1 \\ 0 \end{pmatrix} \otimes \begin{pmatrix} 1 \\ 0 \end{pmatrix} \otimes \begin{pmatrix} 1 \\ 0 \end{pmatrix}, \tag{2.5.2}$$

and hence belong to $G_2$ (and, incidentally, $f_{ijk} = \eta^T \Gamma_{[i} \Gamma_j \Gamma_{k]} \eta$, with the $\Gamma$-matrices as defined in (2.4.2)). Note that this leads to the same representation for the $G_2$-holonomies, since the 8-dimensional spinor-representation of $SO(7)$ splits in a singlet and a 7-dimensional representation of $G_2$.

All $G_2$-representations fall into one conjugacy-class, meaning that all have a trivial centre, and all are simply connected, $\pi_1(G_2) = 0$ (see e.g. [12]). The previously stated theorem guarantees that a non-trivial periodic flat $G_2$ connection exists, with holonomies as above. As in $SO(7)$, this vacuum breaks all of $G_2$, and so gives one new vacuum state. This extra vacuum state adds up with the $r_{G_2} + 1 = 3$ "old" states associated to the constant gauge potentials belonging to the Cartan subalgebra, to make the total vacuum state counting in accordance with the result $\text{Tr}(-1)^F = h(G_2) = 4$.

These results for orthogonal groups and $G_2$ suggest that also supersymmetric Yang-Mills theories with the other exceptional groups as gauge group should have $h(G)$ vacua. The $h(G)$ vacua may be found from different components in the moduli space, according to the formula [55]

$$h(G) = \sum_i (r_i + 1) \tag{2.5.3}$$

where the sum runs over all components of the moduli space, and $r_i$ is the rank of the unbroken subgroup on component $i$. Equation (2.5.3) is satisfied for the unitary and symplectic groups, where the moduli space consists of one single component. For the orthogonal groups and for $G_2$, with a moduli space consisting of two components, the formula holds as well. We will see in the next chapters that (2.5.3) is also satisfied for the remaining exceptional groups, with the number of components of the moduli space being larger than two.

# 3 Non-trivial flat connections on the 3–torus

Inclusion of the extra vacua for gauge theories with orthogonal gauge groups, as well as for the case of the exceptional group $G_2$, leads to a resolution of the problems with the calculation of the Witten index. This strongly suggest that the presence of extra vacua may also solve the problem for the remaining exceptional groups $F_4$, $E_6$, $E_7$ and $E_8$. A simple comparison of numbers already shows that for these cases the vacuumstructure might be more involved. As an example consider $E_8$ whose rank $r$ is 8, but whose dual Coxeter number equals 30. The discussion in the previous section suggests that a particular component of the moduli space can contribute at most $r + 1 = 9$ to the index, hence there should be at least 4 components to reach 30. This is of course an optimistic estimate, and we will find no less than 12 components for the moduli space of $E_8$ flat connections on the 3–torus.

The indicated exceptional groups do not allow a formulation in terms of D-branes, nor do they have a simple relation to any of the classical groups. Also a brute force computation is not attractive (the smallest non-trivial representation of $E_8$ consists of $248 \times 248$-matrices). So we need new ideas. The right idea involves a variant of the construction of twisted boundary conditions, and was found more or less simultaneously by a number of groups [28] [29] [26] [6]. In this chapter we describe the approach of the first two references, which may be described as pragmatic, but nevertheless leads to the correct results. The remaining two references are more mathematical and formal. These will be discussed in the next chapter.

In this thesis we use a different normalisation for roots, group generators and related quantities than in [28] [29]. This was done to conform with the conventions as used in [6] [26], and the remainder of this thesis. The different normalisations for the algebra do not affect the group elements constructed in [28] [29], and hence the results stay the same. The reader is referred to appendix A for the new conventions.

## 3.1 holonomies and vacua

We recall some relevant fact from the previous chapter. For any flat periodic connection on a torus, one can define a set of holonomies, Wilson loops along nontrivial cycles of the torus,

$$\Omega_k = P\exp\left\{i \int_0^{L_k} A_k(x)dx^k\right\}, \quad (3.1.1)$$

where $k = 1, 2, 3$ labels the holonomy corresponding to the cycle wrapping around the $x, y, z$ direction respectively, $L_k$ is the length of the cycle.

The fact that the connection is flat implies that the holonomies commute. If one is only working with fields that take values in a representation of the gauge group that does not faithfully represent the centre (implying that the representation is not simply connected), it is possible to impose commutativity up to an element of the centre [23][24]. Working in a representation of the group that is simply connected, one can easily distinguish between commut-ativity, and commutativity up to a centre element (twisted boundary conditions). We mention these points because twisted boundary conditions will form an essential ingredient in what follows, although in a somewhat unexpected way. As shown in the previous chapter, the fact that holonomies commute in a simply connected representation, is sufficient for the existence of a corresponding flat connection. Thus, we can use the holonomies (3.1.1) to characterise the flat connections.

The holonomies transform covariantly ($\Omega'_k = g\Omega_k g^{-1}$) under periodic gauge transformations, and hence their traces are gauge invariant. For the trivial class of solutions $A_\mu = c^a_\mu H_a$, the holonomies $\Omega_k = \exp(ic^a_k H_a L_k)$ are on the so-called maximal torus (obtained by exponentiating the CSA). To find a different solution, one has to find holonomies that commute, but do not lie on a maximal torus. The corresponding flat connections are no longer of a simple form in these cases.

For $T^2$, using the complex structure on the 2-torus, it is possible to prove (as sketched in a footnote in [55]) that the moduli space of flat connections is connected, and hence the trivial class is the only class of solutions. For $T^3$ this is not the case, but using the result for $T^2$ one can arrange that any two out of the three holonomies are exponentials of elements of the CSA. The third holonomy is therefore the crucial one: either it can be written as the exponent of an element of the CSA, and we have a trivial solution, or it cannot, and one has a non-trivial solution.

The local parameters of the moduli space can be found by perturbing the holonomies $\Omega_k$ around a solution, demanding they still commute (so that the perturbations also lead to admissible vacua), see eqs. (2.4.6) and (2.4.7). For those generators $T^a$ that commute with each of the $\Omega_k$, eq. (2.4.7) does not lead to any restrictions on the corresponding $\alpha^a_k$, and these generators correspond to admissible perturbations. They will generate a group which is unbroken by the holonomies, and this is the group that is relevant in the calculation of $\text{Tr}(-1)^F$. We now show that the generators that do not commute with all holonomies correspond to global gauge transformations. We will see in the following that one can choose a basis $T^a$ for the Lie algebra such that (see e.g. (3.3.11) (3.4.9) (3.5.8))

$$\text{Ad}(\Omega_k)T^a = \Omega_k T^a \Omega_k^{-1} = \exp(\frac{2\pi i n^a_k}{N})T^a \quad n^a_k, N \in \mathbb{Z} \quad \forall k,a. \tag{3.1.2}$$

In that case the condition (2.4.7) reads

$$\alpha^a_k(\exp(-\frac{2\pi i n^a_l}{N}) - 1)\Omega_k\Omega_l T^a = \alpha^a_l(\exp(-\frac{2\pi i n^a_k}{N}) - 1)\Omega_k\Omega_l T^a, \tag{3.1.3}$$

Restricting ourselves to generators that do not commute with all holonomies, we define

$$\beta^a = \frac{\alpha^a_k}{1 - \exp(-\frac{2\pi i n^a_k}{N})} \tag{3.1.4}$$

for $k$ such that $[\Omega_k, T^a] \neq 0$. Equation (3.1.3) implies that $\beta^a$ is independent of $k$, whereas (2.4.7) implies that $\alpha^a_k = 0$ if $[\Omega_k, T^a] = 0$, so we can write

$$\alpha^a_k \Omega_k T^a = \beta^a [\Omega_k, T^a]. \tag{3.1.5}$$

We are thus left with a global gauge transformation, generalising the result below eq. (2.4.8) for groups other than $SO(7)$.

## 3.2 method of construction

Our construction is based on a variant of the so-called multi-twisted boundary conditions, as considered first in [10], involving subgroups of the gauge group.

We restrict ourselves to Yang-Mills theories with a compact, simple and simply connected gauge-group $G$. We will be interested in subgroups $PG'_N$ whose universal covering $\widetilde{PG}'_N$ is the product of $N$ factors $\tilde{G}'$. The representation of $PG'_N$ will in general not be irreducible, nor are its irreducible components in the same congruence class.

The *global* structure of our subgroup $PG'_N$ will *not* be that of a direct product group. For the realisation of twisted boundary conditions, it is necessary that a non-trivial discrete central subgroup has been divided out; for multi-twisted boundary conditions this discrete central subgroup is diagonal. As a relevant example of such a subgroup, consider $SO(4)$ which is locally $SU(2) \times SU(2)$, but has global structure $(SU(2) \times SU(2))/\mathbb{Z}_2$, where $\mathbb{Z}_2$ is the diagonal subgroup in the $\mathbb{Z}_2 \times \mathbb{Z}_2$ centre of $SU(2) \times SU(2)$.

For the discussion here we will restrict ourselves to a subgroup $PG'_2$ (with universal covering $\tilde{G}'^2$), since the generalisation to $PG'_N$ will be obvious. We assume that $\tilde{G}'$ allows a non-trivial centre $Z$, and that the global subgroup $PG'_2$ is a representation of $(\tilde{G}' \times \tilde{G}')/Z_{diag}$ where $Z_{diag} \cong Z$ is the diagonal subgroup of the centre $Z \times Z$ of $\tilde{G}'^2$. Now select two elements $P_1$, $Q_1$ generated by the Lie algebra of the first $\tilde{G}'$ factor ($\tilde{G}'_1$), and two elements $P_2$, $Q_2$ generated by the Lie algebra of the second $\tilde{G}'$ factor ($\tilde{G}'_2$), such that they commute up to a non-trivial element of the centre $Z$ of $\tilde{G}'_i$:

$$P_1 Q_1 = z_1 Q_1 P_1 \quad (3.2.1)$$
$$P_2 Q_2 = z_2 Q_2 P_2 \quad (3.2.2)$$

For irreducible representations of $\tilde{G}'_i$, $z_i$ is a root of unity, for a reducible representation $z_i$ is a diagonal matrix, with on the diagonal the centre elements appropriate for the different irreducible components. Since we only deal with representations of $(\tilde{G}' \times \tilde{G}')/Z_{diag}$, $z_1$ and $z_2$ are actually elements of the central subgroup $(Z \times Z)/Z_{diag} \cong Z$. By picking $P_i$ and $Q_i$ in a specific way, we can thus arrange that (3.2.1, 3.2.2) are satisfied with the additional condition

$$z_1 = (z_2)^{-1} \equiv z \quad (3.2.3)$$

We now have

$$P = P_1 P_2 \quad Q = Q_1 Q_2 \quad \Rightarrow PQ = QP \quad (3.2.4)$$

It is possible[1] and convenient to embed the $PG'_2$ subgroup in such a way that its maximal torus $T'$ is a subgroup of a maximal torus $T$ of $G$. If the $PG'_2$-subgroup has the same rank as $G$, then the tori $T$ and $T'$ coincide. If this is not the case, then there are multiple ways to extend $T'$ to a maximal torus $T$ of $G$.

We can choose the elements $P_1$ and $P_2$ to lie on the torus $T'$. Since the maximal torus is Abelian, it immediately follows that neither $Q_1$ nor $Q_2$ is on $T'$, and neither is their product $Q$ (since $Q$ does not commute with either of the $P_i$). We will now construct a third element $P'$ by "twisting" one of the $G'$'s with respect to the other $G'$

$$P' = Q_1^n P Q_1^{-n} = Q_1^n P_1 Q_1^{-n} P_2 = P_1 Q_2^{-n} P_2 Q_2^n = z^{-n} P \qquad (3.2.5)$$

---

[1]It was proven by Dynkin [17] that if a Lie algebra $\mathcal{L}_G$ has a subalgebra $\mathcal{L}_{G'}$, then the CSA of $\mathcal{L}'_G$ can be chosen to be contained in the CSA of $\mathcal{L}_G$ (by applying a suitable automorphism of $\mathcal{L}_G$). Upon exponentiation, one finds that the maximal torus of the group $G'$, generated by $\mathcal{L}_{G'}$, is contained in the maximal torus of the group $G$, generated by $\mathcal{L}_G$.

We can define a set of holonomies by setting $\Omega_1 = P$, $\Omega_2 = P'$ and $\Omega_3 = Q$. We will always assume $P'$ and $P$ to be different, which is essential for finding non-trivial vacua. This limits $n$ to a finite set, since there exists some $n$ for which $z^{-n} = 1$. We also should not allow $z$ to be an element of the centre of $G$, since this will also imply that the connection defined by the holonomies $P$, $P'$ and $Q$ is trivial (we can arrange that $P$ and $Q$ are on a maximal torus of $G$, and then, since the centre of $G$ is also on this maximal torus, $P'$ must be on this maximal torus). Since $P$ is on the maximal torus $T'$, and $z$ is an element of the centre of $PG_2'$, and the centre is generated by the CSA, $P'$ is on the maximal torus $T'$. By assumption the torus $T'$ is contained in a maximal torus $T$ of $G$. To define a non-trivial flat connection $Q$ should not be on any maximal torus $T$ of $G$. If $Q$ is on a maximal torus of $G$, then the maximal Abelian subgroup that commutes with $P$, $P'$ and $Q$ is a maximal torus.

Since the construction involves subgroups $\tilde{G}'$ with a non-trivial centre, one may always take a subgroup of $\tilde{G}'$ that is a product of unitary groups[2]. Henceforth, we shall always assume $\tilde{G}'$ to be a product of $SU(N)$'s. Thus we may take $\tilde{G}' = SU(N_1)^{n_1} \times SU(N_2)^{n_2} \times \cdots$ ($N_i$ different, $n_i$ positive integers), which has as centre $(\mathbb{Z}_{N_1})^{n_1} \times (\mathbb{Z}_{N_2})^{n_2} \times \cdots$.

Although not strictly necessary, extremely useful for our calculations is the concept of a diagonal subgroup. It is possible to construct a diagonal subgroup $D$ (with universal covering $\tilde{D} = \tilde{G}'$) in $PG_N'$ as follows: construct a Lie algebra $\mathcal{L}$ for $\tilde{G}'$, consisting of elements $T^a$. Then the Lie algebra for $(PG_N'$ has the structure $\mathcal{L}_1 \oplus \mathcal{L}_2 \oplus \cdots \oplus \mathcal{L}_N$ with each of the $\mathcal{L}_i \cong \mathcal{L}$. Hence we can write $T_i^a$, for the generator from $\mathcal{L}_i$ that corresponds to $T^a$ under an isomorphism mapping $\mathcal{L}$ to $\mathcal{L}_i$. The diagonal subgroup of $PG_N'$ is then constructed by taking as generators $T_1^a + T_2^a + \cdots + T_N^a$. This construction is not unique, there are many isomorphisms from $\mathcal{L}_i$ to $\mathcal{L}$, and these will give different diagonal subgroups.

We now take $P_1, P_2, Q_1, Q_2$ as before. The $P_i$ are elements of the maximal torus of $\tilde{G}_i'$, so we can write $P_i = \exp(ih_i)$ with $h_i$ an element of the CSA of $\mathcal{L}_i$, and similarly $Q_i = \exp(ie_i)$ with $e_i$ *not* in the CSA. The elements $P = P_1 P_2 = \exp(i(h_1 + h_2))$ and $Q = Q_1 Q_2 = \exp(i(e_1 + e_2))$ are then elements of a diagonal group $D$, as constructed in the above. Conjugating with $Q_1^n$ will produce $Q_1^n P Q_1^{-n} = P'$ and $Q_1^n Q Q_1^{-n} = Q$, which are elements of a diagonal subgroup $D'$, isomorphic to $D$ (the isomorphism being given by $D \leftrightarrow Q_1^n D Q_1^{-n}$).

## 3.3 Constructions based on $\mathbb{Z}_2$-twist

In this section we specialise to $\tilde{G}' = SU(2)$. We will start by developing the relevant tools for this subgroup. After that we will give an overview of groups in which our construction can be realised. These include $SO(N)$ where the result from [55] is reproduced, and $G_2$, where we rederive the result from section 2.5.

---

[2]If some simple subgroup with non-trivial centre is not $SU(N)$, its centre is either $\mathbb{Z}_2$ ($Sp(n)$, $SO(2n+1)$ and $E_7$), $\mathbb{Z}_4$ ($SO(4n+2)$), $\mathbb{Z}_2 \times \mathbb{Z}_2$ ($SO(4n)$) or $\mathbb{Z}_3$ ($E_6$). In these cases the centre is contained in an $SU(2)$, $SU(4)$, $SU(2) \times SU(2)$ or $SU(3)$-subgroup, respectively

We will use the following convention for the $su(2)$ algebra:

$$[L_3, L_+] = L_+ \qquad [L_3, L_-] = -L_- \qquad [L_+, L_-] = 2L_3 \qquad (3.3.1)$$

We have $L_3^\dagger = L_3$ and $L_+^\dagger = L_-$. With these conventions the eigenvalues of $L_3$, $(L_+ + L_-)/2$ and $i(L_+ - L_-)/2$ are half-integers for representations that exponentiate to $SU(2)$, and integers for representations that exponentiate to $SO(3)$.

In $SU(2)$ we will be looking for elements $p$ and $q$ such that

$$pq = -qp \qquad (3.3.2)$$

The standard convention is to take:

$$p = \begin{pmatrix} i & 0 \\ 0 & -i \end{pmatrix} = \exp\frac{\pi i}{2}\begin{pmatrix} 1 & 0 \\ 0 & -1 \end{pmatrix} \qquad q = \begin{pmatrix} 0 & i \\ i & 0 \end{pmatrix} = \exp\frac{i\pi}{2}\begin{pmatrix} 0 & 1 \\ 1 & 0 \end{pmatrix} \qquad (3.3.3)$$

In terms of generators this is $p = \exp(i\pi L_3)$ and $q = \exp(i\pi(L_+ + L_-)/2)$ (When lifted to $SO(3)$, the elements are $p = \text{diag}(-1,-1,1)$ and $q = \text{diag}(1,-1,-1)$, which commute). The commutation relations of $p,q$ with the algebra elements are easily determined

$$\begin{array}{rclcrcl} pL_3 & = & L_3 p & \quad & qL_3 & = & -L_3 q \\ pL_+ & = & -L_+ p & \quad & qL_+ & = & L_- q \\ pL_- & = & -L_- p & \quad & qL_- & = & L_+ q \end{array} \qquad (3.3.4)$$

Note that $q$ induces the Weyl reflection on the root lattice. This has a nice analogon for twist in $SU(N), N > 2$, as we will see in the subsequent sections.

The elements $P_i$ will take the role of $p$ in the above, the elements $Q_i$ take the role of $q$. Now notice that the condition (3.2.3) implies that the diagonal group $D$ that contains $P = P_1 P_2$ and $Q = Q_1 Q_2$ is actually an $SO(3)$ (it has a trivial centre). This can be seen as follows: the diagonal subgroup-construction provides a homomorphism from any of the $G' \cong SU(2)$ factors to the diagonal subgroup $D$. Under this homomorphism the non-trivial centre element of $SU(2)$ is mapped to the identity in $D$ (because $P_i \to P$, $Q_i \to Q$, we have $P_i Q_i P_i^{-1} Q_i^{-1} = z \to PQP^{-1}Q^{-1} = 1$).

One of the issues we did not address so far is the fact that we require the holonomies $\Omega_i$ ($P$, $P'$ and $Q$) to commute in a simply connected representation (otherwise the theorem of section 2.2 might not be applicable). In fact we will show that they commute in *any* representation, which seems more general, but is equivalent to the previous statement by the theory of compact Lie groups. For the $SU(2)$-based construction described in this paper, a sufficient condition for the commuting of all holonomies is that $D$ is an $SO(3)$-group (since this will imply that $P$ and $Q$ commute, and from this it follows that $P'$ and $Q$ commute. $P$ and $P'$ commute by construction). To determine whether $D$ is an $SO(3)$ subgroup, we construct its algebra.

As remarked in footnote 1 on page 37, it is always possible to choose an embedding of a subgroup such that its CSA is contained in the CSA of the group that contains the subgroup, so let $L_3 = h_\zeta$ for some $h_\zeta$ in the CSA of $G$. The eigenvalues of $h_\zeta$ are determined by taking inner products with the weights of $G$. We want $L_3$ to be an $SO(3)$-generator, and therefore its eigenvalues should be integers. Therefore $\langle \lambda, \zeta \rangle$ should be integer for any weight $\lambda$. Since any weight is expressible as a linear combination of fundamental weights and simple roots with integer coefficients, the condition that $L_3$ should have integer eigenvalues can be translated to

$$\langle \alpha_i, \zeta \rangle \in \mathbb{Z} \tag{3.3.5}$$
$$\langle \Lambda_i, \zeta \rangle \in \mathbb{Z} \tag{3.3.6}$$

where $\alpha_i$ are the simple roots of $G$, and $\Lambda_i$ are its fundamental weights. However, if the first of these two conditions is satisfied, then the second condition is satisfied for some weights, namely those fundamental weights with $\Lambda_i = \sum_k q_k \alpha_k$ with $q_k$ integer (we call these integer weights). Hence the second condition only has to be checked for what we will call non-integer weights. A list of these is contained in our appendix B.

We will want to compute products like $pTp^{-1}$, $qTq^{-1}$, where $T$ are generators of the group $G$ we are working in, and $p$ and $q$ are as in the previous paragraph. We will be looking at $SO(3)$-subgroups, and the generators of $G$ split into $SO(3)$ representations. $pTp^{-1}$ is easily calculated. With $L_3 = h_\zeta$ we have

$$ph_\alpha p^{-1} = h_\alpha \qquad pe_\alpha p^{-1} = \exp(i\pi \langle \alpha, \zeta \rangle) e_\alpha \tag{3.3.7}$$

Note that this implies that the generators either commute or anticommute with $p$.

There is also a nice and easy way to compute $qTq^{-1}$. First decompose the representation of $G$ into irreducible representations of $SO(3)$. In each irrep, construct the normalised eigenvectors $\psi_\lambda$ of $L_3$: $L_3 \psi_\lambda = \lambda \psi_\lambda$. It then follows that

$$L_3 q \psi_\lambda = -q L_3 \psi_\lambda = -\lambda q \psi_\lambda \tag{3.3.8}$$

Hence we conclude $q \psi_\lambda = \phi_\lambda \psi_{-\lambda}$ (where $\phi_\lambda$ is a phase factor). In fact, since we are only dealing with $SO(3)$ representations of $q$, we know that the eigenvalues of $q$ should be $\pm 1$ and hence $q^2 = 1$, meaning that $\phi_\lambda \phi_{-\lambda} = 1$. Moreover from $qL_+ = L_- q$ we easily find that[3]

$$\phi_\lambda \psi_{-(\lambda+1)} = \phi_{\lambda+1} \psi_{-(\lambda+1)} \tag{3.3.9}$$

So, $\phi_\lambda = \phi_{\lambda+1} = \phi$, independent of $\lambda$, and $\phi^2 = 1$. It is now trivial to construct eigenvectors and eigenvalues for $q$:

$$\begin{aligned}\psi_\lambda + \psi_{-\lambda} &\to \text{eigenvalue } \phi \\ \psi_\lambda - \psi_{-\lambda} &\to \text{eigenvalue } -\phi\end{aligned} \tag{3.3.10}$$

Now remember once more that only $SO(3)$ representations occur, meaning that $\lambda \in \mathbb{Z}$. If the dimension of our representation is $2n+1$, then the eigenvalue $\phi$ occurs $n+1$ times, and $-\phi$ $n$ times. The determinant of the matrix is then $(-)^n \phi^{2n+1} = (-)^n \phi$. The determinant should however be 1 since the matrix has been obtained by exponentiating a traceless generator.

---

[3] We use the Condon-Shortley phase convention: $L_+ \psi_\lambda^j = \sqrt{(j-\lambda)(j+\lambda+1)} \psi_{\lambda+1}^j$ and $L_- \psi_\lambda^j = \sqrt{(j+\lambda)(j-\lambda+1)} \psi_{\lambda-1}^j$, where $j$ is the usual angular momentum number, related to the dimension $d$ of the representation by $j = (d-1)/2$

Hence we find $\phi = (-)^n$, and the action of $q$ on any representation of $SO(3)$ is fully determined, and since by assumption $G$ splits into $SO(3)$ irreps, the action of $q$ in $G$ is completely determined. Most of the above is also valid for $SU(2)$-irreps, but there are two differences: one finds $\phi^2 = -1$, and it is not possible to determine whether $\phi = \mathrm{i}$ or $-\mathrm{i}$. This ambiguity comes from the centre of $SU(2)$, which we are unable to detect since we are trying to determine $q$ from commutation relations alone. Notice that these considerations imply that for an appropriately chosen basis of generators $T$ of $G$ that (compare to (3.1.2))

$$pT^a p^{-1} = \pm T^a \quad qT^a q^{-1} = \pm T^a \qquad (3.3.11)$$

We will now discuss the cases in which our construction actually gives a non-trivial flat connection. Our conventions concerning roots and weights can be found in our appendices. For the decomposition of groups into subgroups, use was made of [33].

Although in each case our construction can be carried out in a subgroup $PG'_2 = SO(4)$, we will often take $PG'_N$ with $\widetilde{PG}'_N = SU(2)^N$ where $N > 2$. This allows us to choose $PG'_N$ to be a regular subgroup [17], that is, the root-lattice of $PG'_N$ is a sublattice of the root-lattice of $G$. This gives an enormous simplification of the calculations. Our methods are not limited to regular subgroups, and we have actually carried out our construction for several irregular embeddings, but we always found the same results as for the regular embeddings. Therefore we will describe only constructions with regular embeddings.

### 3.3.1  $G_2$

Our first example will be the non-trivial flat connection in $G_2$, already described in section 2.5. We will treat this example in full detail to clarify our methods. $G_2$, being the group of lowest rank that possesses non-trivial flat connections on $T^3$, is the simplest from the point of view of our construction (unlike the constructions in the previous chapter that are simpler for orthogonal groups).

$G_2$ possesses an $SO(4) \cong (SU(2) \times SU(2))/\mathbb{Z}_2$ subgroup. The first $SU(2)$-factor can be taken to be generated by $h_{\alpha_1}, e_{\alpha_1}$ and $e_{-\alpha_1}$, the second one is then generated by $h_{\alpha_1+2\alpha_2}, e_{\alpha_1+2\alpha_2}$ and $e_{-(\alpha_1+2\alpha_2)}$. We label the two $SU(2)$ factors $SU(2)_{1,2}$, and normalise their generators such that they satisfy the algebra (3.3.1):

$$\begin{array}{ccc} SU(2)_1 & \times & SU(2)_2 \\ l^1_3 = h_{\alpha_1}/2 & & l^2_3 = 3h_{\alpha_1+2\alpha_2}/2 \\ l^1_+ = e_{\alpha_1}/\sqrt{2} & & l^2_+ = \sqrt{3/2}\, e_{\alpha_1+2\alpha_2} \end{array} \qquad (3.3.12)$$

The diagonal subgroup is now easily constructed, being generated by

$$\begin{aligned} L^D_3 &= l^1_3 + l^2_3 = h_{2\alpha_1+3\alpha_2} \\ L^D_+ &= l^1_+ + l^2_+ = (e_{\alpha_1} + \sqrt{3}\, e_{\alpha_1+2\alpha_2})/\sqrt{2} \end{aligned} \qquad (3.3.13)$$

The diagonal $SU(2)$ turns out to be an $SO(3)$. To check this we construct the weights for the $su(2)$-representation. These are $\langle 2\alpha_1 + 3\alpha_2, \lambda \rangle$ where $\lambda$ are the weights of $G_2$. However, since any weight of $G_2$ is of the form $p\alpha_1 + q\alpha_2$, with $p,q$ integer, it is sufficient to compute $\langle 2\alpha_1 + 3\alpha_2, \alpha_1 \rangle = 1$ and $\langle 2\alpha_1 + 3\alpha_2, \alpha_2 \rangle = 0$. We find that the weights of the $su(2)$ are always integer, no matter what representation of $G_2$ we use, and hence the $su(2)$ is actually an $so(3)$. We find the following decompositions for the fundamental and adjoint irreps:

$$\begin{array}{rcccl} G_2 & \to & SU(2) \times SU(2) & \to & SU(2)/\mathbb{Z}_2 \cong SO(3) \\ \mathbf{7} & \to & (\mathbf{1,3}) \oplus (\mathbf{2,2}) & \to & \mathbf{3} \oplus \mathbf{3} \oplus \mathbf{1} \\ \mathbf{14} & \to & (\mathbf{3,1}) \oplus (\mathbf{2,4}) \oplus (\mathbf{1,3}) & \to & \mathbf{3} \oplus \mathbf{5} \oplus \mathbf{3} \oplus \mathbf{3} \end{array} \quad (3.3.14)$$

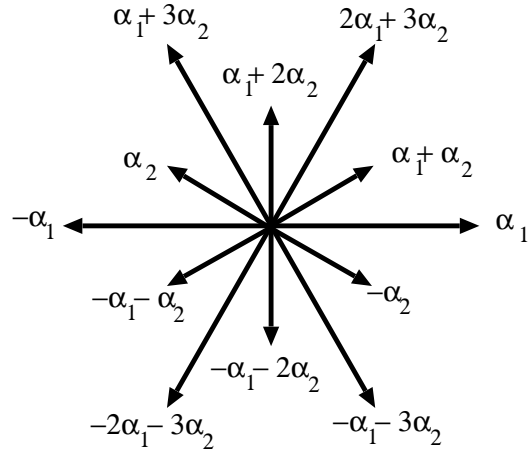

Figure 3-1. the root diagram for $G_2$

It is easy to construct the elements $P$ and $Q$: Take

$$P = \exp(i\pi L_3^D) \quad Q = \exp(\frac{\pi i}{2}(L_+^D + L_-^D)) \quad (3.3.15)$$

Now we wish to obtain $P'$. Conjugation with $Q_1 = \exp(\frac{\pi i}{2}(l_+^1 + l_-^1))$ generates the Weyl reflection in the first of the two $SU(2)$-factors:

$$l_3^1 \leftrightarrow -l_3^1 \qquad l_+^1 \leftrightarrow l_-^1 \quad (3.3.16)$$

This will take the diagonal subgroup $D$ to a diagonal subgroup $D'$ generated by:

$$\begin{aligned} L_3^{D'} &= -l_3^1 + l_3^2 &= h_{\alpha_1 + 3\alpha_2} \\ L_+^{D'} &= l_-^1 + l_+^2 &= (e_{-\alpha_1} + \sqrt{3} e_{\alpha_1 + 2\alpha_2})\sqrt{2} \end{aligned} \quad (3.3.17)$$

Since $L_+^{D'} + L_-^{D'} = L_+^D + L_-^D$, $Q$ is also an element of $D'$. Constructing $P'$ proceeds as in the above:

$$P' = \exp(i\pi L_3^{D'}) \quad (3.3.18)$$

$P$, $P'$ and $Q$ commute by construction. It is also clear that the flat connection implied by $\Omega_1 = P$, $\Omega_2 = P'$ and $\Omega_3 = Q$ is non-trivial, since the maximal torus of $G_2$ is simply the direct product of the tori of $D$ and $D'$, and $Q$ is not on either one.

To calculate the unbroken subgroup, we calculate the commutators of generators with the $\Omega_i$'s. As explained before only generators that commute with all three $\Omega_i$'s will be relevant. Therefore the most efficient way to proceed is to first compute the commutators with $\Omega_1 = P$ and $\Omega_2 = P'$, since these are the easiest, and only compute the commutator with $\Omega_3 = Q$ for those generators that commute with both $P$ and $P'$.

We can use the results of the beginning of this section. If there we substitute $L_3^D$ for $L_3$, and $L_\pm^D$ for $L_\pm$, then it is clear we should identify $P$ with $p$, and $Q$ with $q$. If we substitute $L_3^{D'}$ for $L_3$, and $L_\pm^{D'}$ for $L_\pm$, then we should identify $P'$ with $p$, and $Q$ with $q$. For the commutators of the algebra with $P$, we use (3.3.7), with $\zeta = 2\alpha_1 + 3\alpha_2$. We find that the CSA commutes with $P$, and $e_\beta$ commutes with $P$ only for $\beta = \pm\alpha_2$ and $\beta = \pm(2\alpha_1 + 3\alpha_2)$. For the commutators of the algebra with $P'$, we use (3.3.7) with $\zeta = \alpha_1 + 3\alpha_2$, from which it follows that $e_\beta$ with $\beta = \pm\alpha_2$ and $\beta = \pm(2\alpha_1 + 3\alpha_2)$ anticommute with $P'$. Hence the only generators that commute with both $P$ and $P'$ are the CSA-generators $h_\beta$.

To determine the effect of conjugation with $Q$ on these, we study the branching of $G_2 \to SO(3)$, where the $SO(3)$ is either $D$ or $D'$. We will take $D$, so the ladder operators are $L_\pm^D$ and the CSA-generator is $L_3^D$. We work with the generators of $G_2$ themselves, and hence we are in the adjoint representation of $G_2$. Eigenvectors of $L_3^D$ are easily found, since $\text{ad}(L_3^D)e_\beta = [h_{2\alpha_1+3\alpha_2}, e_\beta] = \langle 2\alpha_1 + 3\alpha_2, \beta\rangle e_\beta$ in our conventions, and , $\text{ad}(L_3^D)h_\beta = 0$, $h_\beta$ and $e_\beta$ are eigenvectors of $\text{ad}(L_3^D)$. We can now split the representation in irreducible components, by the standard procedure of looking for highest eigenvalues, and then applying the ladder operators $L_\pm^D$ to complete a representation, constructing the orthogonal complement etc.. We find that both CSA-generators have eigenvalue 0 in a **3** irrep of $SO(3)$. We should thus identify each CSA-generator to $\psi_0$ in a 3-dimensional representation of $SO(3)$, and using the results of the beginning of this section we find that, since $q\psi_0 = -\psi_0$, $Qh_\beta Q^{-1} = \text{Ad}(Q)h_\beta = -h_\beta$. The CSA generators thus anticommute with $Q$. Thus there is no generator that commutes with all $\Omega_i$, and, as already established in section 2.5, the vacuum implied by these holonomies is isolated, and there is only a discrete unbroken subgroup.

Finally we will give a matrix representation of the holonomies. The **7** irrep of $G_2$ has as its weights $0, \alpha_1 + \alpha_2, \alpha_1 + 2\alpha_2, \alpha_2, -(\alpha_1 + \alpha_2), -(\alpha_1 + 2\alpha_2), -\alpha_2$. Using this ordering of weights, we find:

$$\begin{aligned}
\Omega_1 = P &= \text{diag}(1, -1, -1, 1, -1, -1, 1) \\
\Omega_2 = P' &= \text{diag}(1, 1, -1, -1, 1, -1, -1) \\
\Omega_3 = Q &= \begin{pmatrix} -1 & 0 & 0 & 0 & 0 & 0 & 0 \\ 0 & 0 & 0 & 0 & -1 & 0 & 0 \\ 0 & 0 & 0 & 0 & 0 & -1 & 0 \\ 0 & 0 & 0 & 0 & 0 & 0 & -1 \\ 0 & -1 & 0 & 0 & 0 & 0 & 0 \\ 0 & 0 & -1 & 0 & 0 & 0 & 0 \\ 0 & 0 & 0 & -1 & 0 & 0 & 0 \end{pmatrix}
\end{aligned} \quad (3.3.19)$$

These can be conjugated to the form used in the previous chapter.

### 3.3.2   $SO(n)$

In $SO(7)$, there exists a subgroup $PG'_3$ with $\widetilde{PG'}_3 = SU(2)^3$. The $SU(2)$-factors can be taken to be generated by subalgebras with elements $h_{\beta_i}, e_{\pm\beta_i}$, with $\beta_1 = (1,-1,0)$ for the first factor, $\beta_2 = (1,1,0)$ for the second factor, and $\beta_3 = (0,0,1)$ for the third. The diagonal subgroup is then (appropriate normalisations for each generator included)

$$\begin{aligned} L_3^D &= l_3^1 + l_3^2 + l_3^3 &&= h_{(1,0,1)} \\ L_+^D &= l_+^1 + l_+^2 + l_+^3 &&= e_{(1,-1,0)}/\sqrt{2} + e_{(1,1,0)}/\sqrt{2} + e_{(0,0,1)} \end{aligned} \quad (3.3.20)$$

Again this is always an $so(3)$-algebra, which can be easily verified by calculating the inner products of $(1,0,-1)$ with the simple roots and non-integer weights. Vector, spin, and adjoint representation branch as follows:

$$\begin{array}{rcccl} SO(7) & \to & (SU(2))^3 & \to & SO(3) \\ 7 & \to & (\mathbf{1,1,3}) \oplus (\mathbf{2,2,1}) & \to & 2(\mathbf{3}) \oplus \mathbf{1} \\ 8 & \to & (\mathbf{1,2,2}) \oplus (\mathbf{2,1,2}) & \to & 2(\mathbf{3}) \oplus 2(\mathbf{1}) \\ 21 & \to & (\mathbf{3,1,1}) \oplus (\mathbf{1,3,1}) & & \\ & & \oplus (\mathbf{1,1,3}) \oplus (\mathbf{2,2,3}) & \to & \mathbf{5} \oplus \mathbf{5}(\mathbf{3}) \oplus \mathbf{1} \end{array} \quad (3.3.21)$$

To construct $D'$, twist the first factor with respect to the other two

$$l_3^1 \leftrightarrow -l_3^1 \qquad l_+^1 \leftrightarrow l_-^1$$

which leads to

$$\begin{aligned} L_3^{D'} &= l_3^1 + l_3^2 + l_3^3 &&= h_{(0,1,1)} \\ L_+^{D'} &= l_+^1 + l_+^2 + l_+^3 &&= e_{-(1,-1,0)}/\sqrt{2} + e_{(1,1,0)}/\sqrt{2} + e_{(0,0,1)} \end{aligned} \quad (3.3.22)$$

The set of holonomies is

$$\Omega_1 = P = \exp(i\pi L_3^D) \quad \Omega_2 = P' = \exp(i\pi L_3^{D'}) \quad \Omega_3 = Q = \exp(\frac{\pi i}{2}(L_+^D + L_-^D)) \quad (3.3.23)$$

In the vector representation, these are equivalent to the holonomies of Witten [55]. There is no generator of the algebra that commutes with all three $\Omega_i$.

    Two remarks are in place here. First, one might think that it is arbitrary which of the $SU(2)$-factors one chooses to twist (in the sense of eq. (3.2.5)). Indeed, twisting the first factor leaving the other two the same, as in the above, will give a result equivalent to twisting the second factor while leaving the other two. However, twisting the third factor with respect to the other two will not work: If one tries

$$l_3^3 \leftrightarrow -l_3^3 \qquad l_+^3 \leftrightarrow l_-^3,$$

one finds

$$P' = \exp(i\pi h_{(1,0,-1)}) = \exp(i\pi h_{(0,0,-2)}) \, P$$

But, noticing that $h_{(0,0,1)}$ is a (correctly normalised) generator of an integral $su(2)$-subalgebra of $so(7)$, we easily find $\exp(i\pi h_{(0,0,-2)}) = \pm 1$ (+1 if the $so(7)$-algebra is in the same congruence class as the vector or adjoint representation, $-1$ if the $so(7)$-irrep is isomorphic to the spin representation). Hence $P$ and $P'$ differ only by an element of the centre of $Spin(7)$. But the connection implied by setting $\Omega_1 = P$, $\Omega_2 = P'$ and $\Omega_3 = Q$ must be trivial, as any element that commutes with $\Omega_1$ commutes with $\Omega_2$, and $\Omega_1$ and $\Omega_3$ do not reduce the rank (if they did they could be used to construct a reduced rank periodic solution on the 2–torus, which does not exist).

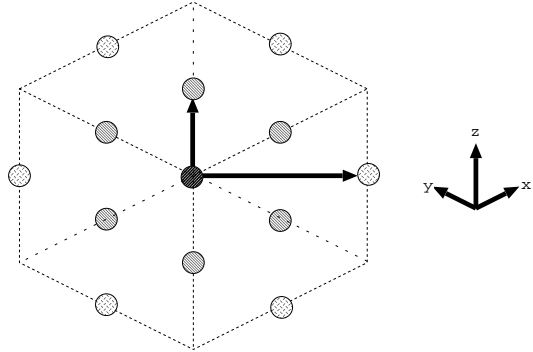

Figure 3-2. projection of the roots of $so(7)$ onto those for $g_2$

As a second remark, we consider the embedding of $G_2$ in $SO(7)$. The $so(7)$-root diagram fits into a cube. To see its $g_2$-subalgebra, project onto the plane orthogonal to a diagonal. Take as diagonal the direction $(-1,-1,1)$. The roots $\beta_2 = (1,1,0)$ and $\beta_3 = (0,0,1)$ will coincide under this projection (they will both project to $(1,1,2)/3$), and we find that the $SO(7)$ result can be understood from the $G_2$-result. This is less relevant for the orthogonal groups, but explains the encountered unbroken subgroups for the construction in the exceptional groups.

In $SO(8)$, there is a subgroup $PG'_4$, with $\widetilde{PG'}_4 = SU(2)^4$. All $SU(2)$-factors can be taken to be generated by algebras with elements $h_{\beta_i}, e_{\pm\beta_i}$, with $\beta_1 = (1,-1,0,0)$, $\beta_2 = (1,1,0,0)$, $\beta_3 = (0,0,1,-1)$ and $\beta_4 = (0,0,1,1)$. The diagonal subgroup is then

$$L_3^D = h_{(1,0,1,0)} \qquad (3.3.24)$$
$$L_+^D = (e_{(1,-1,0,0)} + e_{(1,1,0,0)} + e_{(0,0,1,-1)} + e_{(0,0,1,1)})/\sqrt{2}$$

It is easy to verify that this is an $so(3)$-algebra. Vector, and adjoint representation branch as follows:

$$\begin{array}{rcccl}
SO(8) & \to & (SU(2))^4 & \to & SO(3) \\
\mathbf{8_v} & \to & (\mathbf{2,2,1,1}) \oplus (\mathbf{1,1,2,2}) & \to & 2(\mathbf{3}) \oplus 2(\mathbf{1}) \quad (3.3.25) \\
\mathbf{28} & \to & (\mathbf{3,1,1,1}) \oplus (\mathbf{1,3,1,1}) \oplus (\mathbf{1,1,3,1}) & & \\
& & \oplus (\mathbf{1,1,1,3}) \oplus (\mathbf{2,2,2,2}) & \to & (\mathbf{5}) \oplus 7(\mathbf{3}) \oplus 2(\mathbf{1})
\end{array}$$

Because of triality, the spin-representations $\mathbf{8_c}$ and $\mathbf{8_s}$ have a branching that differs only from the one for $\mathbf{8_v}$ in $SU(2)^4$ by permutations of the $\mathbf{2}$'s and $\mathbf{1}$'s, their $SO(3)$-content is the same.

Twist the first factor to construct $D'$:

$$\begin{aligned} L_3^{D'} &= h_{(0,1,1,0)} \\ L_+^{D'} &= (e_{-(1,-1,0,0)} + e_{(1,1,0,0)} + e_{(0,0,1,-1)} + e_{(0,0,1,1)})/\sqrt{2} \end{aligned} \qquad (3.3.26)$$

The holonomies are then constructed in the usual way (3.3.23). The unbroken subgroup is again discrete. The projection of $SO(8)$ onto $SO(7)$ is trivial, and we find that we can understand the $SO(8)$-result from the $G_2$-result.

It is clear that the $SO(7)$-example and the $SO(8)$-example can both be embedded in $SO(9)$. Notice however, that both in the embedding of the $SO(7)$-example, and the embedding of the $SO(8)$-example we find

$$L_3^D = h_{(1,0,1,0)} \quad L_3^{D'} = h_{(0,1,1,0)}$$

The vectors $(1,0,1,0)$ and $(0,1,1,0)$ are called "defining vectors" for $D$ and $D'$, and there is a theorem by Dynkin [17], that two representations of $SO(3)$ are equivalent, if their defining vectors are equivalent. Hence the $SO(7)$ and $SO(8)$ embeddings are equivalent, and will not lead to different results. One finds a $U(1) \cong SO(2)$ unbroken subgroup, but, as explained in [55, 27], the actual unbroken subgroup is $O(2)$, because of discrete symmetries that are invisible in our approach.

In a similar way, the non-trivial flat connections for $SO(N)$ with $N > 9$ can be constructed. For $N$ sufficiently large, there are multiple ways of embedding an $SU(2)^M$ with a diagonal $SO(3)$. Although we know no simple way of proving this in our approach (their defining vectors need not be equivalent, for example), the analysis of [55] shows that these can never lead to new results, other than an $SO(7)$ or $SO(8)$ embedding will do. Note the chain of subgroups

$$SO(N) \to SO(N-1) \to \cdots \to SO(7) \to G_2 \qquad (3.3.27)$$

that shows that non-trivial flat connections in $SO(N)$ can be derived from its subgroup $G_2$. The connected component of the maximal unbroken subgroup is $SO(N-7)$, as can be understood from the fact that $SO(N)$ branches into $G_2 \times SO(N-7)$.

### 3.3.3 $F_4$

The easiest way to proceed in $F_4$ is by using the fact that the $so(7)$ root lattice is a sublattice of the $f_4$ root lattice. We can use $su(2)$-algebra's with elements $h_{\beta_i}, e_{\pm \beta_i}$, with $\beta_1 = (1,-1,0,0)$, $\beta_2 = (1,1,0,0)$, and $\beta_3 = (0,0,1,0)$. The two diagonal subgroups and holonomies are constructed in the standard way.

$$\begin{aligned} L_3^D &= l_3^1 + l_3^2 + l_3^3 &= h_{(1,0,1,0)} \\ L_+^D &= l_+^1 + l_+^2 + l_+^3 &= e_{(1,-1,0,0)}/\sqrt{2} + e_{(1,1,0,0)}/\sqrt{2} + e_{(0,0,1,0)} \end{aligned} \qquad (3.3.28)$$

$$\begin{aligned} L_3^{D'} &= l_3^1 + l_3^2 + l_3^3 &= h_{(0,1,1,0)} \\ L_+^{D'} &= l_+^1 + l_+^2 + l_+^3 &= e_{-(1,-1,0,0)}/\sqrt{2} + e_{(1,1,0,0)}/\sqrt{2} + e_{(0,0,1,0)} \end{aligned} \qquad (3.3.29)$$

Again these are found to be $so(3)$-algebra's. The holonomies are as in (3.3.23). Calculating the (connected component of the) unbroken subgroup, one finds that it is $SO(3)$. This can be understood from the branching of $F_4$ into $G_2 \times SO(3)$.

$$\begin{aligned} F_4 &\to G_2 \times SO(3) \\ 26 &\to (7,3) \oplus (1,5) \\ 52 &\to (14,1) \oplus (7,5) \oplus (1,3) \end{aligned} \qquad (3.3.30)$$

Note that here it is important that our construction fits into $G_2$. Starting from $SO(7)$ might lead to the wrong expectation that the unbroken subgroup would be $SO(2)$, since $F_4$ branches into $SO(7) \times SO(2)$.

### 3.3.4 $E_6$, $E_7$, $E_8$

In $E_6$ we use that the $so(8)$ root lattice is a sublattice of the $e_6$ root lattice. We use $su(2)$-algebra's with elements $h_{\beta_i}, e_{\pm\beta_i}$, with $\beta_1 = (0,1,-1,0,0,0)$, $\beta_2 = (0,1,1,0,0,0)$, and $\beta_3 = (0,0,0,1,-1,0)$, and $\beta_4 = (0,0,0,1,1,0)$. The two diagonal subgroups are:

$$\begin{aligned} L_3^D &= l_3^1 + l_3^2 + l_3^3 = h_{(0,1,0,1,0,0)} \\ L_+^D &= l_+^1 + l_+^2 + l_+^3 = (e_{(0,1,-1,0,0,0)} + e_{(0,1,1,0,0,0)} + \\ &\qquad\qquad e_{(0,0,0,1,-1,0)} + e_{(0,0,0,1,1,0)})/\sqrt{2} \end{aligned} \qquad (3.3.31)$$

$$\begin{aligned} L_3^{D'} &= l_3^1 + l_3^2 + l_3^3 = h_{(0,0,1,1,0,0)} \\ L_+^D &= l_+^1 + l_+^2 + l_+^3 = (e_{-(0,1,-1,0,0,0)} + e_{(0,1,1,0,0,0)} + \\ &\qquad\qquad e_{(0,0,0,1,-1,0)} + e_{(0,0,0,1,1,0)})/\sqrt{2} \end{aligned} \qquad (3.3.32)$$

These are $so(3)$-algebra's. The holonomies are as in (3.3.23). The (connected component of the) unbroken subgroup is found to be $SU(3)$. This can be understood from the branching of $E_6$ into $G_2 \times SU(3)$.

$$\begin{aligned} E_6 &\to G_2 \times SU(3) \\ 27 &\to (7, \bar{3}) \oplus (1, 6) \\ 78 &\to (14, 1) \oplus (7, 8) \oplus (1, 8) \end{aligned} \qquad (3.3.33)$$

The $E_6$-example can be trivially embedded in $E_7$, using that the $e_6$ root lattice is a sublattice of the $e_7$ root lattice. We add an extra zero to the vectors, and obtain

$$\begin{aligned} L_3^D &= l_3^1 + l_3^2 + l_3^3 = h_{(0,1,0,1,0,0,0)} \\ L_+^D &= l_+^1 + l_+^2 + l_+^3 = (e_{(0,1,-1,0,0,0,0)} + e_{(0,1,1,0,0,0,0)} + \\ &\qquad\qquad e_{(0,0,0,1,-1,0,0)} + e_{(0,0,0,1,1,0,0)})/\sqrt{2} \end{aligned} \qquad (3.3.34)$$

$$\begin{aligned} L_3^{D'} &= l_3^1 + l_3^2 + l_3^3 = h_{(0,0,1,1,0,0)} \\ L_+^D &= l_+^1 + l_+^2 + l_+^3 = (e_{-(0,1,-1,0,0,0,0)} + e_{(0,1,1,0,0,0,0)} + \\ &\qquad\qquad e_{(0,0,0,1,-1,0,0)} + e_{(0,0,0,1,1,0,0)})/\sqrt{2} \end{aligned} \qquad (3.3.35)$$

These are $so(3)$-algebra's. The holonomies are as in (3.3.23). The (connected component of the) unbroken subgroup is a simple group of 21 generators. It takes a little more work to show that it is $Sp(3)$ (and not $SO(7)$). This can be understood from the branching of $E_7$ into[4] $G_2 \times Sp(3)$

$$\begin{aligned} E_7 &\to G_2 \times Sp(3) \\ \mathbf{56} &\to (\mathbf{7},\mathbf{6}) \oplus (\mathbf{1},\mathbf{14}) \\ \mathbf{133} &\to (\mathbf{14},\mathbf{1}) \oplus (\mathbf{7},\mathbf{14'}) \oplus (\mathbf{1},\mathbf{21}) \end{aligned} \quad (3.3.36)$$

The $e_8$ root lattice contains as a sublattice the $e_7$ root lattice. Again we add an extra zero to the vectors, to obtain

$$\begin{aligned} L_3^D &= l_3^1 + l_3^2 + l_3^3 = h_{(0,1,0,1,0,0,0,0)} \\ L_+^D &= l_+^1 + l_+^2 + l_+^3 = (e_{(0,1,-1,0,0,0,0,0)} + e_{(0,1,1,0,0,0,0,0)} + \\ &\qquad\qquad e_{(0,0,0,1,-1,0,0,0)} + e_{(0,0,0,1,1,0,0,0)})/\sqrt{2} \end{aligned} \quad (3.3.37)$$

$$\begin{aligned} L_3^{D'} &= l_3^1 + l_3^2 + l_3^3 = h_{(0,0,1,1,0,0,0,0)} \\ L_+^D &= l_+^1 + l_+^2 + l_+^3 = (e_{-(0,1,-1,0,0,0,0,0)} + e_{(0,1,1,0,0,0,0,0)} + \\ &\qquad\qquad e_{(0,0,0,1,-1,0,0,0)} + e_{(0,0,0,1,1,0,0,0)})/\sqrt{2} \end{aligned} \quad (3.3.38)$$

These are $so(3)$-algebra's. The holonomies are as in (3.3.23). The (connected component of the) unbroken subgroup is $F_4$. This can be understood from the branching of $E_8$ into $G_2 \times F_4$.

$$\begin{aligned} E_8 &\to G_2 \times F_4 \\ \mathbf{248} &\to (\mathbf{14},\mathbf{1}) \oplus (\mathbf{7},\mathbf{26}) \oplus (\mathbf{1},\mathbf{52}) \end{aligned} \quad (3.3.39)$$

## 3.4 Constructions based on $\mathbb{Z}_3$-twist

We will now develop the relevant tools for the case $\tilde{G}' = SU(3)$. Twist in an $SU(3)$-subgroup can be realised in all the exceptional groups except $G_2$. $F_4$ will be discussed in quite some detail.

Like in section 3.3 where $\tilde{G}' = SU(2)$ was discussed, our construction can always be carried out in a subgroup $SU(3)^2$, but we will often take $SU(3)^M$ with $M > 2$, as this allows us to choose it to be a regular subgroup. This will give important simplifications in the calculations.

We will take the canonical form (in the conventions of the appendices A and B, with the modification that we will use capital $E$ and $H$ to be able to distinguish the subgroup generated by these from the original group, whose generators we keep on denoting by $h_\alpha$ and $e_\alpha$):

$$\begin{aligned} &[H_{\alpha_1}, E_{\alpha_1}] = 2E_{\alpha_1} \quad &[H_{\alpha_1}, E_{\alpha_2}] = -E_{\alpha_1} \quad &[H_{\alpha_1}, E_{\alpha_1+\alpha_2}] = E_{\alpha_1+\alpha_2} \\ &[H_{\alpha_2}, E_{\alpha_1}] = -E_{\alpha_1} \quad &[H_{\alpha_2}, E_{\alpha_2}] = 2E_{\alpha_2} \quad &[H_{\alpha_2}, E_{\alpha_1+\alpha_2}] = E_{\alpha_1+\alpha_2} \\ &[E_{\alpha_1}, E_{-\alpha_1}] = 2H_{\alpha_1} \quad &[E_{\alpha_2}, E_{-\alpha_2}] = 2H_{\alpha_2} \quad &[E_{\alpha_1}, E_{\alpha_2}] = E_{\alpha_1+\alpha_2} \\ &[E_{\alpha_1+\alpha_2}, E_{-\alpha_1-\alpha_2}] = 2H_{\alpha_1+\alpha_2} = 2H_{\alpha_1} + 2H_{\alpha_2} \end{aligned} \quad (3.4.1)$$

---

[4]In the decompositions given, both 14-dimensional representations of $Sp(3)$ are present. By **14**, we denote the representation with Dynkin labels (001), while **14'** is the irrep with Dynkin labels (010) (the simple roots of $Sp(3)$ are ordered such that the longest one appears on the right) [33].

All other relations can be found by conjugation, and the Jacobi identity.

In $SU(3)$ we look for two matrices satisfying

$$pq = \exp(\frac{2\pi i}{3})qp \qquad (3.4.2)$$

$p$ and $q$ will commute when lifted to $SU(3)/\mathbb{Z}_3$. We take:

$$p = \begin{pmatrix} \exp(\frac{2\pi i}{3}) & 0 & 0 \\ 0 & 1 & 0 \\ 0 & 0 & \exp(-\frac{2\pi i}{3}) \end{pmatrix}, \qquad q = \begin{pmatrix} 0 & 1 & 0 \\ 0 & 0 & 1 \\ 1 & 0 & 0 \end{pmatrix} \qquad (3.4.3)$$

In terms of generators this is

$$p = \exp\left(\frac{2\pi i}{3} H_{\alpha_1+\alpha_2}\right) \qquad (3.4.4)$$

$$q = \exp\left(\frac{2\pi}{3\sqrt{3}}(E_{\alpha_1} + E_{\alpha_2} + E_{-\alpha_1-\alpha_2} - E_{-\alpha_1} - E_{-\alpha_2} - E_{\alpha_1+\alpha_2})\right)$$

The commutation relations of $p$ and $q$ with the group generators are most easily calculated in a specific representation. One finds

$$\begin{array}{ll} pH_\alpha = H_\alpha p, & qH_\alpha = H_{R\alpha}q, \\ pE_\alpha = \exp(\frac{2\pi i}{3}\langle\alpha,\alpha_1+\alpha_2\rangle)E_\alpha p, & qE_\alpha = E_{R\alpha}q \end{array} \qquad (3.4.5)$$

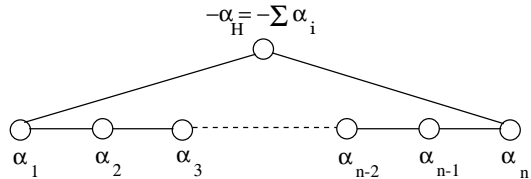

Figure 3-3. The extended Dynkin diagram for $SU(N+1)$

The action of the rotation $R$ (This is a genuine rotation over $2\pi/3$ in the root diagram) is fully determined by its action on the simple roots. $R$ is an element of the Weyl group: It is the composition of the Weyl reflection generated by $\alpha_2$, followed by the reflection generated by $\alpha_1$. The effect of $R$ is

$$R: \quad \alpha_1 \to \alpha_2 \to -(\alpha_1+\alpha_2) \to \alpha_1 \qquad (3.4.6)$$

A nice mnemonic for the action of $R$ is provided by the "extended Dynkin diagram" [17] of $SU(3)$. The extended Dynkin diagram of an algebra consists of its Dynkin diagram, extended with one more root, $-\alpha_H$, with $\alpha_H$ the highest root of the algebra. The node representing the highest root is then connected to the diagram via the standard rules. For $su(3)$, $-\alpha_H = -(\alpha_1+\alpha_2)$ and the extended diagram consists of 3 nodes, connected to form a cycle. The action of $R$ is nothing but a rotation of this cycle by one step. For twist in the higher unitary groups we will see the same feature: the action of the analogon of the element $q$ can be represented by rotating the (cyclic) extended Dynkin diagram of the unitary group by one step.

We now turn to the computation of the unbroken subgroup. We wish to compute the combinations $pT^a p^{-1}$ and $qT^a q^{-1}$, with $T^a$ the generators of $G$. We will generalise our methods for the $SU(2)$-case to $SU(3)$, but this is complicated by a crucial difference between $SU(2)$ and $SU(3)$: For $SU(2)$ all weights in an irreducible representation have multiplicity 1, whereas this is not true for $SU(3)$-irreps. However, we will need only three irreps of $SU(3)$: the **1** and **10**, which do contain only simple weights, and the **8**, which is equivalent to the adjoint. Note that these are all irreps of $SU(3)/\mathbb{Z}_3$.

We proceed as in the $SU(2)$-case. $pT^a p^{-1}$ is easily calculated, with $p = \exp(ih_\zeta)$ we have

$$ph_\alpha p^{-1} = h_\alpha \qquad pe_\beta p^{-1} = \exp(i\langle \zeta, \beta \rangle) e_\beta \qquad (3.4.7)$$

To calculate $qT^a q^{-1}$ we start by decomposing the representation of the group $G$ into irreducible representations of $SU(3)/\mathbb{Z}_3$. First we construct the normalised eigenvectors $\psi_\lambda$ of the CSA: $H_\alpha \psi_\lambda = \langle \alpha, \lambda \rangle \psi_\lambda$. It then follows that

$$H_{R\alpha} q \psi_\lambda = q H_\alpha \psi_\lambda = \langle \alpha, \lambda \rangle q \psi_\lambda = \langle R\alpha, R\lambda \rangle q \psi_\lambda \qquad (3.4.8)$$

This means that $q\psi_\lambda$ is an eigenvector of the CSA with weight $R\lambda$. If all weights are simple, as in the **1** and **10**, then everything is easily solvable, as in the $SU(2)$ case: $q\psi_\lambda = \phi_\lambda \psi_{R\lambda}$ with $\phi_\lambda$ a phase. From $q^3 = 1$, we find that $\phi_\lambda \phi_{R\lambda} \phi_{R^2\lambda} = 1$. With the ladder operators one shows that $\phi_\lambda$ is actually independent of $\lambda$ (in a suitable phase convention). The following three combinations are eigenvectors:

$$\begin{aligned}
\psi_\lambda + \psi_{R\lambda} + \psi_{R^2\lambda} &\to \text{eigenvalue } \phi \\
\psi_\lambda + \exp(\tfrac{4\pi i}{3})\psi_{R\lambda} + \exp(\tfrac{2\pi i}{3})\psi_{R^2\lambda} &\to \text{eigenvalue } \exp(\tfrac{2\pi i}{3})\phi \\
\psi_\lambda + \exp(\tfrac{2\pi i}{3})\psi_{R\lambda} + \exp(\tfrac{4\pi i}{3})\psi_{R^2\lambda} &\to \text{eigenvalue } \exp(\tfrac{4\pi i}{3})\phi
\end{aligned}$$

Note that formally we have to turn to the complexification of the real Lie-algebra for this to make sense. For $\lambda = 0$ the last two of these combinations are zero. We find that each triple of eigenvectors contributes $\phi^3 = 1$ to the determinant, and since all weights, apart from the zero weight, occur in triples we find that the determinant is equal to $\phi^{3n+1} = \phi = 1$.

When not all weights have multiplicity 1, the above discussion for the non-zero weights applies anyway. It is the zero weight that causes the problems. The eigenvectors with weight zero form a subspace that will be mapped into itself, our previous methods fail, and we know no easy way to determine the action of $q$ on any vector of this subspace. For the **8**, the solution is nevertheless easy to find by realising that this is the adjoint. The zero weights of the adjoint representation are the generators of the CSA, the $H_\alpha$, on which we already know the action of $q$, and the eigenvectors of $q$ are

$$\begin{aligned}
H_\alpha + H_{R\alpha} + H_{R^2\alpha} &= 0 \\
H_\alpha + \exp(\tfrac{4\pi i}{3})H_{R\alpha} + \exp(\tfrac{2\pi i}{3})H_{R^2\alpha} &\to \text{eigenvalue } \exp(\tfrac{2\pi i}{3}) \\
H_\alpha + \exp(\tfrac{2\pi i}{3})H_{R\alpha} + \exp(\tfrac{4\pi i}{3})H_{R^2\alpha} &\to \text{eigenvalue } \exp(\tfrac{4\pi i}{3})
\end{aligned}$$

We can now construct the action of $q$ on any representation of $G$ that splits into singlets, octets and decuplets. As a final remark we note that these considerations imply that

$$pT^a p^{-1} = \exp(\frac{2\pi i n}{3}) T^a, \qquad qT^a q^{-1} = \exp(\frac{2\pi i m}{3}) T^a \qquad (3.4.9)$$

For the $su(3)$-subalgebra of an algebra $\mathcal{L}_G$, we define the CSA generators $H_{\alpha'_1}$ and $H_{\alpha'_2}$ associated to the simple roots of $su(3)$. Set $H_{\alpha'_1} = h_{\beta_1}$, $H_{\alpha'_2} = h_{\beta_2}$ for $h_{\beta_i} \in \mathcal{L}_G$. We wish to check how $\mathcal{L}_G$ branches in representation of $su(3)$. A way to do this is to project the weights $\lambda$ of $\mathcal{L}_G$ to the subspace spanned by $\beta_1$ and $\beta_2$. The projected weights $\lambda^{proj}$ have an expansion of the form

$$\lambda^{proj} = c_1 \beta_1 + c_2 \beta_2 \qquad (3.4.10)$$

The coefficients $c_i$ may be found by taking the inner product with the fundamental coweights $\omega_i$ of $su(3)$. These are identical to the fundamental weights $\omega_i = \Lambda'_i$ of $su(3)$, as $su(3)$ is simply laced. The fundamental weights can be expressed in the roots as

$$\Lambda'_1 = \frac{2}{3}\beta_1 + \frac{1}{3}\beta_2 \qquad \Lambda'_2 = \frac{1}{3}\beta_1 + \frac{2}{3}\beta_2.$$

The $c_i$ may now be found by computing

$$c_i = \langle \lambda^{proj}, \Lambda'_i \rangle = \langle \lambda, \Lambda'_i \rangle \qquad (3.4.11)$$

The last equality hold because the $\Lambda'_i$ are orthogonal to $\lambda - \lambda^{proj}$.

The weights of a representation of the group $G$ as

$$\lambda = \sum_i (n_i \Lambda_i + m_i \alpha_i) \quad \text{with } n_i, m_i \in \mathbb{Z}$$

with $\Lambda_i$ the fundamental weights of $G$ and $\alpha_i$ its simple roots. We will demand that the representation of $G$ branches into representations of $SU(3)/\mathbb{Z}_3$. This implies that the $c_i$ are integers, a condition that can be translated to

$$\langle \alpha_i, \frac{2}{3}\beta_1 + \frac{1}{3}\beta_2 \rangle, \quad \langle \alpha_i, \frac{1}{3}\beta_1 + \frac{2}{3}\beta_2 \rangle, \quad \langle \Lambda_i, \frac{2}{3}\beta_1 + \frac{1}{3}\beta_2 \rangle, \quad \langle \Lambda_i, \frac{1}{3}\beta_1 + \frac{2}{3}\beta_2 \rangle \quad \in \mathbb{Z} \quad (3.4.12)$$

### 3.4.1 $F_4$

$F_4$ possesses a $PG'_2$-subgroup with $\widetilde{PG'}_2 = SU(3)^2$. The roots of the subalgebra are contained in the $f_4$ root lattice. An $su(3)$-sublattice is completely determined by making a choice for its simple roots. We take for the first $su(3)$ factor the two roots $(1,-1,0,0)$ and $(0,1,-1,0)$, which are of equal length and have an angle in-between of $2\pi/3$ and can therefore be used as simple roots for $su(3)$. For the second $su(3)$ factor we take as simple roots $(0,0,0,1)$ and $(1,1,1,-1)/2$. These two are orthogonal to the root vectors of the first $su(3)$. To construct the $su(3)$ algebra's, we need to include appropriate normalisation factors. The first $SU(3)$-factor is generated by:

$$\begin{aligned} H^1_{\alpha_1} &= h_{(1,-1,0,0)} & H^1_{\alpha_2} &= h_{(0,1,-1,0)} \\ E^1_{\alpha_1} &= e_{(1,-1,0,0)} & E^1_{\alpha_2} &= e_{(0,1,-1,0)} & E^1_{\alpha_1+\alpha_2} &= e_{(1,0,-1,0)} \end{aligned} \qquad (3.4.13)$$

The second $SU(3)$ factor is generated by

$$H^1_{\alpha_1} = 2h_{(0,0,0,1)} \quad H^1_{\alpha_2} = 2h_{(1,1,1,-1)/2} \quad (3.4.14)$$
$$E^1_{\alpha_1} = \sqrt{2}e_{(0,0,0,1)} \quad E^1_{\alpha_2} = \sqrt{2}e_{(1,1,1,-1)/2} \quad E^1_{\alpha_1+\alpha_2} = \sqrt{2}e_{(1,1,1,1)/2}$$

The diagonal subalgebra $D$ is then easily constructed:

$$\begin{aligned}
H^D_{\alpha_1} &= & H^1_{\alpha_1} + H^2_{\alpha_1} &= h_{(1,-1,0,2)} \\
H^D_{\alpha_2} &= & H^1_{\alpha_2} + H^2_{\alpha_2} &= h_{(1,2,0,-1)} \\
E^D_{\alpha_1} &= & E^1_{\alpha_1} + E^2_{\alpha_1} &= e_{(1,-1,0,0)} + \sqrt{2}e_{(0,0,0,1)} \\
E^D_{\alpha_2} &= & E^1_{\alpha_2} + E^2_{\alpha_2} &= e_{(0,1,-1,0)} + \sqrt{2}e_{(1,1,1,-1)/2} \\
E^D_{\alpha_1+\alpha_2} &= & E^1_{\alpha_1+\alpha_2} + E^2_{\alpha_1+\alpha_2} &= e_{(1,0,-1,0)} + \sqrt{2}e_{(1,1,1,1)/2}
\end{aligned} \quad (3.4.15)$$

By checking the conditions (3.4.12) we see that this algebra generates an $SU(3)/\mathbb{Z}_3$-group. We have $\Lambda'_1 = (1,0,0,1)$ and $\Lambda'_2 = (1,1,0,0)$, and the inner product of these vectors with the simple roots of $F_4$ all give integers. $F_4$ has no non-integer fundamental weights. The decomposition of $F_4$ into the diagonal $SU(3)/\mathbb{Z}_3$ is as follows:

$$\begin{array}{cccccc}
F_4 & \to & SU(3) \times SU(3) & \to & SU(3)/\mathbb{Z}_3 & \\
\mathbf{26} & \to & (\mathbf{3},\bar{\mathbf{3}}) \oplus (\bar{\mathbf{3}},\mathbf{3}) \oplus (\mathbf{8},\mathbf{1}) & \to & 3(\mathbf{8}) \oplus 2(\mathbf{1}) & (3.4.16) \\
\mathbf{52} & \to & (\mathbf{8},\mathbf{1}) \oplus (\mathbf{6},\mathbf{3}) \oplus (\bar{\mathbf{6}},\bar{\mathbf{3}}) \oplus (\mathbf{1},\mathbf{8}) & \to & (\mathbf{10}) \oplus (\overline{\mathbf{10}}) \oplus 4(\mathbf{8}) &
\end{array}$$

Now look for two matrices that commute, but cannot be written as exponentials of generators of $D$ that commute. From (3.4.4)

$$P = \exp\left(\frac{2\pi i}{3} H^D_{\alpha_1+\alpha_2}\right) \quad (3.4.17)$$
$$Q = \exp\left(\frac{2\pi}{3\sqrt{3}}(E^D_{\alpha_1} + E^D_{\alpha_2} + E^D_{-\alpha_1-\alpha_2} - E^D_{-\alpha_1} - E^D_{-\alpha_2} - E^D_{\alpha_1+\alpha_2})\right)$$

Splitting the generators of the diagonal group as $H^D_\alpha = H^1_\alpha + H^2_\alpha$ and $E^D_\alpha = E^1_\alpha + E^2_\alpha$ reveals the product structure of $P$ and $Q$, which decompose accordingly as $P = P_1 P_2$, and $Q = Q_1 Q_2$. That $P$ and $Q$ commute follows from the fact that they take the role of the elements $p$ and $q$ constructed previously, and that they are elements of an $SU(3)/\mathbb{Z}_3$-subgroup. Now we wish to construct $P'$. Our methods for $SU(2)$ have to be generalised in such a way, that it produces a subgroup $D'$ that is isomorphic to $D$, such that $Q$ is an element of both $D$ and $D'$. The appropriate generalisation has the form of a rotation of one of the two factors

$$H^1_\alpha \to H^1_{R\alpha} \quad E^1_\alpha \to E^1_{R\alpha} \quad (3.4.18)$$

where $R$ is the rotation (3.4.6). The second $SU(3)$ will be left as it is. We now construct the diagonal group $D'$:

$$H^{D'}_{\alpha_1} = H^1_{\alpha_2} + H^2_{\alpha_1} = h_{(0,1,-1,2)}$$

$$
\begin{aligned}
H^{D'}_{\alpha_2} &= H^1_{-\alpha_1-\alpha_2} + H^2_{\alpha_2} &= h_{(0,1,2,-1)} \\
E^{D'}_{\alpha_1} &= E^1_{\alpha_2} + E^2_{\alpha_1} &= e_{(0,1,-1,0)} + \sqrt{2}e_{(0,0,0,1)} \\
E^{D'}_{\alpha_2} &= E^1_{-\alpha_1-\alpha_2} + E^2_{\alpha_2} &= e_{(-1,0,1,0)} + \sqrt{2}e_{(1,1,1,-1)/2} \\
E^{D'}_{\alpha_1+\alpha_2} &= E^1_{-\alpha_1} + E^2_{\alpha_1+\alpha_2} &= e_{(-1,1,0,0)} + \sqrt{2}e_{(1,1,1,1)/2}
\end{aligned}
\quad (3.4.19)
$$

We set $P'$ to be

$$
P' = \exp(\frac{2\pi i}{3} H^{D'}_{\alpha_1+\alpha_2}) \tag{3.4.20}
$$

$P$, $P'$ and $Q$ commute by construction. Since $F_4$ is simply connected, the gauge connection implied by the holonomies $\Omega_1 = P$, $\Omega_2 = P'$ and $\Omega_3 = Q$ is flat and non-trivial.

Now we calculate the unbroken subgroup: first find the generators that commute with $P$, of these check how many commute with $P'$, and finally compute the commutator with $Q$ of the generators that commute with both $P$ and $P'$. These computations can be done by using our previous results and substituting $P$ for $p$ and $Q$ for $q$ in the first step, and in a second step $P'$ for $p$ and $Q$ for $q$. The only $e_{\pm\beta}$ that commute with $P$ have

$$
\beta \in \{(0,0,1,0),(1,1,0,0),(1,0,0,1),(0,1,0,-1), \tfrac{(1,-1,1,-1)}{2}, \tfrac{(1,-1,-1,-1)}{2}\}
$$

However, none of these commute with $P'$. Hence the CSA-generators are the only ones that commute with both $P$ and $P'$. The effect of $Q$ on the CSA can be computed by studying the branching of $F_4$ into $SU(3)/\mathbb{Z}_3$. Careful examination shows that all CSA-generators have weight zero for the **8**'s in the decomposition. There is no combination of zero weights for the **8** that is invariant under the adjoint action of $Q$ and hence no generator commutes with $P$, $P'$ and $Q$. The unbroken subgroup is therefore discrete.

We therefore have found a *new* flat connection: Neither the trivial flat connection nor the flat connection constructed in section 3.3 can be deformed to a flat connection that has a discrete unbroken subgroup. Note that instead of the rotation (3.4.18) we could also have rotated by

$$
H^1_\alpha \to H^1_{R^2\alpha} \qquad E^1_\alpha \to E^1_{R^2\alpha} \tag{3.4.21}
$$

Repeating the steps for the construction of $D'$, this gives another diagonal subgroup, that we will call $D''$. It is also possible to construct an element $P'' = \exp(\frac{2\pi i}{3} H^{D''}_{\alpha_1+\alpha_2})$. $P''$ commutes with $P$, $P'$, and $Q$, and can be used in the construction of non-trivial flat connections. The flat connection corresponding to $\Omega_1 = P$, $\Omega_2 = P'$ and $\Omega_3 = Q$ is not equivalent to the one corresponding to $\Omega_1 = P$, $\Omega_2 = P''$ and $\Omega_3 = Q$. These two flat connections both have discrete unbroken subgroups.

We remind the reader that the non-trivial flat connection in $SO(7)$ (see section 2.4) was characterised by simultaneously diagonalising the three holonomies in the 7 dimensional vector representation. The 7 triples $((\Omega_1)_{ii},(\Omega_2)_{ii},(\Omega_3)_{ii})$ are the triples $(\pm 1, \pm 1, \pm 1)$ with at least one $-1$. In Witten's original construction, the triples $(\pm 1, \pm 1, \pm 1)$ correspond to the positions of D-branes at orientifold fixed points [55].

There is a remarkable resemblance between this result and the $SU(3)$-based non-trivial flat connection in $F_4$. We can diagonalise the holonomies constructed here. On the diagonal we have the eigenvalues $\exp(2\pi i n/3)$. Take now the triples

$$(\exp(\frac{2\pi i n_1}{3}), \exp(\frac{2\pi i n_2}{3}), \exp(\frac{2\pi i n_3}{3})) \qquad n_i \in \mathbb{Z} \tag{3.4.22}$$

and exclude the triple $(1,1,1)$. There are $3^3 - 1 = 26$ distinct triples, and this is precisely the dimension of the fundamental irrep of $F_4$. Indeed, constructing the elements $P$, $P'$ and $Q$ in the **26** of $F_4$ and diagonalising, we find that the triples of diagonal elements $(P_{ii}, P'_{ii}, Q_{ii})$ are precisely the 26 triples mentioned above. It is an intriguing question whether this too can be related to an M-theory construction.

### 3.4.2   $E_6$, $E_7$, $E_8$

In $E_6$ there is a $SU(3)^3$-subgroup that is suitable for our purposes. The $su(3)$-factors have root vectors

$$\begin{aligned}
su(3)_1: \quad & \alpha_1^1 = (0,0,0,0,1,1), \quad & \alpha_2^1 = (\sqrt{3},-1,-1,-1,-1,-1)/2; \\
su(3)_2: \quad & \alpha_1^2 = (0,0,1,-1,0,0)/, \quad & \alpha_2^2 = (0,1,-1,0,0,0); \\
su(3)_3: \quad & \alpha_1^3 = (0,0,0,0,1,-1), \quad & \alpha_2^3 = (-\sqrt{3},-1,-1,-1,-1,1)/2.
\end{aligned} \tag{3.4.23}$$

The relevant geometrical properties are easily verified. To construct the $su(3)$ algebra's, we need to include appropriate normalisation factors. A full exposition would be very space consuming, so we limit ourselves to a few points.

The diagonal subalgebra $D$ is generated by:

$$\begin{aligned}
H^D_{\alpha_1} &= H^1_{\alpha_1} + H^2_{\alpha_1} + H^3_{\alpha_1} & = h_{(0,0,1,-1,2,0)} \\
H^D_{\alpha_2} &= H^1_{\alpha_2} + H^2_{\alpha_2} + H^3_{\alpha_1} & = h_{(0,0,-2,-1,-1,0)} \\
E^D_{\alpha_1} &= E^1_{\alpha_1} + E^2_{\alpha_1} + E^3_{\alpha_1} & = e_{\alpha_1^1} + e_{\alpha_1^2} + e_{\alpha_1^3} \\
E^D_{\alpha_2} &= E^1_{\alpha_2} + E^2_{\alpha_2} + E^3_{\alpha_2} & = e_{\alpha_2^1} + e_{\alpha_2^2} + e_{\alpha_2^3} \\
E^D_{\alpha_1+\alpha_2} &= E^1_{\alpha_1+\alpha_2} + E^2_{\alpha_1+\alpha_2} + E^3_{\alpha_1+\alpha_2} & = e_{\alpha_1^1+\alpha_2^1} + e_{\alpha_1^2+\alpha_2^2} + e_{\alpha_1^3+\alpha_2^3}
\end{aligned} \tag{3.4.24}$$

Checking the conditions (3.4.12), we see that this algebra generates an $SU(3)/\mathbb{Z}_3$-group. This subgroup corresponds to the decompositions

$$\begin{aligned}
E_6 \quad &\to \quad SU(3) \times SU(3) \times SU(3) \quad &&\to \quad SU(3)/\mathbb{Z}_3 \\
\mathbf{27} \quad &\to \quad (\mathbf{3},\bar{\mathbf{3}},\mathbf{1}) \oplus (\bar{\mathbf{3}},\mathbf{1},\mathbf{3}) \oplus (\mathbf{1},\mathbf{3},\bar{\mathbf{3}}) \quad &&\to \quad 3(\mathbf{8}) \oplus 3(\mathbf{1}) \\
\mathbf{78} \quad &\to \quad (\mathbf{8},\mathbf{1},\mathbf{1}) \oplus (\mathbf{1},\mathbf{8},\mathbf{1}) \oplus (\mathbf{1},\mathbf{1},\mathbf{8}) \\
& \quad \oplus (\mathbf{3},\mathbf{3},\mathbf{3}) \oplus (\bar{\mathbf{3}},\bar{\mathbf{3}},\bar{\mathbf{3}}) \quad &&\to \quad (\mathbf{10}) \oplus (\overline{\mathbf{10}}) \oplus 7(\mathbf{8}) \oplus 2(\mathbf{1})
\end{aligned} \tag{3.4.25}$$

The diagonal subalgebra's $D'$ and $D''$ are obtained by rotating one of the three $SU(3)$ factors by (3.4.18) resp. (3.4.21). The holonomies are then $\Omega_1 = P$, $\Omega_2 = P'$ and $\Omega_3 = Q$ with

$$\begin{aligned}
P &= \exp(\tfrac{2\pi i}{3} H^D_{\alpha_1+\alpha_2}) & P' &= \exp(\tfrac{2\pi i}{3} H^{D'}_{\alpha_1+\alpha_2}) \\
Q &= \exp(\tfrac{2\pi}{3\sqrt{3}}(E^D_{\alpha_1} + E^D_{\alpha_2} + E^D_{-\alpha_1-\alpha_2} - E^D_{-\alpha_1} - E^D_{-\alpha_2} - E^D_{\alpha_1+\alpha_2}))
\end{aligned} \tag{3.4.26}$$

Again the subgroup $D''$ gives rise to an element $P''$, and the exchange $P' \leftrightarrow P''$ produces a non-equivalent flat connection with the unbroken discrete subgroup.

These $E_6$-flat connections are essentially the same as the ones for $F_4$, as can be understood from the decomposition.

$$
\begin{aligned}
E_6 &\to F_4 \\
\mathbf{27} &\to \mathbf{26} \oplus \mathbf{1} \\
\mathbf{78} &\to \mathbf{52} \oplus \mathbf{26}
\end{aligned}
\tag{3.4.27}
$$

Like the embedding of $G_2$ in $SO(7)$ of section 3.3, it is not hard to make this explicit. Take the roots of $E_6$ (in the conventions of appendix B) and set the first and last component of each vector to zero. This projection gives the roots of $F_4$. Applying the same projection to the roots (3.4.23), one finds that the $SU(3)^3$ subgroup of $E_6$ projects onto an $SU(3)^2$ subgroup of $F_4$.

We furthermore note that, concerning the remark in the previous paragraph on the triples (3.4.22), that in the fundamental $\mathbf{27}$ irrep of $E_6$, the holonomies can be constructed from these triples with the triple $(1,1,1)$ *included*.

Because the $E_6$-lattice is a sub-lattice of the $E_7$-lattice, it is not hard to embed the previous $E_6$ result into $E_7$. Working out all details, one finds two inequivalent non-trivial flat connections with an unbroken $SU(2)$, precisely as one would naively expect from the decomposition of $E_7$ in $F_4 \times SU(2)$

$$
\begin{aligned}
E_7 &\to F_4 \times SU(2) \\
\mathbf{56} &\to (\mathbf{26}, \mathbf{2}) \oplus (\mathbf{1}, \mathbf{4}) \\
\mathbf{133} &\to (\mathbf{52}, \mathbf{1}) \oplus (\mathbf{26}, \mathbf{3}) \oplus (\mathbf{1}, \mathbf{3})
\end{aligned}
\tag{3.4.28}
$$

As the $E_6$-lattice is also a sublattice of the lattice of $E_8$, our construction is easily extended to $E_8$. One finds two inequivalent non-trivial flat connections with unbroken subgroup $G_2$, in accordance to the decomposition

$$
\begin{aligned}
E_8 &\to G_2 \times F_4 \\
\mathbf{248} &\to (\mathbf{14}, \mathbf{1}) \oplus (\mathbf{7}, \mathbf{26}) \oplus (\mathbf{1}, \mathbf{52})
\end{aligned}
\tag{3.4.29}
$$

## 3.5 Constructions based on $\mathbb{Z}_4$-twist

It is also possible to use $SU(4)$ for a construction of non-trivial flat connections. However, we need a subtle modification with respect to the $SU(2)$ and $SU(3)$ constructions. After developing the relevant tools for $SU(4)$, we will construct the explicit realisation of this construction in the groups $E_7$ and $E_8$.

We will take the canonical form, where again we will use $H_\alpha$ and $E_\alpha$ for the generators of the subgroup, and $h_\alpha$ and $e_\alpha$ for the generators of the original group.

In $SU(4)$ we look for two matrices satisfying

$$
pq = \exp(\frac{2\pi i}{4}) qp \tag{3.5.1}
$$

We take:

$$p = \begin{pmatrix} \exp(\frac{3\pi i}{4}) & 0 & 0 & 0 \\ 0 & \exp(\frac{\pi i}{4}) & 0 & 0 \\ 0 & 0 & \exp(-\frac{\pi i}{4}) & 0 \\ 0 & 0 & 0 & \exp(-\frac{3\pi i}{4}) \end{pmatrix} \quad q = \begin{pmatrix} 0 & 1 & 0 & 0 \\ 0 & 0 & 1 & 0 \\ 0 & 0 & 0 & 1 \\ 1 & 0 & 0 & 0 \end{pmatrix} \exp(\frac{\pi i}{4}) \quad (3.5.2)$$

In terms of generators this is

$$p = \exp\left(\frac{\pi i}{4} H_{3\alpha_1 + 4\alpha_2 + 3\alpha_3}\right) \quad (3.5.3)$$

$$q = \exp\left(\frac{\pi i}{4}((1-i)(E_{\alpha_1} + E_{\alpha_2} + E_{\alpha_3} + E_{-\alpha_1-\alpha_2-\alpha_3}) - (E_{\alpha_1+\alpha_2} + E_{\alpha_2+\alpha_3})\right. \quad (3.5.4)$$

$$\left. -(E_{-\alpha_1-\alpha_2} + E_{-\alpha_2-\alpha_3}) + (1+i)(E_{-\alpha_1} + E_{-\alpha_2} + E_{-\alpha_3} + E_{\alpha_1+\alpha_2+\alpha_3}))\right)$$

The commutation relations of $p$ and $q$ with the group generators are most easily calculated in a specific representation. One finds

$$\begin{array}{ll} p H_\alpha = H_\alpha p, & q H_\alpha = H_{R\alpha} q, \\ p E_\alpha = \exp(\frac{\pi i}{4}\langle \alpha, 3\alpha_1 + 4\alpha_2 + 3\alpha_4\rangle) E_\alpha p, & q E_\alpha = E_{R\alpha} q. \end{array} \quad (3.5.5)$$

The action of the rotation $R$ (which is in this case a combination of a genuine rotation and a reflection) is fully determined by its action on the simple roots.

$$R: \quad \alpha_1 \to \alpha_2 \to \alpha_3 \to -(\alpha_1 + \alpha_2 + \alpha_3) \to \alpha_1 \quad (3.5.6)$$

Just like in the $SU(3)$-case this is nicely visualised by a rotation of the extended Dynkin diagram of $SU(4)$. $R$ is an element of the Weyl group: It is the composition of the Weyl reflection generated by $\alpha_3$, followed by the reflection generated by $\alpha_2$ and the reflection generated by $\alpha_1$. It is obvious that $R^4 = 1$, but on the vectors that are a multiple of $\alpha_1 + \alpha_3$ $R$ acts as a reflection, and we even have $R^2(\alpha_1 + \alpha_3) = \alpha_1 + \alpha_3$.

We want to compute the subgroup commuting with $p$ and $q$. We will determine the effect on the generators $T^a$. To compute of the combinations $pT^a p^{-1}$ and $qT^a q^{-1}$, we will need several representations of $SU(4)$. For the construction in the fundamental **56** irrep of $E_7$ we need the **1**, **15** and **6** of $SU(4)$. For the construction in the adjoints of $E_7$ and $E_8$ (the **133** resp. **248**) we furthermore need the **10** and **20** of $SU(4)$. Note that the **6** and **10** still contain a non-trivial $\mathbb{Z}_2$ centre, which is a subgroup of the $\mathbb{Z}_4$ centre of their simply connected covering $SU(4)$. To compensate for the non-trivial $\mathbb{Z}_2$, we will use the $SU(2)$ factor in the decomposition $E_7 \to SU(4) \times SU(4) \times SU(2)$. We will need only the one-, two- and three dimensional representation of $SU(2)$, and relevant facts about these representations can be found in section 3.3.

The calculation of $pT^a p^{-1}$ proceeds in the same way as in (3.4.7). For $qT^a q^{-1}$ we start again by decomposing the representation of the group $G$ into irreducible representations of $SU(4)$. Construct the normalised eigenvectors $\psi_\lambda$ of the CSA: $H_\alpha \psi_\lambda = \langle \alpha, \lambda \rangle \psi_\lambda$. It then follows that

$$H_{R\alpha} q \psi_\lambda = q H_\alpha \psi_\lambda = \langle \alpha, \lambda \rangle q \psi_\lambda = \langle R\alpha, R\lambda \rangle q \psi_\lambda \quad (3.5.7)$$

This means that $q\psi_\lambda$ is an eigenvector of the CSA with weight $R\lambda$. For simple weights, we have $q\psi_\lambda = \phi_\lambda \psi_{R\lambda}$ with $\phi_\lambda$ a phase. Because all representations we consider are representations of $SU(4)/\mathbb{Z}_2$, we have $q^4 = 1$ and we find that $\phi_\lambda \phi_{R\lambda} \phi_{R^2\lambda} \phi_{R^3\lambda} = 1$. With the ladder operators one shows that $\phi_\lambda$ is actually independent of $\lambda$ (in a suitable phase convention). For $\lambda \neq 0$ the following four combinations are eigenvectors:

$$\begin{aligned}
\psi_\lambda + \psi_{R\lambda} + \psi_{R^2\lambda} + \psi_{R^3\lambda} &\to \text{eigenvalue } \phi \\
\psi_\lambda - i\psi_{R\lambda} - \psi_{R^2\lambda} + i\psi_{R^3\lambda} &\to \text{eigenvalue } i\phi \\
\psi_\lambda - \psi_{R\lambda} + \psi_{R^2\lambda} - \psi_{R^3\lambda} &\to \text{eigenvalue } -\phi \\
\psi_\lambda + i\psi_{R\lambda} - \psi_{R^2\lambda} - i\psi_{R^3\lambda} &\to \text{eigenvalue } -i\phi
\end{aligned}$$

Note that for weights $\lambda$ with $R^2\lambda = \lambda$, two of these combinations are zero.

The **6** and **10** do not posses zero weights, and by checking the eigenvalues we find that the phase $\phi = \pm 1$. The above considerations thus determine the action of $q$ up to an overall sign. In the actual constructions for $E_7$ and $E_8$, this sign will be compensated for by another sign coming from an $SU(2)$-subgroup.

The **15** is the adjoint of $SU(4)$. Among the fifteen weights there are 12 non-zero weights, that decompose in 3 4-cycles under the action of $q$. The remaining three zero weights correspond to the generators of the CSA, the $H_\alpha$, on which we already know the action of $q$. The eigenvectors of $q$ are

$$\begin{aligned}
H_\alpha + H_{R\alpha} + H_{R^2\alpha} + H_{R^3\alpha} &= 0 \\
H_\alpha - iH_{R\alpha} - H_{R^2\alpha} + iH_{R^3\alpha} &\to \text{eigenvalue } i \\
H_\alpha - H_{R\alpha} + H_{R^2\alpha} - H_{R^3\alpha} &\to \text{eigenvalue } -1 \\
H_\alpha + iH_{R\alpha} - H_{R^2\alpha} - iH_{R^3\alpha} &\to \text{eigenvalue } -i
\end{aligned}$$

The last irrep we wish to consider is the **20**. The easiest way to reconstruct the action of $q$ in this irrep, is to use the decomposition of the tensor product of two **6**'s: $\mathbf{6} \otimes \mathbf{6} = \mathbf{1} \oplus \mathbf{15} \oplus \mathbf{20}$. From this decomposition we quickly find that the **20** has 18 non-zero-weights, which under the action of $q$ decompose in 4 4-cycles and one 2-cycle. The two zero weights form a 2-cycle.

We note that these considerations imply that

$$pT^a p^{-1} = \exp\left(\frac{2\pi i n}{4}\right) T^a, \quad qT^a q^{-1} = \exp\left(\frac{2\pi i m}{4}\right) T^a \tag{3.5.8}$$

To compute the representations of $SU(4)$ we follow the same route as for $SU(3)$. Define the CSA generators $H_{\alpha'_1}$, $H_{\alpha'_2}$ and $H_{\alpha'_3}$ associated to the simple roots of $su(4)$. Set $H_{\alpha'_i} = h_{\beta_i}$, for $h_{\beta_i} \in \mathcal{L}_G$. To determine how $\mathcal{L}_G$ branches in representation of $su(3)$ project the weights $\lambda$ of $\mathcal{L}_G$ to the subspace spanned by $\beta_1$ and $\beta_2$. The projected weights $\lambda^{proj}$ have an expansion of the form

$$\lambda^{proj} = c_1\beta_1 + c_2\beta_2 + c_3\beta_3 \tag{3.5.9}$$

The coefficients $c_i$ may be found by taking the inner product with the fundamental weights $\Lambda'_i$ of $su(4)$ (as $su(4)$ is simply laced, the fundamental weights equal the fundamental coweights). The fundamental weights can be expressed in the roots as

$$\Lambda'_1 = \frac{3}{4}\beta_1 + \frac{1}{2}\beta_2 + \frac{1}{4}\beta_3 \qquad \Lambda'_2 = \frac{1}{2}\beta_1 + \beta_2 + \frac{1}{2}\beta_3 \qquad \Lambda'_3 = \frac{1}{4}\beta_1 + \frac{1}{2}\beta_2 + \frac{3}{4}\beta_3.$$

The $c_i$ may again be found by computing

$$c_i = \langle \lambda^{proj}, \Lambda'_i \rangle = \langle \lambda, \Lambda'_i \rangle \tag{3.5.10}$$

### 3.5.1 $E_7$, $E_8$

As mentioned before, we need the subgroup of $E_7$ whose universal cover is $SU(2) \times SU(4) \times SU(4)$. This subgroup can be chosen to be regular, such that its lattice is a sublattice of the $E_7$-lattice. We take as $SU(2)$-root:

$$(\sqrt{2}, 0, 0, 0, 0, 0, 0) \tag{3.5.11}$$

For the first $SU(4)$ factor that we will label $SU(4)_1$, we take as roots:

$$\alpha_1^1 = (0,0,0,0,0,1,1) \quad \alpha_2^1 = (0,0,0,0,1,-1,0) \quad \alpha_3^1 = (0,0,0,0,0,1,-1) \tag{3.5.12}$$

The second $SU(4)$-factor ($SU(4)_2$) has as roots:

$$\alpha_1^2 = (0,0,1,1,0,0,0) \quad \alpha_2^2 = (0,1,-1,0,0,0,0) \quad \alpha_3^2 = (0,0,1,-1,0,0,0) \tag{3.5.13}$$

With these roots we have the following branching rules for $E_7$ into $SU(2) \times SU(4) \times SU(4)$

$$\begin{aligned}
E_7 &\to SU(2) \times SU(4) \times SU(4) \\
56 &\to (1,4,\bar{4}) \oplus (1,\bar{4},4) \oplus (2,6,1) \oplus (2,1,6) \\
133 &\to (1,15,1) \oplus (1,1,15) \oplus (3,1,1) \oplus (1,6,6) \oplus (2,4,4) \oplus (2,\bar{4},\bar{4})
\end{aligned} \tag{3.5.14}$$

In the $SU(4) \times SU(4)$-part we construct as usual a diagonal subgroup $D$. We only list the Cartan generators:

$$\begin{aligned}
H^D_{\alpha_1} &= h_{(0,0,1,1,0,1,1)} \\
H^D_{\alpha_2} &= h_{(0,1,-1,0,1,-1,0)} \\
H^D_{\alpha_3} &= h_{(0,0,1,-1,0,1,-1)}
\end{aligned} \tag{3.5.15}$$

The $SU(2)$-factor has as its generators

$$L_3 = h_{(\sqrt{2},0,0,0,0,0,0)}/2 \quad L_+ = e_{(\sqrt{2},0,0,0,0,0,0)}/\sqrt{2} \tag{3.5.16}$$

We compute the inner products of the weights of $E_7$ with $(1/\sqrt{2}, 0, 0, 0, 0, 0, 0)$ to find the $SU(2)$-representations (if the inner product is integer, the corresponding weight belongs to an $SO(3)$ representation, otherwise it is a genuine $SU(2)$ representation with $\mathbb{Z}_2$-centre [28]). To check the $SU(4)$ irreps, it is sufficient to compute $c_1$ (see (3.5.10)) for the diagonal group. $c_1$ is found by computing the inner product of the weights of $E_7$ with $\frac{3}{4}\alpha_1^D + \frac{1}{2}\alpha_2^D + \frac{1}{4}\alpha_3^D = (0,1,1,1,1,1,1)/2$ (where $\alpha_i^D = \alpha_i^1 + \alpha_i^2$ are the simple roots for the diagonal group). Defining $c'_1 = c_1$ mod 1, there are 4 possibilities. If $c'_1 = 3/4$ the weight belongs to a representation of the fundamental $SU(4)$, for $c'_1 = 1/4$ it belongs to an irrep that is congruent to the complex conjugate of the fundamental, for $c_1' = 1/2$ the irrep is congruent to $SU(4)/\mathbb{Z}_2$, and if $c_1 = 0$ the irrep is congruent to $SU(4)/\mathbb{Z}_4$. Careful examination shows that all weights of $E_7$ fall into two categories: One category is formed by weights that are in an irrep of $(SU(4)/\mathbb{Z}_4) \times SO(3)$, while the second category consist of weights that are in an faithful irrep of $(SU(4)/\mathbb{Z}_2) \times SU(2)$. We now construct the elements

$$P_{SU(4)} = \exp\left(\frac{\pi i}{4} H^D_{3\alpha_1+4\alpha_2+3\alpha_3}\right) \tag{3.5.17}$$

$$Q_{SU(4)} = \exp\frac{\pi i}{4}((1-i)(E^D_{\alpha_1} + E^D_{\alpha_2} + E^D_{\alpha_3} + E^D_{-\alpha_1-\alpha_2-\alpha_3}) - (E^D_{\alpha_1+\alpha_2} + E^D_{\alpha_2+\alpha_3}) \tag{3.5.18}$$
$$-(E^D_{-\alpha_1-\alpha_2} + E^D_{-\alpha_2-\alpha_3}) + (1+i)(E^D_{-\alpha_1} + E^D_{-\alpha_2} + E^D_{-\alpha_3} + E^D_{\alpha_1+\alpha_2+\alpha_3}))$$

in the diagonal $SU(4)$-group, and

$$P_{SU(2)} = \exp(i\pi L_3) \qquad Q_{SU(2)} = \exp\left(\frac{i\pi}{2}(L_+ + L_-)\right) \tag{3.5.19}$$

in the $SU(2)$ group. For a weight $\lambda$ with corresponding eigenvector $\psi_\lambda$, that belongs to an $(SU(4)/\mathbb{Z}_4) \times SO(3)$-irrep, we have

$$P_{SU(4)} Q_{SU(4)} \psi_\lambda = Q_{SU(4)} P_{SU(4)} \psi_\lambda \qquad P_{SU(2)} Q_{SU(2)} \psi_\lambda = Q_{SU(2)} P_{SU(2)} \psi_\lambda$$

If the weight $\lambda$ belongs to a faithful $(SU(4)/\mathbb{Z}_2) \times SU(2)$-irrep, we have

$$P_{SU(4)} Q_{SU(4)} \psi_\lambda = -Q_{SU(4)} P_{SU(4)} \psi_\lambda \qquad P_{SU(2)} Q_{SU(2)} \psi_\lambda = -Q_{SU(2)} P_{SU(2)} \psi_\lambda$$

The products

$$P = P_{SU(2)} P_{SU(4)} = \exp(\frac{\pi i}{2} h_{(\sqrt{2},2,1,0,2,1,0)}), \qquad Q = Q_{SU(2)} Q_{SU(4)} \tag{3.5.20}$$

are commuting elements of $E_7$, since they commute on each weight of any representation. We construct another diagonal subgroup $D'$ in $SU(4) \times SU(4)$ as usual, by applying the rotation

$$R : \alpha_1 \to \alpha_2 \to \alpha_3 \to -(\alpha_1 + \alpha_2 + \alpha_3) \to \alpha_1, \tag{3.5.21}$$

to the first $SU(4)$ factor and then reconstructing the diagonal subgroup. In this diagonal subgroup we construct an element

$$P'_{SU(4)} = \exp(\frac{\pi i}{4} H^{D'}_{3\alpha_1+4\alpha_2+3\alpha_3}) \tag{3.5.22}$$

The same considerations as above lead to the conclusion that the element

$$P' = P_{SU(2)} P'_{SU(4)} = \exp(\frac{\pi i}{4} h_{(\sqrt{2},0,-1,-2,2,1,0)}) \tag{3.5.23}$$

commutes with $P$ and $Q$, and hence we have three candidate holonomies $\Omega_1 = P$, $\Omega_2 = P'$ and $\Omega_3 = Q$.

Next we calculate the unbroken subgroup. Computing the commutators of the generators of $E_7$ with $P$ and $P'$, we find that only the CSA-generators of $E_7$ commute with both $P$ and $P'$. These CSA generators correspond to zero weights in the representation of the $SU(2) \times SU(4) \times SU(4)$ subgroup, and from previous considerations we know that there is no combination of zero weights of the adjoints of $SU(4)$ or $SU(2)$ that commutes with $Q$. Hence we find that there is no generator that commutes with $P$, $P'$ and $Q$, and the unbroken subgroup is at most discrete.

Instead of the rotation $R$ given by (3.5.21), one could also apply the rotations $R^2$ or $R^3$ to construct elements that we will call $P''$ resp. $P'''$. The holonomies $\Omega_1 = P$, $\Omega_2 = P'''$ and $\Omega_3 = Q$ also only commute with a discrete subgroup of $E_7$, and give a non-trivial flat connection inequivalent to the one implied by $\Omega_1 = P$, $\Omega_2 = P'$ and $\Omega_3 = Q$. The non-trivial flat connection implied by $\Omega_1 = P$, $\Omega_2 = P''$ and $\Omega_3 = Q$ gives a bigger unbroken subgroup. The unbroken symmetry group can be calculated to be $U(1)^3$, and this non-trivial flat connection is actually a gauge deformation of the flat connection constructed in section 3.3, based on twist in $SU(2)$. Therefore the construction based on twist in $SU(2) \times SU(4) \times SU(4)$ gives two new vacua with discrete symmetry group, and a gauge deformation of a vacuum configuration that we have encountered before.

We wish to point out that also the $SU(4)$-based construction in $E_7$ fits into a pattern. We already remarked that the $G_2$ and $SO(7)$ non-trivial flat connection can be described by the 7 triples $(\pm 1, \pm 1, \pm 1)$ with at least one $-1$, and that the $SU(3)$-based construction in $F_4$ can be characterised by the 26 triples (3.4.22) with at least one element not equal to 1. For the $SU(4)$-based construction, the eigenvalues that appear on the diagonal of the diagonalised holonomies are $i^n$. Consider now the triples $(i^{n_1}, i^{n_2}, i^{n_3})$ with $n_j \in \mathbb{Z}$. There are $4^3 = 64$ distinct triples. Now exclude all triples that are not of order 4, by which we mean that we demand each triple to contain at least one $i$ or $-i$. We are then left with $64 - 2^3 = 56$ triples. 56 is precisely the dimension of the fundamental irrep of $E_7$, and indeed, constructing the holonomies in this representation and diagonalising, we find that the triples $((\Omega_1)_{ii}, (\Omega_2)_{ii}, (\Omega_3)_{ii})$ are precisely the triples mentioned above.

The $SU(2) \times SU(4) \times SU(4)$ construction of $E_7$ can be easily embedded in $E_8$ by using the fact that the $E_7$-lattice is a sublattice of the $E_8$-lattice. More specific, we can take as $SU(2)$-root:

$$(1, 1, 0, 0, 0, 0, 0, 0) \qquad (3.5.24)$$

One $SU(4)$ factor has as roots:

$$\begin{aligned}
\alpha_1 &= (0, 0, 0, 0, 0, 0, 1, 1) \\
\alpha_2 &= (0, 0, 0, 0, 0, 1, -1, 0) \\
\alpha_3 &= (0, 0, 0, 0, 0, 0, 1, -1)
\end{aligned} \qquad (3.5.25)$$

A second $SU(4)$-factor is generated by the roots:

$$\begin{aligned}
\alpha_1 &= (0, 0, 0, 1, 1, 0, 0, 0) \\
\alpha_2 &= (0, 0, 1, -1, 0, 0, 0, 0) \\
\alpha_3 &= (0, 0, 0, 1, -1, 0, 0, 0)
\end{aligned} \qquad (3.5.26)$$

With this information it is trivial to copy the $E_7$ construction. Like in the $E_7$-case, one finds 3 vacua. For two of these, the subgroup unbroken by the $E_7$ holonomies is $SU(2)$. This is to be expected, since $E_8$ decomposes into $E_7 \times SU(2)$:

$$\begin{aligned} E_8 &\to E_7 \times SU(2) \\ \mathbf{248} &\to (\mathbf{133},\mathbf{1}) \oplus (\mathbf{56},\mathbf{2}) \oplus (\mathbf{1},\mathbf{3}) \end{aligned} \qquad (3.5.27)$$

The third vacuum is, as for $E_7$, a gauge deformation of the vacuum based on twist in $SU(2)$ constructed in section 3.3.

## 3.6 Constructions based on $\mathbb{Z}_5$-twist

A construction based on the group $SU(5)$ is also possible, but only in the largest exceptional group $E_8$. We will again use $H_\alpha$ and $E_\alpha$ for the subgroup, and $h_\alpha$ and $e_\alpha$) for the original group.

In $SU(5)$ we look for two matrices satisfying

$$pq = \exp(\frac{2\pi i}{5})qp \qquad (3.6.1)$$

We take:

$$p = \begin{pmatrix} \exp(\frac{4\pi i}{5}) & 0 & 0 & 0 & 0 \\ 0 & \exp(\frac{2\pi i}{5}) & 0 & 0 & 0 \\ 0 & 0 & 1 & 0 & 0 \\ 0 & 0 & 0 & \exp(-\frac{2\pi i}{5}) & 0 \\ 0 & 0 & 0 & 0 & \exp(-\frac{4\pi i}{5}) \end{pmatrix}, \quad q = \begin{pmatrix} 0 & 1 & 0 & 0 & 0 \\ 0 & 0 & 1 & 0 & 0 \\ 0 & 0 & 0 & 1 & 0 \\ 0 & 0 & 0 & 0 & 1 \\ 1 & 0 & 0 & 0 & 0 \end{pmatrix}$$
(3.6.2)

In terms of generators this is

$$p = \exp\left(\frac{\pi i}{5} H_{4\alpha_1+6\alpha_2+6\alpha_3+4\alpha_4}\right) \qquad (3.6.3)$$

$$\begin{aligned} q = \exp\Big(\frac{2\pi i}{25}(a(E_{\alpha_1}+E_{\alpha_2}+E_{\alpha_3}+E_{\alpha_4}+E_{-(\alpha_1+\alpha_2+\alpha_3+\alpha_4)})+ \\ b(E_{\alpha_1+\alpha_2}+E_{\alpha_2+\alpha_3}+E_{\alpha_3+\alpha_4}+E_{-(\alpha_1+\alpha_2+\alpha_3)}+E_{-(\alpha_2+\alpha_3+\alpha_4)}) \\ +\text{complex conjugate})\Big) \end{aligned} \qquad (3.6.4)$$

with

$$a = \left[\exp\left(-\frac{2\pi i}{5}\right) - \exp\left(\frac{2\pi i}{5}\right)\right] + 2\left[\exp\left(-\frac{4\pi i}{5}\right) - \exp\left(\frac{4\pi i}{5}\right)\right] \qquad (3.6.5)$$

$$b = \left[\exp\left(-\frac{4\pi i}{5}\right) - \exp\left(\frac{4\pi i}{5}\right)\right] + 2\left[\exp\left(\frac{2\pi i}{5}\right) - \exp\left(-\frac{2\pi i}{5}\right)\right] \qquad (3.6.6)$$

The commutation relations of $p$ and $q$ with the group generators are most easily calculated in a specific representation. One finds

$$pH_\alpha = H_\alpha p, \qquad\qquad qH_\alpha = H_{R\alpha}q,$$
$$pE_\alpha = \exp(\tfrac{\pi i}{5}\langle\alpha, 4\alpha_1+6\alpha_2+6\alpha_3+4\alpha_4\rangle)E_\alpha p, \qquad qE_\alpha = E_{R\alpha}q. \qquad (3.6.7)$$

The action of the rotation $R$ (which is in this case a genuine rotation) is fully determined by its action on the simple roots.

$$R: \qquad \alpha_1 \to \alpha_2 \to \alpha_3 \to \alpha_4 \to -(\alpha_1+\alpha_2+\alpha_3+\alpha_4) \to \alpha_1 \qquad (3.6.8)$$

Again this can be conveniently depicted by a permutation of the roots of the extended Dynkin diagram of $SU(5)$. $R$ is an element of the Weyl group: It is the composition of the Weyl reflection generated by $\alpha_4$, followed by the reflections generated by $\alpha_3$, $\alpha_2$, and $\alpha_1$.

For the calculation of the unbroken subgroup we will only need the action of $q$ on the zero weights of the adjoint. These are the CSA-generators, on which the action of $q$ is easily diagonalised. The only fact we will need is that there is no linear combination of CSA-generators that commutes with $q$.

### 3.6.1 $E_8$

Only $E_8$ allows a suitable $PG'_2$-subgroup with $\widetilde{PG'}_2 = SU(5) \times SU(5)$. We take as root vectors for the first $SU(5)$-subgroup

$$\alpha^1_1 = (\tfrac{1}{2}, -\tfrac{1}{2}, -\tfrac{1}{2}, -\tfrac{1}{2}, -\tfrac{1}{2}, -\tfrac{1}{2}, -\tfrac{1}{2}, -\tfrac{1}{2}), \qquad \alpha^1_2 = (0,0,0,0,0,0,1,1), \qquad (3.6.9)$$
$$\alpha^1_3 = (0,0,0,0,0,1,-1,0), \qquad \alpha^1_4 = (0,0,0,0,0,0,1,-1).$$

while the second $SU(5)$-factor will have roots

$$\alpha^2_1 = (0,0,0,1,-1,0,0,0), \qquad \alpha^2_2 = (0,0,1,-1,0,0,0,0), \qquad (3.6.10)$$
$$\alpha^2_3 = (0,1,-1,0,0,0,0,0), \qquad \alpha^2_4 = (-1,-1,0,0,0,0,0,0).$$

The decomposition of $E_8$ into this $SU(5) \times SU(5)$ for the adjoint is given by[5]

$$E_8 \to SU(5) \times SU(5)$$
$$\mathbf{248} \to (\mathbf{24,1}) \oplus (\mathbf{1,24}) \oplus (\mathbf{10,5}) \oplus (\mathbf{5,\overline{10}}) \oplus (\mathbf{\bar{5},10}) \oplus (\mathbf{\overline{10},\bar{5}}) \qquad (3.6.11)$$

The congruence classes of the irreducible components of $SU(5) \times SU(5)$ are then[1] $(0,0)$ with multiplicity 2, and $(3,1)$, $(1,2)$, $(4,3)$ and $(2,4)$ all with multiplicity 1. We note that for each congruence class $(x, y)$ we have $x + 2y \mod 5 = 0$. Since all $E_8$ representations are isomorphic, this must automatically hold for any $E_8$ representation.

---

[5]We will label the irrep with Dynkin labels (1000) as **5** and the irrep with labels (0010) as **10**. We define the congruence class of the representation with Dynkin labels $(n_1, n_2, n_3, n_4)$ to be labelled by the integer $n_1 + 2n_2 + 3n_3 + 4n_4 \mod 5$.

We now construct the elements

$$P_i = \exp\left(\frac{\pi i}{5} H_{4\alpha_1^i + 6\alpha_2^i + 6\alpha_3^i + 4\alpha_4^i}\right) \tag{3.6.12}$$

$$Q_i = \exp\Big(\frac{2\pi i}{25}(a(E_{\alpha_1^i} + E_{\alpha_2^i} + E_{\alpha_3^i} + E_{\alpha_4^i} + E_{-(\alpha_1^i+\alpha_2^i+\alpha_3^i+\alpha_4^i)}) + \tag{3.6.13}$$
$$b(E_{\alpha_1^i+\alpha_2^i} + E_{\alpha_2^i+\alpha_3^i} + E_{\alpha_3^i+\alpha_4^i} + E_{-(\alpha_1^i+\alpha_2^i+\alpha_3^i)} + E_{-(\alpha_2^i+\alpha_3^i+\alpha_4^i)})$$
$$+\text{complex conjugate})\Big)$$

When acting on an eigenvector $\psi_\lambda$, corresponding to a weight $\lambda$ that belongs to an irrep with $SU(5) \times SU(5)$-congruence class labelled by $(n_1, n_2)$, $P_i$ and $Q_i$ obey the following commutation rule

$$P_i Q_i \psi_\lambda = \exp\left(\frac{2\pi i n_i}{5}\right) Q_i P_i \psi_\lambda \tag{3.6.14}$$

From the above observation on congruence classes it follows that for any weight $\lambda$ of $E_8$ we have

$$P_1 Q_1 (P_2)^2 Q_2 \psi_\lambda = \exp\left(\frac{2\pi i (n_1 + 2n_2)}{5}\right) Q_1 P_1 Q_2 (P_2)^2 \psi_\lambda = Q_1 P_1 Q_2 (P_2)^2 \psi_\lambda \tag{3.6.15}$$

Hence the elements

$$P = P_1(P_2)^2, \qquad Q = Q_1 Q_2 \tag{3.6.16}$$

commute. A third commuting element is constructed in the standard way:

$$P' = Q_1^{-n} P Q_1^n = Q_2^n P Q_2^{-n} \tag{3.6.17}$$

$n$ can range from 1 to 4, and therefore there are 4 different flat connections constructed this way. We will work out the case where $n = 1$.

In that case we have

$$P = \exp\left(\frac{\pi i}{5} h_{(-6,2,-2,-6,-10,4,2,0)}\right) \tag{3.6.18}$$

$$P' = \exp\left(\frac{\pi i}{5} h_{(-2,-2,-6,-10,6,4,2,0)}\right) \tag{3.6.19}$$

After a somewhat tedious calculation we find that none of the ladder operators $e_\beta$ of $E_8$ commutes with both $P$ and $P'$, so we are only left with the 8 Cartan generators. We have to check whether these commute with $Q$. $Q$ has been constructed in such a way that it is an element of the diagonal $SU(5)$-subgroup, so we can apply our previous methods. The elements of the Cartan subalgebra are also the CSA elements for $SU(5) \times SU(5)$, and we know that none of these are invariant under commutation with $Q$ ($Q$ takes the role of $q$ in the diagonal subgroup of $SU(5) \times SU(5)$). Hence no group generator commutes with $P$, $P'$ and $Q$ simultaneously, and the unbroken subgroup is at most discrete.

For the $SU(5)$ based construction, the eigenvalues that appear in the diagonalised holonomies are $\exp\left(\frac{2\pi i n}{5}\right)$. Consider now the triples

$$(\exp(\frac{2\pi i n_1}{5}), \exp(\frac{2\pi i n_2}{5}), \exp(\frac{2\pi i n_3}{5})) \qquad n_i \in \mathbb{Z} \tag{3.6.20}$$

If one excludes the triple $(1,1,1)$, there are $5^3 - 1 = 124$ distinct triples of this type. Unlike the cases we considered previously, 124 is not the dimension of any representation of $E_8$. The smallest non-trivial representation of $E_8$ is the 248-dimensional adjoint. Constructing the holonomies in the adjoint and diagonalising, one finds that the 248 triples $(P_{ii}, P'_{ii}, Q_{ii})$ consist of precisely two sets of the above 124 triples.

## 3.7　Constructions based on $\mathbb{Z}_6$-twist

At first sight all possibilities seem to be exhausted, since constructions with $SU(N)$, $N > 5$ are impossible because the exceptional groups are not big enough to contain more than one $SU(6)$-factor. There is however still one more possibility, based on our previous constructions based on $SU(2)$ and $SU(3)$. It is easy to see that $E_8$ allows the $SU(2)$ non-trivial flat connection to be realised simultaneously with the $SU(3)$-flat connection.

### 3.7.1　$E_8$

According to the decomposition (3.4.29) $E_8$ allows a $G_2 \times F_4$ subgroup (in this case we are dealing with a subgroup that is a genuine product). In section 3.3 we constructed holonomies for a non-trivial flat connection in $G_2$, that we named $P$, $P'$ and $Q$. Here we will rename them to $P_{G_2}$, $P'_{G_2}$ and $Q_{G_2}$ to avoid confusion. $F_4$ allows different types of non-trivial flat connections. We will use the non-trivial flat connections constructed in section 3.4.1, and will rename the holonomies $P$, $P'$, $P''$ and $Q$ constructed there to $P_{F_4}$, $P'_{F_4}$, $P''_{F_4}$ and $Q_{F_4}$. Decomposing $E_8$ into its $G_2 \times F_4$ subgroup, elements of this subgroup can be denoted by a pair of elements $(g_{G_2}, g_{F_4})$ with $g_{G_2} \in G_2$ and $g_{F_4} \in F_4$. Therefore $(P_{G_2}, P_{F_4})$ and $(Q_{G_2}, Q_{F_4})$ represent commuting elements of $E_8$. A third commuting element can be constructed by applying the standard procedure of rotating group factors to both elements of the pair to construct the element $(P'_{G_2}, P'_{F_4})$. The rotation can be applied multiple times to construct multiple triples. We have the following possibilities:

$$\Omega_1 = (P_{G_2}, P_{F_4}) \quad \Omega_2 = (P'_{G_2}, P'_{F_4}) \quad \Omega_3 = (Q_{G_2}, Q_{F_4}) \tag{3.7.1}$$
$$\Omega_1 = (P_{G_2}, P_{F_4}) \quad \Omega_2 = (P_{G_2}, P''_{F_4}) \quad \Omega_3 = (Q_{G_2}, Q_{F_4}) \tag{3.7.2}$$
$$\Omega_1 = (P_{G_2}, P_{F_4}) \quad \Omega_2 = (P'_{G_2}, P_{F_4}) \quad \Omega_3 = (Q_{G_2}, Q_{F_4}) \tag{3.7.3}$$
$$\Omega_1 = (P_{G_2}, P_{F_4}) \quad \Omega_2 = (P_{G_2}, P'_{F_4}) \quad \Omega_3 = (Q_{G_2}, Q_{F_4}) \tag{3.7.4}$$
$$\Omega_1 = (P_{G_2}, P_{F_4}) \quad \Omega_2 = (P'_{G_2}, P''_{F_4}) \quad \Omega_3 = (Q_{G_2}, Q_{F_4}) \tag{3.7.5}$$

However, not all of these are new. The flat connection implied by the holonomies (3.7.2) has twice the element $P_{G_2}$ in the $G_2$-factor: This means that in the $G_2$ subgroup the connection can be deformed to a trivial one, and only the $F_4$ part is non-trivial. This is a deformation of one of the flat connections based on $SU(3)$-twist that was already constructed previously. Similarly the flat connection implied by the holonomies (3.7.4) is deformable to the other flat connection based on $SU(3)$-twist. The flat connection implied by (3.7.3) is trivial in its $F_4$ factor, and only non-trivial in its $G_2$ part. It is therefore a deformation of the flat connection in $E_8$ that was constructed in section 3.3. The remaining two flat connections are new. We note that, to commute a generator with the $\Omega_i$, we can commute it first with the $G_2$-element and then with the $F_4$-element. We can choose a basis of generators such that they commute or anti-commute with the $G_2$-elements, and they commute up to a $\mathbb{Z}_3$-element with the $F_4$-elements. We conclude that to commute with the $\Omega_i$ a generator has to commute with both the $G_2$ and the $F_4$-elements. The generators that commute with the $G_2$ part of the holonomies generate the $F_4$ subgroup in the decomposition $G_2 \times F_4$. The $F_4$ part of the holonomies breaks this $F_4$ group completely. Hence only discrete symmetries commute with the holonomies constructed here.

## 3.8 Counting vacua

We will label the different vacua by an integer related to the centre of the appropriate $SU(N)$ embeddings required for twisting. Hence the integer is taken to be $N$ for the $SU(N)$ based construction, and 6 for the $SU(2) \times SU(3)$ construction. Obviously we reserve the label 1 for the trivial component.

The essence of the $SU(2)$ based construction is the existence of a suitable subgroup $PG'_N$ within a $G_2$ subgroup of $G$. The non-trivial flat connections arise through the decomposition of $G$ into $G_2 \times H$, with $H$ the maximal subgroup commuting with $G_2$. It is the CSA of $H$ that determines the deformations of the non-trivial $G_2$ connection as embedded in $G$, fixing the dimension of this connected vacuum component (rank($H$)). Since $H$ commutes with the $G_2$ subgroup containing the holonomies, it also plays the role of the maximal unbroken subgroup, apart from some global discrete symmetries. Compare the situation to Witten's D-brane construction [55].

Our construction presented here goes via the chains

$$
\begin{array}{ccccc}
E_8 & \to & E_7 & \to & E_6 \\
& & & & \searrow \\
& SO(2N) & \to & SO(8) & \\
& & & & \searrow \\
& & & F_4 & \to & SO(7) & \to & G_2 \\
& & & \nearrow & \\
& SO(2N+1) & & &
\end{array}
\tag{3.8.1}
$$

It is this general feature that repeats itself for the $SU(N > 2)$ based constructions. For $SU(3)$ the role of $G_2$ is played by $F_4$, with the embedding chain $E_8 \to E_7 \to E_6 \to F_4$. For $SU(4)$ the role of $G_2$ is played by $E_7$, with chain $E_8 \to E_7$, and finally for $SU(5)$ and $SU(2) \times SU(3)$, $E_8$ stands alone.

Our results are summarised in table (3-1), which presents the connected component of the maximal unbroken subgroup for each vacuum-type.

So far we have concentrated on the classical gauge fields. For a calculation of the Witten index $\text{Tr}(-)^F$, these should be quantised, and the fermions should be included. The computation is essentially the same as in chapter 2 and [52] [55]: Each vacuum component implies a unique bosonic vacuum, and fermions can be added in $r'+1$ ways, where $r'$ is the dimension of the vacuum component, which is equal to the rank of the unbroken subgroup. Thus each vacuum component contributes $r'+1$ to $\text{Tr}(-)^F$. In table 3-2, we list $r'+1$, where $r'$ is the rank of the group listed in table 3-1. The columns labelled by $3,4,5,6$ contain more entry's to indicate that there is more than one vacuum component for each type. Finally, table 3-2 also lists the dual Coxeter number for each group. It is easy to verify that (2.5.3) is satisfied.

| Group    | Vacuum-type |           |         |       |         |         |
| -------- | ----------- | --------- | ------- | ----- | ------- | ------- |
| $G$      | 1           | 2         | 3       | 4     | 5       | 6       |
| $SU(N)$    | $SU(N)$     |           |         |       |         |         |
| $Sp(N)$    | $Sp(N)$     |           |         |       |         |         |
| $SO(2N+1)$ | $SO(2N+1)$  | $SO(2N-6)$ |         |       |         |         |
| $SO(2N)$   | $SO(2N)$    | $SO(2N-7)$ |         |       |         |         |
| $G_2$      | $G_2$       | discrete  |         |       |         |         |
| $F_4$      | $F_4$       | $SO(3)$   | discrete |       |         |         |
| $E_6$      | $E_6$       | $SU(3)$   | discrete |       |         |         |
| $E_7$      | $E_7$       | $Sp(3)$   | $SU(2)$ | discrete |         |         |
| $E_8$      | $E_8$       | $F_4$     | $G_2$   | $SU(2)$ | discrete | discrete |

Table 3-1. The connected part of the maximal unbroken subgroups

| Group     |        | Vacuum-type |     |       |       |           |       |
| --------- | ------ | ----------- | --- | ----- | ----- | --------- | ----- |
| $G$       | $h$    | 1           | 2   | 3     | 4     | 5         | 6     |
| $SU(N)$   | $N$    | $N$         |     |       |       |           |       |
| $Sp(N)$   | $N+1$  | $N+1$       |     |       |       |           |       |
| $SO(2N+1)$ | $2N-1$ | $N+1$       | $N-2$ |       |       |           |       |
| $SO(2N)$  | $2N-2$ | $N+1$       | $N-3$ |       |       |           |       |
| $G_2$     | 4      | 3           | 1   |       |       |           |       |
| $F_4$     | 9      | 5           | 2   | (1+1) |       |           |       |
| $E_6$     | 12     | 7           | 3   | (1+1) |       |           |       |
| $E_7$     | 18     | 8           | 4   | (2+2) | (1+1) |           |       |
| $E_8$     | 30     | 9           | 5   | (3+3) | (2+2) | (1+1+1+1) | (1+1) |

Table 3-2. Contributions to $\text{Tr}(-1)^F$

# 4 Almost commuting triples

In chapter 3 it was shown that for gauge theories on a 3–torus, sufficient vacuum solutions exist for all exceptional groups. In a more formal approach, one may also prove that there are no more solutions than the ones already constructed. In this chapter we will review parts of [26] [6], and point out the relations with our work as described in the previous chapter. In [6] also the generalisation to general boundary conditions was undertaken. We will reproduce parts of their analysis for gauge theories on a 3–torus with classical groups in the chapter 5, using a different approach. In [6], a method for calculating the Chern-Simons invariant was developed, as will be reviewed in section 4.4. This invariant will play a role in the considerations of chapter 6.

## 4.1 Reduction of the rank for gauge theories on an $n$–torus

In chapter 3 the starting point was a 2–torus with two commuting holonomies. A third holonomy commuting with the first two was then constructed. In [6] [26] an opposite viewpoint was taken. First one holonomy is constructed with a special property, and then the remaining two. This is motivated below.

A configuration of gauge fields at an $n$-torus with commuting holonomies can be constructed with all gauge fields taking values in the CSA of the gauge group. A maximal Abelian subgroup commuting with the holonomies may then be obtained by exponentiating the elements of the CSA. The centraliser of the holonomies has then the same rank as the original gauge group.

We are interested here in configurations of gauge fields such that the rank of the subgroup that commutes with the holonomies is smaller than the rank of the original gauge group. Not all holonomies can be chosen to be on a maximal torus in this case.

To describe the various possibilities, we consider compactification of a gauge theory with gauge group $G$ on an $n+1$-torus, such that $n$ holonomies can be chosen on a maximal torus of $G$, and the last one cannot. We may now write $T^{n+1} = S^1 \times T^n$. There is a holonomy $\Omega$ along the circle we have split of, and we will take it to be one of the holonomies that lies on a maximal torus of $G$. The $n$ holonomies along the directions of $T^n$ commute with $\Omega$, and hence lie in the centraliser $Z(\Omega)$ of $\Omega$. Of these $n$ holonomies, $n-1$ can be put on a maximal torus of $Z(\Omega)$, but one cannot, due to our starting assumption. We have not specified the size of the circle, and we may consider shrinking it to zero size. Dimensional reduction then leads us to a gauge theory with gauge group $Z(\Omega)$ compactified on a $T^n$, leading us back to our starting point, with the torus dimension $n+1$ replaced by $n$, and the gauge group $G$ replaced by $Z(\Omega)$, which has the same rank as $G$. Repeating these steps we finally arrive at a gauge theory on a circle, with some gauge group $G'$, and a holonomy $\Omega'$ that cannot be put on the maximal torus.

A single holonomy that is an element of the part of the gauge group $G'$ that is connected to the identity can always be put on a maximal torus of $G'$. So $\Omega'$ is not part of the connected part $G'_c$ of the gauge group. Upon traversing the circle with holonomy $\Omega'$, we certainly return to a point with the same local physics. Therefore $\Omega'$ must specify an automorphism of $G'$ acting by conjugation on the group itself. In particular it maps the connected part $G'_c$ of the group to itself (as it maps the identity, which is an element of the connected part to itself). As $\Omega'$ is an element of a disconnected part of the gauge group, it is an outer automorphism of $G'_c$.

As an example of a theory with rank reduction we consider a theory with $G'_c = SO(2) = U(1)$ compactified on a circle $S_b$, which we will use as base manifold for a bundle. The group manifold of $SO(2)$ is a circle $S_f$, which will be the fibre. The circle has as a reflection symmetry $R$. It is thus possible to demand that upon traversing the circle $S_b$, the fibre $S_f$ is reflected. The reflection symmetry is not an element of $SO(2)$, but nevertheless an automorphism. The structure group of this bundle is not $SO(2)$, but a semi-direct product of $SO(2)$ with the $\mathbb{Z}_2$-group generated by the reflection $R$. In this specific case this group has a name, it is $O(2)$. The element $R$ can be represented as $\text{diag}(1,-1)$, since

$$\begin{pmatrix} 1 & 0 \\ 0 & -1 \end{pmatrix} \begin{pmatrix} \cos\phi & -\sin\phi \\ \sin\phi & \cos\phi \end{pmatrix} \begin{pmatrix} 1 & 0 \\ 0 & -1 \end{pmatrix} = \begin{pmatrix} \cos(-\phi) & -\sin(-\phi) \\ \sin(-\phi) & \cos(-\phi) \end{pmatrix} \quad (4.1.1)$$

The coordinate $\phi$ on the circle is indeed reflected. In a $U(1)$ formulation $R$ corresponds to complex conjugation $C$, as this acts as $\exp i\phi \to \exp i(-\phi)$.

The connected part of $O(2)$, which is $SO(2)$, has rank 1. The centraliser of $\text{diag}(1,-1)$ in $O(2)$ is $O(1) \times O(1) = \mathbb{Z}_2 \times \mathbb{Z}_2$. Being a discrete group, it has rank 0, and we have indeed succeeded to reduce the rank.

This procedure can be used on any group that is disconnected, or possesses an outer automorphism. Examples of outer automorphisms can be found for the simple compact non-Abelian Lie-groups $SU(n)$, $SO(2n)$ and $E_6$. Products of groups that contain several isomorphic factors $G_i$ posses outer automorphisms that permute the $G_i$ amongst each other. The latter possibility is well known in string theory for the case of the $E_8 \times E_8$ gauge group of the corresponding heterotic string. The outer automorphism exchanging the two $E_8$ factors can be used in a holonomy when compactifying the $E_8 \times E_8$ heterotic string on a circle [8], to give the CHL-string [7].

An outer automorphism does not map every representation to itself. Therefore for any representation present in the theory, one should require also its image under the automorphism to be present, otherwise the automorphism is not a symmetry of the theory.

We now return to our original theme, which was compactification of gauge theory on a torus with reduction of the rank. Consider a gauge theory on a 2–torus with gauge group $G$ (which we will now assume to be connected). We write the 2–torus as a product of two circles $S^1_a \times S^1_b$, around which there are commuting holonomies $\Omega_a$, resp. $\Omega_b$. When we assume that $\Omega_a$ is an element of the maximal torus of $G$, the centraliser $Z(\Omega_a)$ of $\Omega_a$ has the same rank as $G$. We may still achieve reduction of the rank if $Z(\Omega_a)$ is not connected, such that we can pick $\Omega_b$ to be an element of the part of the $Z(\Omega_a)$ that is not connected to the identity. This is the situation that occurs in 't Hoofts twisted boundary conditions [23] [24].

As an example consider an $SO(3)$ gauge theory on the 2–torus. The holonomy $\Omega_a$ will be set to $\text{diag}(-1,-1,1)$. The centraliser of $\Omega_a$ is a subgroup of $SO(3)$ that is isomorphic to $O(2)$, as it has an $O(2)$-block in the upper left corner, completed by $\pm 1$ on the third diagonal position to make the total determinant 1. From our previous example we know how to reduce the rank of an $O(2)$-theory. Therefore we set $\Omega_b = \mathrm{diag}(1, -1, -1)$. The centraliser of $\Omega_a$ and $\Omega_b$ is a discrete group which has rank 0, which is smaller than the rank 1 of $SO(3)$.

As we require an element from a connected group $G$ to have a disconnected centraliser, $G$ should be non-Abelian. Assume now that $G$ is simple. A theorem by Bott (as quoted in [26] [6]) states that the centraliser of any element from a simple and simply connected group is connected, and hence $G$ should be a non-simply connected group. Possibilities for these are $SU(n)$, $Sp(n)$, $SO(n)$, $E_6$ and $E_7$ quotiented by a non-trivial subgroup $Z$ of their centre. The possible representations of the gauge group are restricted to those representations that represent the elements of $Z$ by the identity.

We now move on to the 3–torus, and assume a simply connected gauge group $G$. The 3–torus may be split in the product $S^1 \times T^2$. Along the circle $S^1$ we assume a holonomy $\Omega$ that is an element of the maximal torus of $G$. From the above discussion for gauge theories on the 2–torus, we know we can achieve rank reduction if the centraliser $Z(\Omega)$ of $\Omega$ is not simply connected. The classification of group elements of simple groups that have a non-simply connected centraliser is the topic of the next section. One finds that $Spin(n \geq 7)$ and all exceptional groups allow such elements.

One may extend the line of reasoning to still higher dimensional tori. For a gauge theory with gauge group $G$ on an $n$-torus $T^n$, one has a set of $n$ holonomies. Assume that no proper subset of these $n$ holonomies has a centraliser that has a rank smaller than the rank of $G$. A set of $n$ holonomies obeying this condition will be called a non-trivial $n$–tuple [26]. In the above we saw examples of non-trivial 1–, 2– and 3–tuples (2– and 3–tuples are also often called pairs and triples. In [6] 2–tuples are called c-pairs, 3–tuples are called commuting triples)

To construct a non-trivial $n$–tuple, one may proceed as follows. Suppose an $n$–tuple is non-trivial, and pick one element $\Omega$ from the $n$–tuple. By definition the remaining $(n-1)$–tuple is trivial in $G$; the centraliser $Z(\Omega)$ of $\Omega$ has the same rank as $G$. The $(n-1)$-tuple forms a set of commuting elements of $Z(\Omega)$, but the centraliser of the $(n-1)$-tuple has a rank smaller than that of $Z(\Omega)$. Therefore the $(n-1)$–tuple is non-trivial in $Z(\Omega)$. Note however that the existence of a non-trivial $(n-1)$-tuple in a subgroup $Z(\Omega)$ is only a necessary condition, but not a sufficient one. A non-trivial $(n-1)$-tuple in $Z(\Omega)$ lifts to a rank reducing $n$-tuple in $G$ by adding $\Omega$ to the $(n-1)$-tuple. It then remains to be investigated whether the $n$-tuple is non-trivial.

For non-trivial 1–, 2– and 3–tuples we avoided this question by successively imposing restrictions on the gauge group. We studied 1–tuples for general groups, 2–tuples for connected groups and 3–tuples for simply connected groups. For $n$–tuples with $n > 3$ there is no successive restriction on the group. However, by the above analysis one can show that the number of possibilities is very limited, and one may perform a case by case study as performed in appendix D of [26]. Here we will just quote their results.

The simple groups containing non-trivial 4–tuples are $Spin(n \geq 15)$. These groups have elements that have as centraliser $Spin(k_1) \times Spin(k_2)$ with both $k_i \geq 7$. By constructing a non-trivial triple in each of the $Spin(k_i)$-factors, and combining this with the holonomy that gave the subgroup one finds a non-trivial quadruple.

For rank reducing theories on the 5–torus, one may take a gauge group $G$ with an element $\Omega$ whose centraliser allows the above construction on the 4–torus. The simple groups $Spin(n \geq 31)$ have elements with centraliser $Spin(k_1) \times Spin(k_2)$ with both $k_i \geq 15$. Constructing a non-trivial quadruple in each factor gives a non-trivial quintuple. The exceptional group $E_8$ also allows a non-trivial quintuple. It has an element with centraliser $Spin(16)$, and in $Spin(16)$ a non-trivial quadruple can be constructed.

One may extend to tori of arbitrary dimension, but only with $Spin(n)$ groups of sufficiently high rank. On an $n$–torus one may achieve rank reduction with the gauge group $Spin(m)$ with $m \geq 2^n - 1$. These groups have elements with centraliser $Spin(k_1) \times Spin(k_2)$ with both $k_i \geq (2^{n-1} - 1)$. In both factors non-trivial $(n-1)$–tuples should be constructed to achieve the desired result.

## 4.2 Non-simply connected subgroups

We now return to the 3–torus. Consider a simple and simply connected group $G$. We need an element $G$ that has as its centraliser a non-simply connected subgroup. The analysis can be done with the aid of the extended Dynkin diagram of the group. The nodes of the extended diagram correspond to the simple roots, and one extended root, which is minus the highest root. We will also need the root integers and coroot integers, defined by the relations

$$g_0 \alpha_0 + g_i \alpha_i = 0 \qquad h_0 \alpha_0^\vee + h_i \alpha_i^\vee = 0$$

with $g_0 = h_0 = 1$. The $g_i$ and $h_i$ are always bigger than zero. As $\alpha_i^\vee = 2\alpha_i/\langle \alpha_i, \alpha_i \rangle$, the $g_i$ and $h_i$ are related by $g_i = 2h_i/\langle \alpha_i, \alpha_i \rangle$. Note that $g_i \alpha_i = h_i \alpha_i^\vee$.

A single element $g$ of the group may, after conjugation, be taken to on the maximal torus, in which case it can be written as

$$g = \exp\{2\pi i h_\beta\}, \tag{4.2.1}$$

with $h_\beta$ an element of the CSA. Define a basis $\omega_i$ for the coweight lattice by $\langle \alpha_i, \omega_j \rangle = \delta_{ij}$, with $\alpha_i$ the simple roots of the group. The coweights form a complete basis, and hence we can expand $\beta = \sum_i s_i \omega_i$, with $s_i = \langle \alpha_i, \beta \rangle$. It is also convenient to define $s_0 = \langle \alpha_0, \beta \rangle + 1$. We see that

$$s_0 + \sum_i s_i g_i = 1 \tag{4.2.2}$$

The group element $g$ does not define a unique $\beta$. Adding a coroot to $\beta$

$$\beta \sim \beta + \alpha^\vee, \tag{4.2.3}$$

leaves $g$ invariant. Conjugation with suitable group elements implements the Weyl reflections, which act on $\beta$ as

$$\beta \sim \beta - \langle \alpha, \beta \rangle \alpha^\vee \tag{4.2.4}$$

This does not leave $g$ invariant but transforms it to an equivalent element. These ambiguities can be fixed by choosing a suitable fundamental domain under all these transformations. Such a fundamental domain is called an alcove. We take without proof the assertion from [26] that all $s_i$ can be chosen such that $s_i \geq 0$. The formula (4.2.2) then implies that all $s_i \leq g_i^{-1}$. This defines a specific alcove, which is called the fundamental alcove.

For a given $\beta$ we may now define the set of indices $I = \{i : s_i = 0\}$. Consider the ladder operators $e_\alpha$ of the Lie-algebra of $G$. The $\alpha$ can be expanded in the simple roots, as $\sum_k c_k \alpha_k$, with all $c_k$ integer. By direct calculation one immediately verifies that only those $e_\alpha$ commute with the group element $g$ that have $c_k = 0$ if $k \notin I$. In particular, $e_{\alpha_k}$ with $\alpha_k$ a simple root only commutes with $g$ if $k \in I$.

From the direct calculation it immediately follows that the centraliser $Z(g)$ of $g$ can be obtained as follows. The semi-simple part $Z_{ss}(g)$ of $Z(g)$ has as its Dynkin diagram the diagram obtained by erasing all nodes $i$ from the extended Dynkin diagram that have $s_i \neq 0$. The full centraliser $Z(g)$ is obtained by adding to the semi-simple part $U(1)$-factors, such that the rank of the centraliser equals the rank of the original group.

We now concentrate on the semi-simple part $Z_{ss}(g)$ of $Z(g)$. The root lattice of $Z_{ss}(g)$ is a sublattice of the root lattice of $G$. It is generated by $\alpha_i$ with $i \in I, i \neq 0$, and, if $s_0$ was zero,

$$\beta_0 = \sum_{i \notin I} g_i \alpha_i \tag{4.2.5}$$

Similarly, the coroot lattice of $Z_{ss}(g)$ may be obtained as a sublattice of the coroot lattice of $G$. This lattice is generated by $\alpha_i^\vee$ with $i \in I, i \neq 0$, and, if $s_0$ was zero,

$$\beta_0^\vee = \sum_{i \notin I} h_i \alpha_i^\vee = \beta_0 \tag{4.2.6}$$

Now suppose $k = \gcd(h_{i \notin I}) > 1$. We have the result

$$\exp\left(\frac{2\pi i n}{k} h_{\beta_0^\vee}\right) = \mathbb{1} \qquad n \in \mathbb{Z} \tag{4.2.7}$$

because $\beta_0^\vee/k$ corresponds to a coroot of the gauge group $G$. $\beta_0^\vee/k$ is however not a coroot of $Z_{ss}(g)$. That it nevertheless gives the identity in $Z(g)$ must mean that it is a coweight, but coweights can only represent the identity if the group is not simply connected. The fundamental cell of the coroot lattice of the semi-simple part of the centraliser contains $k$ of these points which are $\beta_0^\vee n/k$ for $n = \{0, \ldots k-1\}$. This implies that the $Z_{ss}(g)$ is $k$-fold connected.

The centraliser obtained this way was found by erasing nodes of the extended Dynkin diagram that had associated to them coroot integers divisible by $k$. To obtain a "minimal" non-simply connected group, one may choose an element that has as the semi-simple part of its centraliser the group represented by all nodes with coroot integers not divisible by $k$. This leads to table 4-1. The entries in the table are familiar, as they are the subgroups used in the previous chapter to construct non-trivial flat connections on the 3–torus. Perhaps the only unfamiliar entry is the last one in the table: $E_8$ with subgroup $SU(6) \times SU(3) \times SU(2)/\mathbb{Z}_6$, whereas we used $SU(3)^2 \times SU(2)^2/\mathbb{Z}_6$ in the previous chapter. Note however that $\mathbb{Z}_6$ is isomorphic to $\mathbb{Z}_2 \times \mathbb{Z}_3$. There exists a $SU(2) \times SU(3)$-subgroup of $SU(6)$, where the factors are embedded such that the $\mathbb{Z}_2$-centre of $SU(2)$ and the $\mathbb{Z}_3$-centre of $SU(3)$ form the $\mathbb{Z}_6$-centre of $SU(6)$ (obtained for example by tensoring the **2** of $SU(2)$ with the **3** of $SU(3)$ which gives a 6-dimensional matrix group that is a subgroup of the **6** of $SU(6)$)

| Group | $k$ | subgroup | rank |
|---|---|---|---|
| $SO(2N+7)$ | 2 | $SU(2)^3/\mathbb{Z}_2$ | $N$ |
| $SO(2N+8)$ | 2 | $SU(2)^4/\mathbb{Z}_2$ | $N$ |
| $G_2$ | 2 | $SU(2)^2/\mathbb{Z}_2$ | 0 |
| $F_4$ | 2 | $SU(2)^3/\mathbb{Z}_2$ | 2 |
| $F_4$ | 3 | $SU(3)^2/\mathbb{Z}_3$ | 0 |
| $E_6$ | 2 | $SU(2)^4/\mathbb{Z}_2$ | 3 |
| $E_6$ | 3 | $SU(3)^3/\mathbb{Z}_3$ | 0 |
| $E_7$ | 2 | $SU(2)^4/\mathbb{Z}_2$ | 3 |
| $E_7$ | 3 | $SU(3)^3/\mathbb{Z}_3$ | 1 |
| $E_7$ | 4 | $SU(4)^2 \times SU(2)/\mathbb{Z}_4$ | 0 |
| $E_8$ | 2 | $SU(2)^4/\mathbb{Z}_2$ | 4 |
| $E_8$ | 3 | $SU(3)^3/\mathbb{Z}_3$ | 2 |
| $E_8$ | 4 | $SU(4)^2 \times SU(2)/\mathbb{Z}_4$ | 1 |
| $E_8$ | 5 | $SU(5)^2/\mathbb{Z}_5$ | 0 |
| $E_8$ | 6 | $SU(6) \times SU(3) \times SU(2)/\mathbb{Z}_6$ | 0 |

Table 4-1. Minimal non-simply connected subgroups

## 4.3 Twisted boundary conditions

We have achieved the goal of classifying all group elements (and therefore the holonomies) that will lead to a non-simply connected centraliser. We wish to impose twisted boundary conditions in this centraliser, which amounts to finding two elements that commute up to an element of the centre. Again the analysis can be conveniently done with the aid of extended Dynkin diagrams.

Consider the elements of the centre of a group $G$. These can all be taken to be in a maximal torus, and hence can be written as in the previous section, as $\exp\{2\pi i h_\beta\}$ with $\beta = \sum_i s_i \omega_i$. In particular, the identity corresponds to $\beta = 0$ (and $s_0 = 1$, $s_i = 0$). If there are additional elements in the centre, they may be easily found from the extended Dynkin diagram: if a single node can be erased from the extended diagram to give the diagram of the group, then there is an element of the centre associated to that node, which can be found by exponentiating $h_{\omega_i}$, $\omega_i$ being the relevant coweight. Note that this implies that the coroot integer associated to this node must equal 1. If there are non-trivial centre elements, then there exists a diagram automorphism of the extended Dynkin diagram such that this diagram automorphism maps the node corresponding to the extended root (which corresponds to the identity) to another node corresponding to an element of the centre. The diagram automorphism corresponds to a symmetry of the root lattice, and therefore of the algebra. As it does involve the extended node, it cannot be a symmetry of the Dynkin diagram itself and therefore not be an outer automorphism. Hence it is an inner automorphism of the root lattice, and therefore it corresponds to some element of the Weyl group, say $\theta$. The Weyl reflections can be implemented by conjugating with suitable group elements, and we will denote the element corresponding to Weyl transformation $\theta$ as $\Theta$. As $\theta$ acts on the diagram, it permutes the set of simple roots and the extended root among each other, and we write $\theta(\alpha_i) = \alpha_{p(i)}$, $p$ standing for the permutation that is induced on the indices. $\theta$ preserves geometrical relations between the roots, and therefore $g_{p(i)} = g_i$

We assume that $\theta$ is non-trivial. Then the fundamental alcove is not mapped to itself under the $\theta$, but it is not hard to show that if $\beta$ is an element of the fundamental alcove, then $\theta(\beta) + \omega_{p(0)}$ is also an element of the fundamental alcove. Write

$$\beta = \sum_{i=1}^{r} s_i \omega_i \qquad \theta(\beta) + \omega_{p(0)} = \sum_{i=1}^{r} s'_i \omega_i \tag{4.3.1}$$

$$s_0 + \sum_i s_i g_i = 1 \qquad s'_0 + \sum_i s'_i g_i = 1 \tag{4.3.2}$$

Now observe that

$$s'_i = <\alpha_i, \theta(\beta) + \omega_{p(0)}> = <\theta^{-1}(\alpha_i), \beta> + <\alpha_i, \omega_{p(0)}> = <\alpha_{p^{-1}(i)}, \beta> + <\alpha_i, \omega_{p(0)}> \tag{4.3.3}$$

for $i \neq 0$ where we used that $\theta$ is an element of the Weyl group, which is a subgroup of the orthogonal group $O(r)$, with $r$ the rank of $G$.

For $i = p(0)$ one finds

$$s'_{p(0)} = <\alpha_0, \beta> + <\alpha_{p(0)}, \omega_{p(0)}> = (s_0 - 1) + 1 = s_0 \tag{4.3.4}$$

For $i = 0$ we have

$$s'_0 - 1 = <\alpha_0, \theta(\beta) + \omega_{p(0)}> = <\alpha_{p^{-1}(0)}, \beta> + <\alpha_0, \omega_{p(0)}> = s_{p^{-1}(0)} - g_{p(0)} \tag{4.3.5}$$

Otherwise we have

$$s'_i = <\alpha_{p^{-1}(i)}, \beta> = s_{p^{-1}(i)} \rightarrow s'_{p(i)} = s_i \tag{4.3.6}$$

As $g_{p(0)} = g_0 = 1$, we may simply summarise the above equations as $s'_{p(i)} = s_i$, where $i$ is now allowed to take any value from $0, 1 \ldots, r$. As $0 \leq s'_{p(i)} \leq g_{p(i)}^{-1} = g_i^{-1}$, all coordinates lie within the allowed range and hence $\theta(\beta) + \omega_{p(0)}$ is in the fundamental alcove.

So the fundamental alcove is mapped to itself, but a generic point in the alcove is not. An invariant point has to obey the condition

$$\theta(\beta) + \omega_{p(0)} = \beta \rightarrow s_i = s_{p(i)} \tag{4.3.7}$$

Invariant points always exist. In the extreme case (4.3.7) implies that all $s_i$ are equal, and then (4.2.2) fixes the values of the $s_i$ giving a unique solution. This may be realised for the groups $SU(n)$. In the generic case invariant points are not unique.

Assume $\beta$ to be an invariant point, and define $X = \exp\{2\pi i h_\beta\}$, and the centre element $Z = \exp\{2\pi i \omega_{p(0)}\}$. Also remembering that the transformation $\theta$ is implemented on the algebra by commuting with $\Theta$, we may exponentiate (4.3.7) to

$$\Theta X Z \Theta^{-1} = X \tag{4.3.8}$$

Now define $Y = Y'\Theta$, with $Y'$ an element of the maximal torus that is invariant under $\Theta$. Then clearly

$$YXZY^{-1} = X \rightarrow XY = ZYX \tag{4.3.9}$$

We have therefore found elements $X$ and $Y$ that commute up to the element $Z$ of the centre. That all elements $X$ and $Y$ satisfying (4.3.9) are up to conjugation equivalent to a pair obtained in the above manner is harder to prove, and we leave this to the references [47] [6].

## 4.4  The Chern-Simons invariant

In this section we will explain how to calculate the Chern-Simons invariant for a given triple, following [6]. It is convenient to switch to form-notation. The gauge field is represented by the 1-form $A = -\mathrm{i}A_\mu \mathrm{d}x^\mu$, and the field strength by the two-form $F = \mathrm{d}A + A^2 = -\frac{\mathrm{i}}{2}F_{\mu\nu}\mathrm{d}x^\mu \mathrm{d}x^\nu$, where we abbreviated $A \wedge A$ to $A^2$. The four form $F^2$ can be written locally as an exact form, as $\mathrm{tr}(F^2) = 16\pi^2 h \mathrm{d}CS(A)$ ($h$ is the dual Coxeter number), where $CS(A)$ is known as the Chern-Simons form

$$CS(A) = -\frac{1}{16\pi^2 h}\mathrm{tr}(A\mathrm{d}A + \frac{2}{3}A^3) \tag{4.4.1}$$

This is a three-form, and can be integrated over the 3–torus.

We cannot not directly calculate the Chern-Simons form, as we have no expression for the gauge field. Instead we will use an indirect method. We will construct a gauge field having the right holonomies in steps. Define the one form

$$A' = g^{-1}(A + \mathrm{d})g \tag{4.4.2}$$

where $g$ is some element of the gauge group $g$, and $A$ is another one form. Then

$$CS(A') - CS(A) = CS(g^{-1}\mathrm{d}g) + \mathrm{d}\alpha_2 \tag{4.4.3}$$

where $\alpha_2$ is a two-form given by

$$\alpha_2 = -\frac{1}{16\pi^2 h}\mathrm{tr}((\mathrm{d}g)g^{-1}A) \tag{4.4.4}$$

Formula (4.4.3) can be found in various places in the literature (e.g. [2]), and can be verified by a brute force calculation, or more advanced methods.

We will now construct a flat connection with the appropriate holonomies. Let the holonomies be $\Omega_1$, $\Omega_2$, and $\Omega_3$. As we saw previously, the centraliser $Z(\Omega_1)$ may not be simply connected. Define $\tilde{Z}(\Omega_1)$ to be the simply connected covering of $Z(\Omega_1)$, and $\tilde{\Omega}_i$ to be the lifting of $\Omega_i$ to $\tilde{Z}(\Omega_1)$. Define $z$ as

$$z = \tilde{\Omega}_2 \tilde{\Omega}_3 \tilde{\Omega}_2^{-1} \tilde{\Omega}_3^{-1} \tag{4.4.5}$$

$z$ and $\tilde{\Omega}_1$ are two commuting elements of $\tilde{Z}(\Omega_1)$ and may therefore be put on the maximal torus. We can then write them as exponentials of elements of the CSA, $z = \exp\{2\pi \mathrm{i} h_\zeta\}$, and $\Omega_1 = \exp\{2\pi \mathrm{i} h_\xi\}$.

Take now a two torus $T$, with coordinates $(x_2, x_3)$ with periods $x_i \sim x_i + 1$. On this 2-torus we define a flat connection $A$ for the gauge group $Z(\Omega_1)$ such that the holonomies are $\Omega_2$ and $\Omega_3$. The holonomies define the boundary conditions on the edges of the square $0 \leq x_i \leq 1$, but there may be obstructions extending to the interior of the square, as $Z(\Omega_1)$ is not simply connected. Define an open disk $D$

$$D = \{(x_2, x_3) \in T : (x_2 - \tfrac{1}{2})^2 + (x_3 - \tfrac{1}{2})^2 < (R+\epsilon)^2\} \tag{4.4.6}$$

with $0 < R < \tfrac{1}{2}$ and $\epsilon > 0$ with the limit $\epsilon \downarrow 0$ understood. Define also $T_0$ as

$$T_0 = \{(x_1, x_2) \in T : (x_2 - \tfrac{1}{2})^2 + (x_3 - \tfrac{1}{2})^2 > (R-\epsilon)^2\} \tag{4.4.7}$$

The intersection $B$ of $D$ and $T_0$ is a circular strip of vanishing width. Along this circle we define the angular coordinate $0 \leq \theta < 2\pi$.

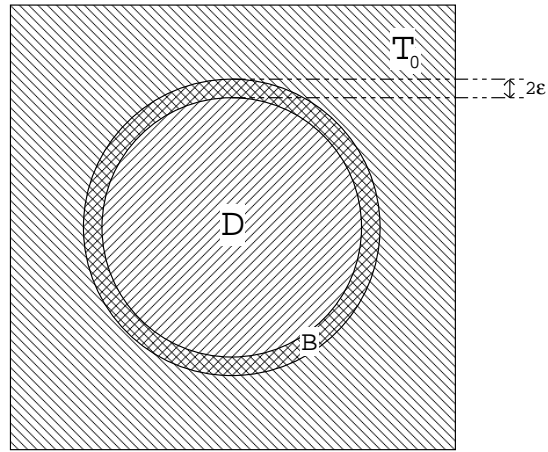

Figure 4-1. Patches $D$, and $T_0$ for the 2–torus

We now define the gauge connection. On $D$ we set $A = 0$. On $T_0$ we define a flat connection such that the holonomies in the $x_2$ and $x_3$-direction correspond to $\Omega_2$ and $\Omega_3$. By a suitable gauge transformation we may take $A$ such that on the edge of $T_0$ it is given by

$$A|_{\partial T_0} = ih_\zeta d\theta \tag{4.4.8}$$

The two definitions of $A$ on the different patches can be glued together, by the transition function $\exp\{i\theta h_\zeta\}$ defined on $B$.

We now move on to the 3–torus $S^1 \times T$, where on $S_1$ we define the coordinate $0 \leq x_1 < 1$. Define the gauge function $g(x_1) = \exp\{2\pi i x_1 h_\xi\}$, and the connection $A'$ over the 3–torus with

$$A'(x_1, x_2, x_3) = g^{-1}(x_1)(A(x_2, x_3) + d)g(x_1) \tag{4.4.9}$$

$A'$ has holonomy $\Omega_1$ in the $x_1$–direction, $\Omega_2$ in the $x_2$–direction and $\Omega_3$ in the $x_3$–direction.

The 3–torus $S^1 \times T^2$ is covered by the patches $S^1 \times D$, and $S^1 \times T_0$. First consider the patch $S^1 \times D$. On this patch we have $A = 0$, $A' = 2\pi i h_\xi dx_1$. By direct calculation one immediately finds $CS(A') = CS(A) = 0$. On the patch $S^1 \times T_0$ we use the formula (4.4.3) to calculate the Chern-Simons form. On this patch we still have $CS(A) = 0$, as $CS(A)$ is a three-form, but $A$ only depends on two coordinates, so the differential $dx_3$ cannot occur. Also $CS(g^{-1}dg) = (2\pi i)^3 CS(h_\xi dx_1) = 0$, and (4.4.3) then reduces to

$$CS(A') = -\frac{1}{16\pi^2 h} d(\text{tr}((dg)g^{-1}A)) \tag{4.4.10}$$

The Chern-Simons invariant reduces to the integral over the surface $S^1 \times \partial T_0$:

$$\int CS(A') = -\frac{1}{16\pi^2 h} \int_0^1 dx_1 \int_0^{2\pi} d\theta \, \text{tr}(2\pi i h_\xi i h_\zeta) = <\zeta, \xi> \tag{4.4.11}$$

For the commuting triples we constructed earlier this number is not hard to calculate. $\Omega_1 = \exp\{2\pi i h_\beta\}$ with $\beta = \sum_i s_i \omega_i$, and hence

$$\xi = \sum_i s_i \omega_i \tag{4.4.12}$$

The centre elements of the non-simply connected subgroup are represented by (see (4.2.6) and (4.2.7))

$$\frac{n}{k}\beta_0^\vee = \frac{n}{k}\sum_{i \notin I} h_i \alpha_i^\vee = \frac{n}{k}\sum_{i \notin I} g_i \alpha_i \tag{4.4.13}$$

These were coweights for the semi-simple part of the centraliser $Z(g)$, and hence our $\zeta = n\beta_0^\vee/k$ for some integer $n$. Then

$$CS(A') = <\xi, \zeta> = \frac{n}{k} < \sum_i s_i \omega_i, \sum_{j \notin I} g_j \alpha_j > \tag{4.4.14}$$

As $0 \in I$, we can use $\langle \omega_i, \alpha_j \rangle = \delta_{ij}$, to obtain

$$CS(A') = \frac{n}{k}\sum_{i \notin I} s_i g_i \tag{4.4.15}$$

Finally, as $I$ was defined to consist of those $i$ for which $s_i = 0$ we may write

$$CS(A') = \frac{n}{k}\sum_i s_i g_i = \frac{n}{k} \tag{4.4.16}$$

which is a fractional number if $n$ is not a multiple of $k$.

    The fractionality of the Chern-Simons invariant has a number of important consequences. First, it provides a check that the various disconnected components of the moduli space of flat connections are indeed disconnected, as the Chern-Simons invariants associated to different components can be shown to be different. Second, it hints at the existence of instanton-like solutions, describing the tunneling between different components in the moduli space. Also the Chern-Simons invariant plays a role in anomaly cancellation for the heterotic string, which is important for the applications discussed in chapter 6.

# 5 Orientifolds and realisations of flat connections

In the case of a compactification of Yang-Mills theory with orthogonal gauge group on a 3–torus with periodic boundary conditions, Witten found a new vacuum solution by means of a construction involving D-branes and orientifolds. The analysis of Borel, Friedman and Morgan shows that if one relaxes the requirement of periodicity, also the other classical groups (unitary, symplectic, orthogonal) allow extra vacuum solutions [6]. All classical groups can be embedded in string theory by including Chan-Paton degrees of freedom, and gauging worldsheet parity for the orthogonal and symplectic cases. In T-dualised descriptions, this will give configurations of orientifolds and D-branes. The solutions obtained by Borel et al. for the 3–torus also allow descriptions of this kind. To construct these, we first consider the T-dual description of a geometrical object, the crosscap. We then move on to compactifications on a 2–torus with twisted boundary conditions, as first proposed by 't Hooft [23] [24], and analysed in detail by Schweigert in [47]. We finally consider compactifications on the 3–torus, reproducing some results from [6].

In this chapter we will not be bothered by the consistency requirements of string theory, which allow only $SO(32)$ as Chan-Paton gauge group. This will be postponed until the next chapter, where we will find that some of the vacua considered here can be used as string theory vacua, but most will suffer from (sometimes subtle) inconsistencies.

## 5.1 The T-dual of a crosscap

We start by considering $U(n)$ theory on a circle. $U(n)$ can be embedded in an open string theory by attaching Chan-Paton charges on the ends of oriented strings. Compactifying this string theory on a circle and applying a T-duality transformation, we obtain a configuration of $n$ D-branes that are transverse to a dual circle, each intersecting the dual circle in one point. The location of the D-branes is controlled by the holonomy $\Omega_1$ along the circle in the original theory. We are interested in configurations with discrete symmetries. The discrete symmetries of the circle are the shift symmetries, shifting the circle by an angle $2\pi q$ with $q$ a rational number, and the order 2 reflection symmetry. Only for specific choices of the original holonomy will the D-brane configuration respect one or some of these symmetries. In this section, we are interested in the reflection symmetry.

The reflection on the circle has two fixed points, which will be taken to be at $X = 0$ and $X = \pi R$ ($X$ being the coordinate along the circle, and $2\pi R$ its circumference). This can always be arranged: in the original theory we had a holonomy in $U(n)$, which is locally equivalent to $U(1) \times SU(n)$. The holonomy for the $U(1)$-factor can be chosen arbitrarily, since it does not couple to anything. In the dual theory this corresponds to an overall translation, which we use to set the coordinates of the fixed points to the above values.

Now compactify in addition on another circle of radius $R'$ with holonomy $\Omega_2$ along this circle. The standard formalism assumes holonomies that can be diagonalised within the group.

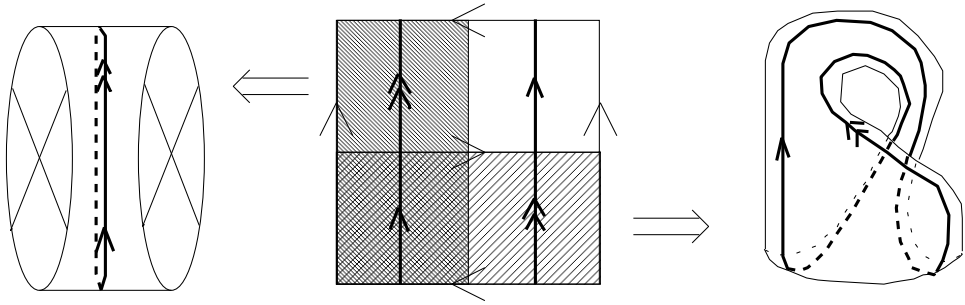

Figure 5-1. The Klein-bottle: a double cover of the Klein-bottle, arrows indicating the direction of identifications (middle); an attempt to draw the standard representation of the Klein-bottle, obtained by taking the lower half of the double cover as fundamental domain (right); the cylinder with two crosscaps, obtained by taking the left half of the fundamental domain (left). We have also drawn an example of a brane in all three pictures (depicted twice on the double cover) as it is positioned after the first T-duality

In case we have the above $\mathbb{Z}_2$ symmetry, we may consider a holonomy that includes the $\mathbb{Z}_2$ reflection. Gluing the circle to a reflected circle upon going around the second cycle, one does not obtain a 2–torus, but a Klein bottle. Instead of a non-trivial line bundle over a circle, we will represent the Klein bottle here as a cylinder of length $\pi R$, circumference $4\pi R'$, bounded by two crosscaps at the end, the crosscap being an identification over half the period of the circle.

The D-branes are wrapped around the cylinder, parallel to the crosscaps. There are two possibilities, controlled by the holonomy $\Omega_1$ in the original theory. D-branes in the bulk (away from the crosscaps) represent branes that were reflected into their image. In this representation, D-brane and image are represented as one brane (which is in this sense a brane pair). In the original theory there can also be D-branes at the fixed point(s) of the $\mathbb{Z}_2$-reflection. In this representation of the Klein-bottle, they are located at the crosscap. Under a smooth deformation of the original holonomy, only even numbers of D-branes can move away from the crosscap. Hence for $U(n)$ with $n$ odd there is at least one brane fixed under the $\mathbb{Z}_2$ reflection and therefore stuck to a crosscap. For $n$ even there are two possibilities: the number of branes at each crosscap is either even or odd. In the latter situation there is at least one brane at each crosscap.

The above is reminiscent of the situation for orientifold planes. For orientifolds, a brane and an image brane on the double cover are mapped to a brane-pair in the orientifold. There is also the possibility of single branes being stuck at an orientifold plane (in the case of $O^-$ planes. By $O^-$ we denote the orientifold plane that gives orthogonal gauge symmetry, and $O^+$ is an orientifold plane that gives symplectic gauge symmetry).

The above configuration can be interpreted in terms of the original gauge theory. $U(n)$ is locally $U(1) \times SU(n)$, and the $U(1)$ background is fixed. For $n > 2$, $SU(n)$ possesses an outer automorphism, which, in a suitable representation, corresponds to complex conjugation. We will be working in the fundamental representation, and denote complex conjugation as $C$ with

action
$$C : U \to U^* \quad U \in SU(n) \tag{5.1.1}$$

One can extend this action to $U(n)$ as $C$ also has a simple action on $U(1)$, and now one may also extend to $n \leq 2$. One normally considers holonomies taking values in the gauge group, which corresponds to combining a translation in space with the action of an inner automorphism (i.e. a conjugation) on the group. One may also consider a holonomy that corresponds to an outer automorphism, and this is precisely what we are doing in the above. The outer automorphism $C$ can be combined with an inner one, say conjugation with a group element $A$. To avoid ambiguities we require that $AC = CA$, which is true if $A$ is real, that is $A \in O(n)$. The holonomy $\Omega_2$ combines the action of $C$ with conjugation with $A$, and we denote it as $\Omega_2 = AC$, with $A$ in the fundamental representation of $U(n)$, and $C$ the operator that implements complex conjugation. The holonomy $\Omega_1$ is an "ordinary" holonomy, and we write $\Omega_1 = B$, with $B$ an element of $U(n)$ in the fundamental representation. $\Omega_1$ should commute with $\Omega_2$, which is solved by taking $B$ commuting with $A$ and $B \in O(n)$. Continuous variation of the $U(1)$-background is incompatible with complex conjugation; in the D-brane picture this corresponds to the fact that a global translation on the D-branes is incompatible with the reflection for generic cases.

By conjugation with $O(n)$ matrices, we may transform $A$ and $B$ to a block diagonal form with $2 \times 2$ blocks of the form
$$\begin{pmatrix} \cos\phi & -\sin\phi \\ \sin\phi & \cos\phi \end{pmatrix} \tag{5.1.2}$$

on the diagonal, and some 1's and $-1$'s as remaining diagonal entries. In the following, we will take $A$ and $B$ to be of this standard form.

We wish to T-dualise the cylinder with the two crosscaps in the direction of the circle. Ignoring the crosscaps one would roughly expect this to lead to a dual theory on a cylinder. The inclusion of the crosscaps can be analysed by examining the symmetries of the original theory.

$A$ and $B$ are elements of the vector representation of $O(n)$, and in particular their eigenvalues occur in pairs: If $\exp i\phi$ is an eigenvalue, then so is $\exp -i\phi$. The ordering is unimportant as there are symmetries that allow the exchange of $\exp i\phi$ and $\exp -i\phi$, for every $\phi$ separately. If $A$ and $B$ where holonomies for an $O(n)$-theory in an orientifold description this symmetry would be simply the orientifold projection itself. This suggests that also for this $U(n)$-theory the dual should be some orientifold.

The radius of the dual theory is expected to be $1/(2R')$, half the "normal" radius. The coordinates of the D-branes in this theory reflect the eigenvalues of the holonomies in the original theory. Naively mapping these onto the dual circle suggests a circle of radius $1/R'$, which seems to lead to a contradiction. The resolution to this paradox lies in the presence of the operator $C$. If $\Omega_i$ are the holonomies for a certain theory, then the holonomies $\Omega'_i = g\Omega_i g^{-1}$ with $g$ some element of $U(n)$ represent the same theory. Consider the set of diagonal matrices with entries $\pm 1, \pm i$ on the diagonal that commute with the $A$ and $B$. Taking $g$ to be a specific element from this set has the effect

$$g : (B, AC) \to (gBg^{-1}, gACg^{-1}) = (B, Ag^2C).$$

This leaves $\Omega_1$ and $\Omega_2$ in standard form, but with $A$ replaced by $Ag^2$. Hence in this construction, $A$ and $Ag^2$ have to be identified. $g^2$ is an element of $O(n)$ that commutes with $A$. By a suitable choice of $g^2$, any eigenvalue $\exp i\phi$ of $A$ can be mapped to $-\exp i\phi = \exp i(\phi + \pi)$. Therefore the periods of the circle and the eigenvalues of the holonomies match, and the dual theory is indeed an orientifold. Now we examine the orientifold planes.

To find maximal symmetry groups we set $B$ to either $\mathbb{1}$ or $-\mathbb{1}$. If we set $A = \mathbb{1}$, then the surviving symmetry group is the subgroup of $U(n)$ that is invariant under $C$, which is $O(n)$. For another maximal symmetry group, assume $n$ to be even for a moment and take $A$ to be of block diagonal form with $2 \times 2$ blocks

$$\begin{pmatrix} 0 & 1 \\ -1 & 0 \end{pmatrix}, \tag{5.1.3}$$

on the diagonal, and call this matrix $J$. The unbroken symmetry group is then the subgroup of $U(n)$ of matrices $U$ that commute with $JC$. $C$ transforms $U \to U^*$, but as $U$ is unitary, $U^* = (U^{-1})^T$, where $T$ is for transposed. We may then rewrite the invariance condition to

$$U^T J U = J \tag{5.1.4}$$

which, together with the unitarity condition defines the symplectic group. It is obvious how to generalise to arbitrary holonomies, and odd $n$: a holonomy with $k$ blocks $\mathrm{diag}(1,1)$ and $k'$ blocks (5.1.3) gives rise to $O(k) \times Sp(k')$-symmetry, completed with some $U(m)$-factors, whenever $m$ eigenvalues not equal to $\pm 1, \pm i$ coincide.

The above $U(n)$ theory on a Klein-bottle is thus T-dual to an orientifold $T^2/\mathbb{Z}_2$, where two of the four orientifold planes are of $O^-$-type and two are of $O^+$-type. The holonomy $B = \pm \mathbb{1}$ distinguishes two parallel configurations of one $O^+$ and one $O^-$-plane, whereas in the theory on the Klein bottle it distinguished the two parallel crosscaps. As a rule of thumb one may therefore state that the dual of the crosscap is a configuration of one $O^+$ and one $O^-$-plane. This fits with the usual charge assignments: opposite charges for the $O^+$ and $O^-$ plane versus no charge for the crosscap. The original theory may have had isolated D-branes at the crosscaps. In the dual theory the isolated branes should be located at the $O^-$-planes, since the $O^+$ planes cannot support isolated branes. Examining the holonomies that will lead to such a situation indeed shows this to be the case.

These ideas are independent of whether D-branes are static in the background, or used as "probes". The above orientifold background is identical to a IIB-orientifold encountered in [55], but consistency requires absence of D-branes. This suggests to regard this model as a "$U(0)$-theory" with a holonomy that includes complex conjugation. Its duality to IIA on a Klein-bottle is obvious from the above. Considering various limits one may also reach other theories discussed in [55] and [14].

We interpreted the Klein-bottle theory as created by combining a translation with an outer automorphism (complex conjugation). Outer automorphisms can always be divided even if not combined with a translation. Dividing a $U(n)$ group by its outer automorphism will give a symplectic or orthogonal theory, where the ambiguity comes from the fact that an outer automorphism may be combined with an inner automorphism to give another outer automorphism. In our case one may consider, instead of $C$, an operator $AC$ with $A$ an element of $U(n)$. One should require $A$ to commute with $C$ and therefore $A \in O(n)$. Consistency also requires that $AC$ acting on the group squares to the identity. The group action on itself is always in the adjoint representation, and hence we have the possibilities $A^*A = (A^{-1})^T A = \pm \mathbb{1}$, so $A$ is either symmetric or antisymmetric. One may now copy a standard textbook derivation [43] to show that this leads to either symplectic or orthogonal groups. The reasoning is parallel to that for orientifolds, so one may interpret the introduction of an orientifold plane as quotienting the gauge group by an outer automorphism. With this point made, which is not stressed in the literature, we may also say that in the above a translation is combined with an orientifold action, as the theory in the last chapter of [55] was originally motivated.

## 5.2 Twist in unitary groups

The $U(n)$-gauge group allows a second form of twist. The circle also has discrete shift symmetries by angles $2\pi q R$, with $q$ a rational number, which can be chosen on the interval $[0,1)$. Choose a configuration of the $n$ D-branes that respects one or some of these shifts. In that case the number $q$ is a multiple of $1/n$. Now compactify on a second circle with a holonomy that includes the shift over $2\pi q R$. This results in a theory on the 2–torus, not with $n$ D-branes, but with $k = \gcd(qn, n)$ branes, wrapped $n/k$ times around the torus. This theory is naturally interpreted as a $U(n)$-theory with twisted boundary conditions [23] [24]. We will not have much new to say on this theory, but mention it for completeness, and to point out some effects that are encountered in other theories as well.

Let $X_1$ and $X_2$ be the coordinates transverse resp. parallel to the branes. Then this 2–torus is $\mathbb{R}^2$ with coordinates $(X_1, X_2)$, quotiented by a lattice generated by the vectors

$$e_1 = 2\pi(qR_1, R_2) \qquad e_2 = 2\pi(R_1, 0) \tag{5.2.1}$$

Now transform to an $SL(2,\mathbb{Z})$-equivalent form. Let $n' = n/k$. Then $n'$ and $qn'$ are integer, and $\gcd(qn', n') = 1$. Hence the equation

$$n'a + qn'b = 1, \qquad a, b \in \mathbb{Z}$$

has a solution, which can be found using Euclid's algorithm. The solution is not unique as $a \to a + mqn'$, $b \to b - mn'$ with integer $m$ gives another solution. Use this arbitrariness to select a $b$ such that $0 \le b < n'$. Then change the fundamental domain of the torus by using the $SL(2,\mathbb{Z})$ transformation

$$\begin{pmatrix} x' \\ y' \end{pmatrix} = \begin{pmatrix} n' & b \\ -qn' & a \end{pmatrix} \begin{pmatrix} x \\ y \end{pmatrix}, \tag{5.2.2}$$

where $(x, y) \in \mathbb{Z}$ are coordinates for the lattice vectors $xe_1 + ye_2$. Under this transformation the basisvectors transform as

$$2\pi(qR_1, R_2) \to 2\pi(0, n'R_2) \qquad 2\pi(R_1, 0) \to 2\pi(R_1/n', bR_2) \tag{5.2.3}$$

On this fundamental domain only $k$ D-branes (which are in a sense configurations of $n'$-tuples of branes) are visible. This is analogous to the two different representations of the Klein-Bottle in the previous section. The set-up is the one considered in [16], which is argued to lead to a Yang-Mills theory on a non-commutative torus in a suitable limit.

We may T-dualise our original theory back to an open string theory (with Neumann boundary conditions) on a torus, using the standard methods [19]. The resulting theory has a non-zero $B$-field (with $B = q$ in appropriate units) in the background, as our original theory does not live on a square torus.

Combining the discrete shift over $2\pi q R$, with the $\mathbb{Z}_2$ reflection does not lead to anything new. The resulting transformation is of the form

$$X \to -X + 2\pi q R$$

which is just a $\mathbb{Z}_2$-reflection, but with other fixed points. This is related to the $U(n)$ theory of the previous section by a trivial translation.

## 5.3  Twist in symplectic groups on the 2–torus

Symplectic groups can be realised in string theories by combining the Chan-Paton construction with the gauging of world sheet parity [43]. This gives a theory of unoriented strings. To complete the description of the theory one has to prescribe how world sheet parity acts on the Chan-Paton matrices. If the reflection of the world sheet is combined with the action of an anti-symmetric matrix on the Chan-Paton indices, the resulting theory will have symplectic gauge symmetry.

Compactifying this theory on a circle of radius $R_1$ and T-dualising leads to an oriented string theory, living on an interval $I = S^1/\mathbb{Z}_2$ of size $(2R_1)^{-1}$, bounded by two $O^+$-planes. For an $Sp(k)$-theory there will be $k$ D-brane pairs distributed along the interval. The two $O^+$-planes do not allow any freely acting shift. We will instead assume that the D-branes are distributed in a configuration that is invariant under the reflection that exchanges the two $O^+$-planes. For odd $k$ one brane-pair is fixed in the middle of the interval. Now compactify on another circle of radius $R_2$ with a holonomy that implements the $\mathbb{Z}_2$-reflection. The resulting compactification manifold is a Möbius strip with an $O^+$-plane as edge. For $k$ even half of the D-branes are exchanged with the other half on going around the circle. For $k$ odd half of $(k-1)/2$ pairs are exchanged with another $(k-1)/2$ and one brane-pair is fixed by the $\mathbb{Z}_2$ reflection. Another representation of the Möbius strip, is a cylinder of diameter $2R_2$ and length $(4R_1)^{-1}$. One end of the cylinder ends in a crosscap, the other end is formed by a single $O^+$-plane. On the cylinder there are $k/2$ D-branes pairs for $k$ even, and $(k-1)/2$ for $k$ odd in which case there is a brane-pair stuck at the crosscap.

This theory is a $U(2k)$-theory as described in section 5.1 with an extra orientifold plane inserted. The mirror symmetry of the orientifold plane turns the Klein-bottle into a Möbius-strip. Take the circle that is dual to the circle of radius $R_1$ and choose coordinates as follows: we will take the orientifold planes at $X = 0$ and $X = \pi/R_1$, and the fixed points of the $\mathbb{Z}_2$-reflection at $X = \pi/2R_1$ and $X = 3\pi/2R_1$. The description from the $U(n)$ theory has to be slightly modified, as the fixed points of the $\mathbb{Z}_2$ are no longer located at $X = 0$ and $X = \pi R$, as before. The action of the $\mathbb{Z}_2$-reflection can be interpreted in the original theory as accomplished by the operator $(-\mathbb{1})C$, which is complex conjugation combined with multiplying by $(-\mathbb{1})$. The symplectic theories one projects onto states invariant under $JC$, with $J$ the matrix

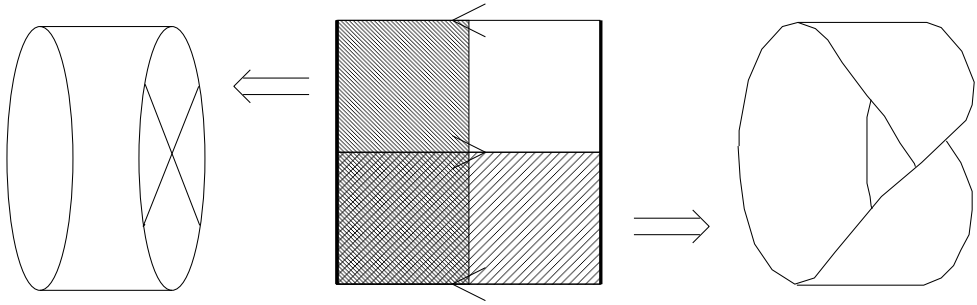

Figure 5-2. The Möbius-strip: a double cover of the Möbius-strip, arrows indicating the direction of identifications, fat lines the edges (middle); the standard representation of the Möbius strip, obtained by taking the lower half of the double cover as fundamental domain (right); the cylinder with one crosscap, obtained by taking the left half as fundamental domain (left)

composed of $2 \times 2$-blocks of the form (5.1.3), and the invariance condition is (5.1.4). In the orientifold projected theory, the operator $(-\mathbb{1})C$ is identified with $(-\mathbb{1})\text{Ad}J$, which has as action "conjugate with $J$ and multiply with $-\mathbb{1}$". Multiplying by $-\mathbb{1}$ is not an outer automorphism of $Sp(k)$ (in fact, the symplectic groups do not posses any outer automorphism at all), and it can be realised by conjugation, as we will show later.

With the appropriate symmetries realised, we can pass from the $U(n)$-theory to the symplectic theory as follows. We argued that the $U(n)$-theory had as its holonomies $(\Omega_1, \Omega_2) = (B, AC)$. Replace the operator $C$ by $(-\mathbb{1})C$, and then perform the orientifold projection. The resulting holonomies are then $(\Omega_1, \Omega_2) = (B, A(-\mathbb{1})\text{Ad}J)$. $\Omega_1$ and $\Omega_2$ do not commute, but anticommute. Their eigenvalues can be read of from $B$ and $A$, but we have to find a way to implement the action of $-\mathbb{1}$.

Anticommutativity of the holonomies is allowed in symplectic theories, provided all representations of $Sp(k)$ have trivial centre (this is the case for all representations one encounters in $Sp(k)$ string perturbation theory. These are the adjoint, which is the symmetric two-tensor; a $k(2k-1)-1$ dimensional representation which is the antisymmetric tensor with an extra singlet removed; and the singlet). This theory may also be analysed by the methods of [47]. Here we will reproduce the results from such an analysis by a different method.

The T-dual theory to the Möbius strip is an orientifold $T^2/\mathbb{Z}_2$, with the size of the $T^2$ being $(2R_1)^{-1} \times (2R_2)^{-1}$ (one fourth of the usual size, compare with [55]). At the four fixed points we find orientifold fixed planes. The original $O^+$-plane splits into two $O^+$-planes intersecting the torus at a point. The crosscap will dualise into one $O^+$-plane and one $O^-$-plane, so we have a total of 3 $O^+$-planes and 1 $O^-$-plane. On the dual we have $k/2$ D-brane pairs at arbitrary positions if $k$ is even. If $k$ is odd, there are $(k-1)/2$ D-branes whose positions can be chosen freely. The remaining brane pair was stuck at the crosscap, so in the dual picture there is an isolated brane at an orientifold plane, which should be the $O^-$.

The corresponding holonomies can be read of as follows. A brane pair in the bulk has two coordinates, and each pair corresponds to four eigenvalues $\lambda_i, -\lambda_i, \lambda_i^{-1}, -\lambda_i^{-1}$ with $\lambda_i = \exp(2\pi i X_i/R_i)$, $X_i$ and $R_i$ being the coordinate and the radius of the corresponding dimension (the $O^-$ plane is located at $(X_1/R_1, X_2/R_2) = (1/4, 1/4)$, the remaining $O^+$ at $(0,0)$, $(1/4, 0)$, $(0, 1/4)$). Corresponding to these eigenvalues we have $2 \times 2$ blocks on the diagonals of the holonomies of the form

$$\begin{pmatrix} \lambda_1 & 0 \\ 0 & -\lambda_1 \end{pmatrix} \quad \begin{pmatrix} 0 & -\lambda_2 \\ -\lambda_2 & 0 \end{pmatrix} \tag{5.3.1}$$

where the left block appears in one of the holonomies and the other, resulting from multiplying a diagonal block with a block of the form (5.1.3) in the other holonomy. There is a second set of blocks with $(\lambda_1, \lambda_2)$ replaced by $(\lambda_1^{-1}, \lambda_2^{-1})$. For a single brane located at the $O^-$ plane we get blocks with $(\lambda_1, \lambda_2) = (i, i)$. One easily verifies that this prescription leads to anticommuting elements in the fundamental representation of the symplectic group.

On the orientifold $T^2/\mathbb{Z}_2$ one should introduce a $B$-field which is half-integer valued. For orthogonal groups this is well known, and it is usually deduced from a path-integral argument [49]. It may also be deduced from duality. The Möbius strip we used may be described as the torus $T^2$, which is $\mathbb{R}^2$ quotiented by the lattice generated by $2\pi(0, 2R_2)$ and $2\pi(R_1/2, R_2)$, quotiented by an orientifold action that takes $(X_1, X_2) \to (-X_1, X_2)$. Omitting the orientifold for a moment, we see that the torus is skew, implying a half-integer value for the $B$-field in its dual [19]. The same reasoning applies to a Möbius strip, where the edge is formed by an $O^-$ instead of $O^+$-plane. This corresponds to an orthogonal theory without vectorstructure, as described in [55], and reproduced by our analysis later.

The resulting orientifolds describe the moduli space of compactifications of $Sp(k)$ theories with twisted boundary conditions. As a check consider the cases $k = 1$ and $k = 2$, since as $Sp(1)/\mathbb{Z}_2 = SU(2)/\mathbb{Z}_2 = SO(3)$, and $Sp(2)/\mathbb{Z}_2 = SO(5)$ these results should be reproduced by other orientifolds. $Sp(1)$ with twist corresponds to an orientifold with 3 $O^+$-planes, and a single D-brane stuck to the $O^-$ plane. The resulting configuration allows no continuous gauge freedom, in accordance with the standard description of $SU(2)$ with twist. The single D-brane at the $O^-$ fixed plane gives $O(1) = \mathbb{Z}_2$ residual symmetry; this should be interpreted as the symmetry of the centre of $SU(2)$ which is the only symmetry of $SU(2)$ that survives the twist.

For $k = 2$ the dual description consists of a single D-brane-pair on the orientifold with 3 $O^+$ and 1 $O^-$-planes. The rank of the unbroken group is 1, and generically it is $U(1)$. At the $O^-$-plane this is enhanced to $O(2)$, while at any of the three $O^+$-planes it is enhanced to $Sp(1) = SU(2)$. We will see in the next section that this nicely agrees with the orientifold description of the $O(5)$ orientifold corresponding to the $\mathbb{Z}_2$-twisted case.

For higher $k$ the analysis is similar. For $k$ even the generic unbroken group is $U(1)^{k/2}$, which can be enhanced to $U(k/2)$ at a generic position at the orientifold, to $Sp(k/2)$ at one of the three $O^+$ planes or $O(k)$ at the $O^-$ point. For $k$ odd this analysis can be copied while replacing $k$ by $k - 1$, with the exception that at the $O^-$ plane $O(k)$ symmetry is possible because of the brane already present there.

## 5.4  Twist in orthogonal groups on the 2–torus

Twist in the orthogonal groups gives a more involved situation and we can distinguish several possibilities. Every orthogonal group has a two-fold cover, so the resulting $Spin$-group has at least a $\mathbb{Z}_2$ centre. Compactification on a two torus with twist in this $\mathbb{Z}_2$ will lead to absence of "spin-structure": fields in the spin representation are not allowed since the holonomies will not commute in this representation.

For $SO(N)$-theories with $N$-odd this is all one can do apart from compactification with periodic boundary conditions. For $N$ even, $SO(N)$ already has a non-trivial centre and the above mentioned $\mathbb{Z}_2$ is just a subgroup of the whole centre. For $N$ divisible by 4, the centre of $Spin(N)$ is $\mathbb{Z}_2 \times \mathbb{Z}_2$. $\mathbb{Z}_2 \times \mathbb{Z}_2$ allows three $\mathbb{Z}_2$ subgroups (basically each of the $\mathbb{Z}_2$ factors, and a diagonal embedding). Of these two are related by the outer automorphism of the $Spin(N)$-groups with $N$ even. Hence there are two options for twisting by a $\mathbb{Z}_2$: The already above mentioned $\mathbb{Z}_2$ leading to compactification without spin structure, and a second one, named "compactification without vector structure". The latter is named so because in this compactification the vector representation is not an allowed one.

For $Spin(N)$ with $N$ even but not divisible by 4, the centre is $\mathbb{Z}_4$. The previously mentioned $\mathbb{Z}_2$ is generated by the order 2 element in $\mathbb{Z}_4$. It is also possible to twist by an element of $\mathbb{Z}_4$ generating the whole centre. We will call this compactification without vector structure, since in this case the vector representation is not an allowed one either. Note however that this twist forbids any representation with a non-trivial centre, so the spin representation should be absent as well. The only representations allowed in this case are conjugate to the adjoint.

The results from this section can also be derived with the methods of [47]. We will follow a different route.

### 5.4.1 No spin structure

Absence of spin-structure does not forbid the vector representation, so one can use an ordinary orientifold $T^2/\mathbb{Z}_2$ with 4 $O^-$-planes. The topological non-triviality has to be treated with the technique of Stiefel-Whitney classes, following the appendix of [55]. Absence of spin-structure implies that the second Stiefel-Whitney class is non-vanishing. We will keep on demanding that the first Stiefel-Whitney class vanishes (which implies $SO(N)$-symmetry, not only $O(N)$), and hence the total Stiefel-Whitney class should be $w = 1 + \omega_2$, with $\omega_2$ the 2-form on the 2-torus. For $SO(N)$ with $N$ odd, this is accomplished by placing 3 branes at the three non-trivial orientifold fixed points, placing $(N-3)/2$ pairs at arbitrary points. For $SO(N)$ with $N$ even one distributes 4 branes over all orientifold fixed points and has $(N/2-2)$ pairs at arbitrary locations. These are the only solutions.

It is instructive to compare the cases of $SO(3)$ and $SO(5)$ without spinstructure to the analysis for $Sp(1)$ and $Sp(2)$ with twist, as $Spin(3) = Sp(1)$ and $Spin(5) = Sp(2)$. For $SO(3)$, 3 isolated D-branes are located at three orientifold planes, and there is no continuous gauge freedom at all, just like in the $Sp(1)$ case. There are discrete symmetries $O(1)^3 = \mathbb{Z}_2^3$, but these just correspond to the 8 diagonal $O(3)$-matrices. Of these, 4 are not elements of $SO(3)$, and of the remaining 4, 3 lift to elements that anticommute with the holonomies in $Spin(3) = Sp(1)$. Only the identity remains, which lifts to the two centre elements of $Sp(1)$, which is how the $\mathbb{Z}_2$ discrete symmetry there is recovered.

For $SO(5)$, we have 3 orientifold planes occupied by one brane each, and a pair of branes at an arbitrary point. Generically the unbroken symmetry is $U(1)$, which can be enhanced to $O(3)$ at any of the three points where an orientifold plane with brane is present. At the remaining orientifold plane $U(1)$ is enhanced to $O(2)$. Stressing again that $SO(3)$ is the double cover of $Sp(1)$, we see that this is exactly the same as the $Sp(2)$ orientifold with twist.

Further easy examples are $SO(4)$ and $SO(6)$ without spin structure. For $SO(4)$ there is no residual gauge symmetry. The $\mathbb{Z}_2$ associated with vector structure acts as twist in both $SU(2)$-factors of $Spin(4)$, eliminating all gauge freedom. $SO(6)$ without spin structure is equivalent to $Spin(6) = SU(4)$ with $\mathbb{Z}_2$-twist. From the above description this gives a rank 1 subgroup, which can be enhanced to $O(3)$ at 4 orientifold-planes. This coincides with the $SU(4)$-description, where $SU(2)$ is a maximal symmetry group.

### 5.4.2 No vector structure

The case of absence of vectorstructure was already analysed by Witten [55] for $O(4N)$. Here we present an analysis from a different point of view, which also nicely extends to the case of $O(4N+2)$.

Compactifying a string theory with orthogonal gauge symmetry $SO(2k)$ on a circle, and T-dualising along this circle, gives a theory on the interval $I = S^1/\mathbb{Z}_2$. The interval is bounded by two $O^-$-planes, and on the interval we have $k$ pairs of D-branes. We will assume that the $O^-$ planes do not contain any isolated D-branes, since if both of them would be occupied we do not have $SO(2k)$ but $O(2k)$-symmetry, and if only one of them would be occupied this would indicate $O(n)$-symmetry with $n$ odd (actually, for $n$ odd the following construction is impossible, which is a reflection of the fact that $SO(n)$ with $n$ odd allows only one kind of twist).

Again the only possible discrete symmetry is $\mathbb{Z}_2$ reflection symmetry, and we will henceforth assume that this is realised. Compactifying on an extra circle, with a holonomy implementing this reflection leads again to a theory on the Möbius strip, this time with $O^-$planes on the boundary. We again go to the representation in which the Möbius strip is a cylinder, bounded on one end by an $O^-$ plane, and on the other end by the crosscap. Notice that for $k$ odd, there is a pair of D-branes fixed by the reflection and, on the cylinder it has to be located at the crosscap. Half the number of the remaining D-branes are visible in this representation. Since we are restricting to $SO(n)$-configurations, we can repeat the whole discussion presented for symplectic groups, with the difference that the orientifold planes we insert here will not give symplectic but orthogonal symmetry. From the geometric picture one may again deduce anticommutativity of the holonomies $\Omega_1$ and $\Omega_2$.

We may now T-dualise as before, and obtain an orientifold with two $O^-$-planes from the original $O^-$-plane, and an $O^+$ and $O^-$-plane from the crosscap. This explains the relation between the IIA-theory on a Möbius strip, that is discussed in [41], and the IIB orientifold in [55], which are both proposed as dual for the CHL-string [7] [8].

We may represent the parts of the holonomies corresponding to branes in the bulk in the same way as in the symplectic case. These can be conjugated to matrices in the real vector representation of $O(n)$. Readers who prefer the real representation of $O(N)$ should substitute for each brane in the bulk the following $4 \times 4$-blocks

$$\begin{pmatrix} \cos\phi_1 & 0 & -\sin\phi_1 & 0 \\ 0 & -\cos\phi_1 & 0 & \sin\phi_1 \\ \sin\phi_1 & 0 & \cos\phi_1 & 0 \\ 0 & -\sin\phi_1 & 0 & -\cos\phi_1 \end{pmatrix} \quad \begin{pmatrix} 0 & -\cos\phi_2 & 0 & \sin\phi_2 \\ -\cos\phi_2 & 0 & \sin\phi_2 & 0 \\ 0 & -\sin\phi_2 & 0 & -\cos\phi_2 \\ -\sin\phi_2 & 0 & -\cos\phi_2 & 0 \end{pmatrix}$$
(5.4.1)

as can be derived straightforwardly. The parameters $\phi_i = X_i/2\pi R_i$ follow from the coordinates of the D-branes on the torus.

For $k$ odd, there was an odd number of D-brane pairs at the crosscap, and hence the $O^-$-plane coming from the crosscap has to contain an odd number of branes. The other two orientifold points also contain isolated branes. This is possible because the even number of branes that should be on the edge of the Möbius strip (to ensure $SO(2k)$-symmetry), may translate into odd numbers of branes at each of the corresponding $O^-$-planes in the dual theory.

Consider a single brane stuck to the $O^-$-plane that came from dualising the crosscap. As can be seen from the geometric picture, this corresponds to a block $\mathrm{diag}(i,-i)$ in the holonomy $\Omega_1$, or equivalently, a block

$$\begin{pmatrix} 0 & -1 \\ 1 & 0 \end{pmatrix}$$

in the real representation of $O(n)$. Demanding anticommutativity with a $2 \times 2$ block in the holonomy $\Omega_2$ leads to the unique solution $\mathrm{diag}(1,-1)$ (in the real representation, up to conjugation with an element of $SO(2)$). One may also consider a single brane stuck to one of the other $O^-$ planes. This defines a block $\mathrm{diag}(1,-1)$ in the holonomy $\Omega_1$. Demanding anticommutativity with a second $2 \times 2$ block leads to two inequivalent possibilities, being

$$\begin{pmatrix} 0 & -1 \\ 1 & 0 \end{pmatrix} \quad \begin{pmatrix} 0 & 1 \\ 1 & 0 \end{pmatrix}$$

These two possibilities correspond to the two $O^-$-planes that came from dualising the original $O^-$-plane. The point is now that occupying 1 or 2 of the $O^-$ planes by an odd number of branes corresponds to holonomies in $O(n)$, but occupying all three at once with an odd number of single branes does give holonomies in $SO(n)$. This also gives the interpretation for the orientifold with 3 $O^-$ planes and one $O^+$-plane where not all of the $O^-$ planes are occupied; these represent $O(k)$ configurations that cannot be represented in $SO(k)$. We see that if we demand $SO(n)$-symmetry, and occupy one $O^-$-plane with an odd number of branes, we have to occupy all $O^-$ planes by an odd number of branes.

On the resulting dual orientifold we have $k/2$ pairs of D-branes if $k$ is even, or $(k-3)/2$ if $k$ is odd. For $k$ even we obtain back the description of [55], with the possibility of $O(k)$ symmetry at 3 planes, and $Sp(k/2)$-symmetry at 1 plane. For $k$ odd we have the possibility of $O(k-2)$ at three planes, and $Sp((k-3)/2)$ at one plane.

It is again instructive to look at a few examples. $k = 1$ is impossible. $k = 2$ corresponds to $SO(4)$-theory without vector structure. Since $Spin(4)$ is $SU(2) \times SU(2)$, this corresponds to twist in one of the $SU(2)$ factors, and arbitrary holonomies in the other $SU(2)$-factor. In the orientifold description, we have the possibility of enhanced symmetries $O(2)$ and $Sp(1)$. $Sp(1) = SU(2)$ obviously corresponds to the unbroken second factor. The $O(2)$'s correspond to situations in which the holonomies in the second factor are $(1, i\sigma^3)$, $(i\sigma^3, 1)$, $(i\sigma^3, i\sigma^3)$. The extra parity transformation is due to the fact that $i\sigma^3$ anticommutes with the elements $i\sigma^{1,2}$, which lifts to the double cover as commutation symmetry.

$k = 3$ gives $SO(6)$ without vector structure. This corresponds to $Spin(6) = SU(4)$ with $\mathbb{Z}_4$-twist. The absence of remaining gauge freedom is completely in agreement with the $SU(4)$ description.

$k = 4$ gives $SO(8)$ without vector structure, which due to triality should be equivalent to $SO(8)$ without spin structure. We certainly do find $Sp(2) = Spin(5)$ symmetry in both descriptions. $O(4)$ is less visible in the above description of $SO(8)$ without spin structure, but this is due to the fact that the parity in $O(4)$ is actually not a real symmetry (compare to the $O(2)$ symmetry found for $k = 2$), and $Spin(4) = SU(2) \times SU(2)$ can also be obtained outside orientifold fixed planes. The asymmetry in the two description is then due to the fact that they represent different projections from the same moduli space.

## 5.5 Commuting n-tuples for orthogonal groups

In this section we will investigate gauge theory with orthogonal gauge group, compactified on higher dimensional tori with periodic boundary conditions.

For orthogonal gauge theory on the 3–torus extra vacua were found by Witten [55]. As described in section 2.4, a configuration with 7 D-branes on the 7 fixed points excluding the origin of an orientifold parametrises a periodic connection for $SO(2N+7)$-theory. Similarly, a configuration with 8 D-branes distributed over all 8 fixed points parametrises a periodic connection of $SO(2N+8)$-theory. On the 3–torus these were the only new solutions. Obviously, these can be trivially embedded on an $n$–torus, by taking 3 holonomies from the 3–torus and $(n-3)$ holonomies equal to the identity. This is however not the only solution.

As a first example consider the non-trivial flat connection of $SO(7)$-theory or $SO(8)$-theory on the 3–torus, and embed the 3–torus in a 4–torus, setting the holonomy along the fourth direction to the identity. The 4–torus has an $SL(4, \mathbb{Z})$ invariance group, that acts non-trivially on the holonomies. If $M_{ij}$ are the entries of the $SL(4, \mathbb{Z})$-matrix, then the holonomies transform as

$$\Omega'_i = \prod_j \Omega_j^{M_{ij}} \tag{5.5.1}$$

An $SL(3, \mathbb{Z})$-subgroup acts trivially on the holonomies, it does not leave the holonomies invariant, but transforms them into gauge equivalent ones. An $SL(4, \mathbb{Z})$ transformation that is not in this subgroup, transforms the holonomies into a set that is *not* gauge equivalent to the original set. An easy way to see this is that it will transform the fourth holonomy, which was the identity, into a non-trivial group element. This can never be accomplished with a gauge transformation. On the Stiefel Whitney class, the $SL(4, \mathbb{Z})$ acts as a simple coordinate transformation. In particular it transform the trivial class into the trivial class, so the $SL(4, \mathbb{Z})$-image of a periodic flat connection is again an allowed solution. In this way already many new solutions can be found.

For the second example consider the non-trivial flat connection of $SO(8)$-theory on the 3–torus. Embed this in a 4–torus, with holonomy $-\mathbb{1}$ along the extra cycle. This surely represents an allowed solution to the vacuum equation, as can also be verified by computing the Stiefel-Whitney class, or writing down an explicit representation. It is also easily verified that this solution is not contained in the previously discussed set. Again one can generate more solutions by acting on this one with $SL(4,\mathbb{Z})$-transformations.

As a matter of fact, each of the above solutions actually represents two solutions, since always two inequivalent liftings to the simply connected $Spin$ groups are possible.

For yet another example we consider $SO(15)$-gauge theory on the 4–torus. In the same fashion as for the 3–torus, using T-duality, a flat connection for this theory is described by a $\mathbb{Z}_2$-invariant configuration of 15 D-branes on $\tilde{T}^4$, which is the covering space of an orientifold $\tilde{T}^4/\mathbb{Z}_2$. There are 16 orientifold planes, corresponding to all possible combinations of eigenvalues $(\pm 1, \pm 1, \pm 1, \pm 1)$ of the four holonomies around the non-trivial cycles. Now distribute the 15 D-branes over the 16 orientifold planes, one brane at each plane, leaving only the orientifold plane that corresponds to the eigenvalues $(1, 1, 1, 1)$ empty. The total Stiefel-Whitney class of this configuration can be easily computed. The easiest way to do this is to divide the branes into two groups, the first group consisting of the 7 branes at $(\pm 1, \pm 1, \pm 1, 1)$, the second group containing the remaining 8 branes. The first 7 branes are in the same configuration that Witten used to prove the existence of a non-trivial flat connection on the 3–torus (embedded in a higher-dimensional orientifold) and hence these do not contribute to the total Stiefel-Whitney class. The remaining 8 also do not contribute, since they actually correspond to the $SO(8)$-configuration described previously. Hence the total Stiefel-Whitney class is trivial, and the bundle is topologically trivial. It is also easily shown that this configuration gives an isolated point in the moduli space of flat $SO(15)$-connections on $T^4$, and that the gauge group is completely broken.

This generalises to still higher dimensional tori: $2^n - 1$ branes distributed over $2^n$ orientifold planes in the covering of the orientifold $\tilde{T}^n/\mathbb{Z}_2$, one brane at each plane but leaving the plane corresponding to $(1,1,\ldots,1)$ empty, corresponds to a topologically trivial flat $SO(2^n - 1)$-connection on $T^n$ (calculation of the total Stiefel-Whitney class of this configuration is trivial, since the $2^n - 1$-branes can be divided in one group of 7, and $2^{n-3} - 1$ groups of 8, where in each group $(n-3)$ eigenvalues of the holonomies are fixed. The group of 7 does not contribute, nor does any of the groups of 8, giving a trivial total Stiefel-Whitney class). Note that these configurations realise the non-trivial $n$-tuples of [26]. To obtain an $SO(2^n)$ connection, one also inserts a D-brane at the orientifold plane corresponding to $(1,1,\ldots,1)$.

A last construction that partially overlaps with the previous examples but also produces new vacua is as follows. Each of the above flat connections corresponds under T-duality to some collection of D-branes, with trivial Stiefel-Whitney class. One can take two such configurations and "add" them, that is superimpose the two. Computation of the Stiefel-Whitney class is modulo two, so one can throw away all the D-brane-pairs that arise in the superposition, and again obtain a flat connection.

As an example, the $SO(7)$-configurations on the 5-torus specified by

$$\begin{aligned}
\Omega_1 &= \text{diag}(\ 1,-1,-1,-1,\ 1,\ 1,-1) \\
\Omega_2 &= \text{diag}(\ 1,-1,\ 1,\ 1,-1,-1,-1) \\
\Omega_3 &= \text{diag}(-1,\ 1,\ 1,-1,\ 1,-1,-1) \\
\Omega_4 &= \mathbb{1} \qquad\qquad\qquad \Omega_5 = \mathbb{1}
\end{aligned}$$

and

$$\begin{aligned}
\Omega_1 &= \mathbb{1} \qquad\qquad\qquad \Omega_2 = \mathbb{1} \\
\Omega_3 &= \text{diag}(\ 1,-1,-1,-1,\ 1,\ 1,-1) \\
\Omega_4 &= \text{diag}(\ 1,-1,\ 1,\ 1,-1,-1,-1) \\
\Omega_5 &= \text{diag}(-1,\ 1,\ 1,-1,\ 1,-1,-1)
\end{aligned}$$

can be added to give

$$\begin{aligned}
\Omega_1 &= \text{diag}(-1,-1,-1,\ 1,\ 1,-1,\ 1,\ 1,\ 1,\ 1,\ 1,\ 1) \\
\Omega_2 &= \text{diag}(-1,\ 1,\ 1,-1,-1,-1,\ 1,\ 1,\ 1,\ 1,\ 1,\ 1) \\
\Omega_3 &= \text{diag}(\ 1,\ 1,-1,\ 1,-1,-1,\ 1,-1,-1,\ 1,\ 1,-1) \\
\Omega_4 &= \text{diag}(\ 1,\ 1,\ 1,\ 1,\ 1,\ 1,\ 1,-1,\ 1,-1,-1,-1) \\
\Omega_5 &= \text{diag}(\ 1,\ 1,\ 1,\ 1,\ 1,\ 1,-1,\ 1,-1,\ 1,-1,-1)
\end{aligned}$$

which determines a flat $SO(12)$ connection on the 5–torus.

## 5.6  $c$-triples for symplectic groups

For symplectic groups on the 3–torus one may also choose non-periodic boundary conditions. This can be done in various ways, but by using $SL(3,\mathbb{Z})$ transformations on the torus, all possibilities are isomorphic to one standard form. We can choose the standard form to have twist between the holonomies in the 1 and 2 direction, and the third holonomy commuting with the former two. Following [6], we call a triple of such holonomies a $c$-triple, where $c$ denotes that the three holonomies only commute up to a (non-trivial) centre element of the gauge group.

From our analysis for twist in symplectic gauge theories on the 2–torus, one easily deduces that the corresponding orientifold description has 6 $O^+$-planes and 2 $O^-$planes. The two planes with 3 $O^+$ and 1 $O^-$, are distinguished by the eigenvalue $\pm 1$ in the third holonomy. Eigenvalues for the third holonomy can be read of in the usual way, with the remark that their multiplicities should be doubled. A configuration for the 2–torus may therefore be imported in either of these planes, corresponding to choosing the third holonomy in $Sp(k)$ to be $\pm\mathbb{1}$, which are the two elements of the centre of $Sp(k)$. One quickly deduces that there are always 2 disconnected possibilities for placing the D-branes in this orientifold background.

First suppose $Sp(k)$ symmetry with $k$ even. For the description with twist on a 3–torus, this should give $k/2$ pairs of D-branes in the above orientifold background. There are two possibilities to distribute the D-branes. First, one can have $k/2$ pairs at arbitrary locations on the orientifold. But one can also split one pair, put one D-brane on one $O^-$-plane and the other on the other $O^-$-plane, and have the remaining $(k/2-1)$-pairs at arbitrary locations.

Both possibilities are legitimate, since $Sp(k)$ is simply connected. The conclusion is thus, that $Sp(k)$-theory on a 3–torus with twisted boundary conditions has a moduli space of 2 components, one with a rank $k/2$ unbroken gauge group, and one with a rank $k/2-1$ unbroken gauge group. We can now perform a Witten index count for this theory, as also performed in [6]: The two components will contribute $k/2+1$ and $k/2$ to the index giving the total value $k+1$ in agreement with both the periodic boundary conditions case, and the infinite volume case [52].

For $k$ is odd the procedure should also be clear. One can place $(k-1)/2$ pairs of D-branes at arbitrary points in the orientifold background. The single D-brane that is left can go on either of the two $O^-$planes. These are inequivalent possibilities, and hence also in this case the moduli space consists of 2 components. Each of these components contributes $(k+1)/2$ to a Witten index calculation, giving also the correct result $k+1$ [6].

Again we check the $k=1$ and $k=2$ cases. According to the above, $Sp(1)$-theory with twist on a 3–torus gives a moduli space of 2 components. On each component the gauge group is completely broken. This is as it should be as $Sp(1) = SU(2)$ which, when compactified on a 3–torus with twist has a moduli space that looks like this. We again have the remaining $O(1) = \mathbb{Z}_2$ symmetry corresponding to the centre of $SU(2)$, which commutes with everything.

Perhaps more interesting is the $Sp(2)$-case, where we have one component for which the gauge group is completely broken, and another where a rank 1 group survives. The rank one gauge group is generically $U(1)$, but can be enhanced to $O(2)$ at two planes or $Sp(1)$ at six other planes. This coincides with the description we will find for $SO(5) = Sp(2)/\mathbb{Z}_2$, without spin structure.

## 5.7 *c*-triples for orthogonal groups

For orthogonal groups on a 3–torus there are more possibilities for the boundary conditions. Like in the case for the 2–torus, we can have absence of either spin- or vectorstructure. By $SL(3,\mathbb{Z})$-transformations on the torus, we can again arrange that the holonomies for the 1 and 2-direction are the ones that do not commute (in the Spin-cover of the group), while the third holonomy does commute with the other two.

In the case of $SO(4n)$ there is however a new possibility. For $Spin(4n)$ the centre of the gauge group is not cyclic but a product of cyclic groups, being $\mathbb{Z}_2 \times \mathbb{Z}_2$. Call the generator of the first $\mathbb{Z}_2$ $z_s$ (with $s$ for spin), and the generator of the second $\mathbb{Z}_2$ $z_c$ ($c$ being the standard notation for the second spin-representation). Also define $z_v = z_s z_c$. This notation is motivated by the fact that identifying $z_v \sim \mathbb{1}$ gives the vector representation.

We can now also impose the following twist conditions on the holonomies:

$$\Omega_1\Omega_2 = z_s \Omega_2 \Omega_1 \qquad \Omega_2\Omega_3 = z_c \Omega_3 \Omega_2 \qquad \Omega_3\Omega_1 = z_v \Omega_1 \Omega_3 \tag{5.7.1}$$

This can be thought of as a standard form. $SL(3,\mathbb{Z})$-transformations result in an isomorphic moduli-space. We will call this case "spin nor vector"-structure, and treat it separately.

### 5.7.1  No spin structure

This is the easiest case, provided we use some previously obtained knowledge. From our description of orthogonal theories on a 2–torus without spin structure, a particular case for the 3–torus can be obtained as follows.

For $SO(k)$ with $k$ odd, place 3 single D-branes at three $O^-$-planes within one plane within the orientifold $T^3/\mathbb{Z}_2$ leaving the fixed plane at the origin empty, and place the others in pairs at arbitrary points at the orientifold. For $k$ even one should place 4 single D-branes at 4 orientifold fixed planes within one plane of $T^3/\mathbb{Z}_2$.

For $k \geq 4$ there is always a second possibility. Remember from section 2.4 and [55] that a configuration of 8 D-branes distributed at all orientifold planes has a trivial Stiefel-Whitney class. We may "add" this orientifold configuration to another as follows. Take a specific configuration of D-branes at the orientifold. This has a certain Stiefel-Whitney class, which can be thought of as providing a topological classification for the configuration. Now adding 8 more D-branes at the orientifold fixed points will not affect the Stiefel-Whitney class. This is so because the Stiefel-Whitney class of the 8 D-branes is trivial, and the Stiefel-Whitney class of the "new" configuration may be simply obtained by multiplying the class of the "old" configuration with that of the added configuration (it is important to realise that Stiefel-Whitney classes are $\mathbb{Z}_2$ valued, and that $-1 = 1$ mod 2, so there is no ordering ambiguity). One may also add or delete any pair of D-branes without affecting the class, also because of its $\mathbb{Z}_2$ nature.

We thus obtain the following possibilities: For $SO(k)$ with $k$ odd, we had 3 single D-branes at three $O^-$-planes. Adding the 8 D-branes and reducing modulo 2, we obtain a configuration of 5 D-branes with the same topological classification as the previous one. The 5 D-branes are precisely at the orientifold planes that were not occupied previously, and in a sense one could speak of a $\mathbb{Z}_2$-complement. One can add pairs of D-branes to again obtain an $SO(k)$ configuration.

For $k$ even one had 4 single D-branes at 4 $O^-$-planes. Taking the $\mathbb{Z}_2$-complement, we get an inequivalent configuration with 4 single D-branes at the other 4 $O^-$-planes, with the same topological classification. Of course, afterwards we must add pairs of D-branes to acquire $SO(k)$.

We will now discuss several cases. $k=3$ corresponds to $SO(3) = SU(2)/Z_2$ with twist on the 3–torus. The $SU(2)$ description has two components. The $SO(3)$-description has also two components, but these cannot be distinguished by their holonomies. In a particular representation, the $SU(2)$ holonomies read

$$\Omega_1 = \mathrm{i}\sigma_3; \qquad \Omega_2 = \mathrm{i}\sigma_1; \qquad \Omega_3 = \pm 1, \qquad (5.7.2)$$

but $\pm 1$ in $SU(2)$ are both projected to the same element of $SO(3)$ being the identity.

$k=4$ gives $SO(4)$ which gives two orientifolds, but some thought will reveal that also in this case there are twice as many components in moduli space. Using that $Spin(4) = SU(2) \times SU(2)$, the no-spin-structure condition amounts to twisting both $SU(2)$ factors simultaneously. For the third holonomy one has then 4 possibilities, being any combination of plus or minus the identity in each $SU(2)$-factor. These 4 possibilities project to only two sets of holonomies in $SO(4)$, and hence two orientifold descriptions.

$k = 5$ gives us $SO(5)$ which is interesting because we should be able to reproduce the $Sp(2)$-results here. $SO(5)$ without spin-structure gives two orientifolds. On one we have 5 fixed D-branes and hence no residual gauge symmetry. On the other we have 3 fixed D-branes and a pair wandering freely. Possible enhanced gauge symmetries are $O(3)$ at three points, and $O(2)$ at five points. However, all but one of the "parity" symmetries (corresponding to elements with det $= -1$) in these $O(n)$ groups are "fake" in the sense that they correspond to elements that anticommute in $Spin(5)$. The remaining $O(2)$ corresponds to the $O(2)$'s we encountered in the $Sp(2)$ case, and the $SO(3)$'s map to the $Sp(1)$-unbroken subgroups in $Sp(2)$. That the multiplicities of these enhanced symmetry groups are only half of those encountered in the $Sp(2)$ description reflects the fact that $SO(5)$ is a double cover of $Sp(2)$, which also translates to the fact that the moduli space of $Sp(2)$-triples is a double cover of the space of $SO(5)$-triples. The moduli space for the gauge theory is the moduli space of $Sp(2)$-triples, as every set of $SO(5)$ holonomies has two inequivalent realisations in terms of gauge fields. Note however that here the number of components in moduli space agrees with the number of orientifolds; the two $SO(5)$-components are double covers of two $Sp(2)$ components, not of 4 $Sp(2)$ components.

$k = 6$ gives $SO(6)$ whose spin cover is $SU(4)$. Here there are two equal dimension components in moduli space, both with a rank 1 gauge group which can be enhanced to $SO(3)$ at 4 points.

It is easy to perform the Witten index count for $k \geq 5$ [6]. In these cases we have always two components of the moduli space and two corresponding orientifold representations. For $k$ even, both components contribute $k/2 - 1$, for a total of $k - 2$. For $k$ odd, one component contributes $(k-1)/2$ whereas the second contributes $(k-3)/2$ for a total of $k - 2$. Of course these answers are as they should be.

### 5.7.2 No vector structure

From the analysis for the 2–torus we deduce that $O(2k)$ without vector structure on a 3–torus corresponds to an orientifold background with 6 $O^-$-planes and 2 $O^+$-planes. Again eigenvalues for the third holonomy can be read off in the usual way, except that their multiplicities should be doubled. One obvious solution to the boundary conditions is to import the solution for the 2–torus here.

For $k$ even we have seen that a particular solution is given by placing all D-brane pairs at arbitrary points. For the second solution we take as before the $\mathbb{Z}_2$-complement. We have not defined how the operation of "$\mathbb{Z}_2$-complement" acts on the $O^+$-planes but this is not hard to guess. Since $O^+$ planes cannot support isolated D-branes, they should remain empty. Hence the second solution has all $O^-$ planes occupied by one D-brane, and $k - 3$ pairs at arbitrary points. A way to see this is as follows. The smallest group for which the configuration with six isolated branes exists is $SO(12)$. One can take the $SO(6)$ holonomies $\Omega_1$ and $\Omega_2$ that gave "no vector structure" on the 2–torus (these are unique up to gauge transformations), to construct the $SO(12)$-holonomies $\Omega_1 \oplus \Omega_1$, $\Omega_2 \oplus \Omega_2$ and $\mathbb{1} \oplus -\mathbb{1}$, where $\mathbb{1}$ stands for the identity in $SO(6)$. That these $SO(12)$-matrices satisfy the required boundary conditions is obvious, and liftings to $Spin(12)$ can be constructed from the liftings of the $SO(6)$-holonomies to $Spin(6) = SU(4)$. That $SO(12)$ is the smallest group allowing these extra solutions can also be deduced with the techniques from [6].

For $k$ is odd we have 3 $O^-$-planes within one plane occupied by D-branes. A priori one has two possible planes, and actually both give distinct solutions. Note that also in these cases the solutions are each others $\mathbb{Z}_2$-complement. For $k$ odd all components of the moduli space are isomorphic, as also follows form the analysis of [6].

$SO(4)$ without vector structure on a 3-torus gives only one solution, since there are simply not enough D-branes to realise the second one. This gives a rank 1 unbroken gauge group which can be enhanced to $Sp(1)$. A naive calculation of the Witten index would give half of the right answer, but as before, the moduli space consists of 2 components that cannot be distinguished by their holonomies in $SO(4)$. We therefore have to multiply the naive value for the index by two, again obtaining the right answer.

$SO(6)$ without vector structure gives two solutions, but from $SU(4)$-analysis one expects four. Again this is due to the fact that inequivalent solutions exist that cannot be distinguished by their holonomies in $SO(6)$. Notice that the gauge group is completely broken, which is as it should be.

A Witten index calculation is straightforward for these theories. For $SO(4N + 2)$ there are always 4 components [6] (projected to two orientifolds) that are all isomorphic. Each orientifold has $2N + 1$ branes on it, of which 3 are stuck. The rest should organise in $N - 1$ pairs, giving a rank $N - 1$ gauge group and contribution $N$ to the Witten index. With 4 components one obtains the total value $4N$, which is indeed the dual Coxeter number for these theories.

For $SO(4N)$ one has 2 components (1 orientifold), where no branes are stuck and $N$ pairs move freely, giving a contribution of $2N + 2$ to the index. For $N \geq 3$ this is not sufficient, but for these cases there exist 2 more components (1 orientifold) having 6 stuck branes, and $N - 3$ pairs at arbitrary points. The total adds up to $2(N + 1) + 2(N - 2) = 4N - 2$, the right answer.

### 5.7.3 Spin nor vector structure

For $SO(4k)$ there is the possibility of holonomies satisfying equation (5.7.1). In some sense this should encompass both the case of no spin- as well as no vector-structure. The orientifold background is as in the case without vector structure, $T^3/\mathbb{Z}_2$ with 2 $O^+$-planes and 6 $O^-$-planes. On top of this $2k$ branes should be distributed.

A clue on the D-brane configuration can be found from T-dualising in the direction of the line connecting the two $O^+$ planes. This gives a theory on the product of a circle and an orientifold $T^2/\mathbb{Z}_2$ with 3 $O^-$'s and 1 $O^+$ plane. This orientifold corresponded to an orthogonal theory without vector structure on the 2–torus. In our case, the gauge group is $SO(4k)$, and we know that if there are branes at an orientifold plane, their number should be even. Translating back to the orientifold $T^3/\mathbb{Z}_2$, the pair of orientifold planes corresponding to one $O^-$-plane in $T^2/\mathbb{Z}_2$ will be occupied either both by an even number of branes, or both by an odd number of branes. This leaves 8 possibilities, 2 of which can be quickly discarded as they correspond to a $SO(4k)$-theory without vector, but with spin structure. The remaining possibilities have thus either 2 or 4 branes stuck at $O^-$ planes, and hence $2k - 2$, resp. $2k - 4$ D-branes in the bulk.

T-dualising in another direction, along a line connecting an $O^+$ and an $O^-$-plane will lead to an orientifold of the form $((T^2/\mathbb{Z}_2) \times S^1)/\mathbb{Z}_2$. One can represent this by an orientifold $T^2/\mathbb{Z}_2$ with 4 $O^-$ planes, quotiented by a $\mathbb{Z}_2$-reflection in one of the points halfway on the line between 2 $O^-$-planes (the multiple possibilities are related by $SL(2,\mathbb{Z})$-transformations). This reflection has a second fixed point. Over the orientifold $(T^2/\mathbb{Z}_2)/\mathbb{Z}_2$ one erects a circle everywhere, except at the two fixed points of the $\mathbb{Z}_2$-reflection where it is replaced by a crosscap. On the orientifold $T^2/\mathbb{Z}_2$ we should have absence of spin structure, meaning that all $O^-$-planes are occupied by an odd number of branes (remember that $4k$ is even). This translates to occupancy of both $O^-$-planes in the quotient $(T^2/\mathbb{Z}_2)/\mathbb{Z}_2$. We now have two possibilities; either there are an odd number of branes at both crosscaps, or there are an even number. For the orientifold $T^3/\mathbb{Z}_2$, these two possibilities translate into the situations with two $O^-$-planes occupied (crosscaps occupied by even number of branes), and four $O^-$-planes occupied (crosscaps occupied by odd number of branes). Thus both possibilities can be realised.

The 3 possibilities of occupying 2 $O^-$-planes are related by $SL(2,\mathbb{Z})$, as are the 3 possibilities of occupying 4 $O^-$ planes. Only one of each set of possibilities solves the boundary conditions (5.7.1) (as these are not $SL(2,\mathbb{Z})$ invariant), and actually, the two possibilities are each others $\mathbb{Z}_2$ complement, as before.

We therefore have two orientifolds representing $SO(4k)$-theory on a 3–torus with spinnor vectorstructure. Each orientifold background is of the form $T^3/\mathbb{Z}_2$, with 6 $O^-$ planes, and 2 $O^+$-planes. One orientifold has 2 $O^-$ planes occupied, and the other has the remaining 4 $O^-$-planes occupied.

$SO(4)$ with spin- nor vector structure has only 2 branes on the dual orientifold, so only one out of the two possibilities mentioned can be realised. This corresponds to 4 components on the moduli space, all consisting of a single point. This can be seen as follows. $Spin(4) = SU(2) \times SU(2)$, and therefore we may write $Spin(4)$ holonomies as $SU(2)$-pairs. The holonomies obeying the boundary conditions are (up to conjugation)

$$\Omega_1 = (i\sigma_3, i\sigma_3) \qquad \Omega_2 = (\pm i\sigma_1, \pm i\sigma_3) \qquad \Omega_3 = (i\sigma_1, i\sigma_1) \qquad (5.7.3)$$

There are four inequivalent possible choices for the signs in $\Omega_2$ (one cannot change a sign in $\Omega_2$ by conjugation without changing some sign in the other holonomies).

For $SO(4k)$ $(k > 1)$ and larger groups each of the two orientifold descriptions represents two components. Each of the two orientifold descriptions again represents 2 components. For $SO(4k)$, the two components contribute $k$, resp. $k-1$ to the Witten index. Taking into account the correct multiplicities, gives the correct answer $4k - 2$ for the Witten index [6].

# 6 New string and M-theory vacua

In the previous chapter we studied orientifold realisations of non-trivial triples. In this chapter we will study consistent string theories, and find that consistency requirements disqualify nearly all orientifolds of the previous chapter as possible string vacua.

The group $E_8 \times E_8$ is realised in 10 dimensional string theory. It is the symmetry group of one of the heterotic string theories. The heterotic $E_8 \times E_8$ theory is a suitable starting point for studying compactifications on a 3–torus, since in this theory all gauge symmetries are realised perturbatively, and are relatively straightforward to check. By duality this also provides clues for realisations of the same symmetries in dual models.

## 6.1 Triples for the heterotic string

### 6.1.1 Asymmetric orbifolds and consistency conditions

The fields living on the world sheet of a closed string can be divided in left and right movers. These are almost independent from each other, making it possible to consider strings where the left-moving fields are different from the right moving ones.

For the heterotic string [21] [22] (in its bosonised description) the right moving fields are 10 bosons and 10 fermions, which is the same as the right moving fields of a closed type II superstring. In the type II superstring the left moving fields are also 10 bosons and 10 fermions. The heterotic string however has a different field content. We may take the left movers to be 26 bosons, which would be the left-moving field content of the bosonic string. In a sense the heterotic string is thus a fusion (heterosis) of a bosonic string and a superstring.

In the superstring and the bosonic string the pairs of left and right moving bosons are interpreted as the 10 resp. 26 coordinates of some embedding space for the string. In the heterotic string this will not work as there is a different number of left and right movers. This may be solved by thinking of 16 of the 26 left-movers as internal coordinates. A consistent theory can be obtained by putting the 16 left-movers on a 16 dimensional torus $\mathbb{R}^{16}/\Lambda$, where consistency requires the lattice $\Lambda$ to be even and self-dual. There are precisely two 16 dimensional even self-dual lattices [36]. These are the lattice $E_8 \oplus E_8$, with $E_8$ denoting the root lattice of the exceptional algebra $E_8$, and the lattice $\Gamma_{16}$, which consists of the root-lattice of $SO(32)$, together with one of the spin weight lattices for this group. Both lattices give rise to a heterotic string theory, usually named the $E_8 \times E_8$ resp. the $Spin(32)/\mathbb{Z}_2$ heterotic string. The remaining pairs of 10 left moving bosons with 10 right moving bosons may then be interpreted as parametrising the 10 dimensional space in which the string lives.

The 16 bosons on the torus take momenta on the dual lattice $\Lambda^*$, which is equal to $\Lambda$ as the lattice is self-dual. The excitations of the string therefore carry quantum numbers corresponding to these hidden momenta. In particular there is a set of 496 massless gauge bosons, that transform in the $(\mathbf{248}, \mathbf{1}) \oplus (\mathbf{1}, \mathbf{248})$ of $E_8 \times E_8$ in the case of the heterotic $E_8 \times E_8$- string, or the 496-dimensional adjoint of $Spin(32)$ for the heterotic $Spin(32)/\mathbb{Z}_2$-string. This gives rise to $E_8 \times E_8$, resp. $Spin(32)$-gauge symmetry in 10 dimensions.

We have seen that both $E_8 \times E_8$ and $Spin(32)$ were among the gauge groups that gave new vacua when used as gauge group for Yang-Mills theory on a 3–torus. It is therefore natural to attempt to realise the new vacua in string theory. We then have to compactify these heterotic string theories on a 3–torus.

Compactifying on a torus makes the momenta in the compact directions discrete. Also new quantum numbers occur, corresponding to the winding of strings around the compact directions. Furthermore, compactification on a torus opens up the possibility of particular non-zero background fields. One of these background fields is the metric on the torus, which is a symmetric tensor. For 2– and higher dimensional tori, one may also choose a non-zero value for the background $B$-field, which is an antisymmetric 2-tensor gauge field appearing in the string theory. Last, and these are the background fields we will be interested in, one may pick a set of holonomies for the gauge fields.

We denote the left and right moving momentum in the compact direction $i$ by $k_{iL}$, resp. $k_{iR}$. The momenta on the internal torus will be denoted by $\mathbf{k}$, which are 16 dimensional lattice vectors of $\Lambda$. The background fields are the metric $g_{ij}$, the anti-symmetric tensor $B_{ij}$, and the gauge fields. For the moment we restrict ourselves to the case that the holonomies for the gauge fields can be taken to be on a maximal torus. One may then set the background gauge fields to take values in the CSA of the group. The elements of the CSA are in turn parametrised by vectors taking values in the root space. One may write the holonomies as $\Omega_i = \exp\{2\pi i h_{\mathbf{a}_i}\}$, associated to a constant background gauge field $A_i = h_{\mathbf{a}_i}/R_i$, $R_i$ being the radius of the corresponding dimension. It is common in the string literature to refer to the dimensionless objects $\mathbf{a}_i$ somewhat imprecisely as "Wilson lines".

We shall be mainly interested in the gauge symmetries, and therefore set $B_{ij} = 0$, and $g_{ij} = \delta_{ij}$. The $\mathbf{a}_i$ are 16-component vectors, taking values in the 16-dimensional torus, which is in a sense interpreted as the maximal torus for the group. The spectrum of the compactified theory can be expressed in the momenta [37] [18]

$$\mathbf{k} = (\mathbf{q} + \sum_i w_i \mathbf{a}_i)\sqrt{\frac{2}{\alpha'}} \tag{6.1.1}$$

$$k_{iL,R} = \frac{n_i - \mathbf{q} \cdot \mathbf{a}_i - \sum_j \frac{w_j}{2}\mathbf{a}_i \cdot \mathbf{a}_j}{R_i} \pm \frac{w_i R_i}{\alpha'} \tag{6.1.2}$$

with $i, j$ running over the compact directions. The $n^i$ and $w^i$ are integers denoting momenta, resp. winding numbers in the compact directions. The vector $\mathbf{q}$ takes values on the 16 dimensional lattice $\Lambda$. For $w_i = 0$ these formulae look just like what one would expect from a theory of particles, with $k_{iL,R}$ as covariant momenta. For strings that wind in the presence of the background gauge fields, non-linearities appear due to the fact that the extra bosons living on the wrapped string world sheet transform under the gauge fields. Especially the correction to $\mathbf{k}$ due to winding strings will be significant below.

The momenta lie on a Lorentzian lattice $\Gamma_{19,3}$ of signature $(19, 3)$ [36]. For two vectors $p$ and $q$ on this lattice, with components $(\mathbf{p}, p_{iL}, p_{iR})$ and $(\mathbf{q}, q_{iL}, q_{iR})$, the innerproduct is given

by
$$(p,q) = \mathbf{p} \cdot \mathbf{q} + p_{iL}q_{iL} - p_{iR}q_{iR} \tag{6.1.3}$$

With this innerproduct one easily verifies that the momentum lattice is even and self-dual.

Thus far we took all holonomies on the maximal torus. The new vacua for gauge theories on the 3–torus were obtained with holonomies that cannot all be put on the maximal torus, and therefore this set-up is too restrictive for our purposes. Remember however how such a vacuum was obtained. For the holonomies three group elements were chosen, such that the first group element leaves an unbroken subgroup that is not-simply connected. In this non-simply connected subgroup we then impose twisted boundary conditions. We may also subdivide this in two steps, the first being to choose an element that leaves a group that has an outer automorphism, and the second step choosing a group element that implements this outer automorphism. In terms of lattices this means that the first two group elements (which can be conjugated simultaneously into a maximal torus) are such that the lattice requires a nontrivial automorphism. These two group elements may be taken as exponentials of elements in the CSA, and it is therefore only for the implementation of the third step that we need something new. The relevant construction is known as an asymmetric orbifold [38] [39].

In a general orbifold construction one quotients by a discrete symmetry of the theory. In an asymmetric orbifold the action of the discrete symmetry is not symmetric in the left and right-moving sectors.

In an asymmetric orbifold, the group elements $g$ of the discrete symmetry act on the momenta as
$$g|P_L, P_R\rangle = e^{2\pi i (P_L \cdot a_L - P_R \cdot a_R)}|\theta_L P_L, \theta_R P_R\rangle \tag{6.1.4}$$

with $P_L$ the left moving momenta $(\mathbf{k}, k_{iL})$, $P_R$ the right moving momenta $k_{iR}$. The $\theta$'s are rotation matrices and $a_{L,R}$ correspond to shift vectors. The rotation matrices act independently on the left and right movers.

In an orbifold construction the theory splits into untwisted and twisted sectors. The untwisted sector describes the states that are left invariant under the symmetry $g$ that is divided out.

The identifications induced by the orbifold group also lead to new, so called twisted sectors. The identifications by the orbifold group lead to the possibility of closed strings that cannot be lifted to closed strings on the space before the quotienting. States from these strings take momenta on a lattice $I^* + a^*$. $I^*$ is the lattice dual to the lattice $I$, which is the sublattice of $\Gamma^{19,3}$ invariant under rotations by the $\theta$'s. The shift $a^*$ is the orthogonal projection of $a$ onto $I$ [38].

The main consistency condition on the asymmetric orbifold comes from level matching [38] (level matching is basically the requirement that the theory is invariant under spatial translations along the string world sheet). If the group element $g$ has order $m$, and $e^{2\pi i r_i/m}$, $i = 1, \ldots, 19$ denote the eigenvalues of $\theta_L$, and $e^{2\pi i s_i/m}$, $i = 1, 2, 3$ denote the eigenvalues of $\theta_R$, then for $m$ odd we should have [38]
$$\sum_i r_i^2 = (ma^*)^2 \bmod m \tag{6.1.5}$$

and for $m$ even

$$\sum_i r_i^2 = (ma^*)^2 \bmod 2m, \qquad \sum_i s_i = 0 \bmod 2 \qquad (6.1.6)$$

The procedure will now be as follows: To construct the heterotic theory corresponding to the triple $(\Omega_1, \Omega_2, \Omega_3)$, we first consider the heterotic string corresponding to $(\Omega_1, \Omega_2, 1)$. The holonomies $\Omega_1$ and $\Omega_2$ correspond to a special locus in the heterotic string moduli space where the lattice $\Gamma^{19,3}$ acquires a non-trivial automorphism (that acts purely within the $\Gamma^{16}$ part corresponding to the $E_8 \oplus E_8$ lattice). From the group theory, we know that the relevant automorphism corresponds to some element of the Weyl group, with order $n$. Next we take an asymmetric orbifold where $\theta_L$ is this automorphism, and $\theta_R$ trivial. The shift vector $a$ will be one period on the spatial 3–torus, divided by $n$. The resulting theory has two holonomies on the maximal torus, while traversing the third cycle (which has become a factor $n$ shorter in the construction) gives a holonomy implementing the Weyl reflection (compare with [32]).

Suppose now we take some commuting triple in one $E_8$ and another commuting triple in the other $E_8$. In terms of the gauge theory, there is no restriction on the allowed triples, but in string theory there is.

We recall some relevant fact from chapter 4. Any element of the group $E_8$ can be conjugated to an element of the form

$$\exp\{2\pi i h_\beta\} \qquad \text{with } \beta = \sum_{j=1}^{8} s_j \omega_j \qquad (6.1.7)$$

where the $\omega_j$ are the fundamental coweights of $E_8$, and the $s_j$ are a set of numbers satisfying $s_j \geq 0$, $\sum_{j=0}^{8} s_j g_j = 1$, with $g_j$ the root integers (this last relation determines the number $s_0$). To obtain the centraliser of this element, one erases from the extended Dynkin diagram all nodes $i$ for which $s_i \neq 0$, and adds $U(1)$'s to complete the rank of the group. The semi-simple part of the centraliser is $n$-fold connected, where $n$ is the greatest common divisor of the coroot integers of the roots that where erased (actually, since we are dealing with a simply laced group, one can drop the distinction between root and coroot, and weight and coweight).

We now consider the $E_8 \times E_8$ theory with a holonomy that is only non-trivial in one of the $E_8$-factors. If we require that the centraliser contains an $m$-fold connected factor with $m \neq 1$, we may never erase the extended root $\alpha_0$ (which has coroot integer 1). Therefore, in the above $s_0 = 0$. $\alpha_0$ will now survive as a root of the subgroup. In the presence of such a holonomy in the compactified heterotic string theory, the momenta $\mathbf{k}$ are of the form

$$\mathbf{k} = \left( \mathbf{q} + w \sum_{j=1}^{8} s_j \omega_j \right) \sqrt{\frac{2}{\alpha'}} \qquad (6.1.8)$$

where $w$ is the winding number in the direction of the holonomy. One may easily calculate

$$< \sum_{j=1}^{8} s_j \omega_j, \alpha_0 > = s_0 - 1 = -1, \qquad (6.1.9)$$

and therefore $\sum_{j=1}^{8} s_j \omega_j$ projected onto the subgroup of which $\alpha_0$ is a node, is minus the (co)weight corresponding to the simple root $\alpha_0$ in the unbroken gauge group. We conclude that in string theory the resulting lattice for the subgroup does not only consist of the lattice given by the group theory, but has extra (co)weight lattices due to the existence of the winding states. It is now easy to check that when these extra weight lattices are added to the group lattice, the lattice of a simply connected group remains.

The conclusion is then that in most cases where gauge theory analysis predicts a non-simply connected group, string theory actually gives a simply connected group. The only way to get to a non-simply connected subgroup is to embed in each $E_8$-factor a Wilson line that breaks the group to a subgroup containing an $m$-fold connected centraliser. In the gauge theory this would result in a centraliser whose semi-simple part has as its fundamental group $\mathbb{Z}_m \times \mathbb{Z}_m$, but the above analysis for the string theory will lead to the conclusion that the semi-simple part has $\pi_1 = \mathbb{Z}_m$. This leaves only the "diagonal" 12 of the original 144 possibilities.

For $E_8 \times E_8$ gauge theory on a 3–torus, the Chern-Simons invariant splits into two contributions, one for each $E_8$-factor. To obtain the total Chern-Simons invariant one may add the two contributions. An analysis for the possible configurations for the string theory shows that these precisely correspond to those cases where the two separate contributions add up to an integer. Non-integrality of the Chern-Simons invariant would present a serious problem for anomaly cancellation in the heterotic string [15]. We do not have to consider the possibility here, as the theory avoids this possibility.

### 6.1.2 Realisation

We will now present an analysis of the $\mathbb{Z}_m$, $m = 2, 3, 4, 5, 6$ orbifolds. The analysis is inspired on the group theory, described in previous chapters, and found in [6] [26] [28] [29]. Since in heterotic string theory (in its bosonised form) the relevant aspects of the gauge group can all be described in a perturbative way, these constructions may shed some light on phenomena in dual theories, where some or all of the relevant physics is non-perturbative.

In chapter 3 we constructed the holonomies in a minimal subgroup and embed them in larger groups. Here we will do the same, but for convenience take the minimal simply laced subgroup. The holonomies for $\mathbb{Z}_m$ orbifolds with $m = 2, 3, 4, 5, 6$ can be embedded in the simply laced groups $SO(8), E_6, E_7, E_8, E_8$. Triples embedded in these subgroups have a number of convenient properties that make them a starting point for a "canonical" construction of the orbifolds. One suitable property is that all three holonomies are conjugate to each other, which implies that they have the same set of eigenvalues (in every representation of the gauge group). Furthermore, one can also show, for example by the techniques of chapter 3, that these eigenvalues are $\exp\{2\pi i k/m\}$, with $k$ an integer. Other convenient properties will arise in the construction.

We will compactify the heterotic string on a 3–torus $T^3 = R^3/\Lambda$, where the lattice $\Lambda$ has an orthogonal basis, but unspecified radii. We start by turning on Wilson lines in the 1- and 2-directions, and will later use the 3-direction for the shift accompanying the orbifold projection. As before we set the metric on the torus $g_{ij} = \delta_{ij}$ (scales are absorbed in the radii, angles are $\pi/2$), and the antisymmetric tensorfield $B_{ij} = 0$ (these values are convenient, but not necessary: the moduli corresponding to these background fields will survive the orbifold projection). The holonomies in the 1 and 2 directions are parametrised by $\mathbf{a}_1, \mathbf{a}_2$. We will set $\mathbf{a}_3$ to zero until further notice. In the formulae for the momenta of the heterotic string compactified with Wilson lines the inner products $\mathbf{a}_1 \cdot \mathbf{a}_2$ appear. Since the Wilson lines at the relevant point in moduli space are conjugate to each other, we have $\mathbf{a}_1^2 = \mathbf{a}_2^2$. Writing $\mathbf{a}_i = (\tilde{a}_i^I, \tilde{a}_i^{II})$ to display the Wilson lines in the "first" (I), and "second" (II) $E_8$ factor, it is convenient to set the Wilson lines to $\mathbf{a}_1 = (\tilde{a}_1, \tilde{a}_1)$, and $\mathbf{a}_2 = (\tilde{a}_2, -\tilde{a}_2)$. This eliminates the inner product $\mathbf{a}_1 \cdot \mathbf{a}_2$ from our formulae. We will use an orbifold projection that is symmetric in both $E_8$-factors. With the conventions for the holonomies this implies that the total gauge field configuration has the Chern-Simons invariant equal to zero. There should be equivalent ways of realising a non-trivial triple, as the two $E_8$-factors are independent, but we find the above conventions the most convenient ones.

The value of $\mathbf{a}_1^2 = \mathbf{a}_2^2$ can be obtained in various ways (which of course all give the same result). In the set-up we have chosen, the holonomy parametrised by $\mathbf{a}_1$ will eliminate only one node from each $E_8$ extended Dynkin diagram, and from the previous discussion it follows that $\mathbf{a}_1$ is of the form $(\omega_j h_j^{-1}, \omega_j h_j^{-1})$, with $\omega_j$ the fundamental (co)weight, and $h_j$ the (co)root integer associated to the node (for some values of $m$ there seem to be more options, but only one corresponds to the embedding of a minimal triple). $\mathbf{a}_i^2$ can now easily be found by using that the weight can be expanded in the simple roots, so $\omega_i = \sum_k p_i^k \alpha_k$. As $\omega_i$ is a fundamental weight, the coefficients $p_i^k$ are simply the entries of the inverse Cartan matrix. It is then trivial to show that $\mathbf{a}_i^2 = 2 p_i^i / (h_i)^2$ (no summation implied). $p_i^i$ is a diagonal element of the inverse Cartan matrix, and for the cases we study we find $\mathbf{a}_i^2 = 2(m-1)/m$.

Putting all these conventions and results together, we find the momenta for the compactified heterotic string *before the orbifold projection*:

$$\mathbf{k} = (\mathbf{q} + \sum_{i=1,2} w_i \mathbf{a}_i) \sqrt{\frac{2}{\alpha'}} \tag{6.1.10}$$

$$k_{iL,R} = \frac{m n_i - m \mathbf{q} \cdot \mathbf{a}_i - w_i(m-1)}{m R_i} \pm \frac{w_i R_i}{\alpha'} \quad i = 1, 2 \tag{6.1.11}$$

$$k_{3L,R} = \frac{n_3}{R_3} \pm \frac{w_3 R_3}{\alpha'} \tag{6.1.12}$$

In the above no summation is implied unless explicitly stated. Due to the special choice of Wilson lines, $\mathbf{q} \cdot \mathbf{a}_i$ is always a multiple of $1/m$, and hence the combination $m n_i - m \mathbf{q} \cdot \mathbf{a}_i - w_i(m-1)$ is always an integer, and actually can take any integer value.

We are now at the point in moduli space where we want to perform the orbifold construction, so we have to make sure that this lattice has the right symmetries and obeys the orbifold consistency conditions. That it has the right symmetries is not entirely obvious due to the behaviour of the string winding states, as explained before. If all $w_i$ would be identically zero, this theory would be the same as the gauge theory, which does have the appropriate symmetries at this point. For checking the symmetries we will need some facts on the orbifold operation. It consists of a shift $a$, and a rotation $\theta_L$, in these cases acting solely on the gauge part of the lattice. Since there is no right rotation we will drop the subscript and write $\theta$ for the rotation, and $\theta(\mathbf{v})$ for the image of the vector $\mathbf{v}$ when rotated by $\theta$. $\theta$ is an element of the Weyl group of $E_8 \times E_8$, which is a discrete subgroup of the orthogonal group $O(16)$ (16 being the rank of $E_8 \times E_8$). Notice that "rotation" in the above also may include reflections. $\theta$, regarded as the holonomy around the third cycle of the torus, should commute with the other Wilson lines. This does not mean that $\theta(\mathbf{a}_i) = \mathbf{a}_i$, but rather implies the weaker condition [6]

$$\theta(\mathbf{a}_i) = \mathbf{a}_i + \mathbf{z}, \qquad (6.1.13)$$

where $\mathbf{z}$ is some lattice vector (for the heterotic string, theories based on lattices that differ by a shift over a lattice vector are identified). There is some ambiguity in the choice of $\mathbf{z}$, but in the cases of non-trivial commuting triples, the lattice vector cannot be chosen to equal zero. $\mathbf{z} = 0$ corresponds to a trivial triple, which would correspond to some Narain compactified-theory. In group theory, this can be seen by exponentiating the relation (6.1.13) to give $yxy^{-1} = Zx$, where $x = \exp\{2\pi i h_{\mathbf{a}_i}\}$, $y$ implements the Weyl reflection $\theta$ by conjugation, and $Z$ is the centre element $\exp\{2\pi i h_{\mathbf{z}}\}$.

To see that the lattice has the right symmetry, we construct the image of a vector with labels $(\mathbf{q}, n_i, w_i, n_3, w_3)$. There should exist an image vector labelled by $(\mathbf{q}', n'_i, w'_i, n'_3, w'_3)$, with $\mathbf{k}' = \theta(\mathbf{q} + w_1 \mathbf{a}_1 + w_2 \mathbf{a}_2)$. We expect existence for generic radii of the space-torus, which implies $w_i = w'_i$ ($i = 1, 2, 3$) and $n_3 = n'_3$, so we have to check the consistency of the equations

$$\mathbf{q}' = \theta(\mathbf{q}) + w_1(\theta(\mathbf{a}_1) - \mathbf{a}_1) + w_2(\theta(\mathbf{a}_2) - \mathbf{a}_2) \qquad (6.1.14)$$
$$n'_i - \mathbf{q}' \cdot \mathbf{a}_i = n_i - \mathbf{q} \cdot \mathbf{a}_i \qquad i = 1, 2 \qquad (6.1.15)$$

The group lattice equation (6.1.14) is consistent by construction, the only thing that has to be verified is that $(\mathbf{q}' - \mathbf{q}) \cdot \mathbf{a}_i$ is an integer for $i = 1, 2$. This is true if $(\theta(\mathbf{q}) - \mathbf{q}) \cdot \mathbf{a}_i$ and $(\theta(\mathbf{a}_i) - \mathbf{a}_i) \cdot \mathbf{a}_j$ are integer.

$(\theta(\mathbf{a}_i) - \mathbf{a}_i) \cdot \mathbf{a}_j$ is actually zero for $i \neq j$, because of the specific choice of $\mathbf{a}_1, \mathbf{a}_2$ and because the transformation $\theta$ is symmetric in both $E_8$ factors. In the above we already remarked that $\theta(\mathbf{a}_i) - \mathbf{a}_i = \mathbf{z}$ for some lattice vector $\mathbf{z}$. Then $\mathbf{a}_i^2 = (\theta(\mathbf{a}_1))^2 = \mathbf{z}^2 + 2\mathbf{a}_i \cdot \mathbf{z} + \mathbf{a}_i^2$, where use was made of the fact that $\theta \in O(16)$. Since $\mathbf{z}$ is on an even lattice, it immediately follows that $\mathbf{a}_i \cdot \mathbf{z}$ is an integer. Finally, rewriting $(\theta(\mathbf{q}) - \mathbf{q}) \cdot \mathbf{a}_i$ as $(\theta^{-1}(\mathbf{a_i}) - \mathbf{a_i}) \cdot \mathbf{q}$, we notice that this is an inner product between two lattice vectors, and hence also integer. Therefore an image point always exists.

For the orbifold consistency conditions (6.1.5), (6.1.6), we need the $r_i$ parametrising the eigenvalues of $\theta$. These can be obtained from group theory. We took $a$ in a particular direction on the spatial 3–torus, which implies that it has the form $a_{iL} = a_{iR}$, and components in the gauge torus directions zero. As $a^*$ is the projection of $a$ on the lattice invariant under $\theta$, and $\theta$ acts only on the gauge torus, one finds that $(a^*)^2$ is always zero. For $m = 2$, $\theta$ has eight nonzero $r_i$, all equal to 1. For $m = 3$, $\theta$ has 12 non-zero $r_i$, 6 equal to 1, and 6 equal to 2. For $m = 4$, $\theta$ has 14 non-zero $r_i$, 4 equal to 1, 6 are equal to 2, and 4 are equal to 3. For $m = 5$ $\theta$ has 16 non-zero $r_i$, 4 equal to 1, 4 equal to 2, 4 equal to 3, and 4 equal to 4. For $m = 6$ finally, there are 16 non-zero $r_i$, 1 with multiplicity 2, 2 with multiplicity 4, 3 with multiplicity 4, 4 with multiplicity 4, and finally 5 having multiplicity 2. In all cases the orbifold consistency conditions are satisfied.

All requirements are fulfilled, hence the asymmetric orbifolds exist and can be constructed.

Having obtained the eigenvalues, it is also possible to calculate the zero-point energies for the twisted sector(s), using that the contribution of each eigenvalue $k$ of $\theta$ to the zero-point energy is [44]

$$\frac{1}{48} - \frac{1}{16}(2\frac{k}{m} - 1)^2 \tag{6.1.16}$$

Each periodic boson contributes $-1/24$. Adding all contributions one finds that the zero point energies are $-1/m$ for the twisted sector(s) of the $\mathbb{Z}_m$-orbifold.

### 6.1.3 Lattices for the orbifolds

Mikhailov [34] defines a lattice for the CHL-string. We will now do the same for the asymmetric orbifolds constructed in the above.

Start with the momenta (6.1.10), (6.1.11), (6.1.12). The momenta in the untwisted sector, after projecting onto invariant states can be described as follows. First define the projection

$$P_\theta = \frac{1}{m} \sum_{i=0}^{m-1} \theta^i \tag{6.1.17}$$

$P_\theta$ projects all lattice vectors onto the space invariant under $\theta$, since $\theta^m = 1$. $P_\theta(\mathbf{a}_i) = 0$ by construction, and we set $P_\theta(\mathbf{q}) = \mathbf{q}_{inv}/\sqrt{m}$ (the extra rescaling with $\sqrt{m}$ will become clear soon, when we will reabsorb it in a rescaling of $\alpha'$). Define $\tilde{n}_i = mn_i - m\mathbf{q} \cdot \mathbf{a}_i - w_i(m-1)$ for $i = 1, 2$. The radii for the 1 and 2-direction are rescaled by defining $R'_i = mR_i$. We also define $\alpha'' = m\alpha'$ (Note that the invariant radii $R_i/\sqrt{\alpha'}$ are only rescaled by a factor $\sqrt{m}$).

We now define a lattice with the vectors

$$\mathbf{v} = (\mathbf{q}_{inv})\sqrt{\frac{2}{\alpha''}} \qquad v_{iL,R} = \frac{\tilde{n}_i}{R'_i} \pm \frac{w_i R'_i}{\alpha''} \quad i = 1, 2 \qquad v_{3L,R} = \frac{n_3}{R_3} \pm \frac{mw_3 R_3}{\alpha''} \tag{6.1.18}$$

The vectors $v_{iL,R}$ and $v_{3L,R}$ form a lattice, which (when rescaled by $\sqrt{\alpha''/2}$) in Mikhailov's language would be called $\Gamma_{2,2} \oplus \Gamma_{1,1}(m)$. This lattice arises in an intermediate step, because we have not yet included the twisted sectors.

Now consider the vectors $\mathbf{v}$ which are the vectors of $\Gamma_8 \oplus \Gamma_8$ projected onto the subspace invariant under $\theta$ (and suitably rescaled). Of course, for every $m$ this defines a different lattice. It can be verified from group theory or explicit calculation that these lattices are $D_4 \oplus D_4$, $A_2 \oplus A_2$, $A_1 \oplus A_1$ for $m = 2, 3, 4$, and the empty lattice for $m = 5, 6$ ($D_4$ is the root lattice of $SO(8)$, $A_2$ is the root-lattice of $SU(3)$, and $A_1$ is the root lattice of $SU(2)$). A note of caution is in place here: We define the lattices $D_4$, $A_2$ and $A_1$ as usual with their roots normalised at length $\sqrt{2}$. For the symmetry groups arising in the gauge theory, the roots of length $\sqrt{2}$ form the *short* roots of non-simply laced algebra's at level 1. For example, at the point in moduli space constructed here the gauge group is $F_4$ for the $m = 2$ case, and $G_2$ for the $m = 3$ case, with the long roots having length 2 and $\sqrt{6}$ respectively. The gauge group $SU(2)$ in the $m = 4$ case has roots of length $\sqrt{8}$ (it's at level 4), the vectors with length $\sqrt{2} = \sqrt{8}/2$ are on the weight lattice of $SU(2)$ (note that, although there is no 4-laced algebra, there is a 4-laced affine Dynkin diagram which turns up in the group description. See Borel et al. [6])

In the twisted sectors, the momenta lie on the lattice $I^*$, which is dual to the lattice $I$ of vectors invariant under $\theta$. We will treat the parts of the lattices that represent the group quantum numbers, and the part that represents the space quantumnumbers separately.

The group parts of the lattices $I$ of invariant vectors can be easily calculated. An invariant vector $\mathbf{v}$ satisfies $P_\theta(\mathbf{v}) = \mathbf{v}$. In the above we defined the lattices of vectors $\sqrt{m}\,P_\theta(\mathbf{v}')$, and named them $D_4 \oplus D_4$, $A_2 \oplus A_2$, $A_1 \oplus A_1$ for $m = 2, 3, 4$. With these definitions, the group parts of the lattices $I$ of invariant vectors should be called $\sqrt{2}(D_4^* \oplus D_4^*)$, $\sqrt{3}(A_2^* \oplus A_2^*)$, $2(A_1^* \oplus A_1^*)$, for $m = 2, 3, 4$. The stars denote the dual lattices, which are the (co)weight lattices. The stars arise because of the definition of $D_4$, $A_2$, $A_1$ in the above, and the fact that we have to keep track of relative orientations. The group parts of the lattices $I^*$ are then $(D_4 \oplus D_4)/\sqrt{2}$, $(A_2 \oplus A_2)/\sqrt{3}$, $(A_1 \oplus A_1)/2$.

To construct the spatial part of the invariant lattice we note that, to be invariant the group vector $\mathbf{q} + \sum w_i \mathbf{a}_i$ has to be on the root lattice, implying that $w_i$ is a multiple of $m$, say $l_i m$ (the $\mathbf{a}_i$ are on the weight lattice of some subgroup, and only multiples of $m\mathbf{a}_i$ are on a root lattice. This can also be seen from the value of $\mathbf{a}_i^2$). Also, if $\mathbf{q} + \sum w_i \mathbf{a}_i$ is on the invariant lattice, then its dot product with either $\mathbf{a}_i$ has to vanish, implying that $\mathbf{q} \cdot \mathbf{a}_i = -2l_i(m-1)$. The spatial momenta on the invariant lattice are thus given by

$$\frac{n_i + l_i(m-1)}{R_i} \pm \frac{m l_i R_i}{\alpha'} \quad i = 1, 2 \qquad \frac{n_3}{R_3} \pm \frac{w_3 R_3}{\alpha'} \tag{6.1.19}$$

Note that $n_i + l_i(m-1)$ can take any integer value, while $l_i m$ is a multiple of $m$. The momenta on the dual to the spatial part of the invariant lattice are then given by the vectors

$$\frac{n_i'}{m R_i} \pm \frac{w_i' R_i}{\alpha'} \quad i = 1, 2 \qquad \frac{n_3'}{R_3} \pm \frac{w_3' R_3}{\alpha'} \tag{6.1.20}$$

We now have constructed $I^*$, we still need the shift $a^*$. The only non-zero components of this vector are in the spatial 3-direction. These are given by multiples of

$$\Delta v_{3L,R} = \pm \frac{R_3}{m\alpha'} \tag{6.1.21}$$

We finally perform the same reparametrisations as in the untwisted sector: we scale the group parts with $\sqrt{m}$, define $R_i' = m R_i$ for $i = 1, 2$, and set $\alpha'' = m\alpha'$. We see that the lattices $I^*$ are simply copies of the lattice for the untwisted sector, confirming a result from appendix A of [38]. The momenta are given by $I^* + a^*$ with the non-zero components of $a^*$ given by $t\Delta v_{3L,R}$ (with $t = 1, \ldots, (m-1)$ labelling the twisted sectors). This implies that the lattices for the untwisted and twisted sectors can be put together to a lattice $\Lambda$, in the process of which the spatial part of the lattice is completed to $\Gamma_{3,3}$.

We only calculated the lattices for very specific orbifolds, with special values of the holonomies and other background fields. One may extend to the general case. First we note that the metric and antisymmetric tensor field did not play any role thus far, and the moduli corresponding to these fields survive the orbifold projection. For the holonomies we took special values, that had

$$P_\theta^\perp(\mathbf{a}_i) = (\mathbb{1} - P_\theta)(\mathbf{a}_i) = \mathbf{a}_i \tag{6.1.22}$$

We may generalise to a more general case with holonomies parametrised by $\mathbf{a}'_i$, provided the projection $P_\theta^\perp$ gives the special value $P_\theta^\perp(\mathbf{a}'_i) = \mathbf{a}_i$. Possible moduli for varying the holonomies are then given by $P_\theta(\mathbf{a}'_i)$. One may use the general formulas from [36] [37] [18] to show that this results in a moduli space that locally has the form

$$\frac{SO(19-\Delta r, 3)}{SO(19-\Delta r) \times SO(3)} \quad (6.1.23)$$

where $\Delta r$ is the rank reduction for the $\mathbb{Z}_m$ orbifold (see also the table below).

We summarise the results found thus far in a table. We also include the standard Narain-compactification (denoted by $\mathbb{Z}_1$), which fits perfectly in the picture when we take trivial holonomies and $m = 1$.

|  | $\Lambda$ | $\Lambda^\perp$ | $\Delta r$ | $E_t$ |
|---|---|---|---|---|
| $\mathbb{Z}_1$ | $\Gamma_{3,3} \oplus E_8 \oplus E_8$ | $\emptyset$ | 0 | $-1$ |
| $\mathbb{Z}_2$ | $\Gamma_{3,3} \oplus D_4 \oplus D_4$ | $D_4 \oplus D_4$ | 8 | $-1/2$ |
| $\mathbb{Z}_3$ | $\Gamma_{3,3} \oplus A_2 \oplus A_2$ | $E_6 \oplus E_6$ | 12 | $-1/3$ |
| $\mathbb{Z}_4$ | $\Gamma_{3,3} \oplus A_1 \oplus A_1$ | $E_7 \oplus E_7$ | 14 | $-1/4$ |
| $\mathbb{Z}_5$ | $\Gamma_{3,3}$ | $E_8 \oplus E_8$ | 16 | $-1/5$ |
| $\mathbb{Z}_6$ | $\Gamma_{3,3}$ | $E_8 \oplus E_8$ | 16 | $-1/6$ |

Table 6-1. Lattices $\Lambda$, complements $\Lambda^\perp$, rank reduction $\Delta r$ and zero-point energies in the twisted sector $E_t$ for $Z_m$ asymmetric orbifolds corresponding to triples.

### 6.1.4 Duality with the CHL-string

One may also consider the possibility of constructing a heterotic $Spin(32)/Z_2$ theory that corresponds to the triple for $Spin(32)$ gauge theory with periodic boundary conditions. Following the same steps as for the $E_8 \times E_8$ theory, one finds that construction of this triple is not possible. We also note that this triple, if it would have existed, would have non-integral Chern-Simons invariant, again confirming that the theory avoids this possibility.

There exists another asymmetric orbifold for the $Spin(32)/\mathbb{Z}_2$ string [32] [55]. Using the fact that the group $Spin(32)/\mathbb{Z}_2$ is not simply connected, one may consider compactifying the theory on a 2–torus with twisted boundary conditions, resulting in a theory without vector structure. The analysis of [6], and the one we presented in chapter 5, indicated multiple components in the moduli space of $Spin(32)$ gauge theory compactified on a 3–torus without vector structure. Again one may deduce from the fact that the topology of the gauge group in string theory differs from the topology found from ordinary gauge field theory due to the presence of winding states, that none of the extra components can be constructed for the $Spin(32)/\mathbb{Z}_2$ string.

Therefore, for the $Spin(32)/\mathbb{Z}_2$ string, the only possibilities for compactification on a 3–torus are the standard Narain-compactification [36], and the one without vector structure from [32] [55], embedding the 2–torus in the 3–torus with a suitable holonomy on the third cycle.

In [32] [55] it was argued that compactification of the $Spin(32)/\mathbb{Z}_2$ string without vector structure is dual to the CHL-compactification of the $E_8 \times E_8$-string. We will show here that also the $E_8 \times E_8$-compactification with a $\mathbb{Z}_2$-triple is dual to these theories.

A derivation of the duality between the $E_8 \times E_8$ and $Spin(32)/\mathbb{Z}_2$ theory can be found in appendix C. We will follow the notation used there, but will replace $R\mathbf{A}$ with $\mathbf{a}$. Consider a heterotic string theory compactified on a circle. The transformation of the momenta of the heterotic string under the inclusion of a holonomy $\mathbf{a}$ is as follows (see eq. (6.1.1) (6.1.2)):

$$W(\mathbf{a})\begin{pmatrix} \mathbf{q} \\ \sqrt{\frac{\alpha'}{2}}(\frac{n}{R}+\frac{wR}{\alpha'}) \\ \sqrt{\frac{\alpha'}{2}}(\frac{n}{R}-\frac{wR}{\alpha'}) \end{pmatrix} = \begin{pmatrix} \mathbf{q}+w\mathbf{a} \\ \sqrt{\frac{\alpha'}{2}}(\frac{n-\mathbf{q}\cdot\mathbf{a}-w\mathbf{a}^2/2}{R}+\frac{wR}{\alpha'}) \\ \sqrt{\frac{\alpha'}{2}}(\frac{n-\mathbf{q}\cdot\mathbf{a}-w\mathbf{a}^2/2}{R}-\frac{wR}{\alpha'}) \end{pmatrix} \quad (6.1.24)$$

In the appendix C we show that the duality transformation between $Spin(32)/\mathbb{Z}_2$ and the $E_8 \times E_8$ string can be characterised by:

- Its form is $W(-\mathbf{a}')uW(\mathbf{a})$, and it maps $\Gamma \oplus \Gamma_{1,1}$ to $\Gamma' \oplus \Gamma'_{1,1}$, with $\Gamma \neq \Gamma'$;
- $\mathbf{a}$ is a root of $G'$, and $2\mathbf{a}$ is a coweight of $G$;
- $\mathbf{a}'$ is a root of $G$, and $2\mathbf{a}'$ is a coweight of $G'$;
- $\mathbf{a} \cdot \mathbf{a}' = -\frac{1}{2}$, and $RR' = \alpha'/2$.

where $\mathbf{a}$ and $\mathbf{a}'$ describe Wilson lines for the two heterotic theories, $R$ and $R'$ are their respective compactification radii, $\Gamma$ and $\Gamma'$ are the lattices $\Gamma_8 \oplus \Gamma_8$ and $\Gamma_{16}$, and $G$ and $G'$ are their associated algebra's $E_8 \times E_8$ and $SO(32)$. Finally $u$ is the transformation that reflects the right moving momentum.

In the example below, we will compactify on a 3–torus, and choose the $\mathbf{a}_i$ to be orthogonal for the different $i$. Then the momenta and winding numbers associated to the dimensions of the spatial torus do not mix any more (as in the momentum formulae for the heterotic string, the off-diagonal contributions from $\mathbf{a}_i \cdot \mathbf{a}_j$, $g_{ij}$ and $B_{ij}$ are all set to 0). We can therefore dualise in each direction separately.

In the example, there are always two holonomies on the maximal torus present. On the third radius of a three-torus we perform an asymmetric orbifold construction, with a shift over half the third circle and a transformation $\theta$ on the group lattice that will be listed explicitly.

We start with a particular example of an $E_8 \times E_8$-string, with background gauge fields

$$\begin{aligned} \mathbf{a}_1 &= (1, 0^{14}, -1,) \\ \mathbf{a}_2 &= (0^4, \frac{1}{2}^4, -\frac{1}{2}^4, 0^4) \\ \theta &: \theta(u_1, \ldots, u_{16}) = -(u_{16}, \ldots, u_1) \end{aligned} \quad (6.1.25)$$

Here the first $E_8$ is in the first 8 positions of the vectors, and the second $E_8$ in the last 8. The transformation $\theta$ swaps the two $E_8$'s. This is the CHL-construction. In this case, both $\mathbf{a}_i$ are invariant under $\theta$ and can actually be smoothly deformed away. The corresponding circles can then be expanded to give the 9 dimensional CHL-string.

The specific form of the $\mathbf{a}_i$ was chosen to allow duality transformations. We dualise in the 1 direction to get the $Spin(32)/\mathbb{Z}_2$ string with holonomies

$$\begin{aligned}
\mathbf{a}_1 &= (-\frac{1}{2}^2, \frac{1}{2}^4, -\frac{1}{2}^2, 0^8) \\
\mathbf{a}_2 &= (0^4, \frac{1}{2}^4, -\frac{1}{2}^4, 0^4) \\
\theta &: \theta(u_1, \ldots, u_{16}) = -(u_{16}, \ldots, u_1)
\end{aligned} \quad (6.1.26)$$

For this case $\mathbf{a}_1$ no longer commutes with $\theta$. The difference $\theta(\mathbf{a}_1) - \mathbf{a}_1$ lies on the spin weight lattice of $Spin(32)$, indicating that we are dealing with a compactification without vector structure. We have reproduced the result from [32] [55]. The "twist" is in the 1-3-direction. $\mathbf{a}_2$ is still invariant under the transformation $\theta$, and can therefore be smoothly deformed away. We can therefore reach 8 dimensions in this situation, but not more.

On this theory we apply another T-duality, this time in the 2 direction, to obtain another $E_8 \times E_8$ string with:

$$\begin{aligned}
\mathbf{a}_1 &= (-\frac{1}{2}^2, \frac{1}{2}^4, -\frac{1}{2}^2, 0^8) \\
\mathbf{a}_2 &= (0^3, 1, -1, 0^9) \\
\theta &: \theta(u_1, \ldots, u_{16}) = -(u_{16}, \ldots, u_1)
\end{aligned} \quad (6.1.27)$$

Now the first $E_8$ is interpreted to be in the positions $1, \ldots, 4, 13, \ldots, 16$, and the second $E_8$ in positions $5, \ldots, 12$. Actually this formulation is not completely accurate, since neither $\mathbf{a}_1$ nor $\mathbf{a}_2$ is invariant under $\theta$, and it is therefore impossible to decompactify any dimension. This is a particular case of a non-trivial triple in $E_8 \times E_8$. Hence we see that the CHL-string, toroidal compactification without vector structure and the non-trivial $\mathbb{Z}_2$-triple are actually one and the same. That they share the same Mikhailov lattice [34], rank reduction and unbroken symmetry groups is now more or less obvious, and also from various proposed dualities it is hard to see how there could be more than one way to divide by a $\mathbb{Z}_2$.

Mikhailov claims that the 7 dimensional CHL-string moduli space has three cusps, which he interprets as "ways for going far away in the moduli space"[34]. It is natural to conjecture that the above three interpretations correspond to these three cusps. Strange is however that Mikhailov claims that two cusps correspond to $Spin(32)$ interpretations and one has a natural $E_8 \times E_8$ interpretation, while the above suggests that it is the other way around, two $E_8 \times E_8$ cusps and one $Spin(32)$. Mikhailov is very sketchy in his interpretation of the cusps, so it is hard to check his arguments.

The other asymmetric orbifolds we constructed (corresponding to $\mathbb{Z}_m$-triples with $m > 2$) can be compactified on additional tori, but probably do not have analogues in dimensions higher than 7. In particular, in the papers [46] and [9], the authors constructed some asymmetric orbifolds in 6 dimensions that are probably compactifications of our 7-dimensional orbifolds.

## 6.2 Orientifold realisations

The $Spin(32)/\mathbb{Z}_2$ heterotic string, and the type I open superstring are conjectured to be one and the same theory [42]. That they appear to be so different is thought to be due to the fact that they describe the theory in different limits that are related by a weak-strong coupling duality. Taking this conjecture to be true, all previously constructed compactifications for the $Spin(32)/\mathbb{Z}_2$ heterotic string should have analogues for the type I string. More dramatically, the compactifications for the type I superstring and its T-duals constructed in chapter 5 that cannot be reproduced by the heterotic string should suffer some kind of inconsistency. There are good indications that the configuration of a 6 dimensional $O^-$ orientifold fixed plane with a single $D_6$ brane on top cannot be embedded in a consistent string theory, invalidating precisely those configurations that are not reproduced by a heterotic string compactification [15].

We also recall at this spot that there were orientifold configurations in chapter 5 that represented multiple components in the moduli space of the gauge theory. These ambiguities should also disappear in the full string theory. The topology of gauge groups as analysed from the heterotic string indicates that this is indeed the case.

In the type I picture the topology of the gauge group is modified by non-perturbative effects. The heterotic string is thought to appear in type I string theory as a soliton, the D-string [42], and its effects cannot be studied from standard string perturbation theory that was the basis for the analysis of chapter 5. There is a remarkable hierarchy of non-perturbative effects in type I theory affecting the topology of the gauge group. In perturbative type I theory all states transform in the adjoint of $O(32)$. In [56] it is argued that instanton effects break $O(32)$ to the adjoint of $SO(32)$. According to [50] [51] [56] there are solitonic particles transforming in a spinor representation, setting the gauge group to $Spin(32)/\mathbb{Z}_2$. And finally in the above we argued that a solitonic string affects the topology of subgroups of $Spin(32)/\mathbb{Z}_2$. The instanton, solitonic particle, and solitonic string all act as obstructions against certain compactifications. They all feature in the paper [56] which tries to explain the existence of these non-perturbative excitations from a unifying framework.

We now return to our original theme, compactifications of string theory on a 3–torus. The $Spin(32)/\mathbb{Z}_2$ heterotic string allowed only one non-trivial compactification on a 3–torus, being one without vector structure. By strong-weak duality this corresponds to a type I compactification without vector structure, with the same holonomies on the 3–torus. From the results of chapter 5 it follows that this is T-dual to IIA string theory on the product of a Möbius strip and a circle. The edge of the Möbius strip is formed by an $O^-$ orientifold plane. This again is dual to a product of a IIB orientifold $T^2/\mathbb{Z}_2$ with a circle. The $\mathbb{Z}_2$ of the orientifold reflects two of the three torus coordinates, and has 4 fixed points on $T^2$, 3 of which correspond to $O^-$ planes, the fourth being an $O^+$ plane. Another T-duality brings us to a IIA orientifold $T^3/\mathbb{Z}_2$ where the $\mathbb{Z}_2$ reflects all torus coordinates, hence gives 8 fixed planes, to be subdivided in 6 $O^-$ planes and 2 $O^+$ planes.

The T-duality that transformed type I theory on a 3–torus without vector structure to IIA theory on the product of the Möbius strip and the circle acted on one of the coordinates of the two-dimensional sub-torus that carried the holonomies that eliminated the vector structure.

On the third circle there is a holonomy taking some value in the maximal torus left invariant by the other two holonomies. One may also T-dualise the type I theory in this direction. This results in a IIA-theory on the product $S^1/\mathbb{Z}_2 \times T^2$. The $\mathbb{Z}_2$ orientifold projection acts on the circle with two fixed points, corresponding to $O^-$ planes. The $T^2$ carries holonomies that encode absence of vector structure.

The last T-duality is more subtle. We know that T-duality on a $T^2$ without vector structure leads to a theory on the Möbius strip, but we have to be careful to include the extra $S^1/\mathbb{Z}_2$ in the right way. A careful analysis leads to a IIB orientifold that may be described as $(T^2/\mathbb{Z}_2^o \times S^1)/\mathbb{Z}_2^t$. We labelled the two $\mathbb{Z}_2$'s appearing here differently, as they have completely different actions. The $\mathbb{Z}_2^o$ acting on $T^2$ is the standard orientifold action acting on the 2–torus. It inverts the two coordinates of the two-torus, and on top of that reflects the worldsheet of the string. The fixed points of $\mathbb{Z}_2^o$ are $O^-$ planes. The action of $\mathbb{Z}_2^t$ is harder to describe. It acts simultaneously on $T^2/\mathbb{Z}_2^o$ and on the circle $S^1$. The action on $S^1$ is a shift over half the period of the $S^1$. The action of $\mathbb{Z}_2^t$ on $T^2/\mathbb{Z}_2^o$ is a reflection in a point (it has a second fixed point because of the various identifications). The presence of $O^-$ planes restricts the possibilities, as the $O^-$ planes should be mapped to each other by $\mathbb{Z}_2^t$. Also the D-brane configuration present should be invariant under the $\mathbb{Z}_2^t$. The point is that in the T-dual to the type I-theory without vector structure, the D-brane configuration on the orientifold $T^2/\mathbb{Z}_2^o$ indeed has this symmetry.

Other T-dualities will take us back to one of the orientifolds considered before.

## 6.3   M-theory and F-theory realisations

This section consists of a brief sketch of extensions of the previous material to M- and F-theory, the theories describing strongly coupled IIA resp. IIB-strings. A more complete account of this material will be published later [15].

It is argued that the strong coupling limit of the 10 dimensional IIA string theory leads to a new 11-dimensional theory, called M-theory [54]. A complete description of M-theory has not been found yet, but there is a large amount of information available on aspects of the theory. The low energy theory of M-theory should be 11-dimensional supergravity [13], the unique maximally supersymmetric theory in 11 dimensions not containing fields with spin bigger than 2. M-theory is thought to be a theory of two-dimensional membranes. Its relation to IIA string theory is by compactification. The IIA theory is obtained as M-theory on a circle, the radius of the circle being proportional to the coupling constant of the IIA theory. For strong coupling the circle is large, and the theory lives in an 11 dimensional space. For weak coupling the circle is very small, and the theory effectively 10 dimensional. The IIA string is thought to be a membrane wrapped around the circle. Also various other extended objects in the IIA theory have an 11 dimensional interpretation.

There is another compactification of M-theory to 10 dimensions. Compactifying the theory on $S^1/\mathbb{Z}_2$ ($\mathbb{Z}_2$ being the reflection on the circle) also leads to a 10 dimensional theory, and the possibility to wrap membranes around $S^1/\mathbb{Z}_2$ suggest that this may also be a string theory. The fixed points of the $\mathbb{Z}_2$ however may lead to problems with the low energy supergravity theory. As it turns out the fixed points indeed lead to anomalies, but they can be successfully cancelled by inserting a 10 dimensional gauge theory with gauge group $E_8$ at the fixed points [25]. The proposal in the cited paper is that M-theory on $S^1/\mathbb{Z}_2$ is the strong coupling limit of the $E_8 \times E_8$ heterotic string. Again the size of the eleventh dimension is interpreted as the strength of the coupling. At large coupling the fixed hyperplanes with the $E_8$ gauge theories are far apart. At small coupling they are a small distance apart, leading to an $E_8 \times E_8$-string theory.

With this information it is almost trivial to obtain the strong coupling limit of the CHL-string. $S^1/\mathbb{Z}_2$ can be represented by an interval, with an $E_8$-gauge theory living at each end. The CHL-string was a compactification of the heterotic $E_8 \times E_8$ string on a circle with a holonomy that interchanges the $E_8$ factors. In the eleven dimensional picture, this implies the presence of an extra circle. Traversing the circle, the two endpoints of the interval should be interchanged. Hence the strong coupling limit of the CHL-string is M-theory on a Möbius strip. Compactifying this theory on an additional 2–torus, one obtains a dual of the 7 dimensional CHL-string. The interpretation as the strong coupling dual of the CHL-string identifies the size of the interval as the coupling constant for the string. Alternatively, one may interpret the size of one of the circles of the additional 2–torus as the coupling constant for a string theory. Moving to weak coupling then means shrinking the circle, and the limit for small circle size is interpreted as IIA theory on a Möbius strip, with an additional circle, which is also a theory we encountered before. $E_8$ is not a gauge group that is encountered in the IIA theory. It is usually argued that $E_8$ arises due to strong coupling effects, and that in this low energy description it is broken to its $SO(16)$-subgroup.

The strong coupling limit for $E_8 \times E_8$ theory with a triple is also not hard to find. The theory lives on a product of $T^3$ with $S^1/\mathbb{Z}_2$. On $T^3$ appropriate holonomies should be included for the $E_8$ theories. In the case of a $\mathbb{Z}_2$ triple, one may again consider shrinking one of the circles of $T^3$ to obtain a weakly coupled IIA string theory. This IIA theory lives on $T^2 \times S^1/\mathbb{Z}_2$, with on the fixed points of the $\mathbb{Z}_2$ $O^-$ planes inserted. On the remaining 2–torus non-trivial holonomies that reduce the rank are included. If one again assumes that the $E_8$'s are broken to $Spin(16)/\mathbb{Z}_2$ subgroups, and realises that this group allows a compactification on a 2–torus without vector structure, then it is clear that this corresponds to one of the theories we discussed before.

In the previous chapter we encountered another IIA theory, living on an orientifold $T^3/\mathbb{Z}_2$ with 6 of the 8 fixed points on the torus occupied by $O^-$-planes, and 2 by $O^+$-planes. The strong coupling limit of this theory is still rather mysterious. The case where all fixed points are occupied by $O^-$ planes, was argued in [54] to correspond to M-theory on a K3. A K3 is a non-trivial 4-dimensional Calabi-Yau manifold. Compactification on K3 breaks half of the supersymmetries of M-theory, which is necessary to make it match with the amount of supersymmetry for heterotic theory on a 3–torus. M-theory compactified on a smooth K3 gives $U(1)^{22}$ as gauge symmetry. Non-Abelian gauge symmetries are argued to occur when the K3 has one or more singularities. Singularities on K3 come from vanishing 2-cycles. The singularities are classified by the intersection matrices for these vanishing 2-cycles. This leads to a classification of singularities that closely resembles the classification of simply laced compact Lie-groups. Membranes can wrap around 2-cycles, and in case the cycle has vanishing size this is argued to lead to extra massless states, and non-Abelian gauge symmetry.

There is then a one to one correspondence between singularities, and possible non-Abelian symmetry groups.

IIA theory, living on an orientifold $T^3/\mathbb{Z}_2$ with 6 $O^-$ planes, and 2 $O^+$ planes closely resembles the above situation. The main differences are that the presence of $O^+$ instead of $O^-$ planes leads to a reduction of the rank of the group, and non-simply laced gauge groups such as $Sp(k)$. In [55] it was argued that the strong coupling limit of this theory should correspond to compactification of M-theory on a K3 with a mysterious so called $D_4 \oplus D_4$ singularity [31], that somehow manages to reduce the rank and produce the non-simply laced groups. The precise mechanism is still unknown. An ordinary $D_4 \oplus D_4$ singularity would give $SO(8) \times SO(8)$ symmetry. It is striking that we argued that $SO(8) \times SO(8)$ is the smallest group that we can use to construct the asymmetric orbifold for the $\mathbb{Z}_2$-triple in the heterotic $E_8 \times E_8$ string. This suggest that also for the other triples a strong coupling limit may exist corresponding to M-theory on K3 with rank reducing singularities. By analogy to the example of [55], we expect the mysterious singularities to be of the type $E_6 \oplus E_6$ for a $\mathbb{Z}_3$-triple, $E_7 \oplus E_7$ for a $\mathbb{Z}_4$-triple, $E_8 \oplus E_8$ for a $\mathbb{Z}_5$-triple and $\mathbb{Z}_6$-triple. This subject is however still under investigation [15].

The idea of M-theory being a strong coupling limit of the IIA theory, with the size of an 11 dimensional circle being the coupling in the IIA theory may make one wonder whether an analogous construction exists for other theories. The strong coupling dual of the IIB theory is thought to be the IIB theory itself. IIB theory contains two types of strings, the fundamental string and a solitonic D-string. These are thought to be interchanged by strong weak coupling duality. One may also consider "dyonic" strings, collective states of fundamental and solitonic strings. Such strings may be labelled by two charges $p$ and $q$, describing a state of $p$ fundamental and $q$ D-strings. All possibilities form a two dimensional lattice, much like the charge lattice that occurs in theories with electrically charged particles as well as magnetic monopoles. In this lattice changes of basis may be made by transforming by an element of $SL(2, \mathbb{Z})$. This $SL(2, \mathbb{Z})$ is conjectured to be a symmetry of the full IIB theory, and it acts also on the coupling by fractional linear transformations.

$SL(2, \mathbb{Z})$ is also known in another context as the modular group of the 2–torus. The 2–torus has a complex structure that transforms under the $SL(2, \mathbb{Z})$ by fractional linear transformations. This led Vafa to propose to interpret the complex coupling constant of the IIB theory as the modular parameter for a 2–torus. One then compactifies IIB theory on a certain base manifold, and erects at each point of this manifold a 2–torus whose modular parameter corresponds to the coupling in the IIB theory. This construction was named F-theory [53]. As an example one may consider the IIB-theory on a 2–torus at constant coupling. According to the prescription one may erect the same 2–torus as a fibre at each point. This results in F-theory on the 4–torus. Things become more interesting if we allow the fibre to become singular. This may happen if in the IIB theory the coupling diverges. Accordingly, at the location of the fibre there should then be a source or sink for the dilaton, and this is interpreted as the position of a D-brane. Coinciding D-branes lead to non-Abelian gauge symmetries. In the F-theory picture this corresponds to a collision of singular fibres. The possible singular fibres are again mathematically classified, and there exists a correspondence between singular fibres and possible non-Abelian symmetry groups.

A particular compactification of F-theory is achieved by compactifying the IIB theory on a 2–sphere. The 2–sphere is not Ricci-flat and therefore not a suitable background for a string theory, as it does not solve the equations of motion. This may be repaired by adding a number of D7-branes transverse to the sphere. In the background of a D7-brane space-time is not flat, and this modifies the curvature. One can achieve a consistent string compactification by adding 24 D7-branes. In the F-theory description we then have a 2–sphere with 24 singular fibres. Mathematically, a 2–sphere with 24 singular fibres gives a subclass of the K3-manifolds, the elliptically fibered K3's. The moduli space classifying elliptically fibered K3's has the same form as the moduli space of Narain compactification of heterotic string theory on a 2–torus. This and other evidence has led to the conjecture that F-theory on an elliptically fibered K3 is dual to heterotic string theory on a 2–torus [53].

Given this duality, it is clear that the Narain compactification of heterotic string theory on a 3–torus should be dual to F-theory on a product of an elliptically fibered K3 with a circle. In the above we considered compactifications of heterotic string theory that led to reduction of the rank of the gauge group. For the case of the CHL-string in 8 dimensions, it was argued in [55] that with a suitable choice of holonomies, there exists a IIB orientifold dual with 3 $O^-$ planes and 1 $O^+$ planes. Another distinguishing feature in this theory is the presence of a non-zero $B$-field on the orientifold. Together with the rank reduction this led to the proposal that this should be dual to F-theory on a K3 with a singular fibre of $D_8$ type, with the property that this singular fibre does not lead to enhanced gauge symmetry. This description was developed in [5] [30], who further emphasised the role of the non-zero B-field over the base of the K3. This is immediately extended to one of the duals for the heterotic string with a $\mathbb{Z}_2$-triple: It should be dual to F-theory on the product of the circle and an elliptically fibered K3, with non-zero B-field over the base.

There exist K3's with extra symmetries; discrete groups mapping the K3 to itself, which leads to another F-theory dual. All possible automorphisms of K3 have been classified by Nikulin [40]. Among the possible automorphisms of K3 there are a $\mathbb{Z}_2$, $\mathbb{Z}_3$, $\mathbb{Z}_4$, $\mathbb{Z}_5$ and $\mathbb{Z}_6$. One may choose a basis of 2–cycles of the K3, such that the $\mathbb{Z}_2$ acts on 8 of the basis cycles, the $\mathbb{Z}_3$ on 12, the $\mathbb{Z}_4$ on 14, and the $\mathbb{Z}_5$ and $\mathbb{Z}_6$ on 16 basis cycles. Recalling that the $\mathbb{Z}_2$, $\mathbb{Z}_3$, $\mathbb{Z}_4$, $\mathbb{Z}_5$ and $\mathbb{Z}_6$ triples led to rank reduction with 8, 12, 14, 16 and 16, there is clearly a connection. Indeed the relevant symmetries can be realised on elliptically fibered K3's. By analogy with the heterotic construction, one may then compactify F-theory on an additional circle, and divide by a $\mathbb{Z}_n$-symmetry, where the $\mathbb{Z}_n$ acts as a shift on the circle and as the Nikulin automorphism on the K3, thus obtaining F-theory on $(K3 \times S^1)/\mathbb{Z}_n$. There exist a few special limits in the K3 moduli space, where K3 may be written as a global orbifold of a 4–torus by a $\mathbb{Z}_n$ symmetry. One of these special K3's is $T^4/\mathbb{Z}_2$, where the $\mathbb{Z}_2$ acts as an inversion of all coordinates. F-theory on $T^4/\mathbb{Z}_2$ is argued to be dual to IIB theory on the orientifold $T^2/\mathbb{Z}_2$ with the fixed points of the $\mathbb{Z}_2$ corresponding to $O^-$ planes [48]. The last IIB-orientifold of the previous section included this orientifold as a sub-manifold of the compactification space. Therefore, this orientifold (with a suitable D-brane configuration) should be dual to F-theory on $(T^4/\mathbb{Z}_2 \times S^1)/\mathbb{Z}_2 = (K3 \times S^1)/\mathbb{Z}_2$.

The whole web of theories that are dual to the CHL-string is depicted in figure 6-1. In this figure $hE8$-stands for the heterotic $E_8 \times E_8$-string, and $hSO$ for the heterotic $Spin(32)/\mathbb{Z}_2$- string. The I, IIA and IIB stand for the corresponding string theories, M and F for M- and F-theory descriptions. Below the theory label we have written the corresponding compactification manifold. "CHL" stands for the special holonomy leading to the CHL-string. The subscript "nv" on tori stands for "no vector structure". On a 3–torus, "triple" denotes the presence of three holonomies leading to reduction of the rank. The symbol $I$ appearing in the compactification manifold stands for $S^1/\mathbb{Z}_2$. The +-signs for IIA on $T^3/\mathbb{Z}_2$ and IIB on $T^2/\mathbb{Z}_2 \times S^1$ denote the presence of two resp. one $O^+$ fixed planes. In one F-theory and one M-theory case we labelled the K3-compactification manifold with $D_8$, resp. $D_4 \oplus D_4$ to remind the reader that there is additional structure in the form of rank reducing singularities.

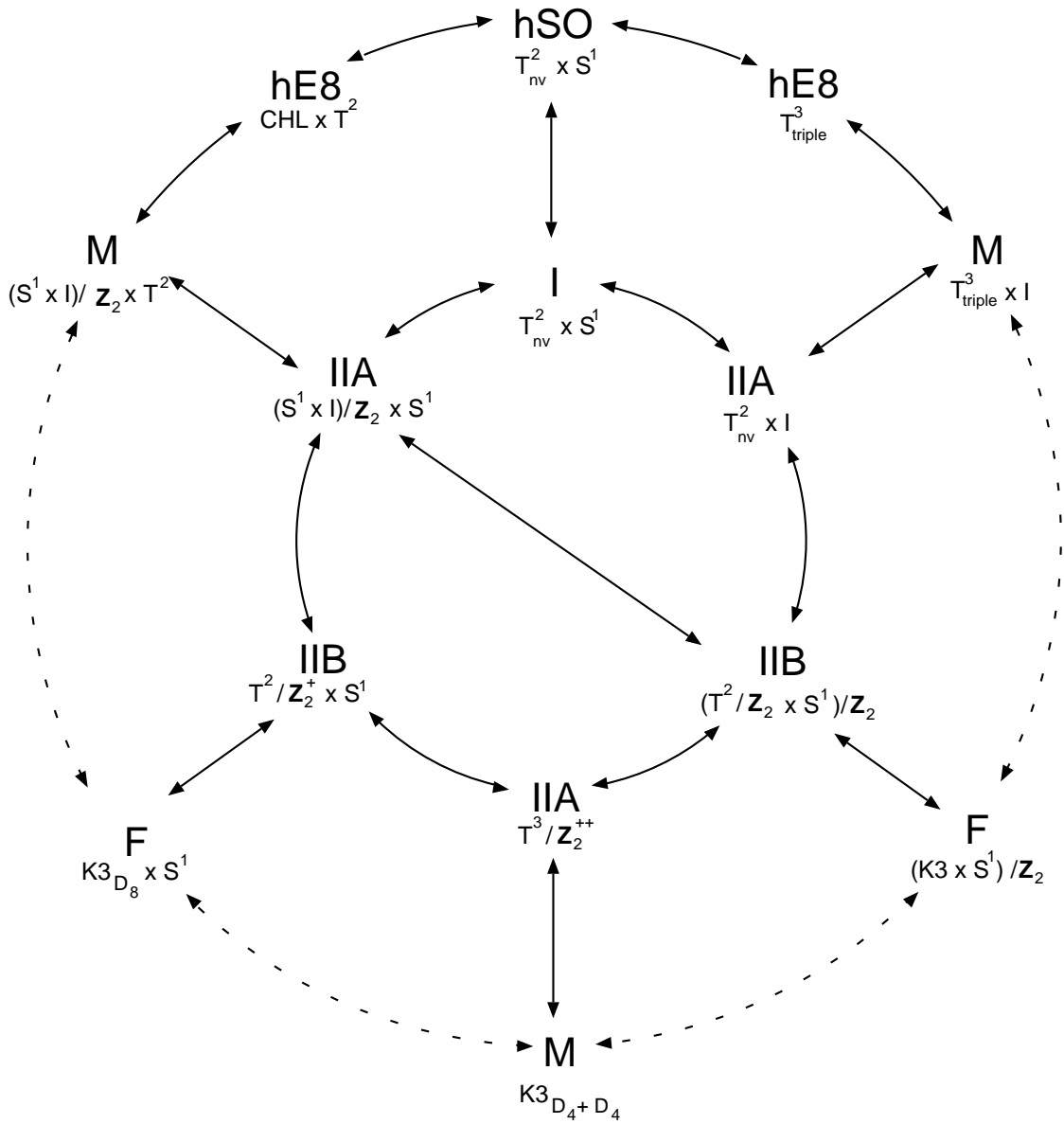

Figure 6-1. The web of relations between the various dual theories to the 7-dimensional CHL-string

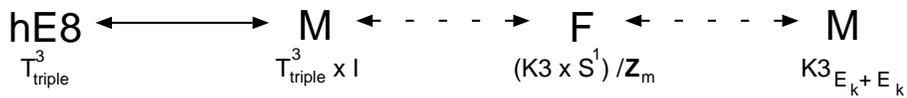

Figure 6-2. Dual theories for $\mathbb{Z}_n$-triples ($n > 2$)

In the case of heterotic theories with $\mathbb{Z}_n$-triples, with $n > 2$, the web of dual theories is less rich and is depicted in figure 6-2. Again one K3 appearing in an M-theory compactification is marked with $E_k + E_k$, to denote the occurrence of an $E_k \oplus E_k$ -singularity that reduces the rank. Note that of the theories appearing in figure $6-2$, the heterotic $E_8 \times E_8$ string gives the only description in terms of weakly coupled strings.

# 7 Conclusions

The moduli space of vacua for Yang-Mills theories compactified on tori turns out to be much richer than was thought a few years ago. For Yang-Mills theory on an $n$–torus with periodic boundary conditions for the gauge fields, one may always obtain a vacuum solution by setting all gauge potentials to be constant elements taking values in the Cartan subalgebra of the gauge group. Recent insights show that for theories with orthogonal and exceptional gauge groups compactified on tori of dimension 3 or higher, there exist vacuum solutions that are not of this form.

The new vacua are valid solutions for pure Yang-Mills theories on tori. Their discovery however was motivated by supersymmetry. For four-dimensional supersymmetric Yang-Mills theories the number of ground states is equal to the Witten index [52], which is argued to be independent of perturbations of the theory. One therefore expects the number of ground states for supersymmetric Yang-Mills theory on a spatial 3–torus with non-compact time direction to be equal to the number of ground states for the same theory in non-compact flat 4-dimensional space-time, as one may continuously deform the theories into each other by varying the volume of the spatial 3–torus. Explicit calculations did not confirm this expectation for supersymmetric Yang-Mills theory with an orthogonal or exceptional gauge group, where the number of vacua in the infinite volume limit seems larger than the number of vacua for the theory on a small 3–torus [52] [1] [11] [35] [45].

A partial resolution of the paradox came from a construction within non-perturbative string theory [55]. The vacua of a gauge theory with an orthogonal gauge group compactified on a torus may be parametrised by the positions of D-branes on an orientifold. Using this construction it can be shown that the gauge theory on the 3–torus admits vacua that were not considered before. Including these extra vacua in the count for the number of vacuum states for the gauge theory on the 3–torus, the discrepancy between the various Witten index calculations disappears. A crucial fact is that the subgroup commuting with the holonomies parametrising the new vacua has a rank that is smaller than that of the original gauge group. For the exceptional group $G_2$, which can be defined as a subgroup of $SO(7)$, one can also show that an extra vacuum state exists, and that including this extra vacuum makes the Witten index calculations agree. These were the topics that were discussed in chapter 2.

With the knowledge that extra vacua solve the Witten index problem for the orthogonal groups and $G_2$, it is natural to conjecture that also for the remaining exceptional groups new vacuum states exist. As these groups can not be studied with a D-brane construction, or as a subgroup of an orthogonal group, new insights are needed. The crucial observation was that new vacua can be constructed by imposing 't Hooft's twisted boundary conditions [23] [24] in appropriate subgroups of the gauge group, as described in chapter 3. With this construction, the results for the orthogonal groups and $G_2$ can be reproduced, and new vacua for the exceptional groups $F_4$, $E_6$, $E_7$ and $E_8$ are found. For the orthogonal groups and $G_2$ the moduli space of vacua consist of 2 components, for $F_4$, $E_6$, $E_7$ and $E_8$ the moduli space consists of resp. 4,4,6 and 12 components. Counting all vacua for the theory on the 3–torus, one obtains the same result as for the infinite volume theory, confirming the constancy of the Witten index.

The subject of compactifications of gauge theory on a 3–torus has been under study by two other groups [6] [26]. In chapter 4 we briefly discussed some elements of their analysis. One may understand the new solutions for the 3–torus from a more general pattern, that gives insight in the possibilities for compactifying a gauge theory on an $n$–torus such that the subgroup of the gauge group that commutes with the holonomies has a rank that is smaller than that of the gauge group itself. This argument leads to a systematic search for the subgroups, in which the twisted boundary conditions can be imposed. We also reviewed how to construct the holonomies in this approach. Finally, a section was devoted to the calculation of the Chern-Simons invariant, which is a topological invariant classifing the various components in the moduli space, as different components have different values for the Chern-Simons invariant.

For theories with orthogonal gauge group and periodic boundary conditions, the new vacua were found by considering D-branes in an orientifold background. We returned to his theme in chapter 5, and showed that all vacua for gauge theories with classical gauge groups on a 2– or 3–torus, with *both* periodic and twisted boundary conditions, can be parametrised by a configuration of D-branes in a certain orientifold background. A crucial element in our analysis is how to apply T-duality to D-branes in the vicinity of a geometrical object called a crosscap, which yields unoriented spaces. We also made a brief excursion to theories with orthogonal gauge groups and periodic boundary conditions on higher dimensional tori.

The string theories with orthogonal gauge groups played an important role in the developments described above. Also exceptional groups occur in string theory. One of the heterotic string theories has as its gauge group $E_8 \times E_8$. Since $E_8$ is one of the groups giving new vacua on the 3–torus, we turned in chapter 6 to compactifications of the heterotic $E_8 \times E_8$ string on a 3–torus. The new vacua are described by asymmetric orbifolds of the heterotic string. Before performing the orbifold construction, we need to carefully examine the gauge symmetries of the theory. This leads to the insight that the topology of the gauge group in string theory is crucially different from the topology found from an analysis of the low energy gauge theory. The difference is due to the presence of winding strings, that have no field theory analogue. These give states transforming in representations of the gauge group different from the ones found from non-winding strings, and modify the topology of the gauge group. The modified topology reduces the number of possibilities for new vacua dramatically. One of the theories constructed this way turns out to be the CHL-string [7] [8]. The new formulation can be related to the traditional one using string duality.

String duality may also be used to obtain other string theories with the same gauge symmetries. In particular, for the CHL-string there are various orientifold descriptions, of which some have been encountered in chapter 5. Strong coupling dualities lead us to M- and F-theory descriptions [54] [53], where the compactification manifold is no longer a torus, but involves the 4-dimensional Calabi-Yau manifold K3. One M-theory description, on a K3 with rank reducing singularities, remains mysterious. M-theory on K3 with a particular rank-reducing singularity has been proposed before [55] as dual for the CHL-string, but in other theories different rank reducing singularities are found, signaling a more general pattern. This is still under investigation. One may hope that resolving this issue will lead to new information on M-theory, and its possible compactifications.

There are many more open ends. The developments described here only concern the (semi-)classical vacua for gauge and string theories. The full quantum dynamics in these theories remains to be explored. The degeneracies of the vacua should be preserved in supersymmetric field theories, because of the Witten index. In non-supersymmetric field theories, non-perturbative effects are likely to lift the degeneracy, selecting a particular vacuum. It is significant that different vacua have different Chern-Simons invariant, which hints at instanton-like excitations describing the tunneling between different vacua. Studying dynamical effects may also take us further away from the extreme limits considered here, being the small volume and infinite volume limits. The constancy of the Witten index suggests that, at least in the supersymmetric theories, also at intermediate volumes a discrete set of vacua may be identified.

Another direction for further research is to classify the vacua for Yang-Mills theories with given gauge group and given boundary conditions on higher than 3-dimensional tori. Some preliminary results for orthogonal groups from chapter 5 indicate that one should expect a multitude of components, but many of these isomorphic. For finding all possibilities the non-trivial $n$-tuples mentioned in chapter 4 may be used as building blocks, but it is a non-trivial problem how to combine them, certainly for the cases with twisted boundary conditions. Results for the higher dimensional tori are relevant for compactifications of 10 dimensional string theories.

# A  Lie algebras: conventions

Let $\mathcal{L}$ be a Lie algebra. A Cartan subalgebra $\mathcal{H}$ is a maximal abelian subalgebra in $\mathcal{L}$. Because all elements in $\mathcal{H}$ commute, they can be simultaneously diagonalised. In particular, in the adjoint representation, the eigenvectors of $\mathcal{H}$ are elements of the Lie algebra. One can write:

$$[h, e_\alpha] = \alpha(h) e_\alpha \quad h \in \mathcal{H} \tag{A.1}$$

The $\alpha(h)$ are linear functionals on the space $\mathcal{H}$, called roots or root vectors. The elements of $\mathcal{H}$ form a vector space. It is possible to associate elements $h_\alpha$ to the functionals $\alpha(h)$ by defining

$$\text{tr}\{\mathbf{ad}(h_\alpha), \mathbf{ad}(h)\} = N\alpha(h) \tag{A.2}$$

with $N$ a normalisation constant to be fixed later (In our articles [28] [29] we set $N = 1$, but in this thesis we will use a different convention. This does not affect the results of [28] [29], which can be found in chapter 3 translated to the conventions defined below). Because of linearity one has

$$h_{p\alpha + q\beta} = p h_\alpha + q h_\beta \tag{A.3}$$

The space $\mathcal{H}^*$ of linear functionals on $\mathcal{H}$ is a vector space. The roots form a (finite) subset of this space. The set of root vectors will be denoted by $\Delta$. One can introduce the notation

$$\text{tr}\{\mathbf{ad}(h_\alpha), \mathbf{ad}(h_\beta)\} = N\alpha(h_\beta) = N\beta(h_\alpha) \equiv N\langle \alpha, \beta \rangle \tag{A.4}$$

The left hand side, and hence the right hand side is clearly symmetric and biliniear. For compact Lie algebra's, (A.4) defines an inner product on the root space. This will be normalised such that the length of the longest roots is always $\sqrt{2}$ (this fixes the normalisation constant $N$). Since it turns out that the roots of the algebra occur in at most two different lengths, one may speak of short and long roots. An algebra that has only roots of one length is called simply laced.

We now have

$$[h_\alpha, e_\beta] = \langle \alpha, \beta \rangle e_\beta \tag{A.5}$$

The $e_\beta$ are normalised such that

$$[e_\alpha, e_{-\alpha}] = 2 h_\alpha \tag{A.6}$$

If $\alpha + \beta \neq 0$:

$$[e_\alpha, e_\beta] = N_{\alpha,\beta} e_{\alpha+\beta} \tag{A.7}$$

(We will never need an explicit form of $N_{\alpha,\beta}$). Hermitean conjugation acts as follows in our conventions

$$h_\alpha^\dagger = h_\alpha \qquad e_\alpha^\dagger = e_{-\alpha} \tag{A.8}$$

One picks a (non-orthogonal) basis of roots $\alpha_i$ such that, if $\alpha_i, \alpha_j$ are in this basis, $\alpha_i - \alpha_j$ is not a root. The roots of such a basis are called "simple". Any root is expressible as $\sum_k c_k \alpha_k$, where the $c_k$ are integers which are either all positive, or all negative. We always denote simple roots by $\alpha_i$, where $i$ is an index or a number. The simple roots of the compact simple Lie algebra's are listed in appendix B. The geometrical relations between the simple roots may be expressed through the Cartan integers

$$n_{ij} = \frac{2\langle \alpha_i, \alpha_j \rangle}{\langle \alpha_j, \alpha_j \rangle} \tag{A.9}$$

The $n_{ij}$ form a matrix, the Cartan matrix, which together with the normalisation fixes all relevant properties of the root system. An easy way of depicting the Cartan matrix is by means of a Dynkin diagram. For this diagram, one draws a node for every value of the index $i$ (the number of dots is thus equal to the rank $r$ of the group). Then every pair of nodes $i$ and $j$ are connected by $n_{ij}n_{ji}$ lines. We will denote the long simple roots by open dots, and the short roots by solid dots.

As any root may be expressed uniquely as $\sum_k c_k \alpha_k$, one may assign a height function $\sum_k c_k$ to any root. For a unique root this function is maximized, and consequently it is called the highest root $\alpha_H$. The lowest root is then $\alpha_0 = -\alpha_H$. When the lowest root is added to the set of simple roots, the resulting set still has the property that $\alpha_i - \alpha_j$ is not a root for any member of the set. By the same rules as for the Dynkin diagram, one may draw a so-called extended Dynkin diagram for this set. A property that this set does not have is linear indepence: There is a relation of the form

$$g_0 \alpha_0 + \sum g_i \alpha_i = 0 \tag{A.10}$$

with the $g_i$ all integers, and $g_0 = 1$. $g = \sum_{i=0}^{r} g_i$ is the Coxeter number, and we see that it is the height of the highest root plus 1.

The coroots are defined as

$$\alpha_i^\vee = \frac{2\alpha_i}{\langle \alpha_i, \alpha_i \rangle} \tag{A.11}$$

The Cartan integers may be written as $n_{ij} = \langle \alpha_i, \alpha_j^\vee \rangle$. One may also define a coroot diagram, following the procedure for a Dynkin diagram, but replacing all roots by coroots. This leads to a diagram that is almost the same as the Dynkin diagram, the exception being that the nodes denoting the long roots become nodes denoting the short coroots, and the nodes denoting short roots become nodes denoting long coroots. Also an extended coroot diagram can be obtained by replacing all roots by coroots for an extended Dynkin diagram. The corresponding set of coroots also does not form an independent set, but obeys a relation of the form

$$h_0 \alpha_0^\vee + \sum h_i \alpha_i^\vee = 0 \tag{A.12}$$

with the integers $h_i$ given by $h_i = g_i <\alpha_i, \alpha_i> /2$, and similar for $h_0$, which turns out always to equal 1. $h_0$ and $h_i$ are the coroot integers, and play a prominent role in chapter 4. $h = \sum_{i=0}^{r} h_i$ is called the dual Coxeter number. Note that for simply laced algebra's, roots and coroots, and therefore Coxeter number and dual Coxeter number may be identified. It can be shown that the normalisation constant $N$ appearing in (A.4) equals twice the dual Coxeter number, and hence

$$\text{tr}\{\mathbf{ad}(h_\alpha), \mathbf{ad}(h_\beta)\} = 2h \langle \alpha, \beta \rangle, \tag{A.13}$$

a result that will be needed in chapter 4.

Because the elements $h_\alpha$ of the CSA always commute, they can be simultaneously diagonalised in any matrix representation. In a specific matrix representation $(h_\alpha)_{ij}$, weights $\lambda_i$ are defined by $(h_\alpha)_{ii} = \langle \alpha, \lambda_i \rangle$ for each $\alpha$. Consequently the number of weights of a representation is equal to its dimension. A weight $\lambda$ of a group is always of the form

$$\lambda = \sum_i n_i \Lambda_i + m_i \alpha_i \qquad n_i, m_i \in \mathbf{Z} \qquad (A.14)$$

where $\Lambda_i$ are the fundamental weights, and $\alpha_i$ the simple roots. The fundamental weights are defined from the simple roots by

$$\langle \Lambda_i, \alpha_j^\vee \rangle = \delta_{ij} \qquad (A.15)$$

The weight lattice is thus the dual lattice to the coroot lattice. The fundamental weights are always of the form $\sum_k q_k \alpha_k$ where the $q_k$ are rational numbers.

The dual lattice to the root lattice is called the coweight lattice. A basis $\omega_i$ for the coweight lattice may be found by taking the dual basis to a basis of simple roots $<\alpha_i, \omega_j> = \delta_{ij}$. If the algebra is simply laced, the weight and coweight lattice may be identified.

Roots and weights have an expansion in the simple roots of the form $\beta = \sum_k c_k \alpha_k$. The coefficients $c_k$ can always be obtained by computing the inner produkt with the coweight $\omega_k$, as

$$c_k = \langle \beta, \omega_k \rangle. \qquad (A.16)$$

# B  Lie algebras: roots and weights

For easy reference we give some quantities for the groups used in this thesis. For $G_2$, we find it easier to work with abstract root vectors, for the orthogonal and other exceptional groups it is more convenient to work with explicit forms for the root vectors. $E_8$, $F_4$ and $G_2$ do not possess non-integer fundamental weights, and hence none are listed. Non-integer fundamental weights of $SU(n)$ and $Sp(n)$ are not listed either since we will not need them.

We use the notation $e_i$ for the unit vector in the $i$-direction. In the non-simply laced algebra's, the solid dots in the Dynkin diagrams denote the shorter roots.

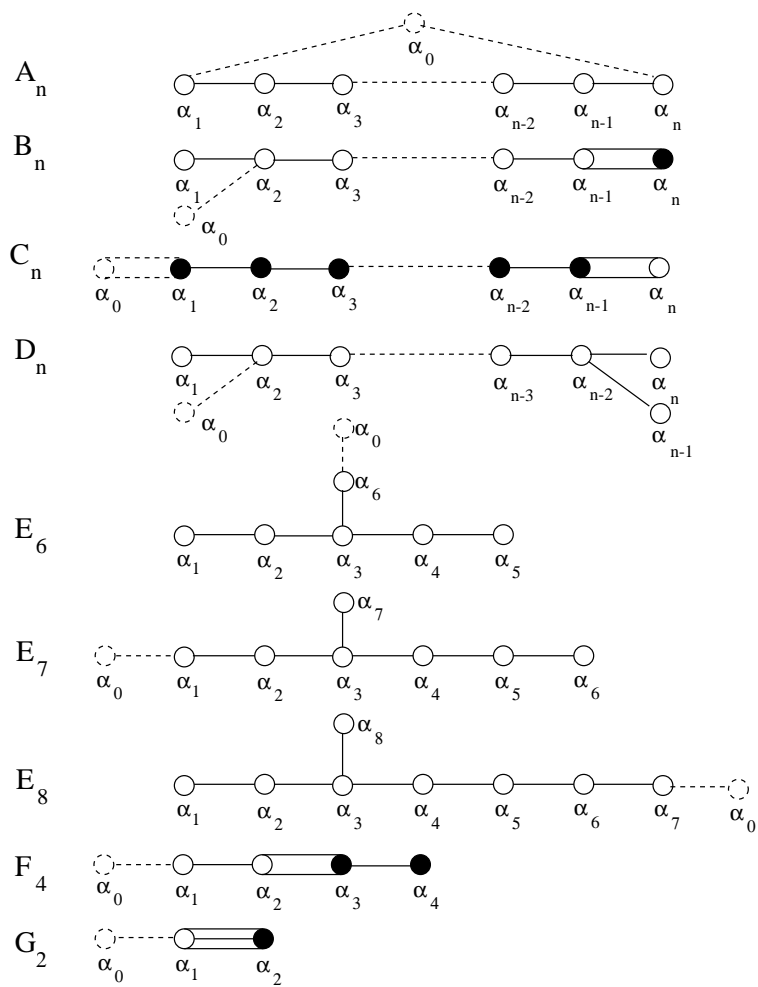

Figure B-1. Extended Dynkin diagrams. The Dynkin diagram is obtained by deleting the node marked by $\alpha_0$ and all lines connected to this node.

### $SU(n)$ $(A_n)$

- coroot integers:
$$h_i = 1 \quad \forall\, i$$

### $SO(2n+1)$ $(B_n)$

- Simple roots, explicit representation:
$$\alpha_i = (e_i - e_{i+1}) \quad (1 \leq i < n) \qquad \alpha_n = e_n$$

- Positive roots:
$$e_k \quad (e_k \pm e_l) \quad (k < l)$$

- Non-integer fundamental weights:
$$\Lambda_n = \sum_{i=1}^{n} e_i/2$$

- coroot integers:
$$h_i = 1 \quad \forall\, i \in \{0, 1, n\} \qquad h_i = 2 \quad \forall\, i \notin \{0, 1, n\}$$

### $Sp(n)$ $(C_n)$

- coroot integers:
$$h_i = 1 \quad \forall\, i$$

### $SO(2n)$ $(D_n)$

- Simple roots, explicit representation:
$$\alpha_i = (e_i - e_{i+1}) \quad (1 \leq i < n) \quad \alpha_n = (e_{n-1} + e_n)$$

- Positive roots:
$$(e_k \pm e_l) \quad (k < l)$$

- Non-integer fundamental weights:
$$\Lambda_{n-1} = (\sum_{i=1}^{n-2} e_i/2 - e_{n-1}/2 - e_n/2)$$
$$\Lambda_n = (\sum_{i=1}^{n-2} e_i/2 - e_{n-1}/2 + e_n/2)$$

- coroot integers:

$$h_i = 1 \quad \forall\, i \in \{0, 1, n-1, n\} \qquad h_i = 2 \quad \forall\, i \notin \{0, 1, n-1, n\}$$

## $E_6$

- Simple roots, explicit representation:

$$\alpha_1 = (e_1\sqrt{3} - e_2 - e_3 - e_4 - e_5 - e_6)/2$$
$$\alpha_2 = (e_5 + e_6) \qquad \alpha_3 = (e_4 - e_5) \qquad \alpha_4 = (e_3 - e_4)$$
$$\alpha_5 = (e_2 - e_3) \qquad \alpha_6 = (e_5 - e_6)$$

- Positive roots:

$$(e_k \pm e_l)\,(1 < k < l) \quad (e_1\sqrt{3} \pm e_2 \pm e_3 \pm e_4 \pm e_5 \pm e_6)/2$$

In the last expression the number of minus-signs should be odd.

- Non-integer fundamental weights:

$$\Lambda_1 = (\tfrac{2}{3}\sqrt{3}e_1)$$
$$\Lambda_2 = (\tfrac{5}{6}\sqrt{3}e_1 + \tfrac{1}{2}(e_2 + e_3 + e_4 + e_5 + e_6))$$
$$\Lambda_4 = (\tfrac{2}{3}\sqrt{3}e_1 + e_2 + e_3)$$
$$\Lambda_5 = (\tfrac{1}{3}\sqrt{3}e_1 + e_2)$$

- coroot integers:

$$h_0 = h_1 = h_5 = 1 \qquad h_2 = h_4 = h_6 = 2 \qquad h_3 = 3$$

## $E_7$

- Simple roots, explicit representation:

$$\alpha_1 = (e_1\sqrt{2} - e_2 - e_3 - e_4 - e_5 - e_6 - e_7)/2$$
$$\alpha_2 = (e_6 + e_7) \qquad \alpha_3 = (e_5 - e_6) \qquad \alpha_4 = (e_4 - e_5)$$
$$\alpha_5 = (e_3 - e_4) \qquad \alpha_6 = (e_2 - e_3) \qquad \alpha_7 = (e_6 - e_7)$$

- Positive roots:

$$e_1\sqrt{2} \quad (e_k \pm e_l)\,(1 < k < l) \quad (e_1\sqrt{2} \pm e_2 \pm e_3 \pm e_4 \pm e_5 \pm e_6 \pm e_7)/2$$

In the last expression the number of minus-signs should be even.

- Non-integer fundamental weights:

$$\Lambda_4 = (\tfrac{3}{2}\sqrt{2}e_1 + e_2 + e_3 + e_4)$$
$$\Lambda_6 = (\tfrac{1}{2}\sqrt{2}e_1 + e_2)$$
$$\Lambda_7 = (\sqrt{2}e_1 + \tfrac{1}{2}(e_2 + e_3 + e_4 + e_5 + e_6 - e_7))$$

- coroot integers:

$$h_0 = h_6 = 1 \quad h_1 = h_5 = h_7 = 2 \quad h_2 = h_4 = 3 \quad h_3 = 4$$

## $E_8$

- Simple roots, explicit representation:

$$\begin{aligned}
\alpha_1 &= (e_1 - e_2 - e_3 - e_4 - e_5 - e_6 - e_7 - e_8)/2 \\
\alpha_2 &= (e_7 + e_8) \quad \alpha_3 = (e_6 - e_7) \quad \alpha_4 = (e_5 - e_6) \\
\alpha_5 &= (e_4 - e_5) \quad \alpha_6 = (e_3 - e_4) \quad \alpha_7 = (e_2 - e_3) \\
\alpha_8 &= (e_7 - e_8)
\end{aligned}$$

- Positive roots:

$$(e_k \pm e_l)\,(k<l) \quad (e_1 \pm e_2 \pm e_3 \pm e_4 \pm e_5 \pm e_6 \pm e_7 \pm e_8)/2$$

In the last expression the number of minus-signs should be odd.

- coroot integers:

$$h_0 = 1 \quad h_1 = h_7 = 2 \quad h_6 = h_8 = 3 \quad h_2 = h_5 = 4 \quad h_4 = 5 \quad h_3 = 6$$

## $F_4$

- Simple roots, explicit representation:

$$\begin{aligned}
\alpha_1 &= (e_2 - e_3) & \alpha_2 &= (e_3 - e_4) \\
\alpha_3 &= (e_4) & \alpha_4 &= (e_1 - e_2 - e_3 - e_4)/2
\end{aligned}$$

- Positive roots:

$$e_k \quad (e_k \pm e_l)\,(k<l) \quad (e_1 \pm e_2 \pm e_3 \pm e_4)/2$$

- coroot integers:

$$h_0 = h_4 = 1 \quad h_1 = h_3 = 2 \quad h_2 = 3$$

## $G_2$

- Positive roots:

$$\alpha_1, \alpha_2, \alpha_1 + \alpha_2, \alpha_1 + 2\alpha_2, \alpha_1 + 3\alpha_2, 2\alpha_1 + 3\alpha_2$$

- coroot integers:

$$h_0 = h_2 = 1 \quad h_2 = 2$$

# C  Heterotic-heterotic duality in 9 dimensions

The heterotic $E_8 \times E_8$-string and $Spin(32)/\mathbb{Z}_2$-string are T-dual to each other when either one is compactified on an $n$–torus [36] [37] [18]. In the following we will rederive the result of [18]. Our main purpose is to point out some facts which are relevant for this thesis.

In the Narain-description of compactification of either heterotic theory on a circle, the momenta lie on the lattice $\Pi_{17,1}$. $\Pi_{17,1}$ is the unique even self-dual Lorentzian lattice of signature (17,1) (up to $O(17,1)$ transformations). One can decompose $\Pi_{17,1}$ as $\Gamma \oplus \Gamma_{1,1}$, where $\Gamma_{1,1}$ is interpreted as the lattice of momenta for one compactified dimension. $\Gamma$ is then a Euclidean Lie algebra lattice, reflecting the gauge symmetry in 10 dimensions. The possible choices are either $\Gamma = \Gamma_8 \oplus \Gamma_8$ with $\Gamma_8$ the $E_8$-root-lattice, or $\Gamma = \Gamma_{16}$, with $\Gamma_{16}$ the composition of the $SO(32)$ root-lattice with one of its spin weight-lattices.

The inclusion of a background gauge field $\mathbf{A}$ in the heterotic theory compactied on a circle with radius $R$, modifies the momenta as follows (we set $\alpha' = 2$ for convenience) [37] [18]

$$\mathbf{k} = (\mathbf{q} + wR\mathbf{A}) \tag{C.1}$$

$$k_{L,R} = \frac{n - \mathbf{q} \cdot R\mathbf{A} - \frac{w}{2}R^2 \mathbf{A} \cdot \mathbf{A}}{R} \pm \frac{wR}{2} \tag{C.2}$$

The $n$ and $w$ are integers denoting momenta, resp. winding numbers in the compact direction. By $\mathbf{A}$ we denote a 16-component Euclidean vector, taking values in the root space of the gauge group. It is thus naturally identified with an element of the Cartan subalgebra of the gauge group, and it corresponds to a holonomy $\exp(2\pi i h_{R\mathbf{A}})$ taking values in a maximal torus.

One can rewrite this as an $SO(17, 1)$-transformation acting on an element of $\Pi_{17,1}$:

$$\begin{pmatrix} \mathbf{k} \\ k_L \\ k_R \end{pmatrix} = \begin{pmatrix} \mathbb{1}_{16} & \mathbf{A} & -\mathbf{A} \\ -\mathbf{A}^T & 1 - \frac{\mathbf{A}^2}{2} & \frac{\mathbf{A}^2}{2} \\ -\mathbf{A}^T & -\frac{\mathbf{A}^2}{2} & 1 + \frac{\mathbf{A}^2}{2} \end{pmatrix} \begin{pmatrix} \mathbf{q} \\ \frac{n}{R} + \frac{wR}{2} \\ \frac{n}{R} - \frac{wR}{2} \end{pmatrix} \equiv W(\mathbf{A}) \begin{pmatrix} \mathbf{q} \\ p_+ \\ p_- \end{pmatrix} \tag{C.3}$$

where we wrote $\mathbf{A}^T$ for the transpose of $\mathbf{A}$, and $\mathbf{A}^2$ for its Euclidean norm. The vector $\mathbf{q}$ takes values in the lattice $\Gamma$, and we define $p_\pm = \frac{n}{R} \pm \frac{wR}{2}$ where $n$ and $w$ are the momentum and winding quantum numbers. The vector with components $(\mathbf{q}, p_+; p_-)$ is an element of $\Gamma \oplus \Gamma_{1,1}$.

Consider the lattice of vectors $W(\mathbf{A})\psi$ with $\psi \in \Gamma \oplus \Gamma_{1,1}$. The vectors in this lattice can also be described as $M\psi'$, with the components of $\psi'$ chosen relative to another basis, $\psi' \in \Gamma' \oplus \Gamma'_{1,1}$, and $M$ an $SO(17,1)$ transformation ($\Gamma'_{1,1}$ is obtained from $\Gamma_{1,1}$ by replacing $R$ by $R'$). To find the coordinate transformation from $\Gamma \oplus \Gamma_{1,1}$ to $\Gamma' \oplus \Gamma'_{1,1}$, we have to find $M$. We will fix its form by a few ansatze. First, we anticipate the fact that the desired duality transformation involves a large $R \leftrightarrow$ small $R$ duality transformation, which can be implemented by the matrix $u = \text{diag}(1^{17}, -1)$. Second, we assume that also in the dual theory a Wilson line is present, which can be implemented in the dual theory by the transformation $W(\mathbf{A})$. We then have to set $M = uW(\mathbf{A}')u$. $M\psi'$ is now given in coordinates appropriate for $\Gamma \oplus \Gamma_{1,1}$. Coordinates appropriate for $\Gamma' \oplus \Gamma'_{1,1}$ involve an extra $u$-transformation. The duality transformation is thus given as: $W(-\mathbf{A}')uW(\mathbf{A})$ which takes $\Gamma \oplus \Gamma_{1,1} \to \Gamma' \oplus \Gamma'_{1,1}$.

The required form of the lattices already fixes most of transformation between them. The remaining part can be fixed by two more ansatze. The first ansatz will be that specific Kaluza-Klein states are mapped to vectors in the root lattice of the gauge group of the dual theory. Specifically, we demand that $(0, \frac{1}{R} + \frac{R}{2}; \frac{1}{R} - \frac{R}{2})$ is mapped to a vector in the Lie algebra lattice with $p_+ = p_- = 0$. This implies that $|R\mathbf{A}|^2 = 2$, that $\mathbf{A} \cdot \mathbf{A}' = -\frac{1}{2}$ and that the image of the above vector is $(R\mathbf{A}, 0, 0)$. As a second ansatz we impose that $(0, \frac{1}{R'} + \frac{R'}{2}; \frac{1}{R'} - \frac{R'}{2})$ is mapped to a vector with $p_+ = p_- = 0$, but now by the inverse transformation. This implies that $|R'\mathbf{A}'|^2 = 2$. This identifies $R\mathbf{A}$ as a root of $\Gamma'$, while $R'\mathbf{A}'$ is a root of $\Gamma$.

The images of vectors of the type $(\mathbf{v}, 0, 0)$ (with $\mathbf{v}$ a lattice vector of $\Gamma$) have the correct form in $\Gamma'_{1,1}$ if and only if $2\mathbf{A} \cdot \mathbf{v}/R'$ is an integer. Thus $2\mathbf{A}/R'$ is a coweight of $\Gamma$. Performing the same computation for the inverse transformation shows that $2\mathbf{A}'/R$ is a coweight of $\Gamma'$.

To complete the basis of $\Gamma \oplus \Gamma_{1,1}$ we check the image of $(0, 1/R, 1/R)$. This image has the correct form if either $RR' = 1$ or $RR' = 2$. $RR' = 2$ leads to a transformation with $\Gamma' = \Gamma$ (it will take either heterotic string theory at $R$ to the same theory at $2/R$). Probably the quickest way to see this is to note that either for $E_8 \times E_8$ or $Spin(32)$, the conditions that $R\mathbf{A}$ is a coweight and has length $\sqrt{2}$ implies that $R\mathbf{A}$ is a root, and the same for $R'\mathbf{A}'$. Hence the Wilson lines are trivial, and this is just a complicated way of stating that the T-dual of any heterotic string theory without Wilson lines is itself, as also noted in [18]. Set $RR' = 1$, so the Wilson lines are non-trivial. This will take one heterotic string theory into the other.

Summarising, the duality transformation found in the above is characterised by:

- Its form is $W(-\mathbf{A}')uW(\mathbf{A})$, and it maps $\Gamma \oplus \Gamma_{1,1}$ to $\Gamma' \oplus \Gamma'_{1,1}$, with $\Gamma \neq \Gamma'$;
- $R\mathbf{A}$ is a root of $\Gamma'$, and $2R\mathbf{A}$ is a coweight of $\Gamma$;
- $R'\mathbf{A}'$ is a root of $\Gamma$, and $2R'\mathbf{A}'$ is a coweight of $\Gamma'$;
- $\mathbf{A} \cdot \mathbf{A}' = -\frac{1}{2}$, and $RR' = 1$.

Notice the symmetry between $\mathbf{A}$ and $\mathbf{A}'$ that is not explicit in most derivations of heterotic-heterotic duality. Especially the observation that $2R\mathbf{A}$ and $2R'\mathbf{A}'$ are coweights of one of the gauge groups will be significant for our applications. For the inverse transformation, simply interchange $R \leftrightarrow R'$ and $\mathbf{A} \leftrightarrow \mathbf{A}'$.

The transformation found in [18] is equivalent to the above transformation. One can check that the requirements on the Wilson lines lead to an unbroken subgroup whose simply connected cover is $Spin(16) \times Spin(16)$. When more dimensions are compactified, explicit duality transformations are harder to find in general. In the cases we will study, things are simplified by a special circumstance. One of the compact directions will have Wilson line $R\mathbf{A}$ and, in the absence of other Wilson lines we could perform the duality transformation to the dual theory with Wilson line $R'\mathbf{A}'$. In the cases we will study there will be additional Wilson lines, say $R_i \mathbf{B}_i$, present, but it will be possible to pick them such that $\mathbf{A} \cdot \mathbf{B}_i = \mathbf{A}' \cdot \mathbf{B}_i = 0$. With these requirements, momenta from different directions will not mix, and one may dualise dimensions separately. The duality transformation will then factorise, and one may dualise a theory with Wilson lines $R\mathbf{A}$, $R_i \mathbf{B}_i$ to a theory with Wilson lines $R'\mathbf{A}'$, $R_i \mathbf{B}_i$.

# Samenvatting

*'Het hoofd van de gemeente keek versuft naar de deur, die zich achter de wetenschapsman sloot.*
*"Wat wilde hij?", vroeg hij zich af. "Hij is tenslotte een geleerde en men kan zijn woorden niet verwaarlozen. Als men ze maar begreep." '*

Burgemeester Dickerdack in "De geweldige wiswassen" - Marten Toonder

## Nieuwe vacua voor Yang-Mills theorie op de 3–torus

Volgens de moderne natuurkunde worden alle fundamentele krachten beschreven door ijktheorieen. Een ijktheorie is niets anders dan een theorie met meer parameters dan vrijheidsgraden. Die extra parameters zijn niet van invloed op fysisch waarneembare grootheden. Ze kunnen daarom variëren, niet alleen globaal, maar zelfs van plek to plek dus lokaal. In het geval van de elektromagnetische kracht, beschreven door de eenvoudigste ijktheorie, bestaat de extra vrijheid uit slechts één parameter. Een ander fundamentele kracht is de zwaartekracht, beschreven door de algemene relativiteitstheorie. Volgens het relativiteitsprincipe mag de fysica niet afhankelijk zijn van de keuze van coördinaten op een ruimte. In de Algemene Relativiteitstheorie is de ijkvrijheid dus zo groot als de vrijheid van keuze voor een coördinatenstelsel op de beschreven ruimte, wat in zekere zin oneindig veel vrije parameters geeft. De overige twee bekende fundamentele krachten, de zwakke en sterke kernkracht, zitten qua complexiteit tussen electromagnetisme en gravitatie in. Ze worden beschreven door niet-Abelse ijktheorieën. Deze theorieën hebben een eindig aantal ijkvrijheidsgraden. Deze vrijheidsgraden vormen een wiskundige structuur, genaamd een groep. "Niet-Abels" betekent dat de elementen van deze groep niet commuteren, dus dat $A \times B \neq B \times A$. Groepentheorie is een goed ontwikkelde tak van de wiskunde, en de groepen die gebruikt kunnen gebruikt worden voor niet-Abelse ijktheorieen zijn bekend. Ze worden onderverdeeld in de unitaire, symplectische, orthogonale en exceptionele groepen. De zwakke kracht wordt beschreven door een model waarin de unitaire groep $SU(2)$ een grote rol speelt. Het model voor de sterke kracht (Quantumchromodynamica, oftewel QCD) heeft $SU(3)$ als (unitaire) ijkgroep. Ook modellen gebaseerd op andere groepen zijn mogelijk, zoals $SU(5)$ (unitair), $SO(10)$ (orthogonaal) en $E_6$ (exceptioneel), die gebruikt worden in zogenaamde geunificeerde theorieën. Deze theorieën pogen de bekende krachten (op de zwaartekracht na) te verenigen in één enkele theorie.

De elektromagnetische kracht en de zwakke kernkracht zijn vrij goed begrepen, en voorspellingen met grote nauwkeurigheid zijn hier mogelijk. Dit ligt anders voor de twee overige krachten, de sterke kracht en de zwaartekracht. Voor de sterke kracht ligt dit in het feit dat de theorie een aantal opmerkelijke eigenschappen bezit. Eén van die eigenschappen is "asymptotische vrijheid", wat betekent dat de krachten die de deeltjes voelen op korte afstand zwak zijn, maar op lange afstand juist zeer sterk. Dit is precies het omgekeerde van wat je ziet bij bijvoorbeeld de elektromagnetische kracht. Als een consequentie daarvan kunnen we het gedrag van bijvoorbeeld quarks (dit zijn de elementaire deeltjes die de sterke kracht voelen) redelijk voorspellen. Maar hoe op een wat grotere schaal een gebonden toestand zoals het proton (dat bestaat uit drie quarks) beschreven moet worden kan niet makkelijk worden afgeleid uit de microscopische beschrijving.

Een manier waarop een theoretisch natuurkundige dit probleem kan ontwijken is door ervoor te zorgen dat er geen lange afstanden zijn. Dit kan bijvoorbeeld door de theorie in een doos van een eindig volume te stoppen. Bepaalde effecten kunnen dan bestudeerd worden door het volume van de doos te variëren. Een andere motivatie voor het bestuderen van theorieën in een eindig volume kan gevonden worden in computersimulaties van natuurkundige systemen. Vanwege het eindige geheugen van een computer moet een te simuleren systeem een eindige grootte hebben.

Het formuleren van een ijktheorie in een eindig volume brengt echter extra problemen met zich mee. De extra ijkvrijheidsgraden moeten op een consistente manier worden ingebouwd, wat inhoudt dat hun gedrag op de randen van het eindige volume moet worden voorgeschreven. Dit kan op meerdere manieren. Deze globale randvoorwaarden zijn echter van invloed op de fysica van het systeem. Om het effect van deze randvoorwaarden te verminderen zouden we het volume groter moeten maken, maar we krijgen dan weer te maken met de sterke krachten die de berekeningen bemoeilijken.

Een andere vereenvoudiging is het weglaten van de materie uit de theorie, waarna een theorie met louter de krachtvoerende deeltjes overblijft. Zo'n theorie heet een Yang-Mills theorie, naar de ontdekkers.

In het begin van de jaren tachtig werd een probleem geconstateerd bij het beschrijven van Yang-Mills theorieën. De motivatie kwam uit supersymmetrische theorieën. Supersymmetrie is een symmetrie tussen bosonen en fermionen, die enorme beperkingen oplegt aan een theorie. De soorten deeltjes en de interacties tussen de verschillende deeltjes worden zwaar beperkt door de eis van supersymmetrie. Desalniettemin zijn vele supersymmetrische theorieën mogelijk, en in het bijzonder bestaan er supersymmetrische generalisaties van niet-Abelse ijktheorieën. Eén interessante vraag is of er in deze theorieën toestanden met energie nul bestaan en zo ja, hoeveel. Deze toestanden zijn belangrijk omdat ze de toestanden van laagste energie zijn (in supersymmetrische theorieën is de energie groter of gelijk aan nul), en omdat ze invariant zijn onder supersymmetrie. Sommige van deze toestanden (ook wel grondtoestanden, of vacua genoemd) zijn bosonisch en andere fermionisch. Het verschil tussen het aantal van bosonische grondtoestanden en het aantal fermionische grondtoestanden is een getal, dat de Witten index wordt genoemd. Voor niet-Abelse ijktheorieën is het aannemelijk dat er geen fermionische vacua zijn, en de index is dan simpel het aantal grondtoestanden. Men beargumenteert nu dat deze index niet verandert als de parameters van de theorie iets worden veranderd.

Het berekenen van het aantal grondtoestanden voor een (supersymmetrische) Yang-Mills theorie is niet eenvoudig. Voor één manier van berekening kan men de bovenstaande procedure volgen: De theorie in een (klein) eindig volume stoppen, in dit geval een 3-dimensionale doos met periodieke randvoorwaarden (ook wel 3–torus genaamd). Berekening van de Witten index geeft nu een bepaald getal. Dit getal zou niet moeten veranderen als de parameters van de theorie worden veranderd. In het bijzonder kan het eindige volume heel groot worden gemaakt. In deze limiet kan een andere berekening van de Witten index worden gedaan. Vergelijking van de twee manieren van berekening leidt nu tot een verrassing: de antwoorden zijn gelijk bij theorieën gebaseerd op de unitaire of symplectische groepen, maar verschillend bij theorieën gebaseerd op orthogonale of exceptionele groepen. Dit is in het bijzonder merkwaardig omdat, afgezien van de groepsstructuur, deze theorieën helemaal niet zoveel van elkaar verschillen.

Het verschil tussen de twee berekeningen bleef lange tijd een raadsel, totdat eind 1997 werd ontdekt dat voor de orthogonale groepen de oude berekening voor een klein volume incorrect is. Er blijken naast de grondtoestanden die men altijd al meenam in de berekening, nog andere grondtoestanden te bestaan, waarvan men zich voor die tijd niet het bestaan realiseerde. Als men deze nieuwe grondtoestanden meetelt, komt het antwoord voor de index in het kleine volume wèl overeen met het antwoord voor het grote volume. Ook is het niet moeilijk om te laten zien dat voor de unitaire en symplectische groepen zulke extra toestanden niet bestaan, wat verklaart dat de berekeningen voor deze gevallen altijd al in overeenstemming waren. Het leek aannemelijk dat ook de exceptionele groepen aanleiding geven tot extra grondtoestanden in een klein volume. Dit kon echter toen nog niet worden aangetoond.

Overigens is supersymmetrie niet relevant voor het bestaan van de nieuwe oplossingen, ze bestaan ook voor Yang-Mills theorieën op de 3–torus zonder supersymmetrie. Voor theorieën zonder supersymmetrie kan het argument van de Witten index echter niet worden toegepast. Deze berekening is een fraai voorbeeld van hoe supersymmetrische theorieën tot inzichten in niet-supersymmetrische theorieën kunnen leiden.

De ontdekking van de nieuwe vacua werd gedaan in de context van snaartheorieën met D-branen. Het oplossen van het probleem voor de exceptionele groepen is niet mogelijk in deze context, en het blijkt nodig de berekening opnieuw te formuleren. Een belangrijk deel van dit proefschrift is hieraan gewijd. Dit leidt tot een beter begrip van de situatie voor orthogonale groepen. Ook voor de exceptionele groepen blijken extra vacua te bestaan op de 3–torus. Hiermee is de titel van dit proefschrift, "Nieuwe vacua voor Yang-Mills theorie op de 3–torus" verklaard. In de hoofdstukken 2, 3 en 4 worden de Witten index berekening en de constructie van de nieuwe vacua besproken. Ook met niet-periodieke randvoorwaarden blijken er nieuwe oplossingen te bestaan.

De overige hoofdstukken van het proefschrift zijn gewijd aan een onderdeel van de theoretische natuurkunde waar de nieuwe oplossingen van belang zijn: snaartheorie (vaak aangeduid met de Engelse naam string theory). Bij vele natuurkundige theorieën zijn de fundamentele objecten puntdeeltjes. In snaartheorieën zijn de bouwstenen draadjes (snaren). Er zijn twee mogelijkheden: de draadjes hebben twee uiteinden (open snaren), of het draadje vormt een gesloten lus (gesloten snaar). Er zijn twee belangrijke verschillen tussen theorieën met snaren en theorieën van puntdeeltjes. Het eerste is dat snaren een zekere ruimtelijke uitgebreidheid hebben, wat er toe leidt dat interacties niet langer in een punt gebeuren, maar uitgesmeerd worden in de tijd-ruimte. Het tweede verschil is een gevolg van de ruimtelijke uitgebreidheid van snaren: in snaren is ook nog een interne structuur aanwezig. Snaren kunnen namelijk in trilling worden gebracht, en de verschillende trillingswijzen (grondtoon en boventonen, denk aan de snaren van muziekinstrumenten) zijn mogelijk. Als men op grotere afstand kijkt, is de uitgebreidheid van de snaar niet meer zichtbaar, maar de energie die in de interne trilling zit geeft het object andere eigenschappen. Zo proberen snaartheorieën het bestaan van verschillende deeltjes te verklaren.

De snaartheorie werd oorspronkelijk (begin jaren 70) opgezet als een mogelijke theorie voor de sterke wisselwerking, maar dit idee werd later verlaten ten gunste van een andere theorie, het eerder genoemde QCD (er bestaan wel vele interessante relaties tussen QCD en snaartheorie, en dit is nog steeds een onderwerp van onderzoek). Men realiseerde zich echter al snel dat snaartheorie wellicht een beschrijving kon geven van de quantumtheorie van de enige kracht die tot nog toe buiten beschouwing is gebleven: de zwaartekracht. Een naïeve formulering van de theorie van quantumzwaartekracht leidt tot grote problemen, vanwege het optreden van oneindigheden in de berekeningen. Die oneindigheden suggereren dat de theorie van zwaartekracht moet worden geherformuleerd voor korte afstanden. Snaartheorie lijkt een geschikte modificatie van de theorie van de zwaartekracht op te leveren. De oneindigheden doen zich in snaartheorie niet meer voor, in essentie omdat interactie tussen snaren niet in een punt plaatsvindt, maar over een eindig oppervlak wordt uitgesmeerd. Een andere gunstige eigenschap is dat ook de eerder besproken niet-Abelse ijksymmetrieën kunnen worden ingebouwd. Dit opent de mogelijkheid dat alle bekende krachten uit een snaartheorie kunnen worden verklaard. Of die mogelijkheid echt gerealiseerd is, is nog steeds een open vraag, maar het verklaart de enorme interesse voor snaartheorieën.

Snaartheorie in zijn meest eenvoudige versie geeft echter aanleiding tot andere inconsistenties. Deze problemen kunnen ten dele worden opgelost door het inbouwen van supersymmetrie in de theorie. Dit geeft de zgn. supersnaartheorieën. De overige inconsistenties kunnen slechts worden geëlimineerd worden door speciale waarden van de parameters te kiezen. Eén van die parameters is het aantal dimensies. Alle bekende consistente supersnaartheorieën vereisen negen ruimtelijke dimensies, plus een tiende tijddimensie.

Er bestaan vijf verschillende supersnaartheorieën. Twee theorieën hebben maximale supersymmetrie, maar geen niet-Abelse ijksymmetrie (althans niet in de storingstheorie). Dit zijn theorieën van gesloten snaren die men IIA en IIB heeft genoemd. De overige drie theorieën hebben wel niet-Abelse ijksymmetrie, en minder supersymmetrie. De strenge consistentie eisen van snaartheorie maakt de keuze voor een ijkgroep echter beperkt: alleen voor $SO(32)$ (een orthogonale groep) en $E_8 \times E_8$ (een produkt van twee exceptionele groepen) is een consistente theorie mogelijk. $SO(32)$ is de ijkgroep van een theorie van open snaren, die type I stringtheorie wordt genoemd. De twee overige theorieën zijn de heterotische theorieën. Dit zijn theorieën van gesloten snaren, en er bestaat een theorie met ijkgroep $SO(32)$ en een andere met ijkgroep $E_8 \times E_8$.

Hoe kan een theorie die in 10 dimensies leeft onze 4-dimensionale ruimte-tijd beschrijven? Er is geen tegenspraak als men aanneemt dat van de 10 dimensies er 6 zeer kleine (te klein om te meten met huidige technieken) afmetingen hebben. Dit is het idee van "compactificatie": een aantal van de 10 dimensies vormen een "doos" van eindig (en zeer klein) volume. Als we 3 van de dimensies nemen en er een doos met periodieke randvoorwaarden van vormen, en ons realiseren dat consistentie vereist dat de ijkgroep orthogonaal of exceptioneel is, is het duidelijk dat de eerder beschreven resultaten van belang zijn voor snaartheorie.

De ontwikkeling van snaartheorie is echter verder gegaan dan we tot nog toe beschreven hebben. Al in de jaren 80 kwam men tot de ontdekking dat er relaties bestonden tussen verschillende theorieën. Als men bijvoorbeeld 1 dimensie periodiek neemt (er dus in feite een cirkel van maakt), blijkt dat de IIA en IIB-theorie dezelfde theorie zijn. Dit is mogelijk omdat gesloten snaren op een cirkel geclassificeerd worden door twee getallen, namelijk een impuls die de beweging op de cirkel beschrijft, en een windingsgetal dat beschrijft hoeveel maal een snaar om de cirkel heen loopt. Het windingsgetal is een geheel getal, en quantummechanica vertelt ons dat ook de impuls een geheel getal is. Nemen we nu een IIA-snaar met impuls $n$ en windingsgetal $w$ op een kleine cirkel, dan blijkt dat we dat ook kunnen interpreteren als een IIB-snaar op een grote cirkel met impuls $w$ en windingsgetal $n$, en vice versa. Ook de twee heterotische theorieen hebben een dergelijke relatie met elkaar. De type I theorie blijkt een zelfde soort relatie te hebben met een andere theorie, maar de grondtoestand van deze nieuwe theorie bevat objecten waarvan de relevantie pas in de jaren 90 begrepen werd: de Dirichlet-branen, of kortweg D-branen. Dit zijn objecten met een ruimtelijke uitgebreidheid, met de eigenschap dat de uiteinden van open snaren erop eindigen. Er zijn D-branen van allerlei dimensies, van puntvormige (D-deeltjes) tot D-branen die de gehele ruimte vullen. Terugredenerend kan men ook beargumenteren dat de type I theorien ook een achtergrond van D-branen heeft, van het type dat de gehele ruimte vult. Ook de IIA en IIB theorieën bevatten diverse soorten D-branen, maar daar maken ze geen deel uit van het vacuum. De type I theorie bevat ook D-branen die geen onderdeel uitmaken van het vacuum, maar wel bij eindige energie kunnen voorkomen.

In de jaren 90 realiseerde men zich dat er nog meer relaties bestaan tussen de verschillende snaartheorieën. Naast de parameters die de vorm van de kleine dimensies beschrijven, heeft elke theorie een parameter die de sterkte van de interacties weergeeft. Het is bijvoorbeeld mogelijk om deze parameter heel groot te maken. In dat geval kan men niet verwachten dat storingstheorie een goede beschrijving geeft van de theorie. Het is toch mogelijk een goede beschrijving te verkrijgen, voornamelijk dankzij supersymmetrie. Naast snaren bevatten de snaartheorieën ook nog andere objecten, waaronder de eerder genoemde D-branen. De massa van zulke objecten hangt op een bepaalde manier van de interactieparameter af, en supersymmetrie garandeert dat deze relatie een eenvoudige vorm heeft die ook geldig blijft als we een storing-theoretische berekening van die massa niet meer kunnen vertrouwen. In het bijzonder kan je laten zien dat als de interactie-parameter groot wordt, deze D-branen heel licht kunnen worden, terwijl de massa van de oorspronkelijke snaren juist erg groot wordt. De theorie met grote interactie parameter is dus een theorie van D-branen. Een aantal snaartheorieën bevat één-dimensionale D-branen, die dan ook wel D-snaren worden genoemd. In de limiet van grote interactieparameter worden zulke theorieen dus beschreven door snaartheorieën! Deze "duale" theorieën zijn niet noodzakelijkerwijs gelijk aan de oorspronkelijke theorie, immers, de D-snaar is niet hetzelfde object als de oorspronkelijke snaar. Het blijken wel altijd bekende supersnaartheorieën. Tezamen met de eerder ontdekte relaties blijken nu alle 5 oorspronkelijke theorieën met elkaar verbonden, en in feite beschrijven ze dus 1 theorie in verschillende limieten.

Niet alle theorieën hebben D-snaren, en voor deze theorieen is de duale theorie dus geen snaartheorie. Beschouwing van deze theorieën leidt tot een nieuwe theorie die niet in 10, maar in 11 dimensies leeft. Dit is niet in tegenspraak met het voorafgaande omdat de grootte van de elfde dimensie gerelateerd is aan de interactieparameter. Is deze klein, dan is de elfde dimensie klein, en de theorie dus zo goed als 10-dimensionaal. Deze theorie, waarvan nog steeds geen volledige formulering bestaat, wordt M-theorie genoemd.

We hebben vermeld dat de IIA en IIB-theorieën geen ijksymmetrie hadden in storingstheorie, maar dat ze wel gerelateerd zijn aan theorieën met ijksymmetrie. Dit impliceert dat de IIA en IIB-theorie wel degelijk ijksymmetrie hebben, en dat deze gerealiseerd moeten zijn op een manier die buiten het bereik van storingstheorie valt. Ook in dit geval blijken D-branen de relevante objecten te zijn tezamen met een constructie die bekend staat als een "orientifold". D-branen en orientifolds zijn ook te definiëren in de niet-supersymmetrische snaartheorieën, en leiden daar tot dezelfde ijksymmetrieën. Als men puur geïnteresseerd is in de ijksymmetrie, is het mogelijk om de consistentie-eisen van snaartheorie te negeren en enkel de configuraties van D-branes en orientifolds te bestuderen. De eerder genoemde nieuwe grondtoestanden voor Yang-Millstheorieën (met periodieke en niet-periodieke randvoorwaarden) moeten dan een vertaling hebben in termen van dit soort configuraties. Deze configuraties zijn het onderwerp van hoofdstuk 5.

Nemen we echter de consistentie eisen van snaartheorie wel in beschouwing, dan wordt al snel duidelijk dat het overgrote deel van de nieuwe oplossingen niet naar de snaartheorie vertaald kan worden. De consistentie eisen gaan echter nog verder dan het eisen van $SO(32)$ of $E_8 \times E_8$ ijksymmetrie: zelfs configuraties die zijn toegestaan in bijvoorbeeld $E_8 \times E_8$-ijktheorie, blijken inconsistent in $E_8 \times E_8$-snaartheorie. Ondanks de eliminatie van vele mogelijkheden, blijken er nog steeds een behoorlijk aantal mogelijkheden realiseerbaar in snaartheorie. Vanwege de relaties tussen de verschillende snaartheorieën en M-theorie, moeten de realiseerbare mogelijkheden een vertaling hebben in de verschillende theorieën. Dit geeft een fascinerend netwerk van relaties tussen verschillende theorieën. Ook blijken de nieuwe oplossingen gerelateerd aan al bekende oplossingen, wat bijdraagt tot een compleet en rijk geschakeerd beeld, dat beschreven wordt in hoofdstuk 6.

# List of publications

- M. Marchevsky, A. Keurentjes, J. Aarts, and P.H. Kes
  Elastic deformations in field-cooled vortex lattices in $NbSe_2$
  *Phys. Rev.* **B57**, (1998) 6061.

- A. Keurentjes, A. Rosly, and A.V. Smilga
  Isolated vacua in supersymmetric Yang-Mills theories
  *Phys. Rev.* **D58** (1998) 081701, hep-th/9805183.

- Arjan Keurentjes
  Non-trivial flat connections on the 3-torus I:
  $G_2$ and the orthogonal groups
  *J. High Energy Phys.* **05** (1999) 001, hep-th/9901154.

- Arjan Keurentjes
  Non-trivial flat connections on the 3-torus II:
  The exceptional groups $F_4$ and $E_{6,7,8}$
  *J. High Energy Phys.*, **05** (1999) 014, hep-th/9902186.

- Arjan Keurentjes
  Flat connections for Yang-Mills theories on the 3-torus
  hep-th/9908164
  Based on a talk given at NATO-ASI and TMR Summer school "Progress in String Theory and M-theory" Cargese May 24 -June 5 1999

- Arjan Keurentjes
  Orientifolds and twisted boundary conditions
  hep-th/0004073
  Submitted to Nuclear Physics B

- J. de Boer, R. Dijkgraaf, K. Hori, A. Keurentjes, J. Morgan, D. Morrison and S. Sethi
  The moduli space of supersymmetric strings in $d \geq 7$
  In preparation

# Curriculum vitae

Ik ben geboren te Bladel en Netersel op 6 Juli 1969. In 1987 behaalde ik het diploma Gymnasium $\beta$ op het Van Maerlantlyceum te Eindhoven. In dat zelfde jaar begon ik de studie sterrenkunde aan de Universiteit Leiden, maar staakte deze later. In de periode 1989-1992 had ik diverse banen. In 1992 startte ik de studie natuurkunde aan de Universiteit Leiden, en in 1993 behaalde ik de propedeuse. Mijn experimentele stage verrichtte ik in de "Metalen"-groep van prof. dr. Peter Kes. In augustus 1996 slaagde ik voor het doctoraal examen natuurkunde. Mijn afstudeeronderzoek betrof een literatuuronderzoek over het onderwerp "Electromagnetic duality", o.l.v. prof. dr. Pierre van Baal. Per 1 september 1996 begon ik mijn promotieonderzoek aan het Instituut-Lorentz voor theoretische natuurkunde te Leiden, o.l.v. prof. dr. Pierre van Baal, aanvankelijk als beurspromovendus, sinds 1998 als Assistent in Opleiding. In 1996 verzorgde ik het werkcollege "Quantumtheorie 1", in 1998 assisteerde ik bij het studentenseminarium "Gravitatie", in 1999 verzorgde ik werkcolleges "Speciale Relativiteitstheorie" en "Quantumfysica 1b" en in 2000 het werkcollege "Quantumtheorie 2". Daarnaast was ik betrokken bij de organisatie van een activiteit om VWO-scholieren met de universitaire opleiding natuurkunde te laten kennismaken. In het kader hiervan werd een hoor- en werkcollege "Speciale relativiteitstheorie voor scholieren" georganiseerd. Deze vond plaats in zowel 1999 als 2000.

Ik heb deelgenomen aan de AIO winterscholen van de Landelijke Onderzoeksschool Theoretische Natuurkunde in januari 1997 (Dalfsen) en 1998 (Nijmegen), en bezocht de NATO-ASI en zomerschool "Confinement, Duality and Non-perturbative Aspects of QCD", in Cambridge (Groot-Brittanië, 1997), de NATO-ASI en TMR zomerschool "Progress in String Theory and M-theory", te Cargese (Frankrijk, 1999) and the "Graduate school on contemporary String Theory and Brane Physics" in Turijn (Italië, 2000).

Vanaf de herfst 2000 zal ik werkzaam zijn als postdoc op een gezamenlijk positie aan de Ecole Normale Supérieure en de Université Pierre et Marie Curie in het kader van het RTN-netwerk "The quantum structure of spacetime and the geometric nature of fundamental interactions".